\documentclass[nofootinbib,preprintnumbers,superscriptaddress,amsmath,amssymb,floatfix]{revtex4-1}
\usepackage{graphicx,setspace,epsfig,color}
\usepackage[letterpaper,dvips,width=9.4in,height=8.5in,includemp=false]{geometry}

\usepackage{exscale}
\usepackage{enumerate}
\usepackage{cancel}
\usepackage{mdframed}
\usepackage{amsmath}
\usepackage[cm]{fullpage}
\usepackage{amssymb}
\usepackage{mathtools}
\usepackage{float}
\usepackage{slashed}
\usepackage{hyperref}
\usepackage{cleveref}
\usepackage{inputenc}
\usepackage{comment}
\usepackage{wrapfig}
\long\def\symbolfootnote[#1]#2{\begingroup%
\def\thefootnote{\fnsymbol{footnote}}\footnote[#1]{#2}\endgroup}

\newcommand{\be}{\begin{equation}}
\newcommand{\ee}{\end{equation}}
\newcommand\bea{\begin{eqnarray}}
\newcommand\eea{\end{eqnarray}} 

\newcommand{\E}

\usepackage{hyperref}
\begin{document}

\title{The Photon in Dense Nuclear Matter: Random Phase Approximation}

\author{Stephan Stetina} 
\email{stetina@uw.edu}
\affiliation{Institut f\"ur Theoretische Physik, Technische Universit\"at Wien,
Wiedner Hauptstrasse 8-10, A-1040 Vienna, Austria}
\affiliation{Institute for Nuclear Theory, University of Washington, Seattle, WA 98195.}

\author{Ermal Rrapaj}
\email{ermalrrapaj@gmail.com}
\affiliation{Department of Physics, University of Washington, Seattle, WA}
\affiliation{Department of Physics, University of Guelph, Guelph, ON N1G 2W1, Canada}

\author{Sanjay Reddy} 
\email{sareddy@uw.edu}
\affiliation{Institute for Nuclear Theory, University of Washington, Seattle, WA 98195.}
\affiliation{Department of Physics, University of Washington, Seattle, WA}

\begin{abstract}
We present a comprehensive and pedagogic discussion of the properties of photons in cold and dense nuclear matter based on the resummed one-loop photon self energy. Correlations among electrons, muons, protons and neutrons in $\beta$ equilibrium that arise as a result of electromagnetic and strong interactions are consistently taken into account within the random phase approximation. Screening effects, damping, and collective excitations are systematically studied in a fully relativistic setup. Our study is relevant to the linear response theory of dense nuclear matter, calculations of transport properties of cold dense matter, and investigations of the production and propagation of hypothetical vector bosons such as the dark photons. 
% \begin{description}
% \item[Usage]
% Secondary publications and information retrieval purposes.
% \item[PACS numbers]
% May be entered using the \verb+\pacs{#1}+ command.
% \item[Structure]
% You may use the \texttt{description} environment to structure your abstract;
% use the optional argument of the \verb+\item+ command to give the category of each item. 
% \end{description}
\end{abstract}

\pacs{Valid PACS appear here}% PACS, the Physics and Astronomy
\maketitle

%\tableofcontents

\section{Introduction}
\label{Intro}
%Excitations in hot or dense systems acquire a finite lifetime due to coupling to other particles and manifest as a finite spectral width. As long as the width is parametrically small compared to its thermal mass at fixed momentum one may consider the in-medium excitation to be a well defined quasi particle. 
Using thermal gauge field theories photon propagation in dense plasmas is well studied, usually in terms of the hard dense loop (HDL) approximation where the photon momentum is assumed to be small compared to the scales associated with the fermions in the plasma, i.e. their chemical potentials $\mu$ and Fermi momenta $k_f$ \cite{Altherr:1992jg} \cite{Altherr:1992mf}. At extreme  densities in deconfined quark matter these methods have been adapted to calculate the photon spectrum in a QCD plasma; see, e.g., Refs. \cite{Litim:2001mv}, \cite{Schmitt:2003aa}. 
In contrast, in cold dense nuclear matter, albeit phenomenologically very relevant to our understanding of matter in neutron stars, photon propagation has received much less attention. In this article, we consider the physical environment present in the core of neutron stars: a degenerate plasma composed of electrons, muons, protons, and neutrons interacting via electromagnetic and strong interactions at densities of around and above nuclear saturation density. We study the photon propagation in detail, utilizing the (relativistic) random phase approximation (RPA), which amounts to resumming one-loop polarization functions to obtain the dressed photon propagator.   \newline

%An emphasis in this work is placed on the distinction of two kinds of quasiparticle excitations in the problem: 
For conditions encountered in neutron stars, nucleons form a strongly interacting Fermi liquid and are regarded as quasiparticles with a finite lifetime whose interactions at low energy are described in terms of Fermi-liquid parameters at low temperature. We shall regard them as \textit{fundamental excitations}, even though their properties are strongly modified by the medium. \textit{Collective excitations} represent long-wavelength oscillations of the density of these quasiparticles in which they remain correlated due to interactions and/or collisions. The propagation of electromagnetic waves with wavelengths that are large compared to the inter-particle distance in the dense medium is interpreted as collective mode. Thus, the photon propagator at low energy will differ substantially from its vacuum counterpart, and its poles encode gauge invariant information about the dispersion relations and damping rates which are determined by the properties of the medium. In addition to the transverse modes which now obtain a mass gap in their dispersion relations,  a new longitudinal mode appears, which, since it owes its existence to the plasma, has been termed \textit{plasmon}. In the following, we shall continue to refer to the transverse mode as the photon and reserve the term plasmon to describe the longitudinal mode. We note that although the plasmon is a well-defined quasi-particle at long-wavelength, as we discuss later, it is strongly damped at short wavelength when it can readily decay into particle-hole excitations.      \newline
%In the following we shall refer to both of these modes as plasma as  and decays rapidly upon approaching the short wavelength limit.  

In dense nuclear matter under neutron star conditions, the vast majority of nucleons are neutrons and one might wonder about their relevance to the electromagnetic response. In free space, and at low momenta, electromagnetic interactions of neutrons are weak as they arise exclusively due to the small neutron magnetic moment. In a dense plasma, on the other hand, photons can couple to neutrons by virtue of an induced interaction due to the polarizability of (electromagnetically and strongly) charged protons. The consistent resummation of this induced interaction to obtain the dressed photon propagator is an important new aspect explored for the first time in this paper. \newline 

A detailed understanding of photon propagation is relevant because in the \textit{space-like} regime (i.e., for photon four-momenta $q^2=q_0^2-\boldsymbol{q}^2 < 0$ ) the photon spectrum determines the dynamics of fermion-fermion scattering in the plasma, which in turn strongly influences transport properties.  In this regard it is particularly important to understand how dynamical screening effects due to Landau damping affect the longitudinal and transverse spectra of the photon in cold degenerate matter. References \cite{Heiselberg:1992ha} and \cite{Heiselberg:1993cr} demonstrate that in a dense QED (or QCD) plasma composed of relativistic fermions the transverse channel dominates electromagnetic interactions. Based on this insight, Ref.\cite{Shternin:2007ee} and \cite{Shternin:2008es} calculate the lepton contribution to thermal conductivity and shear viscosity in neutron star cores (see  \cite{Schmitt:2017efp} for a recent review), ignoring the effect of induced electron-neutron scattering. Reference \cite{Bertoni:2014soa} shows that induced interactions have the potential to substantially  modify these results, and we shall investigate their impact on longitudinal and transverse spectra in detail in this work. We defer the discussion of transport coefficients in this setup to an upcoming publication where we aim for a consistent calculation of the electron and muon contributions to the thermal and electrical conductivity, and shear viscosity using the dynamic photon propagator derived in this work \cite{Reddy2017}. Space-like properties of photons and plasmons also play a role in determining the neutrino scattering rates at high density which are relevant to our understanding of the core-collapse supernovae mechanism, nucleosynthesis and neutrino signals from galactic supernovae  \cite{Horowitz:1991pg,Reddy:1998hb}.    \newline

In the \textit{time-like} ($q^2=q_0^2-\boldsymbol{q}^2 > 0$) region, the low energy spectrum of the photon is relevant for the calculation of the neutrino emissivity since neutrino pairs are produced by the decay of photons and plasmons \cite{Braaten:1993jw,Baldo:2008pb}.  It is also relevant to studies of the production and propagation of hypothetical dark gauge bosons such as the dark photon that can mix with the ordinary photon both in vacuum and in the medium \cite{Batell:2014yra}, \cite{Chang:2016ntp}. While one-loop effects strongly influence the photon and plasmon mass, Landau damping due to particle-hole excitations is absent for these kinematics. Further, under degenerate conditions, the decay to electron-positron pairs of a virtual time-like photon is suppressed exponentially by the large electron chemical potential in dense nuclear matter. A proper description of life-time effects of photons with energies below the electron-positron threshold in the medium rely on the inclusion Compton scattering and inverse nucleon-nucleon Bremsstrahlung process which appear at two-loop order in QED and will be discussed in a future publication.\newline

This paper is organized as follows: Sec.~\ref{sec:QED} reviews the calculation of the linear response within the relativistic RPA in the presence of electromagnetic interactions. We first consider a degenerate plasma comprised of a single fermion species. We compare full results to the hard dense loop approximation which is widely used in literature, study real and imaginary parts of the one-loop polarization functions as well as collective excitations. This serves as a useful preparation to  Sec. \ref{sec:RPAnucl} where strong interactions are introduced: Sec. \ref{sub:Skyrme} elaborates on the derivation of the quasi particle properties from Fermi liquid theory, Sec. \ref{subsec:EMP} discusses the photon spectrum in a multicomponent plasma  composed of electrons, muons, and protons, and finally Sec. \ref{subsec:EMPN} considers the impact of induced interactions with neutrons. Throughout this paper we use natural units $\hbar=c=k_b=1$ and the electric charge $e^2=4\pi\alpha_f$ where $\alpha_f=1/137$ is the fine structure constant and a mostly negative metric convention $g^{\mu\nu}=\textrm{diag}(1,-1,-1,-1)$.
\section{Preliminaries: Photon propagator, linear response and RPA}
\label{sec:QED}
\subsection{Photon propagator and linear response}
\noindent We work in Coulomb gauge subject to the gauge fixing condition $\boldsymbol{\nabla}\cdot\boldsymbol{A}=0$. At zero temperature and in the rest frame of the heat bath, the free Feynman propagator is given by
\be
D^{\mu\nu}(q)=-\frac{1}{q^{2}+i\epsilon}G^{\mu\nu}(q)\,,\hspace{1cm}G^{\mu\nu}(q)=g^{\mu\,0}g^{\nu\,0}\frac{q^2}{\boldsymbol{q}^2}+P_{\perp}^{\mu\nu}\,,\label{eq:PhotonGauge}
\ee
with the usual definition of the transverse projector
\be
P_{\perp}^{ij}=\left(\delta^{ij}-\hat{\boldsymbol{q}}^{i}\hat{\boldsymbol{q}}^{j}\right) \,,\hspace{1cm} P_{\perp}^{00}=P_{\perp}^{0i}=0\,.\label{eq:PPerp}
\ee
The photon self-energy is defined as the difference of the (inverse) dressed and free propagators
\begin{equation}
\Pi^{\mu\nu}(q)=\tilde{D}_{\mu\nu}^{-1}-D_{\mu\nu}^{-1}\,,
\label{eq:SEdef}
\end{equation}
and constrained by current conservation $q_\mu\, \Pi^{\mu\nu}=0$.
As a result it is expressible in terms of two scalar functions $F(q)$ and $G(q)$
\begin{equation}
\Pi^{\mu\nu}(q)=F(q)\,P_{L}^{\mu\nu}+G(q)\,P_{\perp}^{\mu\nu}\,,\label{eq:PiFree}
\end{equation}
where the transverse projector is the same as above while the longitudinal projector reads
\be 
P_L^{\mu\nu}=\frac{q^\mu q^\nu}{q^2} -g^{\mu\nu}-P_\perp^{\mu\nu}\,.\label{eq:PLong}
\ee 
$P_L$ and $P_\perp$ are four transverse to $q^{\mu}$ while $P_{L}^{\mu\nu}$ is three
longitudinal and $P_{\perp}^{\mu\nu}$
is three transverse to $\boldsymbol{q}$ \cite{Weldon:1982aq}. $F$ and $G$ can consequently be related to $\Pi_{00}$ and the trace $\Pi_{\,\,\mu}^{\mu}$ via
\be \label{eq:FG}
F(q)=\frac{q^2}{\boldsymbol{q^2}}\,\Pi_{00}(q)\,,\hspace{1cm}\,G(q)=\frac{1}{2}\left(\delta^{ij}-\hat{q}^{i}\hat{q}^{j}\right)\Pi_{ij}=-\frac{1}{2}\left(\Pi_{\,\,\,\mu}^{\mu}(q)+F(q)\right)\,.
\ee
Multiplying Eq. \ref{eq:SEdef}  with $\tilde{D}$ and $D$ one obtains the Dyson equation
\be
\tilde{D}^{\mu\nu}=D^{\mu\nu}-D^{\mu\alpha}\,\Pi_{\alpha\beta}\,\tilde{D}^{\beta\nu}\label{eq:Dyson}\,,
\ee
which self consistently determines $\tilde{D}$. We may again write the resummed photon propagator in terms of two scalar functions, say $R(q)$ and $S(q)$: 
\be
\tilde{D}^{\mu\nu}(q)=R(q)\,g^{\mu\,0}g^{\nu\,0}+S(q)\,P_{\perp}^{\mu\nu}\,.\label{eq:PropResum}
\ee
Inserting \ref{eq:PhotonGauge},  \ref{eq:PiFree} and \ref{eq:PropResum} into
\ref{eq:Dyson} and solving for $R$ and $S$ one obtains the closed form solutions
\begin{equation}
R(q)=\frac{q^{2}}{\boldsymbol{q}^{2}}\frac{1}{F(q)-q^{2}}=\frac{1}{\Pi_{00}(q)-\boldsymbol{q}^{2}}\,,\hspace{1cm}S(q)=\frac{1}{G(q)-q^{2}}\,.\label{eq:RSresummed}
\end{equation}  
Observe that in the absence of medium effects $\Pi_{00}=G=0$  only the transverse propagator exhibits propagating modes with dispersions $q_{0}=\left|\boldsymbol{q}\right|$ as it should be. This is not the case for covariant gauges where additional unphysical degrees of freedom are to be canceled by the Faddeev-Popov determinant in the partition function. In the low frequency limit the photon spectrum is nevertheless gauge invariant; see Ref. \cite{Kraemmer:2003gd} for a detailed discussion. As an alternative to the direct calculation of the photon propagator according to \ref{eq:SEdef} one may also first resum the polarization tensor  
\be\label{eq:PolResum}
\tilde{\Pi}^{\mu\nu}=\Pi^{\mu\nu}+\Pi^{\mu\alpha}\,D_{\alpha\beta}\,\tilde{\Pi}^{\beta\nu}\,.
\ee
After projecting with $P_L$ and $P_\perp$, the closed form solutions of the polarization functions $\tilde{\Pi}^{\mu\nu}=\tilde{F}P_L^{\mu\nu}+\tilde{G} P_\perp^{\mu\nu}$ read
\begin{equation}
\tilde{F}(q)=\frac{F(q)}{1- (1/q^{2}) F(q)}\,,\hspace{1cm}\tilde{\Pi}_{00}(q)=\frac{\Pi_{00}(q)}{1-(1/\boldsymbol{q}^2)\Pi_{00}(q)}\,,\hspace{1cm}\tilde{G}(q)=\frac{G(q)}{1-(1/q^{2}) G(q)}\,,\label{eq:FGresummed}
\end{equation}  
with $\tilde F = (q^2 / \boldsymbol{q}^2 ) \,\tilde{\Pi}_{00}$. Note that additional factors of $q^2/\boldsymbol{q}^2$ from relation \ref{eq:PhotonGauge} drop out in the evaluation of $\tilde{F}$ and the resummed polarization tensor remains gauge invariant. Inserting the above results for $\tilde{\Pi}$ into the Dyson equation for the photon propagator \ref{eq:Dyson} (whereby $\tilde{D}$ on the right hand side has to be replaced by $D$), one readily recovers \ref{eq:RSresummed}. For the simple case of a one-loop resummation, both approaches to obtain the full photon propagator are depicted in Fig. \ref{fig:Resum}. Once it has been calculated, the spectral representation of longitudinal and transverse photons can be extracted via \cite{Pisarski:1989cs}
\be
\rho_{L}(q)  =  \frac{1}{\pi}\,\text{Im}\, R(q)\,,\hspace{1cm}
\rho_{\perp}(q)  =  \frac{1}{\pi}\,\text{Im}\, S(q)\,.\label{eq:Spectral}
\ee 
The photon propagator and polarization tensor are often discussed in the context of linear response theory and it is instructive to make this relationship explicit. The linear response of an observable $\hat{X}$ to an arbitrary dynamical variable $\hat{Y}$ coupling linearly to an external (classical) source  $\phi_{cl}$ is given by
\be\label{eq:linRes} 
\delta X(x) =-i\int d^4x^{\prime}\,\left\langle\left[\hat{X}(x^{\prime}),\hat{Y}(x)\right]\right\rangle\,\phi_{cl}(x^{\prime})
\ee
where $\delta X$ is defined as the difference of the ensemble averages evaluated in the interacting system with and without an external source
\be\label{eq:response} 
\delta X=\left\langle \hat{X} \right\rangle_\Phi  - \left\langle \hat{X} \right\rangle\,. 
\ee
If we ask for the response of the fermion current to an external photon field we chose $\hat{X}=\hat{Y}=\hat{j}^\mu$, a linear interaction of the form  $\mathcal{H}=A_{cl}^{\mu}\,\hat{j}_\mu$,  and obtain the answer (in momentum space)
\be\label{eq:LRcurrent} 
\delta j^{\mu}(\omega\,,\boldsymbol{k})=A_{cl,\,\nu}(\omega\,,\boldsymbol{k})\,\,\tilde{\Pi}_{R}^{\mu\nu}(\omega\,,\boldsymbol{k})\,,
\ee
where $\tilde{\Pi}_{R}^{\mu\nu}$ is the \textit{full retarded} current-current correlation function, Eq. \ref{eq:FGresummed}. Alternatively, we can ask for the response of the photon field to an external current. In this case, $\hat{X}=\hat{Y}=\hat{A}^\mu$    
with $\mathcal{H}=j_{cl}^{\mu}\,\hat{A}_\mu$ and one obtains
\be\label{eq:LRphoton} 
\delta A^{\mu}(\omega\,,\boldsymbol{k})=-j_{cl,\,\nu}(\omega\,,\boldsymbol{k})\,\,\tilde{D}_{R}^{\mu\nu}(\omega\,,\boldsymbol{k})\,,
\ee 
where $\tilde{D}_{R}$ denotes the  \textit{full retarded} photon propagator, Eq. \ref{eq:PropResum}.  The closed form relation \ref{eq:SEdef} is then merely the statement that introducing a charged test-particle (associated with a classical field $A^\mu_{\textrm{cl}}$) to an interacting system induces a current $j_\mu$, which in turn acts as a source for an internal field $A^\mu_i$. The response $\Pi$ to an external electromagnetic field is often denoted by $\chi$ and termed \textit{generalized susceptibility}. Its explicit calculation will be the content of the next subsection. To conclude the current discussion we employ Eq. \ref{eq:LRcurrent} to calculate the static ($q_0=0$) limit of the \textit{density-density} correlation function $\Pi^{00}$ 
\be \label{eq:MDstatic}
\Pi^{00}=\frac{\delta n}{\delta A^0}\rightarrow\frac{\partial n}{\partial\mu}=\frac{\mu\,k_f}{\pi^2}= \frac{m_D^2}{e^2}\,,
\ee
where the free energy density of relativistic fermions at zero temperature 
\be 
\frac{\partial\mu}{\partial n}=\frac{\partial^{2}}{\partial n^{2}}\,\mathcal{E}_{\text{kin}}\,,\hspace{1cm}\mathcal{E}_{\text{kin}} = \frac{1}{\pi^{2}}\int_{0}^{k_{f}}d\boldsymbol{k}\,\boldsymbol{k}^{2}\,\sqrt{\boldsymbol{k}^{2}+m^{2}}\,,
\ee
and  $k_f=(3\pi^2\,n)^{1/3}$ have been used. Eq. \ref{eq:MDstatic} makes use of the fact that $\delta A^0$ couples to the density $j^0$ in the same way as a chemical potential $\mu$ associated with the fermions. The result yields the standard definition of the \textit{Debye mass} (except for a factor of $e^2$ which appears in a loop calculation in field theory). The susceptibility in the static limit is thus determined by mere thermodynamics.
\subsection{Random phase approximation}
\label{subsec:RPA}
\begin{figure}
\includegraphics[scale=1.11]{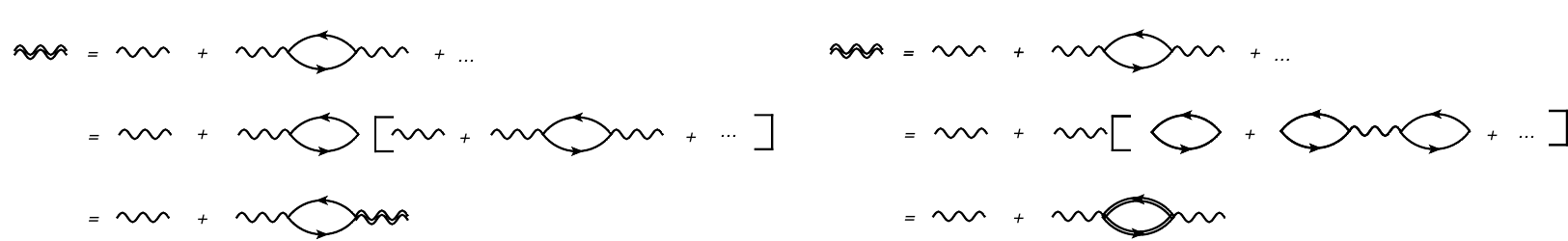}
\caption{\label{fig:Resum} 
\noindent Identical one-loop resummations of the photon propagator. Left: direct resummation according to Eq. \ref{eq:SEdef}. Right: resummation of the polarization function according to Eq. \ref{eq:PolResum} and subsequent insertion into the photon propagator.}
\end{figure}
\begin{figure}
\includegraphics[scale=0.61]{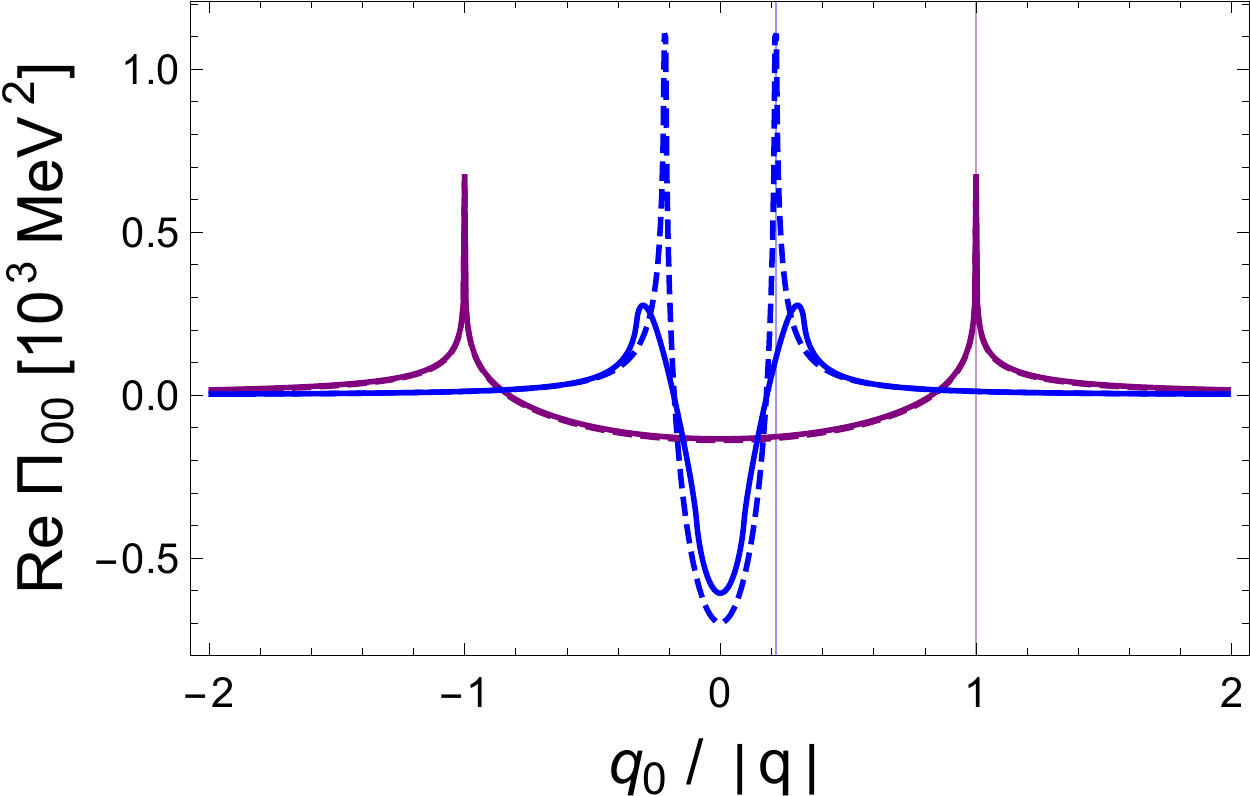}~~~~~~\includegraphics[scale=0.6]{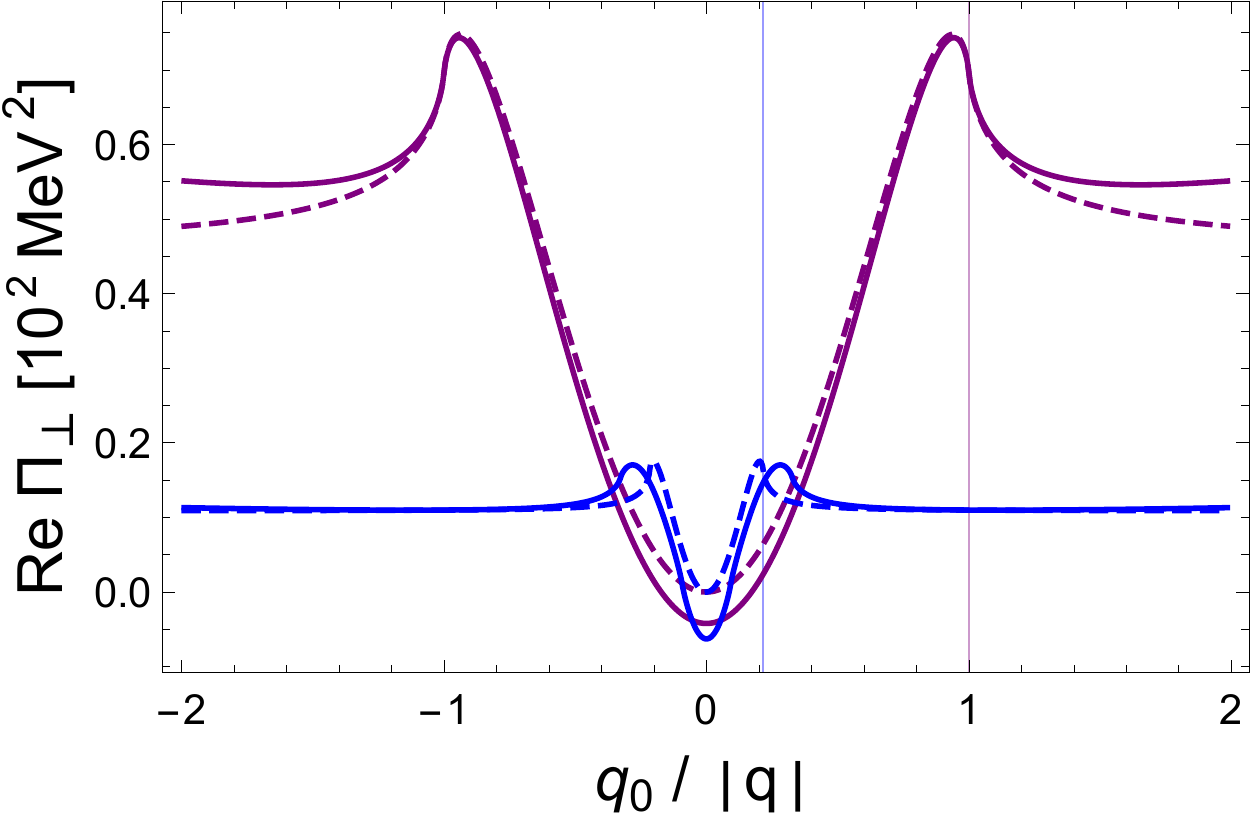}
\caption{\label{fig:RealPi} Real parts of $\Pi_{00}$ and $\Pi_{\perp}$. Purple lines correspond to the case where the fermions in the plasma are electrons, blue lines correspond to the case where they are protons, and dashed lines correspond to the respective HDL approximations. Thin vertical lines indicate the values $q_0=v_f \left|\boldsymbol{q}\right|$ for each particle species. Effective chemical potential and mass of the proton  as well as the chemical potential of the electron are determined at nuclear saturation density in $\beta$ equilibrium,  using NRAPR (non-relativistic Akmal, Pandharipande and Ravenhall  \cite{Steiner:2004fi}) Skyrme type interactions (see Sec. ~\ref{subsec:Fermi} for details): $m_p^{\star}=575\,\textrm{MeV}$ , $m_e=0.5\,\textrm{MeV}$,  $\mu_p^{\star}=589\,\textrm{MeV}$, $\mu_e=122\,\textrm{MeV}$. The momenta $\left| \boldsymbol{q} \right|$ in both cases are fixed at 25$\%$ of the chemical potential and $q_0$ is plotted in between values of $\pm2\left| \boldsymbol{q} \right|$. The real parts of the polarization functions are even functions of the frequency. Despite the relatively large chosen value of $\left| \boldsymbol{q} \right|$, the HDL approximations work very well for electrons. In the presence of protons, one finds the typical behavior of non-relativistic  polarization functions (``Lindhard functions"), which display a much smaller and smoother peak. Hard dense loops do not capture this feature. The transverse component is in general less well approximated by HDL.}
\end{figure}

\noindent Numerous approximations to $\tilde{\Pi}$ exist in the literature. In the Thomas-Fermi model, the susceptibility is approximated by  its value in the static limit. The random phase approximation (RPA) is the simplest approach to a dynamical susceptibility. It consists of a resummation of one-loop polarization functions as outlined in Fig. \ref{fig:Resum}. Calculations of the relativistic photon polarization tensor at finite temperature and density can be found in numerous textbooks; see e.g., Refs. \cite{Kapusta:2006pm}, \cite{Bellac:2011kqa}. 
Renormalization effects are of secondary importance for our purposes and we shall work at fixed coupling $e$. In the degenerate limit the relativistic response function has first been studied by Jancovici \cite{Jancovici:1962jg} who was able to derive analytic expressions for $\Pi$ to leading order in $\alpha_f$.
We carry out finite temperature calculations employing the real time formalism (RTF) in Keldysh representation \cite{Chou:1984es} \cite{Landsman:1986uw}. Propagators are formulated as 2x2 matrices, in the space spanned by particle/thermal ghosts, and most conveniently combined into retarded, advanced and symmetric propagators where only the latter depends on thermal distribution functions \cite{Carrington:1997sq}
\begin{subequations}
\bea
S_{R/A}(q) & = & \frac{\cancel{q} +m}{q^{2}-m^{2}\pm i\textrm{sgn}(q_{0})\epsilon}\,,\label{eq:SRetAdv}\\[2ex]
S_{S}(q) & = & -2\pi i\left(\cancel{q}+m\right)\,(1-2N_{f}(q_{0}))\delta(q^{2}-m^{2})\,,\label{eq:SSym}
\eea
\end{subequations}
with the Fermi distribution function at finite density 
\be
N_{f}(q_{0})=\Theta(q_{0})\,n_{f}(q_{0}-\mu)\,+\Theta(-q_{0})\,n_{f}(-q_{0}+\mu)\,.\label{eq:NFull}
\ee
In terms of these propagators the retarded photon self energy reads 
\be 
\Pi^{\mu\nu}_R(q)=\frac{-ie^{2}}{2}\int\frac{d^{4}k}{(2\pi)^{4}}\left\{ \text{Tr}\left[\gamma^{\mu}S_{S}(k)\gamma^{\nu}S_{R}(k+q)\right]+\text{Tr}\left[\gamma^{\mu}S_{A}(k)\gamma^{\nu}S_{S}(k+q)\right]\right\} \,.\label{eq:PIDefinition}
\ee
Details on the calculation are provided in Appendix~\ref{sec:PI}, here we quote the final result for $\Pi_{00}$ and $\Pi_{\,\,\mu}^{\mu}$ 
\vspace{1mm}
\bea \label{eq:PI0Trace}
\Pi_R(q) & = & -e^{2}\,\int\frac{d^{3}\boldsymbol{k}}{(2\pi)^{3}}\sum_{\xi=\pm}\,\xi\,\frac{1}{(q_{0}+i\epsilon)+\xi(\epsilon_{\boldsymbol{k}}+\epsilon_{\boldsymbol{k}^{\prime}})}\left(C-\frac{K}{\epsilon_{\boldsymbol{k}}\epsilon_{\boldsymbol{k}^{\prime}}}\right)\left[1-n_{f}^{-\xi}(\epsilon_{\boldsymbol{k}})-n_{f}^{\xi}(\epsilon_{\boldsymbol{k}^{\prime}})\right]\\[2ex]
 &  & +e^{2}\,\int\frac{d^{3}\boldsymbol{k}}{(2\pi)^{3}}\sum_{\xi=\pm}\,\xi\,\frac{1}{(q_{0}+i\epsilon)+\xi(\epsilon_{\boldsymbol{k}}-\epsilon_{\boldsymbol{k}^{\prime}})}\left(C+\frac{K}{\epsilon_{\boldsymbol{k}}\epsilon_{\boldsymbol{k}^{\prime}}}\right)\left[n_{f}^{-\xi}(\epsilon_{\boldsymbol{k}})-n_{f}^{-\xi}(\epsilon_{\boldsymbol{k}^{\prime}})\right]\,,\nonumber\\
 \nonumber
\eea
where the coefficients for $\Pi_{00}$ ($\Pi_{\,\,\mu}^{\mu}$) are $C=1$ ($C=-2$) and $K=\epsilon_{\boldsymbol{k}}+\boldsymbol{k}\cdot\boldsymbol{q}$ [$K=2(\epsilon_{\boldsymbol{k}}+\boldsymbol{k}\cdot\boldsymbol{q}+m^{2})$] with the usual relativistic dispersions $\epsilon_{\boldsymbol{k}}=\sqrt{k^{2}+m^{2}}$ and the thermal distributions $n_{f}^{\pm}=n_{f}(\epsilon_{\boldsymbol{k}}\pm\mu)$. 
In the ultrarelativistic limit,  $K/(\epsilon_{\boldsymbol{k}}\epsilon_{\boldsymbol{k}^{\prime}})$ reduces to $\hat{\boldsymbol{k}}\cdot\hat{\boldsymbol{k}}^{\prime}$  ($2\,\hat{\boldsymbol{k}}\cdot\hat{\boldsymbol{k}}^{\prime}$) for $\Pi_{00}$  ($\Pi_{\,\,\mu}^{\mu}$). Unless stated otherwise we are interested in retarded quantities and drop the subscript ``R" in what follows. Before we turn to the RPA resummation, we calculate the hard dense loop (HDL) approximation of Eq. ~\ref{eq:PI0Trace}; see also Refs. ~\cite{Altherr:1992jg} and ~\cite{Altherr:1992mf}. The first step consists of taking the zero temperature limit by replacing Fermi distributions with step functions $n_f^{\pm}\rightarrow\theta(\mp\mu-\epsilon_{\boldsymbol{k}})$. One then assumes that the four-momentum of the photon is small compared to the ``hard"  scales associated with the fermions, i.e., their chemical potentials $\mu$ and Fermi momenta $k_f$. In this limit, the angular integrals of ~\ref{eq:PI0Trace} can be performed analytically and one obtains the following approximations to ~\ref{eq:FG} (see Appendix ~\ref{subsec:HDL} for details):
\vspace*{1mm}
\bea
F_{\textrm{HDL}}(q) & = & (1-\frac{q_{0}^{2}}{\boldsymbol{q}^{2}})\frac{e^{2}\mu k_{f}}{\pi^{2}}\left[1-\frac{1}{2}\frac{\mu q_{0}}{k_{f}\left|\boldsymbol{q}\right|}\text{log}\left(\frac{\mu q_{0}+k_{f}\left|\boldsymbol{q}\right|}{\mu q_{0}-k_{f}\left|\boldsymbol{q}\right|}\right)\right]\,,\label{eq:FHDL}\\[3ex]
G_{\textrm{HDL}}(q) & = & \frac{1}{2}\frac{e^{2}\mu k_{f}}{\pi^{2}}\frac{q_{0}}{\left|\boldsymbol{q}\right|}\left[\frac{q_{0}}{\left|\boldsymbol{q}\right|}+\frac{1}{2}\frac{k_{f}}{\mu}\left(1-\left(\frac{\mu q_{0}}{k_{f}\left|\boldsymbol{q}\right|}\right)^{2}\right)\text{log}\left(\frac{\mu q_{0}+k_{f}\left|\boldsymbol{q}\right|}{\mu q_{0}-k_{f}\left|\boldsymbol{q}\right|}\right)\right]\,,\label{eq:GHDL}
\eea
\noindent with the relativistic Fermi momentum $k_{f}=\sqrt{\mu^{2}-m^{2}}$. In the retarded case, the identification $q_{0}\rightarrow q_0+i\epsilon$ is implied. The HDL expressions are intriguing in the sense that, despite representing an expansion that treats $q_0$ and $\left|\boldsymbol{q}\right|$ on equal footing, they contain both variables to all orders.
Hard loop resummation techniques have been developed to establish a consistent perturbative treatment of gauge theories at finite temperatures and densities which goes beyond the scope of this work. It is nevertheless  interesting to investigate under which conditions results ~\ref{eq:FHDL} and ~\ref{eq:GHDL} are reliable approximations of the photon polarization. If the fermions in the plasma are very light (e.g., electrons), the HDL condition $\left|\boldsymbol{q}\right|\ll k_f$ is easily satisfied in degenerate matter and the HDL results approximate the longitudinal polarization functions remarkably well. The transverse component deviates more strongly with increasing momentum $\left|\boldsymbol{q}\right|$; see also Ref. \cite{Peshier:1998dy}. Decomposition ~\ref{eq:FG} shows that the deviation in the transverse piece originates from the admixture of the trace, for which HDL approximation and full result coincide only close to the light cone. In fact it was shown that  the HDL in-medium dispersions of the photon are exact when one is interested in on-shell photons \cite{Braaten:1993jw}. If the fermions in the plasma are heavy (e.g., nucleons) the HDL condition is violated even for small momenta. The real parts of the polarization functions typically show a peak in the vicinity of $q_0 = v_f \left| \boldsymbol{q} \right|$,  where $v_f=k_f/\mu$ is the relativistic Fermi velocity. For sufficiently heavy fermions, this peak is far away from the light-cone and one should not rely on HDL approximations. The real parts of the full and approximated polarization functions are shown in Fig. \ref{fig:RealPi} for fixed $\left| \boldsymbol{q} \right|$ as a function of $q_0$. 
\subsection{Single fermion species}
\label{subsec:Electrons}
\noindent To set the stage, we first discuss the case of a degenerate QED plasma composed of a single fermion species, say electrons, and outline the calculation of the dressed photon propagator. While a QED plasma at high temperatures and vanishing fermion chemical potential is manifestly charge neutral, one might wonder about the same issue in the opposite limit of low temperatures and high densities. In the multi species case, charge neutrality is achieved by adjusting the chemical potentials of electrons, muons, and protons accordingly. To give some physical meaning to the single species case, one may think of a so called ``jellium" model, where a background distribution of positive charges is determined such that it renders the resulting ground state charge neutral. In its simplest implementation, one assumes a uniform charge distribution which couples to the electron density; see e.g., \cite{FetterWalecka}. Alternatively, one may consider a periodic lattice of positively charged ions. In the first case, the spectrum of excitation includes (quasi-) particle-hole excitations and collective modes which are both modified by the uniform charge background \cite{Baldo:2008pb} \cite{Kobyakov:2017dbl}. This is the scenario relevant for homogeneous matter in the core of neutron stars. In the latter case, the localized charge distribution breaks translational invariance and as a result gives rise to additional low frequency modes, the lattice phonons \cite{Cirigliano:2011tj} \cite{Chamel:2012ix}. This is the scenario relevant for the crust of neutron stars. We shall ignore the issue of charge neutrality for the moment and view this section as a mere preparation for the more complicated multi species case.   
\subsubsection{Dispersion relations and static screening}
\label{subsubsec:RPe}
\begin{figure}
\includegraphics[scale=0.6]{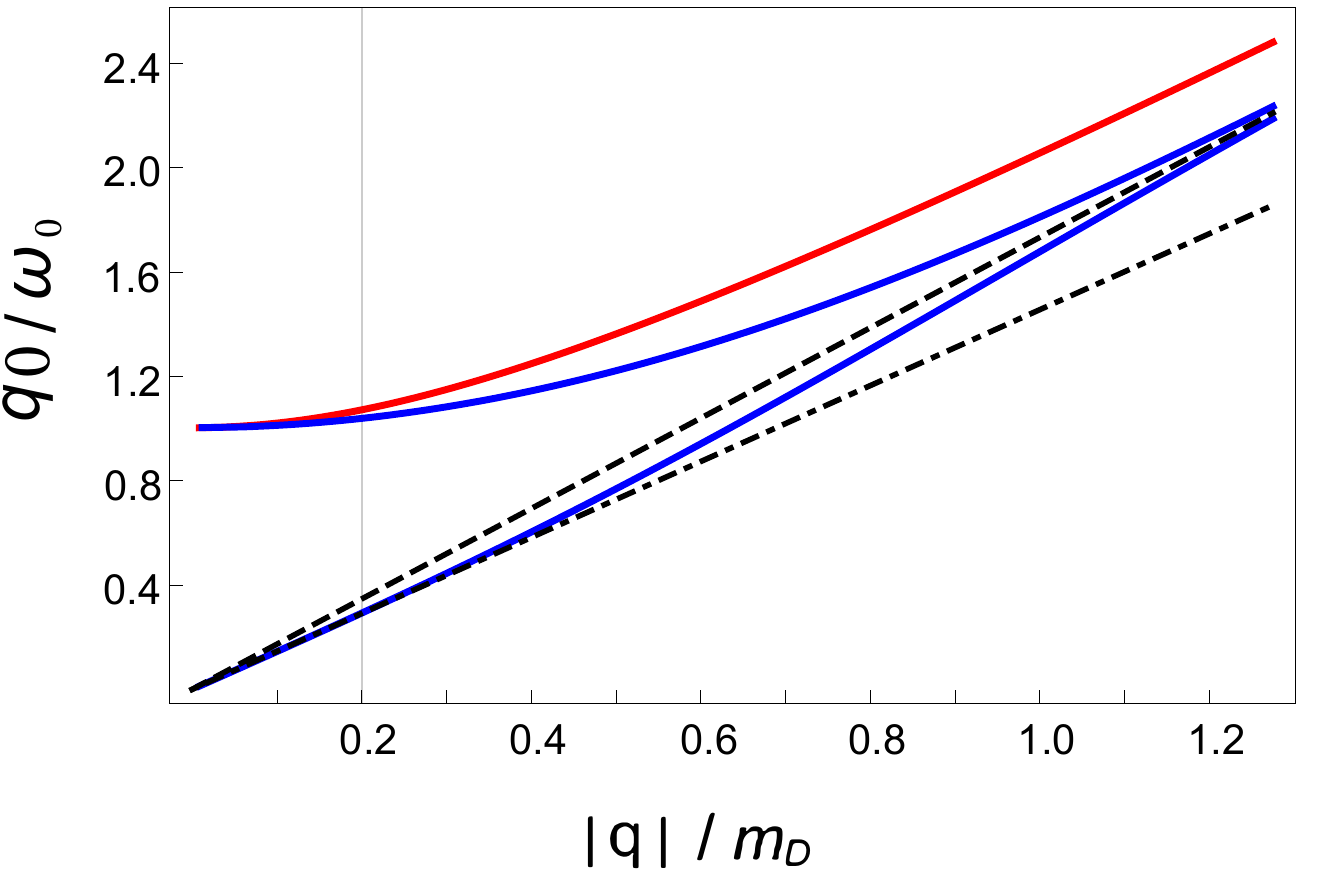}~~~~~\includegraphics[scale=0.6]{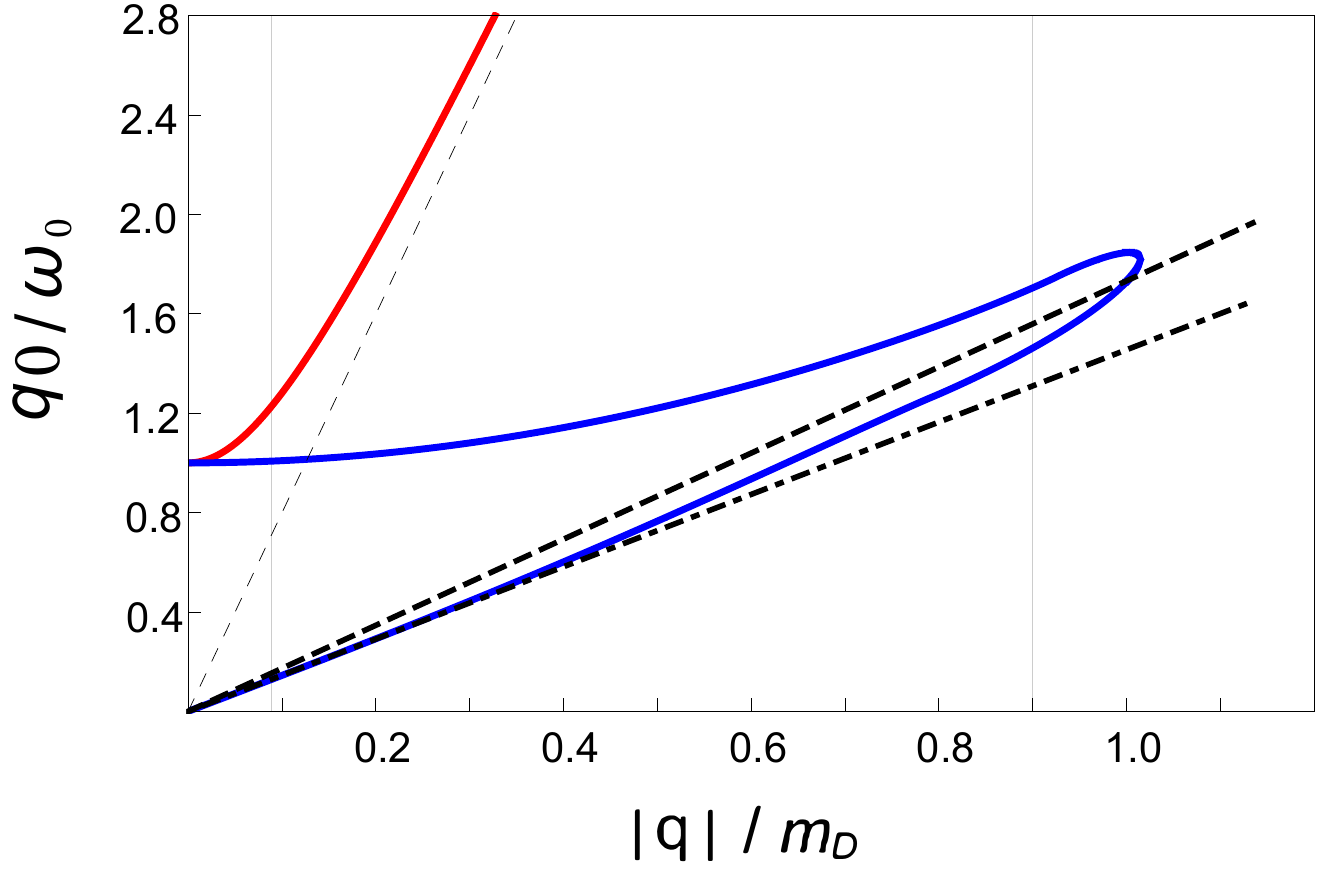}
\setlength{\belowcaptionskip}{-1mm}
\caption{\label{fig:SingleModes} Zeros of the real parts of the longitudinal (blue) and transverse (red) photon propagator in a degenerate plasma composed of electrons (left) and protons  (right). Same parameters as in Fig. \ref{fig:RealPi}. Dashed lines  correspond to $q_0=v_f\,\left|\boldsymbol{q}\right|$ (the gray dashed line in the right figure corresponds to the light cone); dot-dashed lines correspond to $q_0=c\,v_f\,\left|\boldsymbol{q}\right|$ where the constant $c$ is obtained by numerically solving Eq. \ref{eq:slope}. In the longitudinal case one encounters a characteristic ``thumb-like" structure that has been reported in several references \cite{Baldo:2008pb}\cite{mcorist_melrose_weise_2007}. Only the upper branch corresponds to a collective mode for as long as it resides outside the Landau damped region, see also Fig. \ref{fig:ImPi}. The electron plasma qualitatively resembles the ultra-relativistic case ($v_f\sim 1$) in which the two longitudinal branches approach the light cone asymptotically. In the proton plasma both longitudinal branches merge at significantly lower momenta. This feature is not captured by the HDL approximation which predicts the existence a longitudinal mode up to much larger momenta, as can be seen from the proton peak in Fig. \ref{fig:RealPi}. The transverse mode obtains a finite thermal mass and remains always time-like. The spectral functions in Fig. \ref{fig:rhoEP} are plotted along the thin gray lines.}
\end{figure}
\noindent  The  dispersion relations of the longitudinal and transverse  modes of the in-medium photon  are obtained by solving  
\bea
\omega_{L}^{2} & = & \boldsymbol{q}^{2}+\textrm{Re}\,F(q_{0}=\omega_{L},\,\boldsymbol{q})\,,\label{eq:wL}\\[1ex]
\omega_{\perp}^{2} & = & \boldsymbol{q}^{2}+\textrm{Re}\,G(q_{0}=\omega_{\perp},\,\boldsymbol{q})\,,\label{eq:wP}
\eea
[or similarly $\boldsymbol{q^2}=\Pi_{00}(\omega_L,\,\boldsymbol{q})$] for $\omega_{L}$ and $\omega_{\perp}$.  In the following, we discuss the solutions to these equations in detail. Relations \ref{eq:wL} and \ref{eq:wP} define poles of the resummed photon propagator if the polarization functions exhibit no imaginary part for given values of $q_0$ and $\left|\boldsymbol{q}\right|$. As we shall see, this is not necessarily the case. In general, there is one solution to the transverse equation \ref{eq:wP},  and  two solutions to the longitudinal equation \ref{eq:wL}, which are roughly located on each side of the line $q_0=v_f \left|\boldsymbol{q}\right|$; see Fig. \ref{fig:SingleModes}. The branch corresponding to the higher photon energy $q_0$ is denoted by $\omega_{L}$ while the lower branch is denoted by $\omega_{<}$. As explained in detail in the next section, the solution of lower energy is subject to Landau damping (causing a finite imaginary part of $F$) and only $\omega_{L}$ constitutes a well defined mode. In general, solutions to ~\ref{eq:wL} and ~\ref{eq:wP} are obtained numerically. For very small momenta, we may employ the HDL approximation, expand ~\ref{eq:FHDL} to leading order in $\boldsymbol{q}$, and solve for the dispersion relations, which give
\be \label{eq:PlasmaL}
\omega_{L}^2=\omega_0^2+\frac{3}{5}v_f^2\,\boldsymbol{q}^2\,,\hspace{1cm} \omega_0^2=\frac{e^2}{3\pi^2}\frac{k_f^3}{\mu}\,,
\ee 
\noindent where  $\omega_0$ is the  usual plasma frequency which one recovers from approaching the static limit of  ~\ref{eq:FHDL} and ~\ref{eq:GHDL} in the \textit{time-like} region $\omega_0=F(q_0\rightarrow0,\,\boldsymbol{q}=\boldsymbol{0})=G(q_0\rightarrow0,\,\boldsymbol{q}=\boldsymbol{0})$. The lower (damped) solution is more subtle to obtain due to the non-analytic behavior of the logarithm in ~\ref{eq:FHDL} at $q_0=0$. Since any space-like solution has to be gapless, we choose the ansatz    
\be 
\omega_{<}=c\,v_f\,\left|\boldsymbol{q}\right|\,,
\ee 
insert it into Eq.~\ref{eq:wL}, and drop quadratic terms in $\boldsymbol{q}$. The constant $c$ is then determined by the equation 
\be \label{eq:slope}
\frac{1}{2}\,c\,\text{Log}\,\left|\frac{1+c}{1-c}\right|=1
\ee
which yields $c\sim0.83$. The slope of the gapless branch is thus determined by the (universal) number $c$ and the mass of the fermions which enters via $k_f$. With increasing momentum, the gapless branch intersects with the line $v_f\,\left|\boldsymbol{q}\right|$ and merges with the real plasmon mode $\omega_{L}$, leading to a ``thumb-like" shape \cite{Baldo:2008pb} \cite{mcorist_melrose_weise_2007}. At  larger momenta, there are no more longitudinal excitations in the spectrum. Any longitudinal mode in a plasma composed of \textit{massive} fermions shares this fate, which is not unexpected: At higher energies when collective effects become negligible, the photon is expected to return to its  vacuum appearance, which is purely transverse \footnote{One might wonder about the ultra-relativistic limit where $v_f=1$ and the light-cone marks the borderline between the (time-like) region where collective modes are undamped and the (space-like) region where Landau damping sets in. In this case, $\omega_{L}$ and $\omega_{<}$ merge for $\left|\boldsymbol{q}\right|\rightarrow\infty$. The longitudinal mode is nevertheless expected to disappear in the free-space limit. Indeed, it was shown ~\cite{Pisarski:1989cs} that the residue of the pole corresponding to the longitudinal mode in ultra-relativistic plasmas becomes exponentially small for large momenta.}. The exact boundary where Landau damping sets in will be evaluated in the next section. It is important to mention that the ``tip of the thumb" is always located in the Landau damped region, avoiding the awkward scenario of a mode with infinite group velocity $v_g=\partial \omega / \partial \left|\boldsymbol{q}\right|$. The transverse mode in the spectrum remains always time-like. In a low momentum expansion, one finds
\be \label{eq:PlasmaP}
\omega_\perp^2=\omega_0^2+\left(1+\frac{1}{5}v_f^2\right)\boldsymbol{q}^2\,.
\ee
In the opposite limit of large frequencies and momenta, the transverse photon acquires a thermal mass such that for $\boldsymbol{q}\rightarrow\infty$ one indeed recovers the dispersion relation of a free propagating mode.  Using the HDL results, one obtains 
\be 
\omega_{\perp\,,\infty}^2=\left(1+\frac{1}{5}v_f^2\right)\,\omega_0^2+\boldsymbol{q}^2\,.
\ee
Approaching the static limit of the polarization functions in the \textit{space-like} region, one obtains the screening mass (Debye mass),  
\be \label{eq:mD}
F(q_0=0,\,\boldsymbol{q}\rightarrow\boldsymbol{0})=m_D^2=3\omega_0^2\,/v_f^2=e^2\mu k_f/\pi^2\,,\hspace{1cm}G(q_0=0)=0\,,
\ee
in agreement with result \ref{eq:MDstatic}. As expected, the normal (i.e., not superconducting) phase exhibits   \textit{static} screening for longitudinal but not for transverse (magnetic) photons. The inverse Debye mass defines the length scale at which the screening operates: At distances smaller than $m_{D}^{-1}$, electromagnetic interactions in the plasma are effectively unscreened. Because of their high masses, nucleons obviously provide a far less effective screening compared to leptons. In the static limit, the longitudinal propagator  ~\ref{eq:RSresummed} turns into $R(q_0=0, \boldsymbol{q})=(\boldsymbol{q}^2+m_D^2)^{-1}$ which describes a \textit{Thomas-Fermi} screened interaction. Note that in the ultra-relativistic limit, the static screening mass can be read off from the HDL expression of the trace ~\ref{eq:HDLTrace}, $\Pi_{\,\,\mu}^{\mu}=-m_{D}^{2}$. In the non-relativistic case ($k_f\ll\mu$), the mass is the highest scale, and the $m^2$ term in ~\ref{eq:HDLTrace} becomes dominant. To conclude this subsection, we compute the static limit of the \textit{resummed} polarization tensor \ref{eq:FGresummed}
\be \label{eq:DebyeCoulomb}
\tilde{m}_D=-\,\tilde{\Pi}_{00}(q_0=0,\,\boldsymbol{q})=\frac{m_D^2}{1+(1/\boldsymbol{q}^2)\,m_D^2}\,.
\ee
When the full RPA results are used, one encounters \textit{dynamical} (frequency dependent) screening, which affects both longitudinal and transverse photons.
\subsubsection{Imaginary parts}
\begin{figure}
\begin{centering}
\includegraphics[scale=1.15]{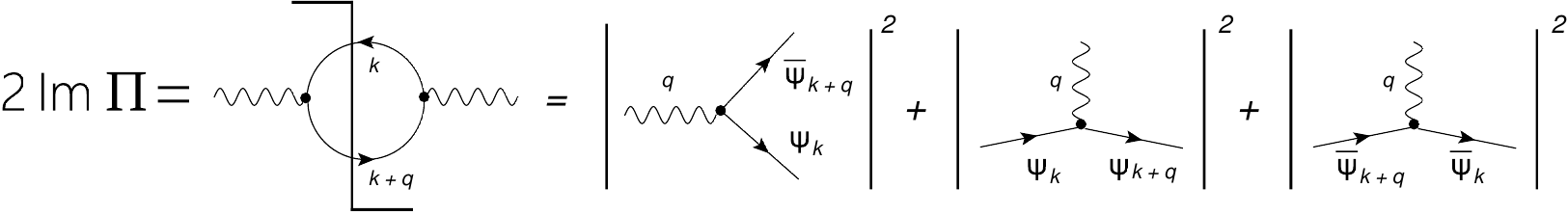}
\caption{\label{fig:CuttingPi}Physical interpretation of the various contributions to the imaginary part of the photon self-energy according to the cutting rules.  The first diagram corresponds to the creation of a fermion/anti-fermion pair while the latter two correspond to Landau damping: particles (and antiparticles) are emitted from or absorbed by the thermal bath. The latter two contributions are particle-hole processes, particle-antiparticle processes are exponentially suppressed in degenerate matter by the large chemical potentials. }
\end{centering}
\end{figure}
\label{subsubsec:IPe}
\begin{figure}
\begin{centering}
\includegraphics[scale=0.69]{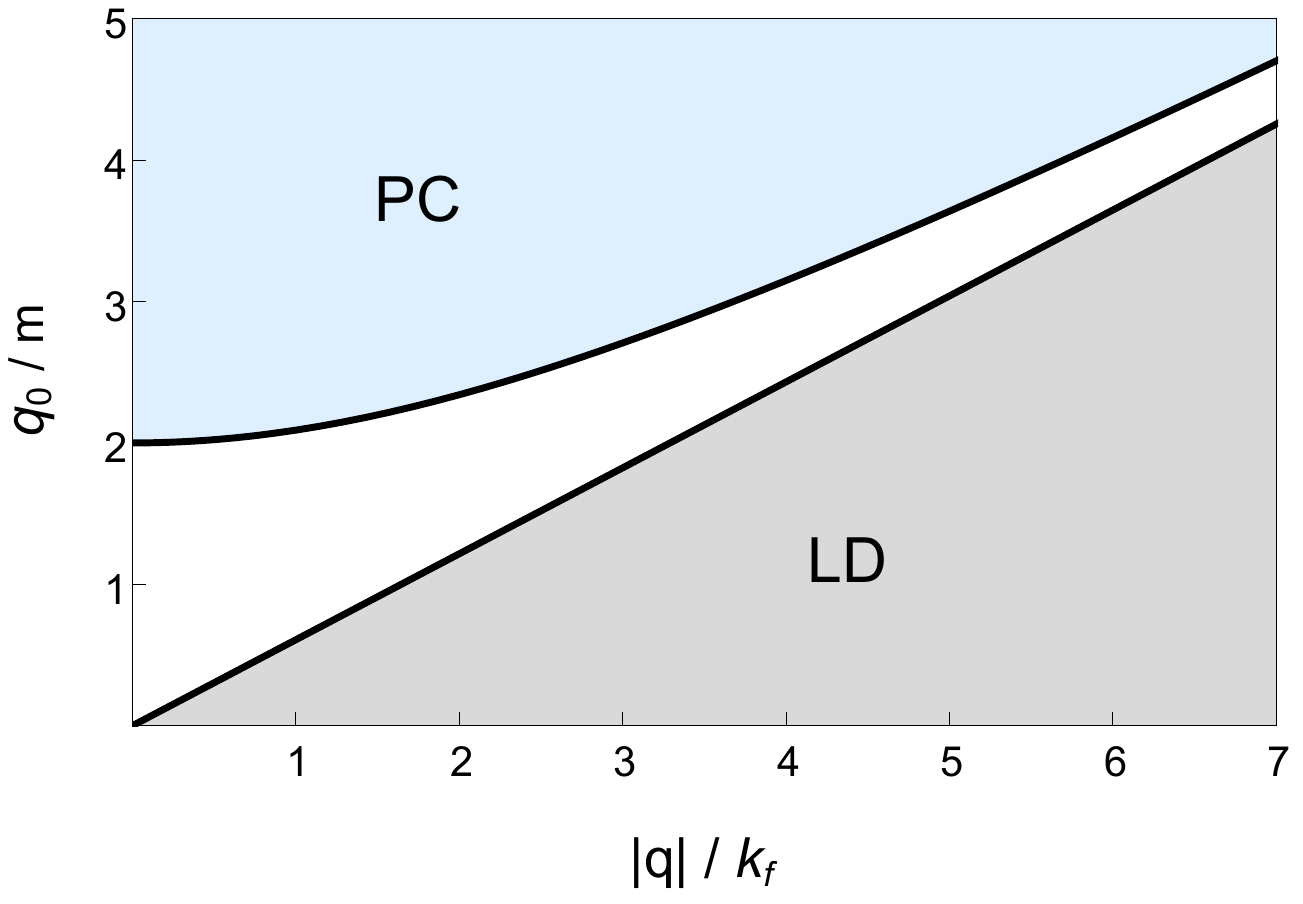}~~~\includegraphics[scale=0.69]{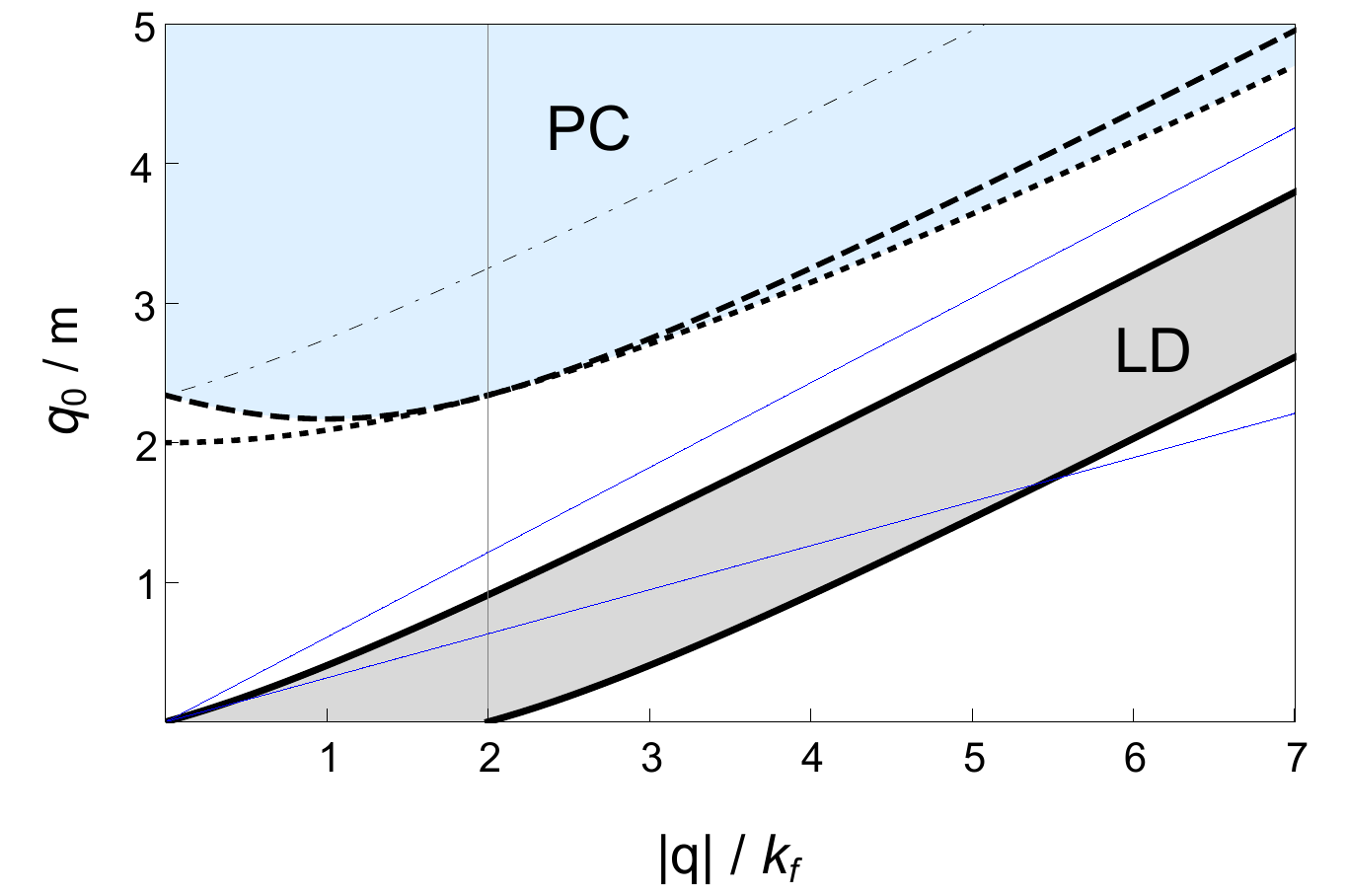}
\par\end{centering}
~\\[2ex]
\begin{centering} \includegraphics[scale=0.7]{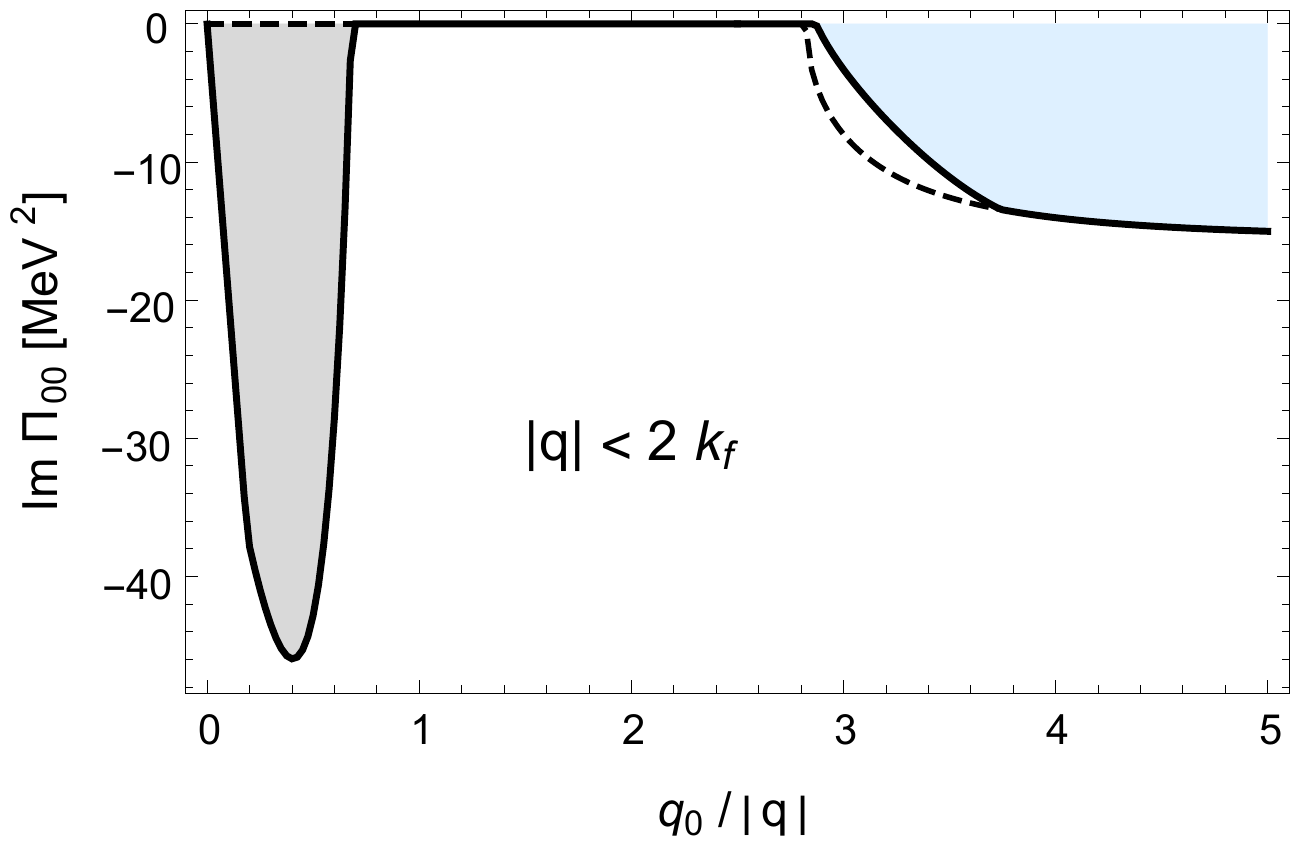}\includegraphics[scale=0.7]{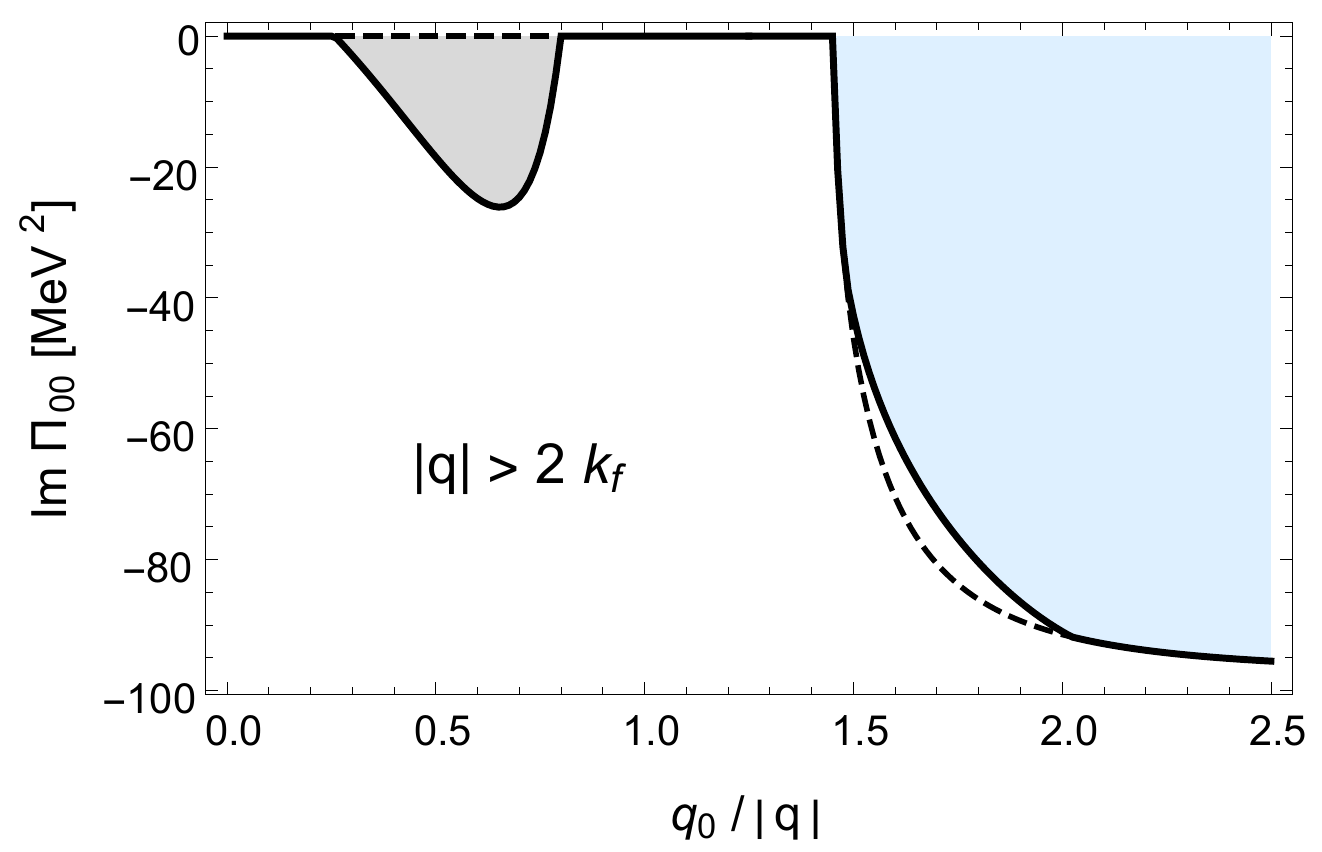}
\end{centering}
\caption{\label{fig:Domains} 
Top left: domains where Landau damping (gray shaded area) and pair creation (blue shaded area) are kinematically allowed. The boundaries are $q_0=\left|\boldsymbol{q}\right|$ and  $q_0=\sqrt{\left|\boldsymbol{q}\right|^2+4m^2}$. Top right: effect of degeneracy on the boundaries. Landau damping is confined in between the two thick solid lines $q_0=-\mu+\sqrt{\mu^2+2 k_f\left|\boldsymbol{q}\right|+\left|\boldsymbol{q}\right|^2}$  and $q_0=-\mu+\sqrt{\mu^2-2 k_f\left|\boldsymbol{q}\right|+\left|\boldsymbol{q}\right|^2}$. The thin vertical line indicates the position of $\left|\boldsymbol{q}\right|=2k_f$. At $q_0 = 0$ there is no more Landau damping for $\left|\boldsymbol{q}\right| > 2k_f$. Pair creation in degenerate matter is determined by the interplay of the two boundaries $q_0=\mu+\sqrt{\mu^2-2 k_f\left|\boldsymbol{q}\right|+\left|\boldsymbol{q}\right|^2}$ (dashed) and $q_0=\sqrt{\left|\boldsymbol{q}\right|^2+4m^2}$ (dotted): For $\left|\boldsymbol{q}\right| < 2k_f$ - despite being kinematically allowed - pair creation is completely suppressed for values of $q_0$ larger than the dotted line and only sets in above the dashed line. At $\left|\boldsymbol{q}\right| = 2k_f$ both boundaries assume the value $q_0=2\mu$ and for $\left|\boldsymbol{q}\right| > 2k_f$ pair creation becomes possible for values of $q_0$ above the dotted line. Finally for values of $q_0$ above the dot-dashed line $q_0=\mu+\sqrt{\mu^2+2 k_f\left|\boldsymbol{q}\right|+\left|\boldsymbol{q}\right|^2}$ the degeneracy suppression has completely faded away and pair creation assumes its vacuum value. The analytic calculation of the imaginary part for each of these regions is presented in Appendix \ref{sub:ImPiDegen}. The thin blue lines indicate the position of the light cone and of $q_0=v_f \left|\boldsymbol{q}\right|$. It is easy to see that the energy $q_0$ required to create a fermion/anti-fermion pair is minimal when the photon has momentum $\left|\boldsymbol{q}\right|=k_f$. \newline
Bottom left: imaginary part of $\Pi_{00}$ for $\left|\boldsymbol{q}\right| < 2k_f$ (left) and $\left|\boldsymbol{q}\right| > 2k_f$ (right). A convenient choice to illustrate the imaginary parts of both regions is to use the muon mass with a muon chemical potential at saturation density of about 122 MeV. In this case the separation of both regions is clearly visible, yet not too large. The value of $2k_f$ calculates to roughly 124 MeV. The momenta are set to $ \left|\boldsymbol{q}\right|$ = 80 MeV (left) and 200 MeV (right). Solid lines represent the total magnitude of the imaginary part, and dashed lines represent the vacuum values. The outline in both cases can be understood from the top right diagram: For $\left|\boldsymbol{q}\right| < 2k_f$, a non zero imaginary part sets in immediately at $q_0=0$ until the boundary of Landau damping is reached. Upon increasing $q_0$, one traverses through a dissipation free region until the threshold for pair creation is reached. Degeneracy effects suppress the imaginary part due to pair creation until one eventually crosses the dot-dashed line in the top right diagram, above which pair creation assumes its unsuppressed vacuum value. For $\left|\boldsymbol{q}\right| > 2 k_f$, the situation is similar except that Landau damping is absent for sufficiently small $q_0$. With increasing momentum, the imaginary part due to pair creation outgrows the imaginary part due to Landau damping. }      
\end{figure} 
\noindent The imaginary part determines the damping $\gamma$ of a given mode. For well defined quasi-particles, we may decompose $q_0=\omega-i\gamma$, assume $\gamma\ll\omega$, and expand the imaginary parts of the poles of \ref{eq:RSresummed} 
\be 
\gamma_L=\textrm{Im}\,\Pi_{00}\left(\frac{\partial\,\textrm{Re}\,\Pi_{00}} {\partial q_0}\right)^{-1}\,,\hspace{1cm}\gamma_\perp=-\frac{\textrm{Im} \,G(q_0,\,\boldsymbol{q})}{2\omega_\perp}\,.
\ee
The imaginary part of the polarization tensor is derived in appendix \ref{sec:ImPI} and reads
\begin{eqnarray}
\text{Im}\Pi^{\mu\nu}(q) & = & \frac{e^{2}}{2}\int\frac{d^{3}\boldsymbol{k}}{(2\pi)^{2}}\frac{1}{2\epsilon_{\boldsymbol{k}}}\frac{1}{2\epsilon_{\boldsymbol{k}^{\prime}}}\sum_{\xi=\pm}\xi\,\left[1-n_{f}^{-\xi}(\epsilon_{\boldsymbol{k}})-n_{f}^{+\xi}(\epsilon_{\boldsymbol{k}^{\prime}})\right]T^{\mu\nu}(k,q)\,\delta(q_{0}-\xi(\epsilon_{\boldsymbol{k}}+\epsilon_{\boldsymbol{k}\prime}))\nonumber \\[2ex]
 & + & \frac{e^{2}}{2}\int\frac{d^{3}\boldsymbol{k}}{(2\pi)^{2}}\frac{1}{2\epsilon_{\boldsymbol{k}}}\frac{1}{2\epsilon_{\boldsymbol{k}^{\prime}}}\sum_{\xi=\pm}\xi\left[n_{f}^{-\xi}(\epsilon_{\boldsymbol{k}})-n_{f}^{-\xi}(\epsilon_{\boldsymbol{k}^{\prime}})\right]T^{\mu\nu}(k,q)\,\delta(q_{0}-\xi(\epsilon_{\boldsymbol{k}}-\epsilon_{\boldsymbol{k}\prime}))\,,\label{eq:ImPi}
\end{eqnarray}
with the trace $T^{\mu\nu}=\text{Tr}\left[\gamma^{\mu}\left(\cancel{k}+m\right)\gamma^{\nu}\left(\cancel{k}^{\prime} +m\right)\right]$. Non-zero imaginary parts emerge whenever the fermions in the loop are put on shell. The resulting on shell conditions are encoded in the delta functions above and indicate the kinematic requirements of the respective one-loop processes: Since $q_{0}+\epsilon_{\boldsymbol{k}}+\epsilon_{\boldsymbol{k}^{\prime}}>0$ for any angle and momentum $\left|\boldsymbol{k}\right|$, the term corresponding to $\xi=-$ in the first line never exhibits a pole. The $\xi=+$ term of the first line develops a pole for $q_{0}=\epsilon_{\boldsymbol{k}}+\epsilon_{\boldsymbol{k}^{\prime}}$ which is fulfilled for time-like photon momenta $q^{2}=q_{0}^{2}-\boldsymbol{q}^{2}\geq\,4m^{2}$, the usual threshold for pair creation. The two terms in the second line develop poles for $q_{0}=\pm\epsilon_{\boldsymbol{k}}\mp\epsilon_{\boldsymbol{k}^{\prime}}$
corresponding to space like photon momenta $q^{2}=q_{0}^{2}-\boldsymbol{q}^{2}\leq0$. Cutting rules ~\cite{Kobes:1985kc} (or unitarity rules; see Ref. \cite{Das:1997gg} for a thorough introduction) relate the imaginary part of the (one-loop) polarization tensor to interaction rates of tree level processes (see also Fig. \ref{fig:CuttingPi} for an illustration): The contribution in the first line of \ref{eq:ImPi} corresponds to the production rate of a fermion/anti-fermion pair $\gamma\rightarrow f\,\bar{f}$, reflected by the mixing of particle and anti-particle distribution functions which can be rearranged to yield the detailed balance factors of the reaction $(1-n_1^{-})(1-n_2^{+})-n_1^{-}n_2^{+}$. The two contributions in the second line of \ref{eq:ImPi} correspond to the rates at which fermions ($\xi=+$) and antifermions ($\xi=-$) scatter [with detailed balance factors $n_1^{\pm}(1-n_2^{\pm})- n_2^{\pm}(1-n_1^{\pm})$ ] where we can dismiss the latter in the degenerate limit. Fermion scattering $\gamma\,f\rightarrow f^{\prime}$  (i.e. the creation of particle-hole pairs) arises due to creation or annihilation of particles during collisions with low momentum photons of the heat bath. It is thus a pure medium effect (observe that in contrast to the first line in \ref{eq:ImPi} there is no vacuum contribution in the second). As a net result the photon looses energy which has been termed \textit{Landau damping}.\newline
In the degenerate limit $\mu - m \gg T$, the fermion gas suppresses Landau damping and pair creation in several regions where it is otherwise kinematically allowed. The domain boundaries become more complicated; see Fig. \ref{fig:Domains} and Appendix \ref{sub:ImPiDegen}. For very low $\left|\boldsymbol{q}\right|$, one obtains simple approximations to the domain boundaries $q_0 = v_f \left|\boldsymbol{q}\right|$ for Landau damping and $q_0 = 2\mu - v_f \left|\boldsymbol{q}\right|$ for pair creation.  Analytic results for arbitrary $q_0$ and $\left|\boldsymbol{q}\right|$ are derived in Appendix \ref{sub:ImPiDegen}. In the HDL limit, the imaginary part can easily be extracted from \ref{eq:FHDL} and \ref{eq:GHDL}
\vspace*{1mm}
\bea
\text{Im}\,\Pi_{\text{HDL},\,R}^{00}(q) & = & -\frac{\pi}{2}m_{D}^{2}\,\frac{\mu}{k_{f}}\frac{q_{0}}{\left|\boldsymbol{q}\right|}\Theta(k_{f}\left|\boldsymbol{q}\right|-\mu \,q_{0})\,,\label{eq:ImP00HDL}\\[3ex]
\text{Im}\,G_{\text{HDL},\,R }(q) & = & -\frac{\pi}{4}m_{D}^{2}\frac{k_{f}q_{0}}{\mu\left|\boldsymbol{q}\right|}\left[1-\left(\frac{\mu\, q_{0}}{k_{f}\left|\boldsymbol{q}\right|}\right)^{2}\right]\Theta(k_{f}\left|\boldsymbol{q}\right|-\mu\, q_{0})\,.\label{eq:ImGHDL}
\eea
As expected, the HDL results indeed exhibit a strict cut-off at $q_0 = v_f \left|\boldsymbol{q}\right| $. A comparison of Landau damping using HDL and full results is shown in Fig. \ref{fig:ImPi}. In addition, we display the solutions $\omega_{L}$ and $\omega_{<}$ over the magnitude of the imaginary part of $\Pi_{00}$. We have previously labeled solutions to \ref{eq:wL} $\omega_{L}$ while they remain in the dissipation free region and $\omega_{<}$ in the Landau damped region. For most of its existence $\omega_{<}$ resides in a region just below the line $q_0 = v_f \left|\boldsymbol{q}\right|$, where it is maximally damped and hence should not be regarded as an actual mode. The value of $\left| \boldsymbol{q} \right| $ at which $\omega_{L}$ ``dives" into the Landau damped region is often referred to as ``cutoff momentum" and has to be determined numerically. \newline
The one-loop calculation obviously leaves  behind gaps in the space spanned by $q_0$ and  $\left|\boldsymbol{q}\right|$ which are covered neither by Landau damping nor pair creation. In these domains, two-loop processes (Compton scattering and Bremsstrahlung) which are kinematically allowed, take over.
\begin{figure}
\begin{centering}
\includegraphics[scale=0.6]{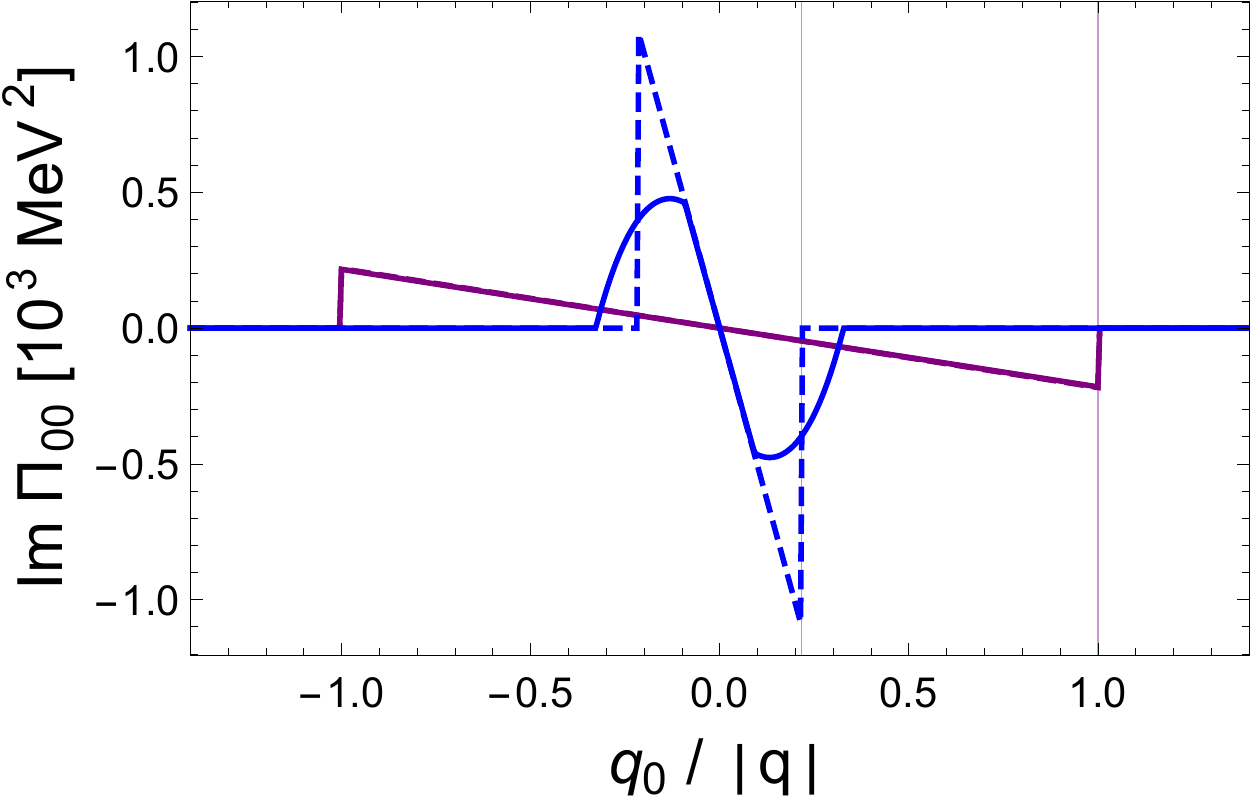}~~~~~~~\includegraphics[scale=0.59]{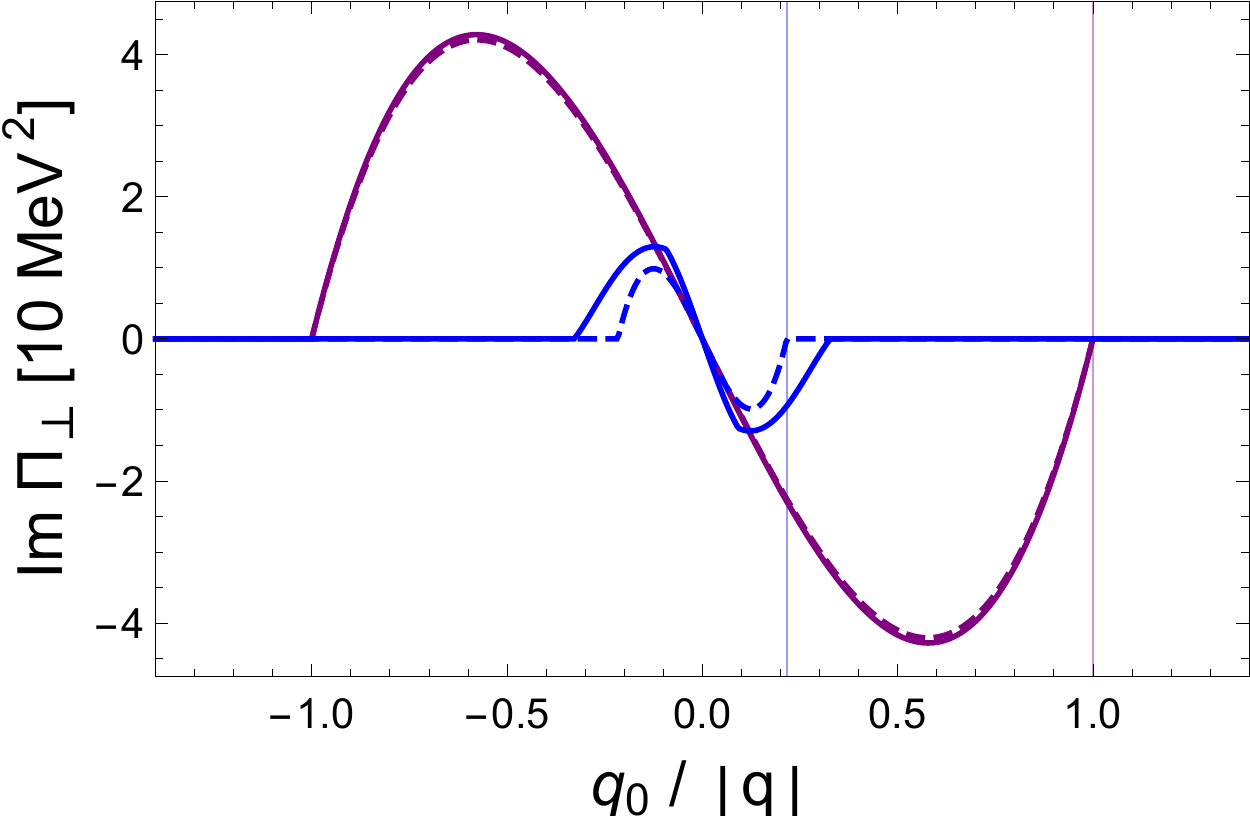}
\par\end{centering}
~\newline
~
\begin{centering} \includegraphics[scale=0.6]{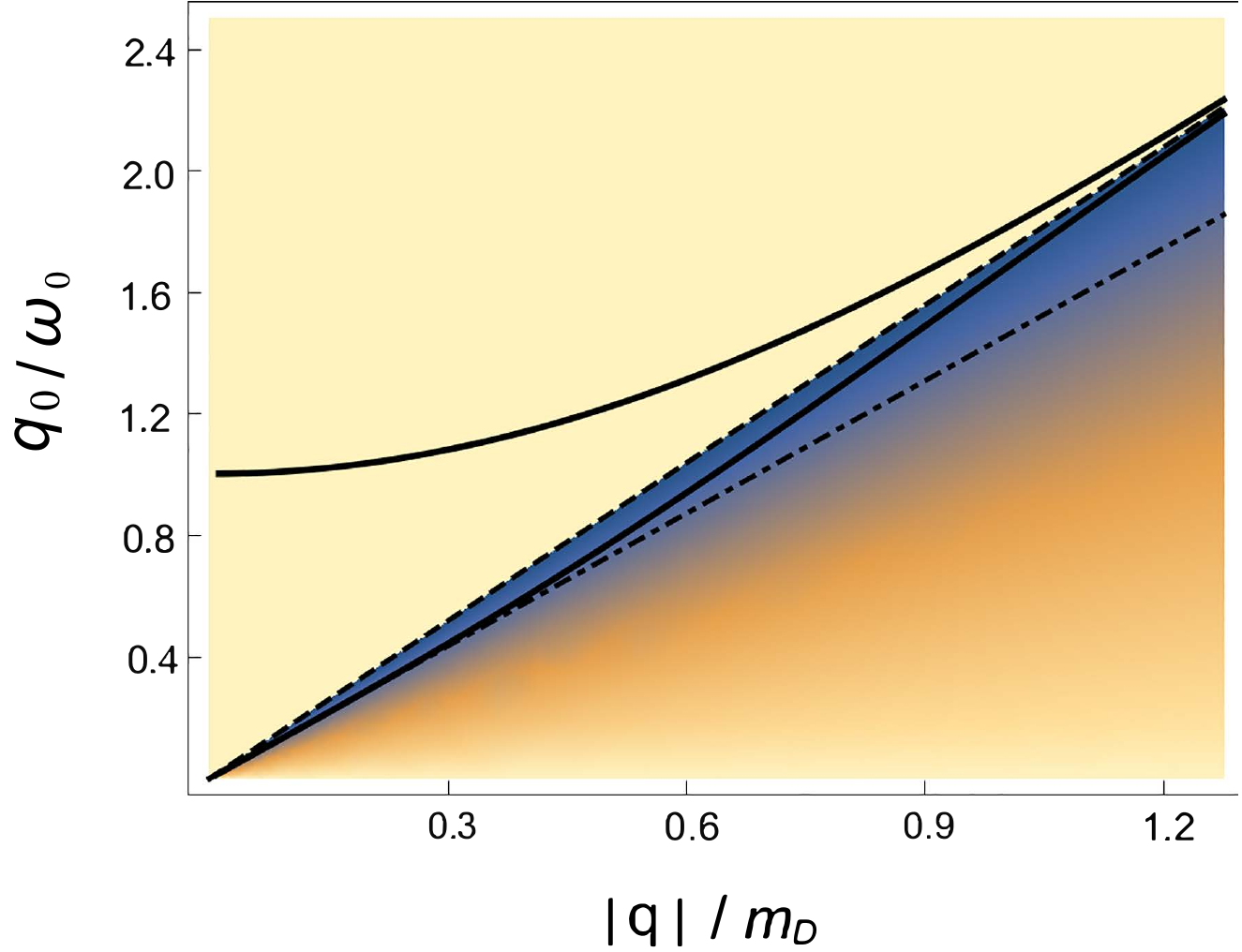}~~~~~\includegraphics[scale=0.59]{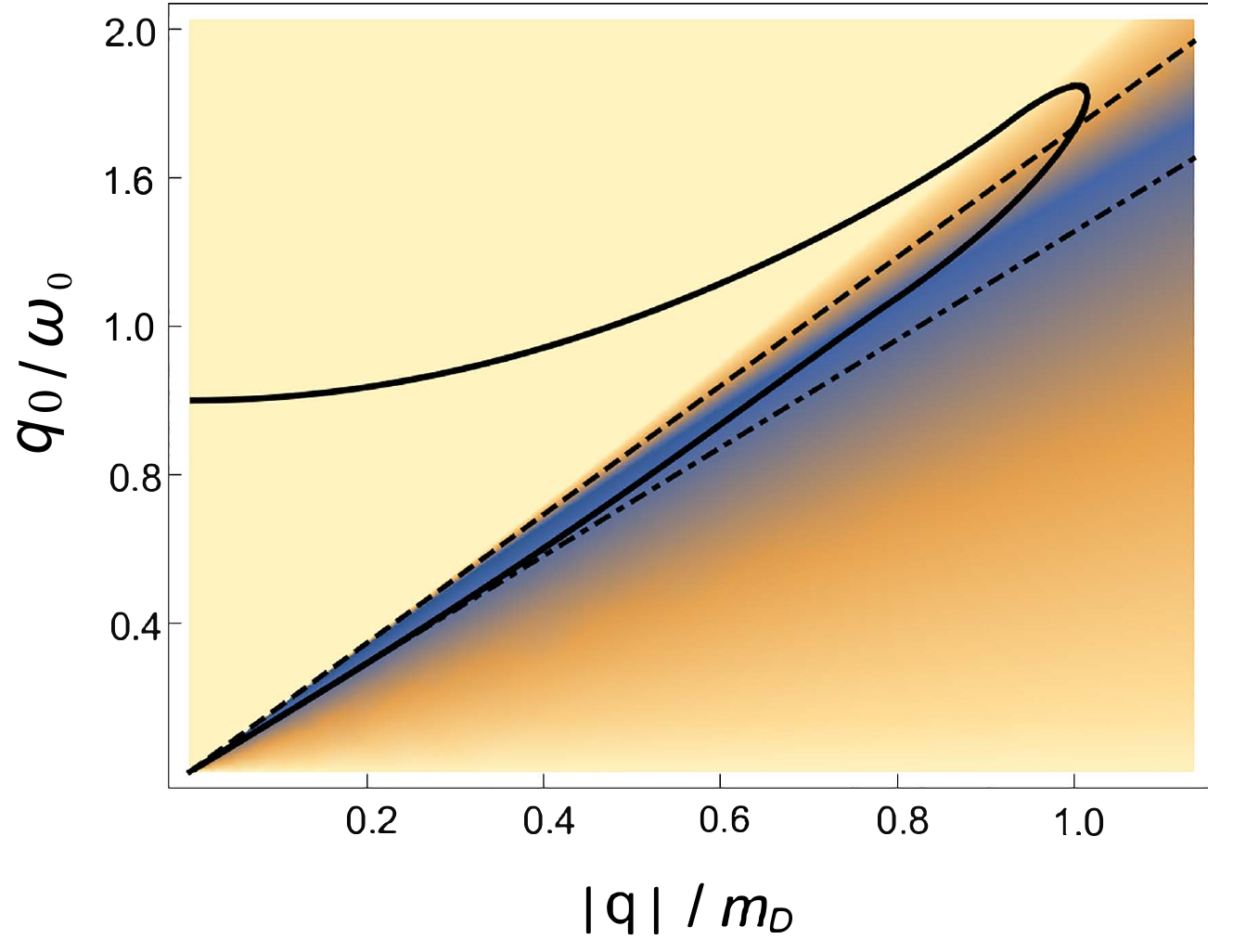}
\end{centering}
\caption{\label{fig:ImPi} 
Top left and right: imaginary parts of $\Pi_{00}$ and $\Pi_{\perp}$. Same parameters as in Fig. \ref{fig:RealPi}. Purple lines correspond to electrons, blue lines correspond to protons, and dashed lines correspond to the respective HDL approximations. Thin vertical lines indicate the positions of $q_0 = v_f\,\left|\boldsymbol{q}\right|$. Imaginary parts of polarization functions are odd functions of $q_0$. Despite the fact that we have chosen a (relatively) large value of $\boldsymbol{q}$ the HDL approximation works very well for electrons, but fails for protons. Analogously to the real parts, the large proton mass leads to a sizable reduction of the magnitude of $\textrm{Im}\,\Pi_{00}$ in the vicinity of $q_0 = v_f \left|\boldsymbol{q}\right|$. Note further that the imaginary parts of $\Pi_{00}$ and $\Pi_{\perp}$  in the presence of protons do not go to zero at  $q_0 = v_f\,\left|\boldsymbol{q}\right|$ [as predicted by \ref{eq:ImP00HDL} and \ref{eq:ImGHDL}] but at a larger value. The correct domain boundaries are worked out in the top right digram of Fig.\ref{fig:Domains}. Using the full RPA results, one finds that the transition to the Landau-damped region is always smooth though very rapid, in particular in the electron case. This phenomenon is further investigated in the bottom two figures: The solutions to \ref{eq:wL} are plotted over the magnitude of the imaginary part in an electron plasma (left) and proton plasma (right), where darker colors indicate larger (negative) values. Dashed and dot-dashed lines correspond to $q_0=v_f\,\left|\boldsymbol{q}\right|$ and $q_0=c v_f\,\left|\boldsymbol{q}\right|$ respectively. In a region where $q_0$  is slightly smaller than  $v_f\,\left|\boldsymbol{q}\right|$, excitations are maximally damped. The position where the undamped mode $\omega_L$ ``dives" into the Landau damped region and loses energy rapidly defines a cutoff momentum $\left|\boldsymbol{q}\right|_\textrm{max}$. Gapless solutions $\omega_{<}$ including the ``tip of the thumb" (the position where $\partial \omega / \partial \left|\boldsymbol{q}\right|$ diverges) are confined to the Landau damped region and hence should not be regarded as actual collective modes. The plasmon mode shares the same fate in a plasma comprised of electrons or protons, but due to the small mass of the electrons the transition to the Landau damped region occurs at much larger momenta.}      
\end{figure}
\subsubsection{Spectral functions}
\label{subsubsec:IPe}
\noindent The spectral functions \ref{eq:Spectral} are obtained from the \textit{resummed retarded propagators} in the limit $\lim_{\epsilon\rightarrow 0}\,\textrm{Im}\,R(q_0+i\epsilon)$ and $\lim_{\epsilon\rightarrow 0}\,\textrm{Im}\,S(q_0+i\epsilon)$. In regions where the photon polarization exhibits no imaginary part the limit is somewhat subtle and requires the expansion  
\be 
\Pi(q_0+i\epsilon)=\Pi(q_0)+i\epsilon\,\partial\,\Pi(q_0)/\partial q_0-(\epsilon^2/2)\,\partial^2\,\Pi(q_0)/\partial q_0^2+\mathcal{O}(\epsilon^3)\,, 
\ee 
where $\Pi$ is $\Pi^{00}$ or $G$. Using the limit expression $\pi\,\delta (x)=\lim_{\epsilon\rightarrow  0}\,\epsilon^2/(x^2+\epsilon^2)$, one obtains the final results
\vspace{1mm}
\bea
\rho_{L}(q) & = & -\frac{1}{\pi}\frac{\text{Im}\,\text{\ensuremath{\Pi}}^{00}}{(\text{Re}\,\Pi^{00} - \boldsymbol{q}^{2})^{2}+(\text{Im}\,\Pi^{00} )^{2}} + \textrm{sgn}(q_0/\left|\boldsymbol{q}\right|)\delta\left. \left(\text{Re}\,\Pi^{00} -\boldsymbol{q}^2 \right) \right|_{\omega_{L}} \,,  \label{SpecL}\\[3ex]
\rho_{\perp}(q) & = & -\frac{1}{\pi}\frac{\text{Im}\,G }{(\text{Re}\,G -q^{2})^{2}+(\text{Im}\,G )^{2}} + \textrm{sgn}(q_0/\left|\boldsymbol{q}\right|)\delta\left(\text{Re}\,G-q^2\right)\,.\label{SpecPerp}
\eea

\begin{figure}
\begin{centering}
\includegraphics[scale=0.55]{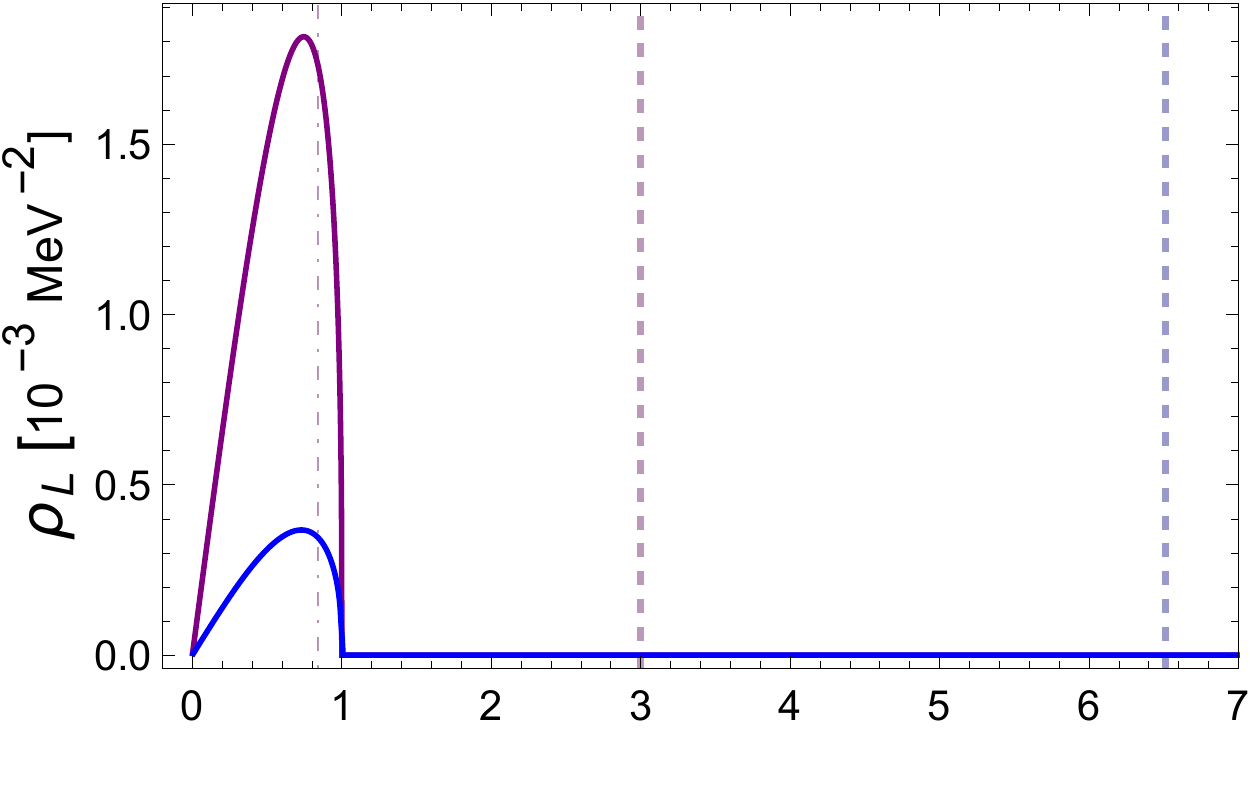}~~~~~~~~~\hspace{0.2cm}\includegraphics[scale=0.54]{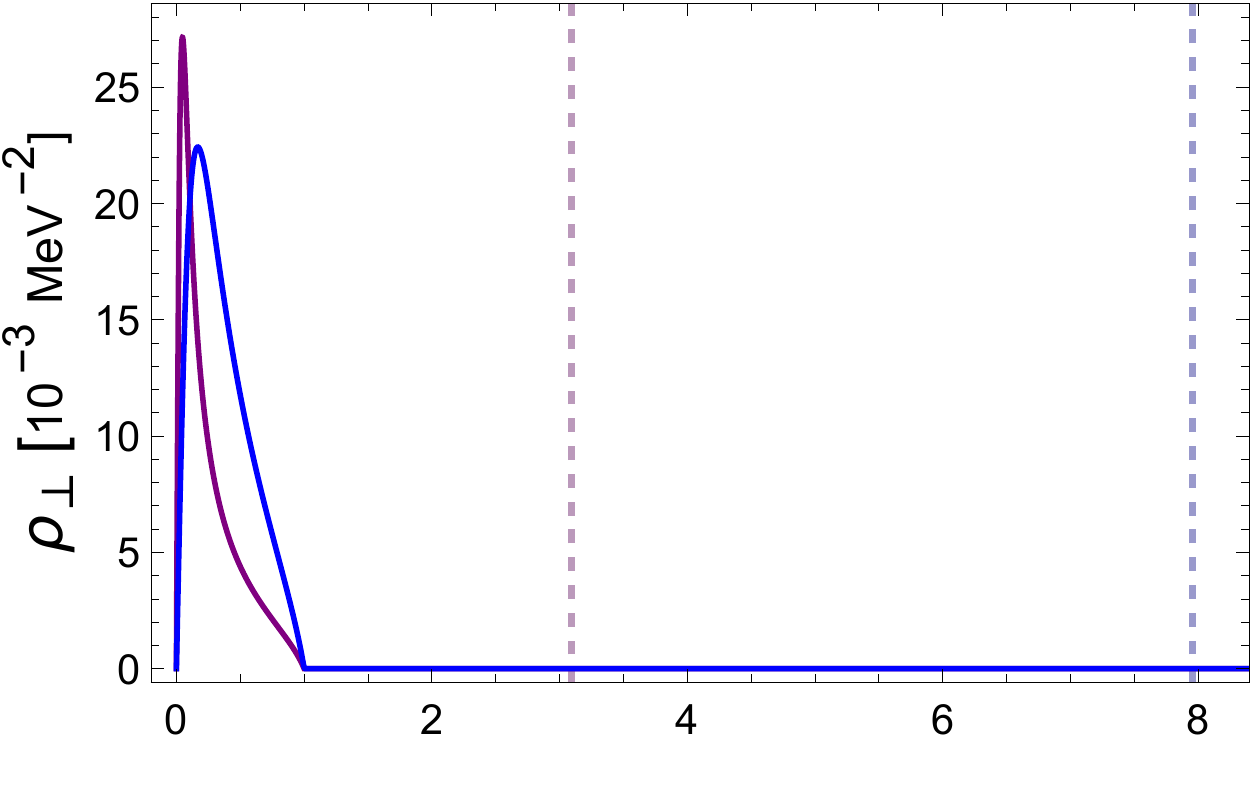}
~
\includegraphics[scale=0.56]{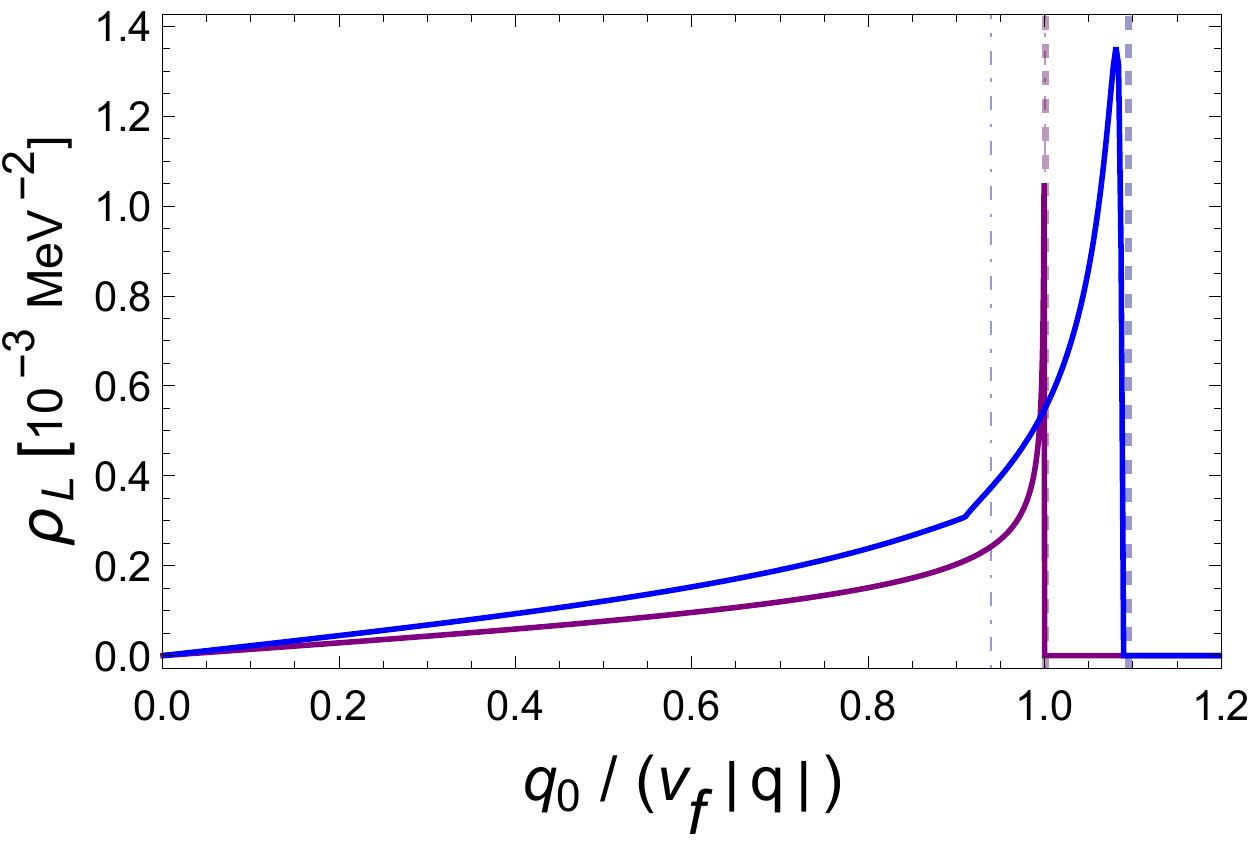}~~~~~~~~\includegraphics[scale=0.57]{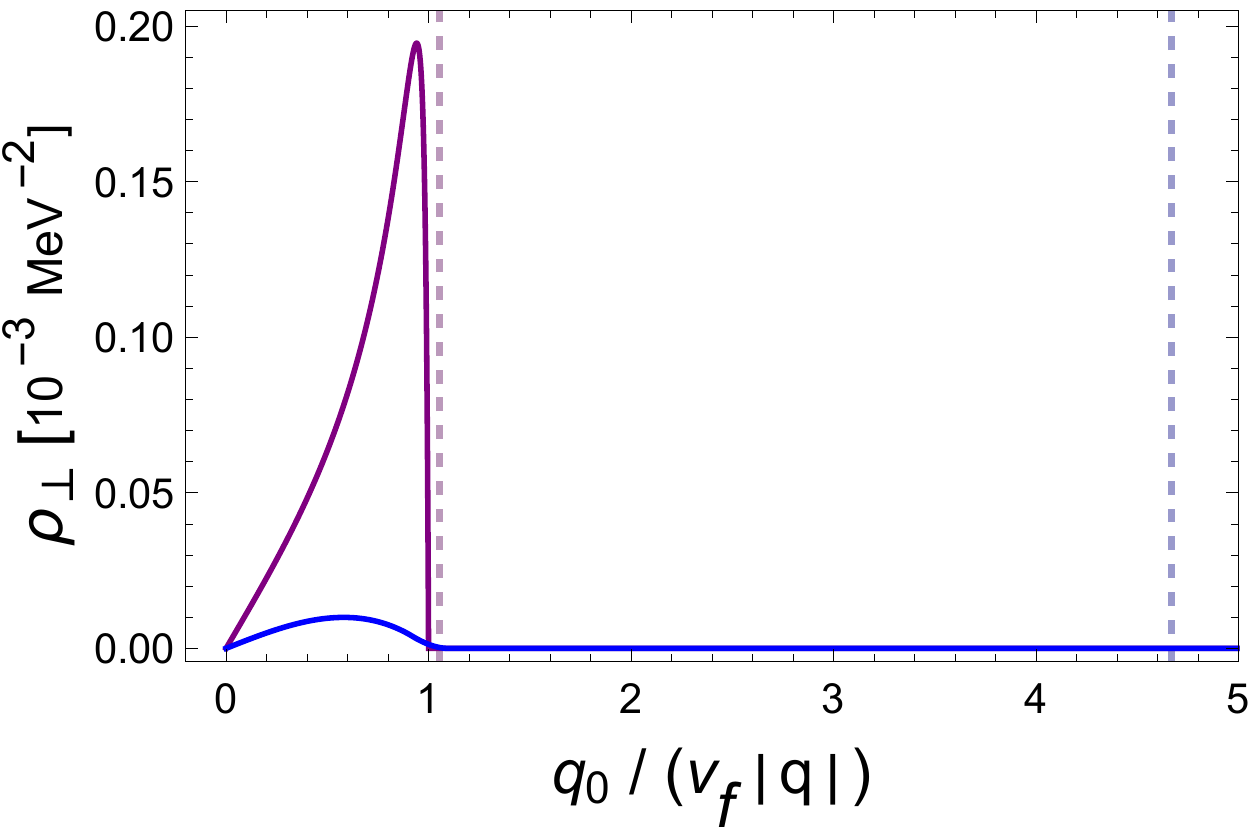}
\end{centering}
\caption{\label{fig:rhoEP} 
\c Longitudinal (left) and transverse (right) low energy photon spectrum in a plasma composed of electrons (purple) and protons (blue). Same parameters as in Fig. \ref{fig:RealPi}. Horizontal axis are normalized over $v_f\,\left|\boldsymbol{q}\right|$, where $v_f$ is the respective Fermi velocity of each particle species. Dashed lines correspond to the positions of the collective modes $\omega_L$ and $\omega_\perp$; dot-dashed lines correspond to positions of the gap-less solutions $\omega_{<}$. The value of $\left|\boldsymbol{q}\right|$ is set to $0.2\,m_{D,\,e}\,\sim\,2.4$ MeV (top) and $0.9\,m_{D,\,p}\,\sim\,$ 23.8 MeV (bottom). These positions are indicated by thin horizontal lines in Fig. \ref{fig:SingleModes} and the spectral functions are plotted along those lines. At sufficiently low momenta the solutions $\omega_{<}$ in the electron and proton cases are both located at a value of roughly $q_0 = 0.83\,v_f\,\left|\boldsymbol{q}\right|$ and hence appear on top of each other in the top right plot. Transverse spectral functions are roughly an order of magnitude larger than longitudinal ones. At higher momenta (second line) the plasmon modes $\omega_L$ are located just outside the Landau-damped region. Note further that the longitudinal spectrum in the proton plasma appears unusually large at the edge. This is due to the fact that the magnitude of the imaginary part is reduced in the vicinity of $\omega_{<}$ at larger momenta (see Fig. \ref{fig:ImPi}, bottom right). If the plasma is charge neutralized by an additional particle species Landau damping due to the other constituent takes over and reduces the magnitude of the spectrum in this region. } 
\end{figure}

\begin{figure}
\begin{centering}
\includegraphics[scale=0.6]{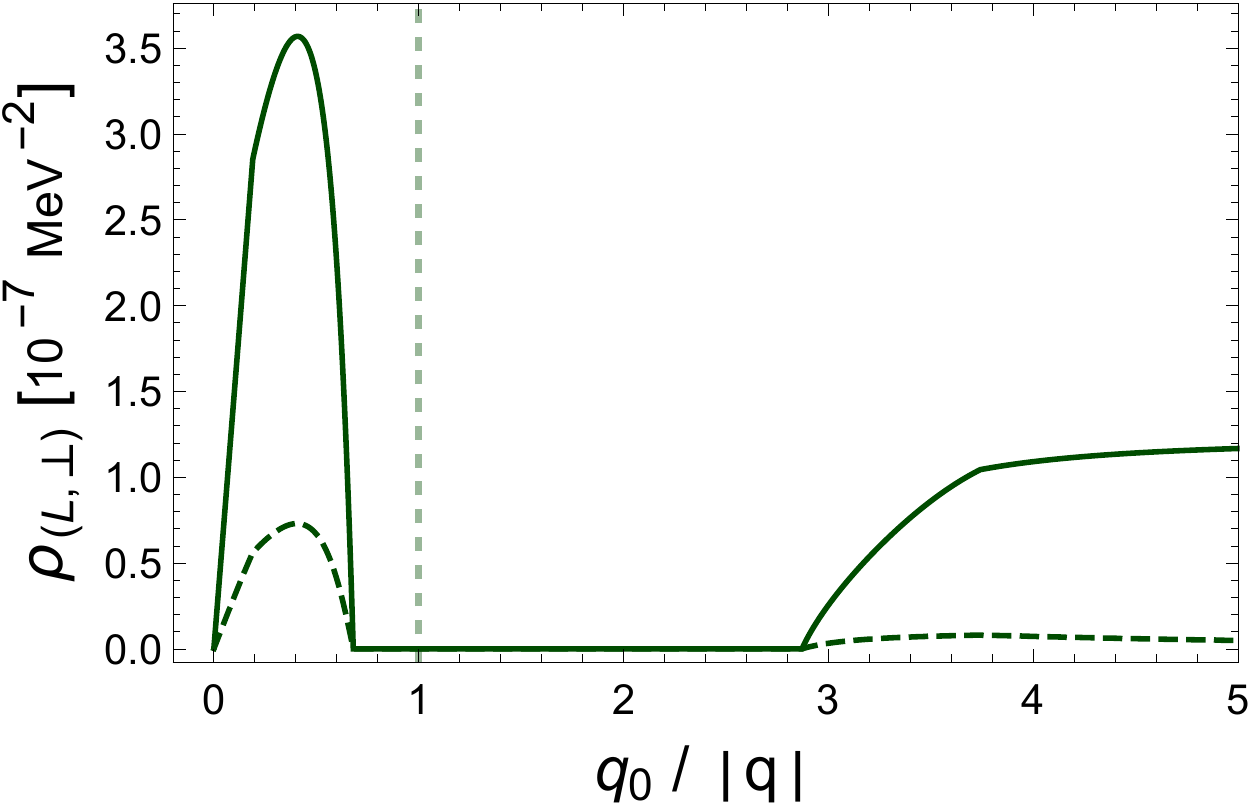}~~~~~~\includegraphics[scale=0.6]{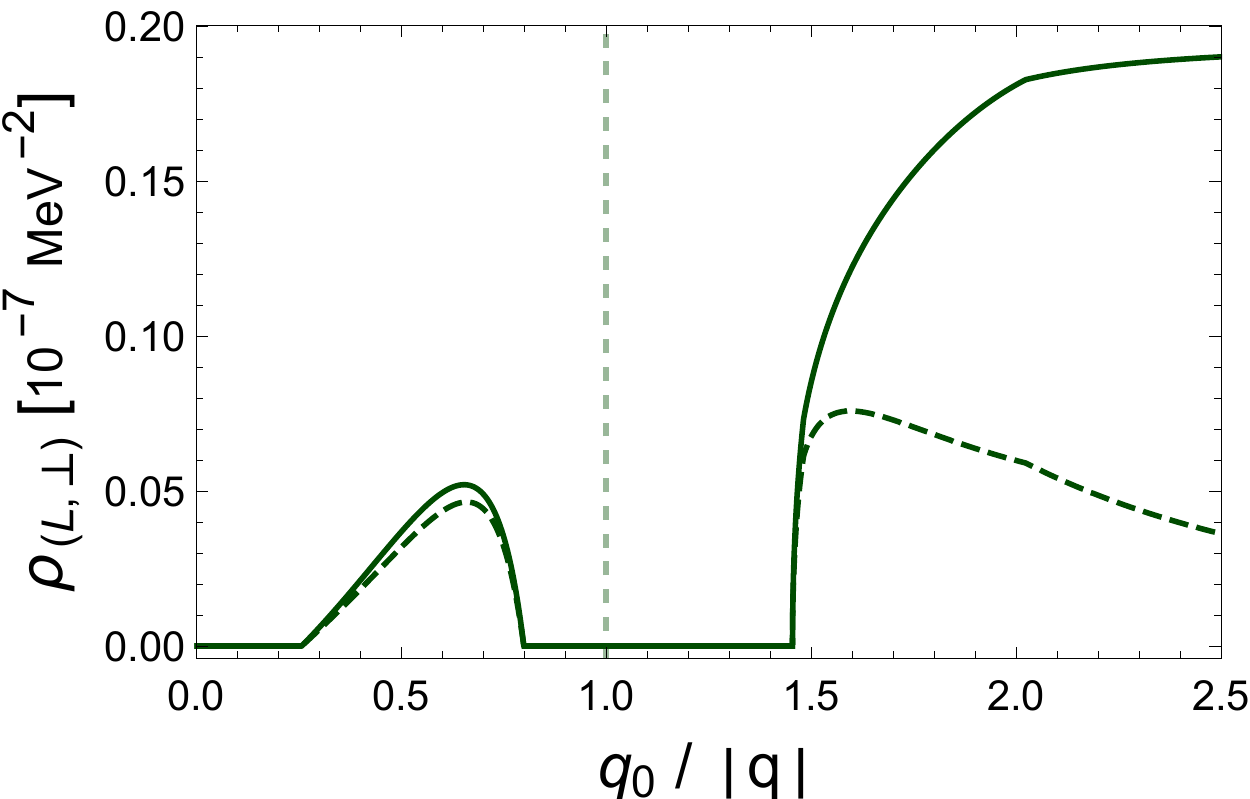}
\par\end{centering}
\caption{\label{fig:rhoMu} 
\c Longitudinal (solid) and transverse (dashed) spectral functions in a muon plasma including regions subject to Landau damping and pair creation. Identical parameters as in Fig. \ref{fig:Domains}: the muon chemical potential is set to $\mu_\mu = 122 MeV$ and the momenta to $ \left|\boldsymbol{q}\right|$ = 80 MeV (left) and 200 MeV (right). At these momenta, plasmon modes are longer present in the spectrum, dashed vertical lines correspond to the position of the (transverse) photon modes located at the light cone in the dissipation free region. With increasing momenta $\left|\boldsymbol{q}\right|$, the spectrum due to pair creation outgrows the spectrum due to Landau damping. Since the chosen values for $\left|\boldsymbol{q}\right|$ are rather large, the corresponding spectral functions are relatively tiny. 
} 
\end{figure}

\vspace{1mm}
\noindent While the first term picks up contributions from Landau damping and pair creation, the second indicates the position of the poles of the resummed propagators corresponding to the collective modes located in the dissipation free region. Therefore, only the pole $\omega_{L}$  develops a (delta) peak in the longitudinal spectrum  while the pole $\omega_{<}$ is located somewhere in the bulk of the spectrum, usually in an area where Landau damping reaches its maximum. Longitudinal and transverse spectral functions are plotted in Fig. \ref{fig:rhoEP}, and including pair creation in Fig. \ref{fig:rhoMu}. For small momenta $\left|\boldsymbol{q}\right|$ transverse spectral functions are considerably larger than longitudinal ones. This is due to lack of static screening for the transverse component \cite{Heiselberg:1992ha} \cite{Heiselberg:1993cr}. For large values of $\left|\boldsymbol{q}\right|$, the situation is reversed.     
\newpage
\newpage
\section{RPA in dense nuclear matter}
\label{sec:RPAnucl}

\noindent Two modifications are necessary to adapt the results of previous sections to the physical environment of dense homogeneous nuclear matter: On one hand, the RPA resummation technique (which describes the long range collective response of the system) has to be generalized to accommodate multiple particle species; on the other hand, medium modifications of the nucleon quasi-particles (due to short-range strong interactions) have to be consistently incorporated. The latter is discussed in Subsec. \ref{subsec:Fermi} (the results of which have at least partially been employed already, e.g., by using effective masses and chemical potentials of protons at saturation density), the latter in Subsecs. \ref{subsec:EMP} and \ref{subsec:EMPN}. While Subsec. \ref{subsec:EMP} discusses a pure QED plasma comprised of electrons, muons and protons Subsec. \ref{subsec:EMPN} additionally introduces  strong interactions whereby the neutron enters the RPA resummation. Electromagnetic interactions of the neutron in vacuum arising from its small magnetic moment are ignored. The results presented here are relevant to the low-energy photon spectrum; at high energies the effects of the nuclear plasma can be expected to be small compared to QED contribution from pair creation processes. When the photon energy is close to being resonant with vector mesons such as $\omega$ and $\rho$, the photon can mix with these excitations in the nuclear medium, but these effects are beyond the scope of our current study.

\subsection{Properties of nuclear matter from Fermi liquid theory}
\label{subsec:Fermi}

\noindent Under neutron star conditions nucleons form a strongly interacting Fermi liquid. We are interested in the quasiparticles properties, in particular the residual particle-hole interactions.  As starting point we employ a Landau energy functional based on Skyrme type interactions taken from Ref. \cite{Chamel:2006rc}, derived under the assumption of spin-averaged, homogeneous nuclear matter 

\begin{equation}
\mathcal{E}[n]=\sum_{T=0,1}\left[\delta_{T,0}\frac{\hbar^{2}}{2m}\tau_{T}+C_{T}^{n}[n]\,n_{T}^{2}+C_{T}^{\tau}n_{T}\tau_{T}+C_{T}^{\boldsymbol{j}}\,\boldsymbol{j}_{T}^{2}\right]\,.\label{eq:EpsChamel}
\end{equation}

\noindent The coefficients $C_{T}^{ {n,\tau,\boldsymbol{j} } }$
are related to standard Skyrme parameters \cite{Chamel:2006rc}. The above functional depends on densities, kinetic energies, and currents, which in turn depend on the quasiparticle occupations via
\begin{equation}\label{eq:functionals}
n_{a}[n_{\boldsymbol{k,}a}]=\int_{k}n_{\boldsymbol{k},a}\,,\hspace{1cm}\tau_{a}[n_{\boldsymbol{k,}a}]=\int_{k}\boldsymbol{k}^{2}n_{\boldsymbol{k},a}\,,\hspace{1cm}\boldsymbol{j}_{a}[n_{\boldsymbol{k,}a}]=\int_{k}\boldsymbol{k}\,n_{\boldsymbol{k},a}\,.
\end{equation}
\noindent with the usual definition $\int_{k}=\int d^{3}\boldsymbol{k}/(2\pi)^{3}$ and flavor index $a = n, p$ (the flavor labels $a$ and $b$ are chosen to avoid confusion with the spatial indices $i$ and $j$). Isoscalar (T=0) and isovector (T=1) densities  are $n_{0}=n=n_{n}+n_{p}$, $n_{1}=n_{n}-n_{p}$ (and similar for $\boldsymbol{j}$ an $\tau$). The last term in \ref{eq:EpsChamel} contains a $\boldsymbol{k}\cdot\boldsymbol{k}^{\prime}\propto\text{cos}\theta$
dependence and consequently corresponds to a $l=1$ contribution ($l$ being the angular momentum) in a partial wave expansion. Quasiparticle dispersions, effective masses, and (density and current) interactions are obtained by taking the derivatives 
\be \label{eq:derivatives}
e_{\boldsymbol{k},a}  =  \frac{\delta\mathcal{E}}{\delta n_{a,\boldsymbol{k}}} = \frac{\hbar^{2}\boldsymbol{k}^{2}}{2m_{a}^{*}}+U_{a}\,,\hspace{1cm}\frac{\hbar^{2}}{2m_{a}^{*}}  :=  \frac{\delta\mathcal{E}}{\delta\tau_{a}}\,,\hspace{1cm}f_{ab}=\frac{\delta^{2}\mathcal{E}}{\delta n_{a,\boldsymbol{k}}\,\delta n_{b,\boldsymbol{k}}}\,,\hspace{1cm}\bar{f}_{ab} \,\delta_{ij} = \frac{\delta^{2}\mathcal{E}}{\delta j_{a}^{i}\,\delta j_{b}^{j}\,.}
\ee
Derivatives with respect to $n_{a,\boldsymbol{k}}$ are \textit{functional} derivatives\footnote{Note that a volume element in momentum space is left over after each functional derivation; i.e., to be precise, one has $\frac{\delta n_{a}}{\delta n_{\boldsymbol{k},b}}=V\,\delta_{ab}$. Ultimately, we are interested in the averaged energies $e_{\boldsymbol{k},a}/V$ and potentials  $f_{a\boldsymbol{k} \boldsymbol{k}^{\prime}}/V^{2}$ and therefore drop factors of $V$ in what follows.}, see Appendix \ref{sub:Skyrme} for further details. Strictly speaking, the interaction potentials obtained via relations \ref{eq:derivatives} are obtained in the static limit and only valid for $f_{ab}(\boldsymbol{q}=\boldsymbol{0})$ and $\bar{f}_{ab}(\boldsymbol{q}=\boldsymbol{0})$, where $\boldsymbol{q} = \boldsymbol{k}-\boldsymbol{k^\prime}$ is the momentum transfer in the scattering of two quasi-particles. Corrections of the order $\boldsymbol{q}^2$ cannot be extracted from functional \ref{eq:EpsChamel}. While this is certainly the biggest setback it is still a reasonable approximation if one is interested in the qualitative impact on the photon spectrum at sufficiently low  momenta. In a partial wave expansion, the static density-density interactions correspond to $l=0$ terms, while current-current interactions as well as effective masses correspond to $l=1$ terms. Note further that in their functional form \ref{eq:functionals} the currents are defined as $\boldsymbol{j}_{a}\sim n_{a}\boldsymbol{p}_{a}$, where $\boldsymbol{p}_{a}$ is the macroscopic momentum. From a standard kinetic definition, one would expect $\boldsymbol{j}_{a}\sim n_{a}\boldsymbol{v}_{a}$ with $\boldsymbol{v_{a}}=\boldsymbol{p}_{a}/\left|\boldsymbol{p}_{a}\right|$. To obtain the proper current-current interactions required for the RPA  $\left(\bar{f}_{ab}\right)^{ij}$ have hence to be multiplied by $k_{f,\,a} \, k_{f,\,b}$. The result is then equivalent to the $l=1$ contribution of the density-density interactions; see Appendix \ref{sub:Skyrme}. In the ground state, i.e., at $\boldsymbol{j}_{i}=\boldsymbol{0}$ and $n_{a,\boldsymbol{k}}=\Theta(k_{f,a}-\left|\boldsymbol{k}\right|)$, energy functional \ref{eq:EpsChamel} evaluates to (after adding the rest mass $m_0$)
\begin{eqnarray}\label{eq:Eps0}
\mathcal{E}_{0} & = & nm_{0}+\frac{1}{2m}\tau_{0}+C_{0}^{\tau}n\tau_{0}+C_{1}^{\tau}n_{1}\tau_{1}+C_{0}^{n}n^{2}+C_{1}^{n}n_{1}^{2}\label{eq:EnergyGround}\\[2ex]
 & = & nm_{0}+\mathcal{E}_{\text{kin}}+\mathcal{E}_{\text{int}}=nm_{0}+\frac{E}{A}\cdot n\,.\nonumber 
\end{eqnarray} 
The result can be expressed in terms of the energy per particle E/A and is therefore directly related to the equation of state. The kinetic terms evaluated in the ground state become
\be
\tau_{a}=\frac{3}{5}(3\pi^{2})^{2/3}n_{a}^{5/3}\,,\hspace{1cm}\tau_{0}=\tau_{p}+\tau_{n}\,,\hspace{1cm}\tau_{1}=\tau_{n}-\tau_{p}\,.
\ee
Energy functional \ref{eq:EpsChamel} is obviously defined non relativistically and so are all quantities derived from it. To incorporate them into a fully relativistic RPA calculation one has to perform a careful matching. A good starting point is to consider the energy density itself. In a fully relativistic setup, the kinetic contributions reads
\bea
\mathcal{E}_{\text{kin,\,rel}} & = & \frac{1}{\pi^{2}}\int_{0}^{k_{f}}d\boldsymbol{k}\,\boldsymbol{k}^{2}\,\sqrt{\boldsymbol{k}^{2}+m^{2}}=\frac{1}{8\pi^{2}}\left[(2k_{f}^{3}+m^{2}k_{f})\sqrt{k_{f}^{2}+m^{2}}-m^{4}\,\text{ln}\left(k_{f}/m+\sqrt{1+k_{f}^{2}/m^{2}}\right)\right]\label{eq:EpsRel}\\[2ex]
 & \simeq & m\,\frac{1}{3\pi^{2}}\,k_{f}^{3}+\frac{1}{m}\frac{1}{10\pi^{2}}\,k_{f}^{5}+\mathcal{O}(m^{-3})=mn + \frac{1}{10\,m}(3\pi^{2})^{2/3}n^{5/3}+\mathcal{O}(m^{-3})
\eea
which to lowest order in a large $m$ expansion precisely reproduces the kinetic pieces of expression \ref{eq:EnergyGround}.  Relativistic corrections to the interaction energy $\mathcal{E}_{\text{int}}$ are at least partially included in the fits of the Skyrme parameters. While it is not entirely consistent to replace the kinetic contribution by its fully relativistic counterpart it should still be a fair approximation as relativistic corrections to the interaction energy are certainly smaller than those to the kinetic energy. In a free Fermi gas, chemical potentials and Debye masses of the fermions are consequently given by 
\be 
\frac{\partial\mathcal{E}_{\text{kin,\,rel}}}{\partial n}=\sqrt{k_{f}^{2}+m^{2}}:=\mu_{rel}\,,\hspace{1cm} \frac{\partial^{2}\mathcal{E}_{\text{kin}}}{\partial n^{2}}=\frac{\pi^{2}}{k_{f}\,\mu_{\,rel}}=\frac{1}{m_{D,\,rel}^{\prime\,2}}\,.
\ee
To indicate that in contrast to its usual definition no factor of $e^2$ is included in the Debye mass, it has been denoted by $m^{\prime\,2}_D$. Single-particle dispersions and energies in the \textit{interacting} case are renormalized by  
\begin{equation}
e_{\boldsymbol{k},\,rel}=\sqrt{\boldsymbol{k}^{2}+m^{*}}+m-m^{*}\,,\hspace{1cm}\mu^{*}_{rel}=\sqrt{k_{f}^{2}+m^{*2}}+m-m^{*}\,,\label{eq:EpsMuRPA}
\end{equation}
to ensure matching with the non-relativistic results \ref{eq:derivatives} in a large $m^*$ expansion (to leading order one expects to find the non relativistic result plus the rest mass rather than the effective mass). Note further that the single-particle potentials $U$ present in \ref{eq:derivatives} have been subtracted. While those potentials are important for the determination of the ground-state properties of nuclear matter (e.g., number densities in $\beta$ equilibrium; see below and in appendix \ref{sub:Skyrme}), they should not be added to the dispersions used in QED loop calculations where fermions are essentially treated as free particles (with effective masses). In thermal distribution functions $n_{f}(e_{\boldsymbol{k}}-\mu)$, the single-particle potentials drop out automatically. In the following, we shall drop the label ``rel" again and generally assume that relativistic quantities are used. \newline      
\subsubsection{Stability and $\beta$ equilibrium}
\label{subsubsec:BetaEq}
\noindent In general, stable homogeneous matter requires a positive curvature of the energy density in a space spanned by $n_{n}$ and $n_{p}$,
\begin{equation}
\frac{\partial^{2}\mathcal{E}_{0}}{\partial n_{n}^{2}}\cdot\frac{\partial^{2}\mathcal{E}_{0}}{\partial n_{p}^{2}}-\frac{\partial^{2}\mathcal{E}_{0}}{\partial n_{n}\partial n_{p}}>0\,.\label{eq:Stability}
\end{equation}
This condition fails below a critical density $n_c$  which marks the onset of the so-called spinodal instability \cite{Baym:1971ax} \cite{Muller:1995ji} \cite{Li:1997ra},  indicating that nuclear matter strives to be in a clustered state. In the models applied here local stability of homogeneous nuclear matter is achieved at densities above 0.6$\,n_{0}$. The screening mass being a second derivative of the energy density is sensible to the instability and diverges upon approaching $n=n_c$ from above. For completeness, it should be mentioned that the above argument involves only strong interactions; the inclusion of electromagnetism turns the first-order phase transition at the spinodal point into a second order phase transition. The screening, however, will in any case diverge at the transition point. There are three regions to be distinguished:
\begin{enumerate}
\item $n<n_{c}$: homogeneous nuclear matter is unstable. 
\item $n_{c}<n<n_{c,\,\mu}$: homogeneous nuclear matter composed of electrons, protons and neutrons under the constraints of $\beta$ equilibrium and charge neutrality, i.e.,
$\mu_{n}-\mu_{p}=\mu_{e}$ and $n_{e}=n_{p}$.
\item  $n_{c,\,\mu}<n$: homogeneous nuclear matter composed of electrons, muons, protons, and neutrons. The critical density for muon onset $n_{c,\,\mu}$ is obtained from $\beta$ equilibrium
under the additional constraints of $n_{\mu}=0$ and $\mu_{e}=m_{\mu}$.
$\beta$ equilibrium and charge neutrality are enforced by $\mu_{n}-\mu_{p}=\mu_{e}=\mu_{\mu}$
and $n_{e}+n_{\mu}=n_{p}$.
\end{enumerate}
As an immediate consequence of the relativistic modifications introduced in the last section, the proton fractions are reduced with increasing densities. This can be seen from the leading relativistic correction to the $\beta$ equilibrium condition 
\begin{equation} \label{eq:protonRel}
\mu_{e}=\mu_{n}-\mu_{p}\sim m_{n}-m_{p}+\frac{k_{f,n}^{2}}{2m_{n}^{*}}\left(1-\frac{k_{f,n}}{4m_{n}^{*}}\right)-\frac{k_{f,p}^{2}}{2m_{p}^{*}}\left(1-\frac{k_{f,p}}{4m_{p}^{*}}\right)+U_{n}-U_{p}+\mathcal{O}(k_{f}^{6}/m^{5})\,.
\end{equation}
In this work, the Skyrme parametrizations recommended in Ref. \cite{Dutra:2012mb} (KDE0v1, SKRA, SQMC700, LNS and NRAPR) are employed. The critical densities calculate on average to $n_c = 0.56\,n_0$ and $n_{c,\,\mu} = 0.75\,n_0$ , i.e. the density region where homogeneous matter is stable but muons are absent is very small. A comparison of relativistic and non-relativistic cases is provided in Appendix \ref{sub:Skyrme}. 
\subsubsection{Static screening in nuclear matter}
\label{subsubsec:StatScreen}
\begin{figure}
\begin{centering}
\includegraphics[scale=0.76]{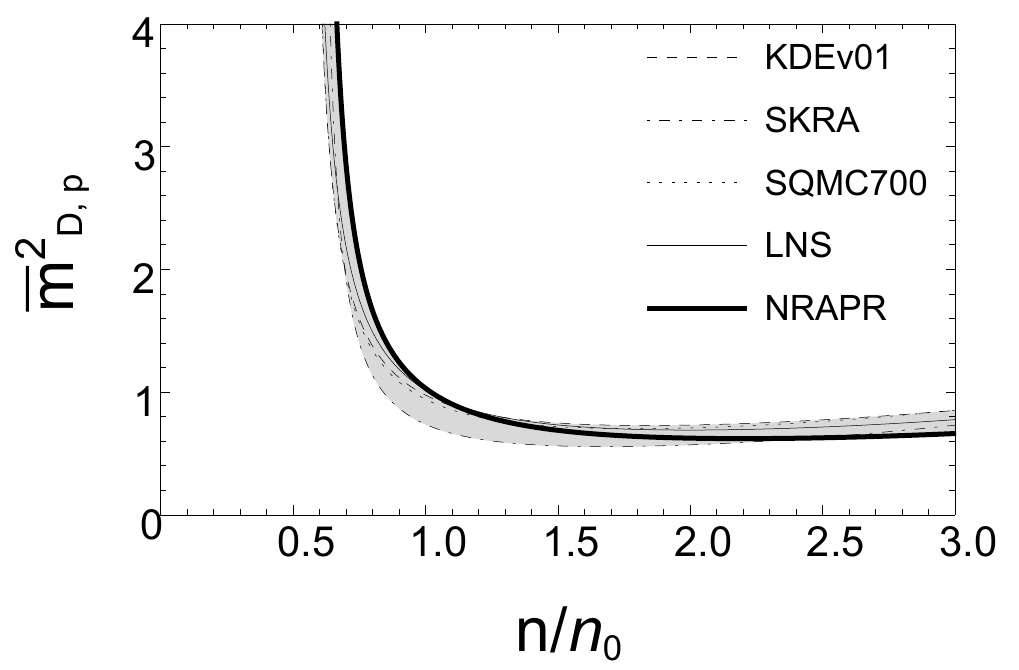}~~~~~~~\includegraphics[scale=0.58]{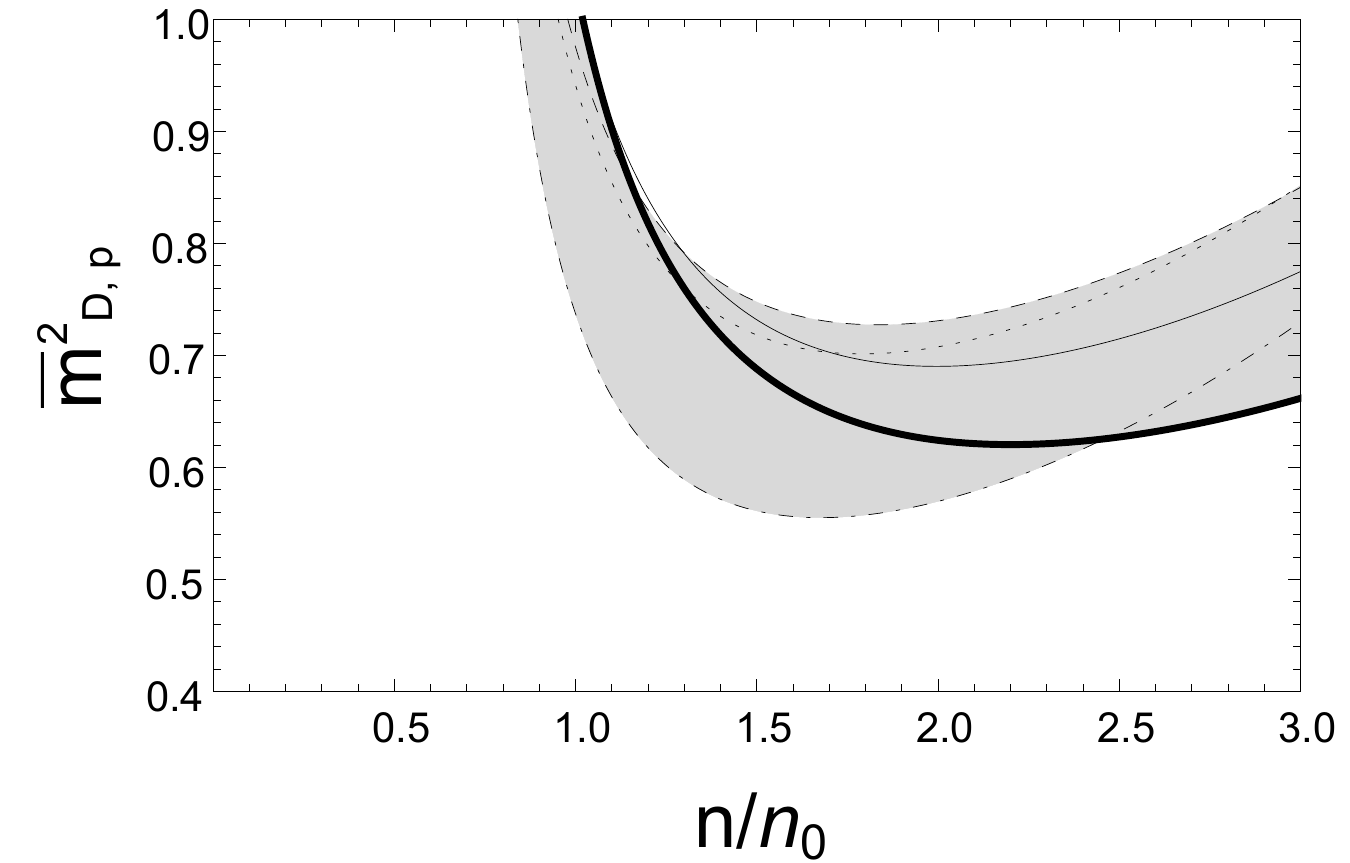}
\par\end{centering}
\caption{\label{fig:NucScreening} 
\c Static screening $\bar{m}_{D\,,p}=\tilde{m}_{D\,,p}/m_{D,\,p}^{\prime}$ in nuclear matter in $\beta$ equilibrium based on various Skyrme forces according to  Eq. \ref{eq:DebyePN}, normalized over the Debye masses in the non-interacting system. Given the considerable differences of effective masses or proton fractions at higher densities (see Appendix \ref{sub:Skyrme}) the results for the screening mass is fairly consistent. The screening diverges close to the spinodal instability and is slightly reduced at higher densities. The plot on the right-hand side displays an enlargment of the one on the left-hand side.} 
\end{figure}

\noindent A quantity of particular interest to this work is the screening mass arising due to strong interactions. Since the induced coupling to neutrons is a core aspect of this study we are especially interested in how proton-neutron interactions modify the screening. Consider for a moment that these interactions are absent. Then, according to its thermodynamic definition, the inverse screening mass resulting from proton-proton interactions calculates to
\begin{eqnarray}\label{eq:DebyePP}
\tilde{m}_{D,\,p}^{\prime\,-2} & = & \left(\frac{\partial\mu_{p}}{\partial n_{p}}\right)=\frac{\partial^{2}\mathcal{E}_{kin}}{\partial n_{p}^{2}}+\frac{\partial^{2}\mathcal{E}_{int}}{\partial n_{p}^{2}}=\frac{1}{m_{D,p}^{\prime\,2}}+V_{pp}\label{eq:ScrenningP}\\[2ex]
\tilde{m}_{D,\,p}^{\prime\,2} & = & \frac{m_{D,p}^{\prime\,2}}{1+m_{D,p}^{\prime\,2}\,V_{pp}}\,,
\end{eqnarray}
with the usual Debye mass of the non-interacting system. The potential $V_{pp}$ is obtained as second (standard) derivative of the (interaction part of the) energy density \ref{eq:EnergyGround} in the ground state with respect to $n_p$. It should not be confused with the actual quasiparticle potentials $f_{ab}$ which are obtained as (functional) derivative of energy functional \ref{eq:EpsChamel} which describes the ground state as well as excited states. Both expressions differ by $l=1$ contributions and this difference is consequently expressible in terms of the effective chemical potentials (or effective masses in the non relativistic case)
\be 
f_{aa}=V_{aa}+\frac{\pi^{2}\hbar^{2}}{k_{f,a}}\left(\frac{1}{\mu_{a}}-\frac{1}{\mu_{a}^{*}}\right)\,,\hspace{1cm}f_{pn}=V_{pn}\,.\label{eq:VFrelation}
\ee
Equation \ref{eq:DebyePP} can alternatively be obtained from the static limit of the resummed polarization tensor $\tilde{m}_D=-\tilde{\Pi}_{00}(q_0=0)$ [see Eq. \ref{eq:DebyeCoulomb} with the Coulomb interaction replaced by $V_{pp}$] and is easy to interpret: Attractive interactions $V_{pp}<0$ increase the screening while repulsive interactions reduce it.  Now switch on interactions with neutrons: $\beta$ equilibrium links the chemical potential of protons with those of neutrons $\mu_{p}=\mu_{p}(\mu_{n})$ and the calculation of the (resummed) static screening requires the calculation of the Jacobian $J=\left|\text{Det}(\partial\mu_{i}/\partial n_{j})\right|$
such that 
\begin{eqnarray} \label{eq:DebyePN}
\tilde{m}_{D,\,p}^{\prime\,2} & = & \frac{1}{J}\,\frac{\partial\mu_{n}}{\partial n_{n}} = \frac{m_{D,p}^{\prime\,2}\left(1+m_{D,n}^{\prime\,2}V_{nn}\right)}{1+m_{D,n}^{\prime\,2}V_{nn}+m_{D,p}^{\prime\,2}V_{pp}+m_{D,p}^{\prime\,2}m_{D,n}^{\prime\,2}(V_{pp}V_{nn}-V_{np}^{2})}\, \\[2ex]
 & = &  \frac{m_{D,p}^{*2}\left(1+m_{D,n}^{*2}f_{nn}\right)}{1+m_{D,n}^{*2}f_{nn}+m_{D,p}^{*2}f_{pp}+m_{D,p}^{*2}m_{D,n}^{*2}(f_{pp}f_{nn}-f_{np}^{2})}\,,
\end{eqnarray}
\\
where relations \ref{eq:VFrelation} and the definition $m_{D}^{*2}=\mu^{*}k_{f}/\pi^{2}$ have been used. The density-dependence of the static nuclear screening is displayed in Fig. \ref{fig:NucScreening}. The denominator of the Jacobian is precisely the stability condition \ref{eq:Stability}. Note further that one recovers expression \ref{eq:DebyePP} when $f_{pn}$ is set to zero. Again Eq. \ref{eq:DebyePN} can alternatively be obtained from the static limit of a resummed polarization tensor similar to Eq.\ref{eq:PolResum}; see Eq. \ref{eq:Screentot} in section \ref{subsec:EMPN}. For illustrative purposes the static screening is obtained from a relativistic mean field model in Appendix \ref{sub:RMF}.
\subsection{QED with Electrons, Muons and Protons}
\label{subsec:EMP}
\begin{figure}
\begin{centering}
\includegraphics[scale=1]{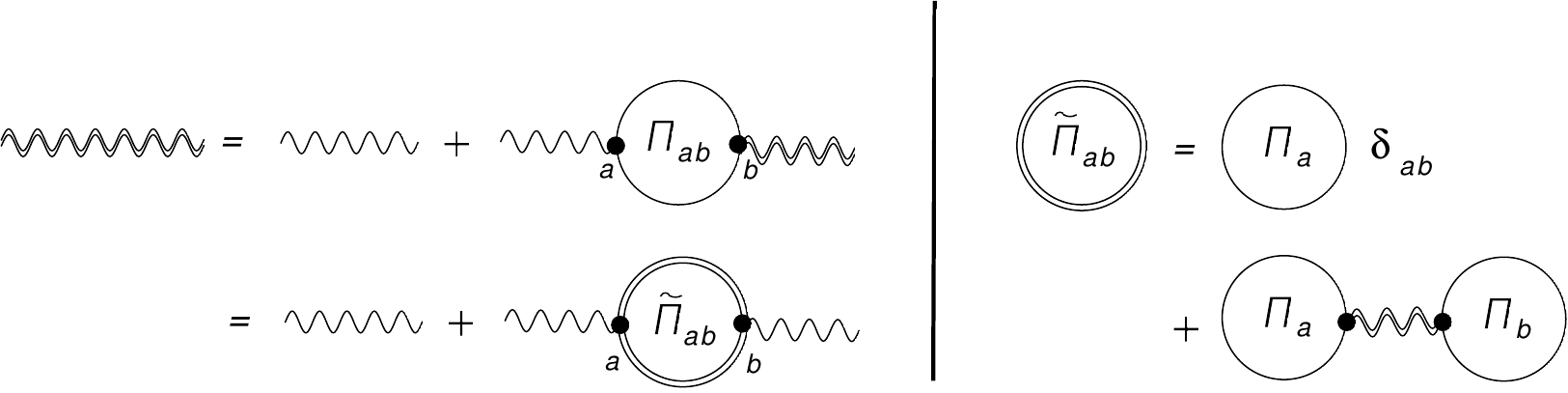}
\setlength{\belowcaptionskip}{30pt}
\caption{\label{fig:DysonMulti}Dyson equation of the photon propagator resumming contributions from all constituents of the plasma interacting via single photon exchange. Vertices carry a flavor index, and the one-loop polarization tensor is diagonal in flavor space. The photon propagator naturally remains a scalar in flavor space, and summing over the indices $a$ and $b$ simply returns the trace $\textrm{Tr}\, \Pi$. As in the single species case Fig. \ref{fig:Resum}, one may alternatively utilize the dressed polarization tensor, Eq. \ref{eq:screenEMP}. To leading order, one recovers the (diagonal) one-loop polarization tensor, and higher order terms describe how the interactions between the various species in the plasma are screened. }
\end{centering}
\end{figure}

\noindent Equipped with the results of Sec. \ref{subsec:Fermi} we extend the single-species RPA to include multiple fermion species. Multi component RPA in the context of nuclear matter is well studied; see e.g., Refs. \cite{Horowitz:1991pg}, \cite{Reddy:1998hb}. In a first approach, we consider pure electromagnetism and ignore the impact of the nuclear potentials $f_{ab}$ on the resummation, (i.e., on the \textit{collective} properties of the system). The polarization tensor is now promoted to a matrix defined in a flavor space with indices $a,\,b=\left\{ e^-,\,\mu^-,\,p^+\right\} $ 
\begin{equation}
\Pi_{ab}^{\mu\nu}=\text{diag}\{\Pi_{e}^{\mu\nu},\,\Pi_{\mu}^{\mu\nu},\,\Pi_{p}^{\mu\nu}\}\,.\label{eq:PiMulti}
\end{equation}
Similarly the generalized photon-fermion vertex reads $V_{a}^{\mu}=ie\,\gamma^{\mu}U_{a}$
where in the present case $U_{a}=(1,\,1,\,-1)$. The photon propagator remains a scalar in flavor space and its Dyson equation can be constructed with the aid of the new vertices according to
\begin{eqnarray}
\tilde{D}^{\mu\nu} & = & D^{\mu\nu}+D^{\mu\alpha}\,U_{a}^{T}\,\left(\Pi_{\alpha\beta}\right)_{ab}\,U_{b}\,\tilde{D}^{\beta\nu}\,\nonumber \\[2ex]
 & = & D^{\mu\nu}+D^{\mu\alpha}\,(\text{Tr}\,\Pi_{\alpha\beta})\,\tilde{D}^{\beta\nu}\,.\label{eq:PropMulti}
\end{eqnarray}
\noindent where the trace is taken in flavor space; see also Fig.
\ref{fig:DysonMulti}. We could have obtained this result again from the definition of the self energy which now reads $(\tilde{D}^{-1})^{\mu\nu}=(D^{-1})^{\mu\nu}+\text{Tr}\,\Pi^{\mu\nu}$.
Observe that we have neglected the neutron entirely in the RPA resummation: Eq. \ref{eq:PropMulti} includes only those fermion loops which couple directly to the photon and as a result the neutron appears neither in \ref{eq:PiMulti} nor in \ref{eq:PropMulti}. Alternatively, one may again ask for the generalization of the resummed polarization tensor Eq. \ref{eq:PolResum}, which reads
\be 
\tilde{\Pi}_{ab}^{\mu\nu}=\Pi_{ab}^{\mu\nu}+\Pi_{ac}^{\mu\alpha}\,\left(V_{\alpha\beta}\right)_{cd}\,\tilde{\Pi}_{db}^{\beta\nu}\,,\label{eq:EPMPolResum}
\ee
where the propagator has been replaced by the interaction matrix $V_{ab}$ which governs the interactions of all the constituents in the plasma. The above equation is the relativistic analog of Eq. 1 in Ref. \cite{Baldo:2008pb}, where it was used to study collective modes in homogeneous neutron star matter.  Considering Coulomb interactions, only $V_{ab}$ is given by 
\be\label{eq:EPMIntMatrix} 
V^{\mu\nu}(q)=-\frac{1}{q^{2}+i\epsilon}\left(\begin{array}{ccc}
1 & 1 & -1\\
1 & 1 & -1\\
-1 & -1 & 1
\end{array}\right)G^{\mu\nu}\,,
\ee
with the definition of $G^{\mu\nu}$ from Eq. \ref{eq:PhotonGauge}. The closed-form solutions of the resummed polarization functions now read 
\be \label{eq:FTildeMatrix}
\tilde{F}_{ab} =  (\delta_{ad}+F_{ac}V_{cd})^{-1}F_{db}\,, 
\ee
with a similar expression for $\tilde{G}_{ab}$. The explicit result \ref{eq:FTildeMatrix} can be written in the instructive form (compare it as well to Eq. 6 in Ref. \cite{Baldo:2008pb})
\be \label{eq:screenEMP}
\tilde{F}=\left(\begin{array}{ccc}
F_e + F_e \frac{1}{\Delta} F_e  & F_{e} \frac{1}{\Delta}F_{\mu} & - F_{e} \frac{1}{\Delta} F_{p}\\[2ex]
F_{\mu} \frac{1}{\Delta} F_{e} & F_\mu +F_\mu \frac{1}{\Delta} F_\mu  & -F_{\mu} \frac{1}{\Delta} F_{p}\\[2ex]
-F_{p} \frac{1}{\Delta} F_{e} & -F_{p} \frac{1}{\Delta} F_{\mu} & F_p+F_p \frac{1}{\Delta} F_p
\end{array}\right)\,,
\ee 
with the determinant 
\be 
\Delta = q^2 - F_e - F_\mu - F_p \,.
\ee
$\tilde{F}_{ab}$ tensor is obviously no longer diagonal in flavor space. It describes how electromagnetic interactions between constituents $a$ and $b$ are screened by \textit{all} constituents of the plasma. At leading order in an $1/q^2$ expansion of $\tilde{F}_{ab}$, only the diagonal terms are nonzero and contain the one-loop polarization functions. Higher order terms describe screened electromagnetic interactions between fermions of species $a$ and $b$: see, e.g., the expansion of $\tilde{F}_{ee}$
\be\label{eq:FeeExpand}
\tilde{F}_{ee}=F_e + F_e \frac{1}{q^2} F_e + F_e\,\frac{1}{q^2} \left( F_e + F_\mu +F_p \right) \frac{1}{q^2} F_e + \mathcal{O}(\frac{1}{q^6})\,.     
\ee One may again use the resummed polarization tensor to obtain the dressed photon propagator 
\be 
\tilde{D} =  D^{\mu\nu}+D^{\mu\alpha}\,U^T_a(\,\tilde{\Pi}_{\alpha\beta})_{ab}\,U_b\,D^{\beta\nu}\,.
\ee 
The relationship between the dressed photon propagator and polarization tensor is illustrated in Fig. \ref{fig:DysonMulti}. After summing over all components of  $U^T \Pi \,U$ one finds 
\be \label{eq:FempFull}
U^T \tilde{F}\, U=\frac{F_e+F_\mu+F_p}{1-(1/q^2)(F_e+F_\mu+F_p)}\,,
\ee
in complete analogy to expressions \ref{eq:FGresummed}.
\subsubsection{Collective modes, damping and spectral densities.}
\label{subsubsec:RPemp}
\begin{figure}
\begin{centering}
\includegraphics[scale=0.75]{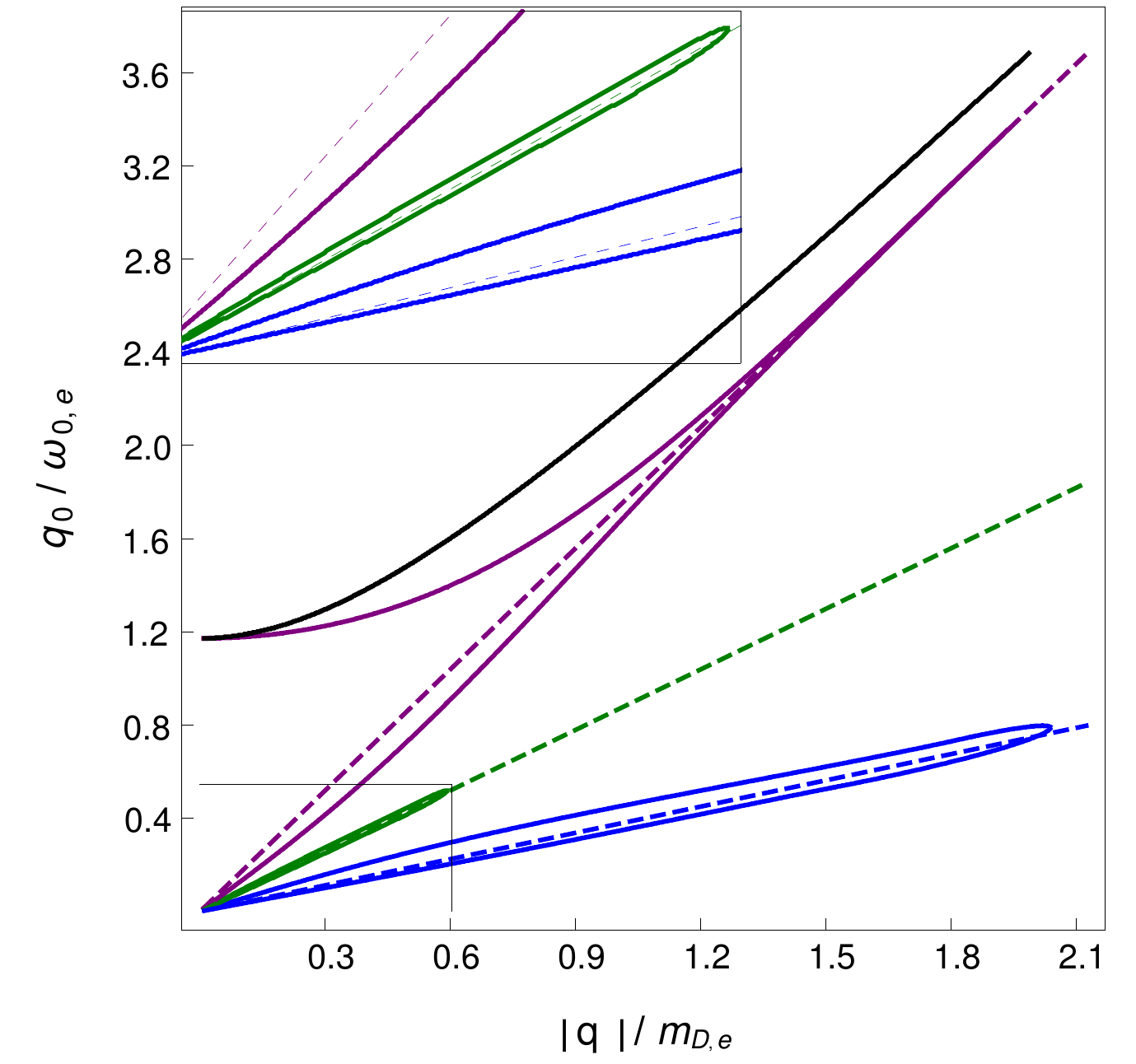}
\par\end{centering}
\caption{\label{fig:EPMmodes} 
Solutions to equations \ref{eq:wLEP} and \ref{eq:wPEP} in a plasma composed of electrons, protons, and muons at saturation density using NRAPR Skyrme forces. Electron and muon chemical potentials are $\mu_e=\mu_\mu=\,122\,\textrm{MeV}$, for protons  one finds $\mu_p=589\,\textrm{MeV}$ and $m^*_p=575\,\textrm{MeV}$.  Blue, purple and green lines correspond to longitudinal proton-like, electron-like, and muon-like solutions respectively. Dashed lines correspond to $q_0 = v_f \left| \boldsymbol{q} \right|$ for each particle species,  the black line corresponds to the transverse mode. In the collisionless limit, the various longitudinal solutions appear completely decoupled from one another. The electron-like solution is the only gapped solution in the spectrum and consequently identified with the plasmon mode of the combined electron, muon, and proton system. Muon-muon interactions are screened by electrons and proton-proton interactions are screened by electrons and muons. As a result, their respective plasmon modes are effectively reduced to sound-like modes.}      
\end{figure}
\begin{figure}
\begin{centering}
\includegraphics[scale=0.49]{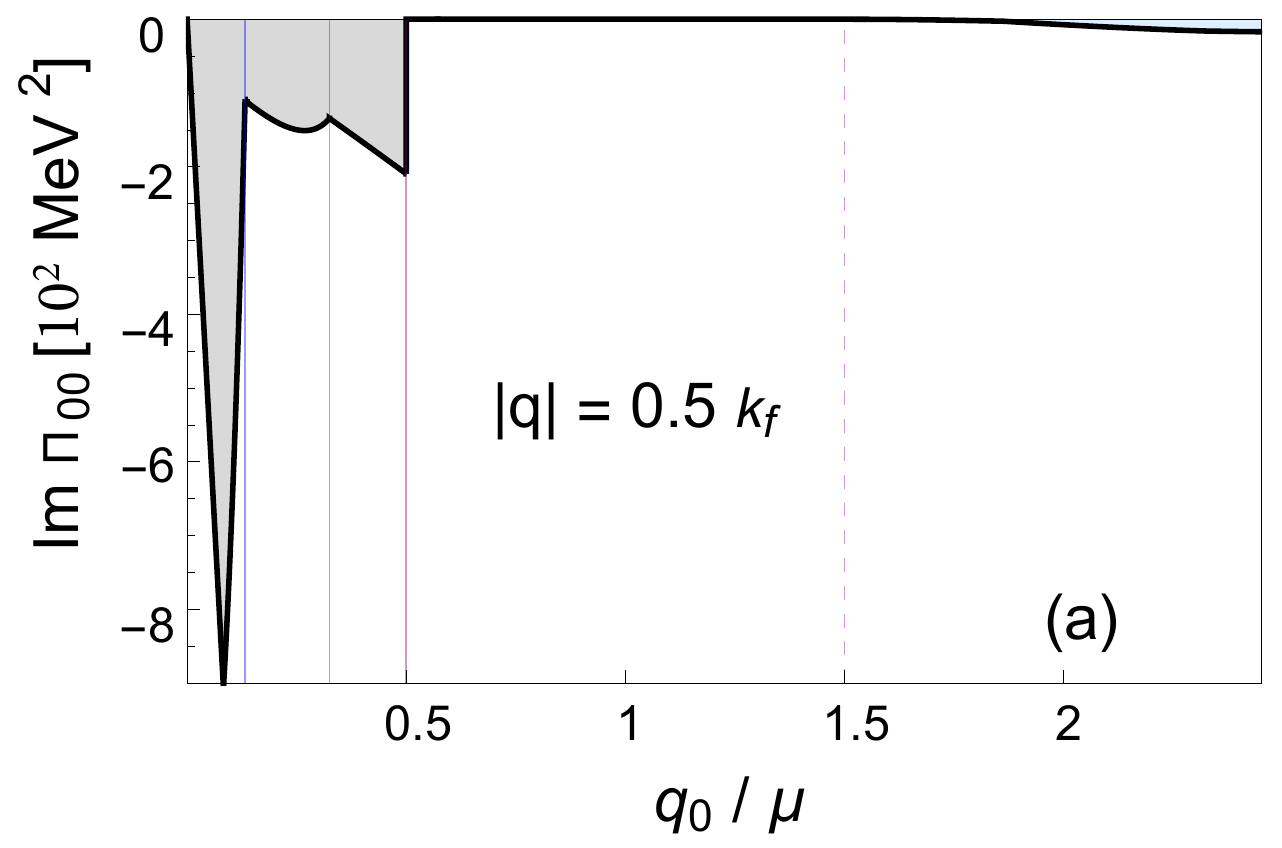}~
\includegraphics[scale=0.48]{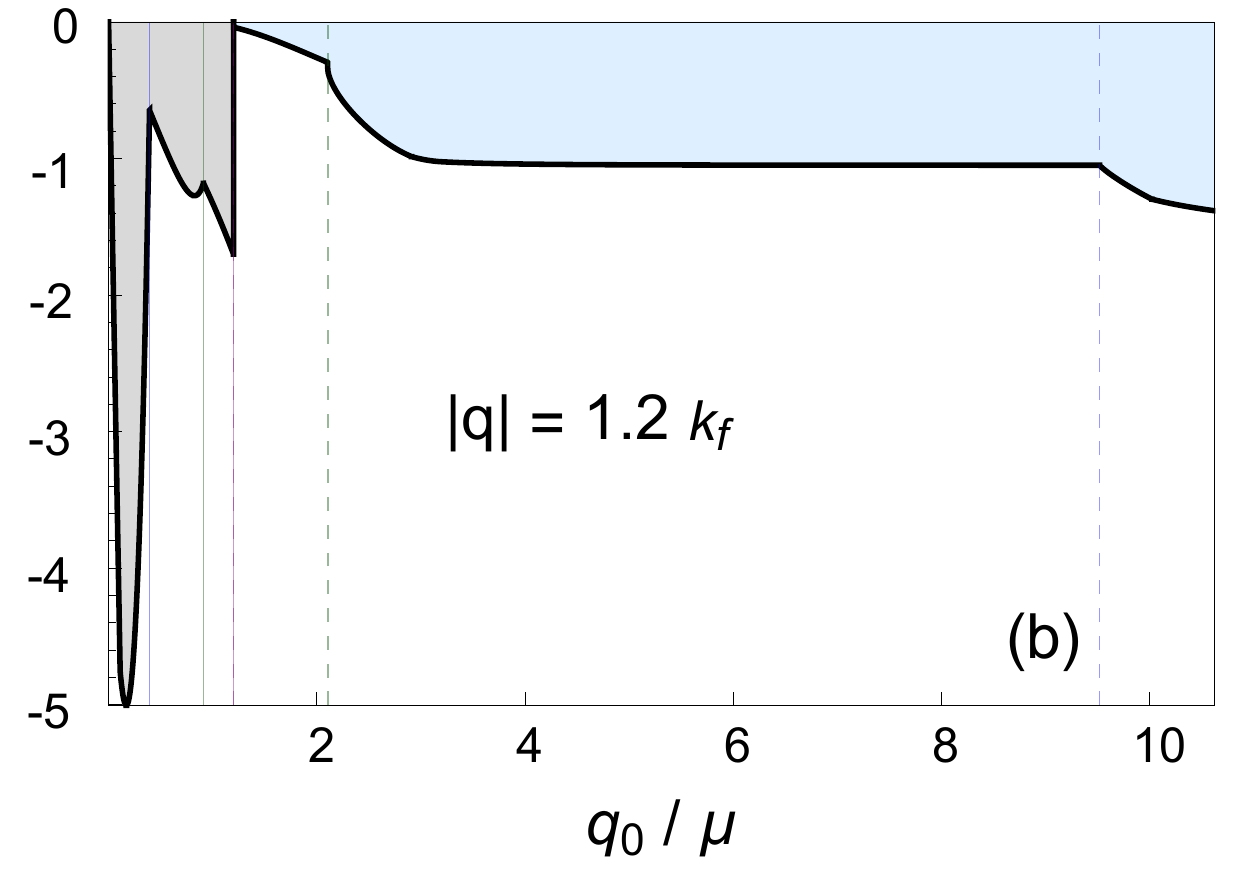}~
\includegraphics[scale=0.48]{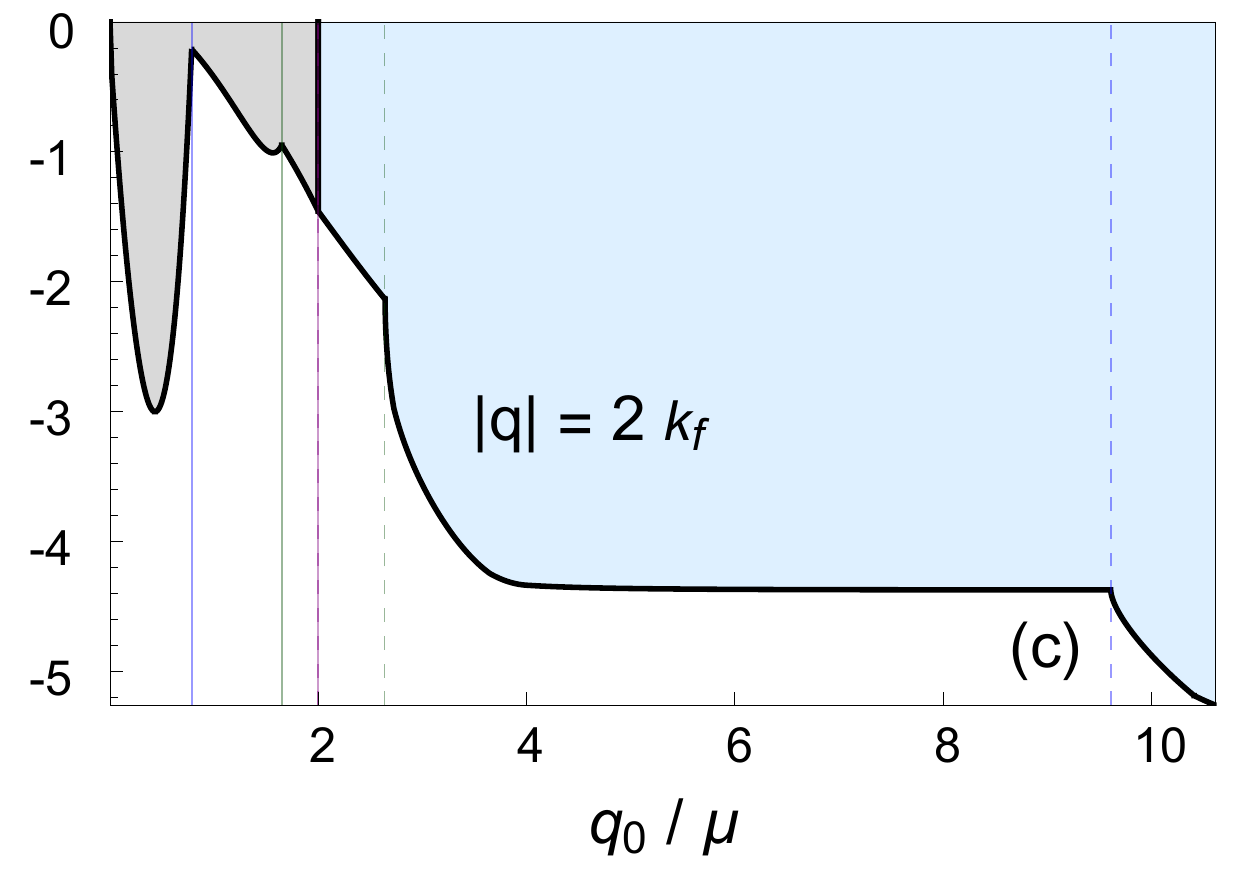}
\par\end{centering}
\caption{\label{fig:EPMPlots} Evolution of the total imaginary part for different values of $\left| \boldsymbol{q} \right|$. Thin vertical lines (with the usual color coding of each constituent) mark the positions of $q_0=-\mu_a+\sqrt{\mu_a^2+2 k_{f,\,a}\left|\boldsymbol{q}\right|+\left|\boldsymbol{q}\right|^2}$ below which Landau damping operates and thin dashed lines the positions $q_0=\mu_a+\sqrt{\mu_a^2-2 k_{f,\,a}\left|\boldsymbol{q}\right|+\left|\boldsymbol{q}\right|^2}$ above which pair creation sets in.  At low $\left| \boldsymbol{q} \right|$ the imaginary part due to Landau damping (gray shaded) is huge compared to the imaginary part due to pair creation. The situation is reversed at high $\left| \boldsymbol{q} \right|$. Owing to the small mass of electrons the dissipation free area squeezed in between Landau damping and $e^+\,e^-$ pair creation becomes tiny for $\left| \boldsymbol{q} \right|> k_{f,\,e}$.}      
\end{figure}

\noindent Collective modes of the multi component plasma are determined from
\bea
\omega_{L}^{2} & = & \boldsymbol{q}^{2}+\textrm{Tr}\,\left[ \textrm{Re}\,F(q_0=\omega_{L},\,\boldsymbol{q})\right] \,,\label{eq:wLEP}\\[3ex]
\omega_{\perp}^{2} & = & \boldsymbol{q}^{2}+\textrm{Tr}\,\left[ \textrm{Re}\,G(q_0=\omega_{\perp},\,\boldsymbol{q})\right] \,.\label{eq:wPEP}
\eea
In the longitudinal case, one finds a fine-structure of solutions which can be identified with the individual responses of the various particle species in the plasma. The existence of these distinct solutions is tied to the fact that the RPA probes the plasma in the collisionless limit characterized by $q_0 \tau >> 1$ where $\tau$ is the lifetime of the Landau quasi-particles (see  \cite{Shen:2013kxa} for a more detailed discussion). In the collisional hydrodynamic regime ($q_0 \tau < 1$), these modes are expected to merge into one collective response of the plasma. Among the longitudinal solutions, one finds the usual gapped plasmon mode $\omega_L$ which is predominantly a collective excitation of electrons. Electrons are very agile and effective in screening  muons and protons in the plasma. As a consequence of the screened muon-muon and proton-proton interactions their respective plasmon modes are reduced to (gapless) sound modes $u_{\mu}$, $u_{p}$ . This phenomenon is well studied in condensed matter theory and the corresponding modes are often termed  ``Bohm-Staver sound modes" \cite{Bohm1950}; see \cite{Baldo:2008pb} for a discussion in the context of nuclear matter. Because of their lower Fermi momenta $k_{f,\,\mu}<k_{f.\,p}$, the muon branch disappears much earlier with increasing momentum $\boldsymbol{q}$. In addition to these three modes, there are the overdamped gapless solutions $\omega_{<,\,e}$, $\omega_{<,\,\mu}$, $\omega_{<,\,p}$. The transverse spectrum $\omega_\perp$ does not exhibit such a fine-structure of modes. The analytical approximations for the transverse and plasmon mode in the limit of small $\boldsymbol{q}$ are
\bea 
\omega_{L,\,e}^2 & = & \sum_a \omega_{0,\,a} 
^2+\frac{3}{5}\frac{1}{\bar{k}_f^5} \left[ \frac{k_{f,\,e}^5\, \mu_\mu \mu_p}{\mu_e^2} + \frac{k_{f,\,\mu}^5\, \mu_e \mu_p}{\mu_\mu^2} +\frac{k_{f,\,p}^5\, \mu_e \mu_\mu}{\mu_p^2}  \right]\boldsymbol{q}^2\,,\label{eq:wLEPM}\\[3ex]
\omega_{\perp}^2 & = &  \sum_a \omega_{0,\,a}^2 +  \frac{1}{\bar{k}^5_f} \left[ k_{f,\,e}^3\mu_\mu \mu_p \left( 1+\frac{1}{5}v_{f,\,e}^2 \right) + k_{f,\,\mu}^3\mu_e \mu_p \left(  1 + \frac{1}{5}v_{f,\,\mu}^2 \right) + k_{f,\,p}^3 \mu_e \mu_\mu \left( 1 + \frac{1}{5}v_{f,\,p}^2 \right) \right] \boldsymbol{q}^2\,,\label{eq:wPEPM} \\[3ex]
\bar{k}_f^5 & = & k^3_{f,\,e}\,\mu_\mu \mu_p + k^3_{f,\,\mu}\,\mu_e \mu_p + k^3_{f,\,p}\,\mu_e \mu_\mu\,.
\eea
The above expressions are easily generalized to an arbitrary number of particle species. Setting $m_e = m_\mu = m_p := m$ and $\mu_e = \mu_\mu = \mu_p := \mu$ one obtains $3\,\omega_0^2$ in the limit $\boldsymbol{q}\rightarrow\boldsymbol{0}$. This is to be expected as $\omega_0^2$ can be expressed in terms of the particle number $\mathcal{N}$,  $\omega_0^2 =e^2 \mathcal{N} / \mu $, and adding three times particles of the same species to the plasma simply triples the magnitude of $\mathcal{N}$. The slope of $\omega_L$ and $\omega_\perp$, on the other hand should not change, and indeed one recovers the $\boldsymbol{q}^2$ terms of Eqs. \ref{eq:PlasmaL} and \ref{eq:PlasmaP}. In the limit $\boldsymbol{q}\rightarrow \infty$ ,the transverse modes become 
\begin{equation}
\omega_{\perp,\,\infty}^2=\boldsymbol{q}^2+\sum_a \omega_{0,\,a}^2 \left( 1 + \frac{1}{5} v_{f,\,a}^2\right)\,.
\end{equation}
The gapless solutions in the small momentum limit may again be obtained by assuming a dispersion of the form $q_0 = c v_f \left| \boldsymbol{q} \right| $ (where it is a matter of taste with respect to which of the Fermi velocities one wants to measure the slopes) and numerically solving Eq. \ref{eq:wLEP} for $c$. At saturation density, one finds the two proton-like solutions $\omega_{<,\,p}\sim 0.89\, v_{f,\,p} \left| \boldsymbol{q} \right|$ , $u_{p} \sim 1.43\, v_{f,\,p} \left| \boldsymbol{q} \right| $, the two muon-like solutions $\omega_{<,\,\mu}\sim 0.95\, v_{f,\,\mu} \left| \boldsymbol{q} \right|$ , $u_{\mu} \sim 1.43\, v_{f,\,\mu} \left| \boldsymbol{q} \right| $  and finally the gapped electron-like mode $\omega_{<,\,e} \sim 0.76\, v_{f,\,e} \left| \boldsymbol{q} \right|$. The complete spectrum is displayed in Fig. \ref{fig:EPMmodes}.\newline
Next we consider static screening. Since the trace $\textrm{Tr}\,\Pi_{00}$  enters the longitudinal photon propagator \ref{eq:PropMulti}, the total screening mass is simply given by the sum of the individual Debye masses. The static limit of the resummed polarization tensor summed over all constituens of the plasma \ref{eq:FempFull} yields
\be \label{eq:ScreenEMP}
\tilde{m}_D^2=\frac{m_{D,\,e}^2+m_{D,\,\mu}^2+m_{D,\,p}^2}{1+(1/\boldsymbol{q}^2)(m_{D,\,e}^2+m_{D,\,\mu}^2+m_{D,\,p}^2)}\,.
\ee
To extract the static screening of Coulomb interactions between two particular particle species one needs to refer to the corresponding entry in matrix \ref{eq:screenEMP}.\newline 
It remains to discuss the imaginary parts  stemming from Landau damping and pair creation of electrons, muons and protons. Fig. \ref{fig:EPMdamping} illustrates the landscape where these processes are allowed in degenerate matter in $\beta$ equilibrium at saturation density. At the one-loop level, one finds two dissipation-free regions characterized by the lightest and the heaviest particles in the system: There is always a gap in between the areas where Landau damping and pair creation of electrons operate, i.e., for 
\bea  
-\mu_e+\sqrt{\mu_e^2+2\, k_{f,\,e}\left|\boldsymbol{q}\right|+\left|\boldsymbol{q}\right|^2} \,< q_0 & < &\, +\mu_e+\sqrt{\mu_e^2-2\, k_{f,\,e}\left|\boldsymbol{q}\right|+\left|\boldsymbol{q}\right|^2}\,,\hspace{1cm} \left|\boldsymbol{q}\right| < k_{f,\,e}\,,\\[3ex] 
-\mu_e+\sqrt{\mu_e^2+2\, k_{f,\,e}\left|\boldsymbol{q}\right|+\left|\boldsymbol{q}\right|^2} \,< q_0 & < &\, \sqrt{\left|\boldsymbol{q}\right|^2+4m_e^2}\,,\hspace{3.6cm} \left|\boldsymbol{q}\right| > k_{f,\,e}\,,
\eea 
\\
even if this gap becomes tiny when the photon momentum exceeds $k_{f,\,e}$; see Fig. \ref{fig:EPMPlots} (b) and (c). The photon modes $\omega_\perp$ reside in this dissipation-free ``canal", and the plasmon mode $\omega_L$ is restricted to the dissipation-free area on the left side of Fig. \ref{fig:EPMdamping}. Because of the small electron mass, the boundaries are very well approximated by $q_0=\left|\boldsymbol{q}\right|$ and  $q_0=2\mu_e-\left|\boldsymbol{q}\right|$. The threshold for $e^+\,e^-$ pair creation assumes a minimum for photons with momenta $\left|\boldsymbol{q}\right| = k_{f,\,e} \sim \mu_e$, where energies of $q_0\gtrsim \mu_e$ are required. At higher momenta, there is no dissipation for energies below the (lower) threshold for Landau damping due to protons
\be 
q_0 <\, -\mu_p+\sqrt{\mu_p^2-2\, k_{f,\,p}\left|\boldsymbol{q}\right|+\left|\boldsymbol{q}\right|^2}\,,\hspace{1cm} \left|\boldsymbol{q}\right| > 2k_{f,\,p}\,.
\ee 
In conclusion, the solutions $u_a$ and $\omega_{<,\,a}$ are all subject to Landau damping due to one or more particle species of the plasma; the question is to which extent. While solutions $u_a$ are damped by the remaining constituents $b\neq a$, solutions $\omega_{<,\,a}$ are additionally damped by their own kind. As a result we shall find that only the solutions $u_a$ represent maxima of the spectrum.\newline The spectral functions generalize in a straight forward manner:
\begin{figure}
\includegraphics[scale=0.80]{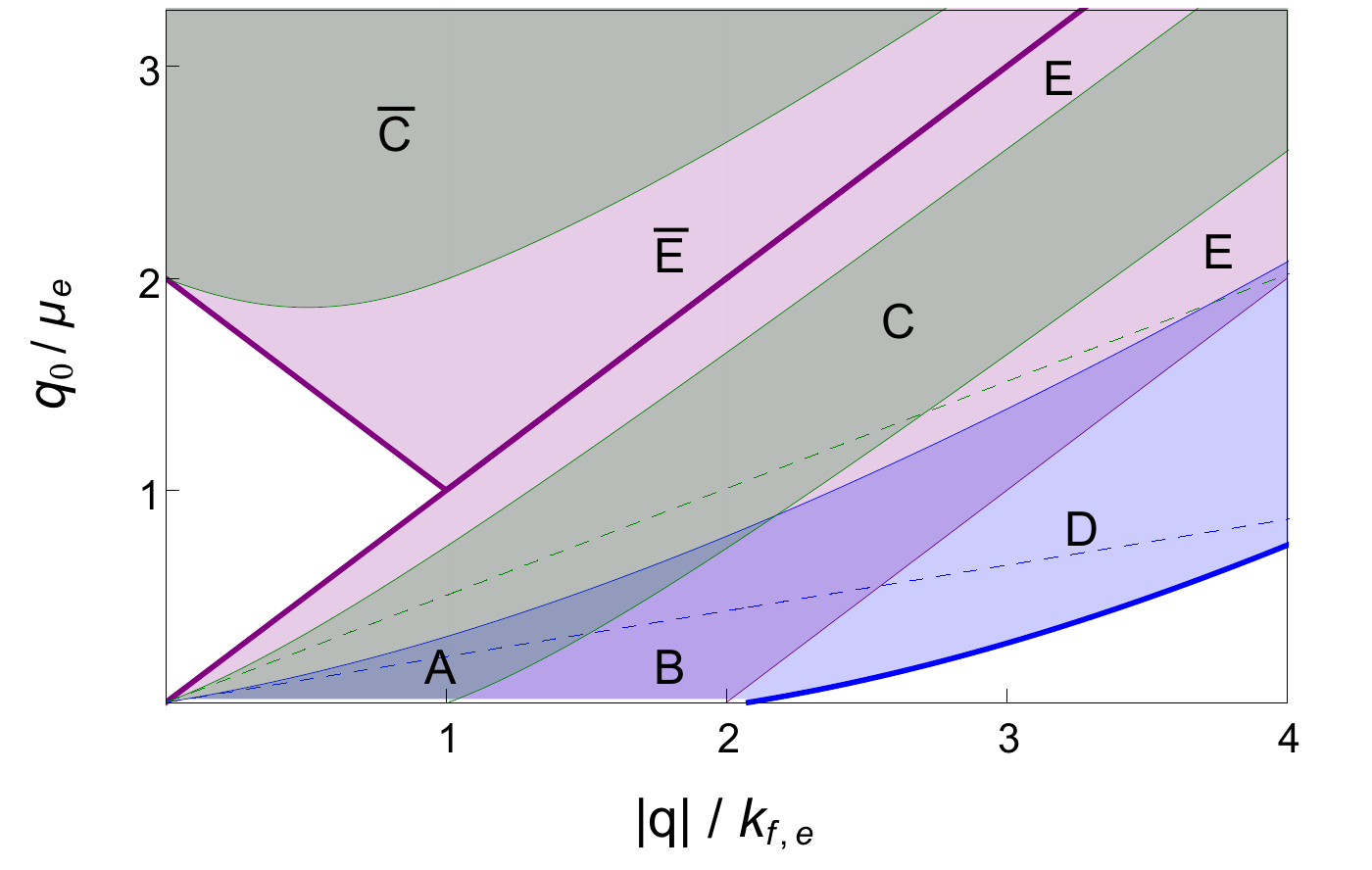}\label{fig:MultiDamping}
\caption{\label{fig:EPMdamping} 
Regions in which Landau damping and pair creation of electrons (purple shaded), muons (green shaded), and protons (blue shaded) operate. White areas indicate dissipation-free regions, thick lines denote the outermost boundaries where dissipation sets in, and thin lines mark the boundaries where dissipation of a particular constituent sets in. Analytic expressions for the boundaries are listed in the caption of Fig. \ref{fig:Domains}. Dashed lines display the positions of $v_{f} \left| \boldsymbol{q} \right|$ for each species (which in the case of electrons overlaps with the thick purple line). The labels are assigned as follows: ($A$) Landau damping due to all three constituents, ($B$) Landau damping due to electrons and protons, ($C$), Landau damping due to electrons and muons, ($D$) Landau damping due to protons, ($E$) Landau damping due to electrons, ($\bar{E}$) electron pair creation and ($\bar{C}$) electron and muon pair creation. Proton pair creation appears for much higher values of $q_0$ and is not displayed in the plot. As long as the fermions in the plasma are massive, there is always a finite gap in between the area where pair creation and Landau damping operate, due to their small mass this gap becomes tiny for electrons.}
\end{figure}
\begin{figure}
\begin{centering}
\textbf{$\left|\boldsymbol{q}\right|=0.3\,m_{D,\,e}\hspace{4cm}\left|\boldsymbol{q}\right|=\,m_{D,\,e}\hspace{4cm}\left|\boldsymbol{q}\right|=1.8\,m_{D,\,e}$}\par\medskip
\includegraphics[scale=0.5]{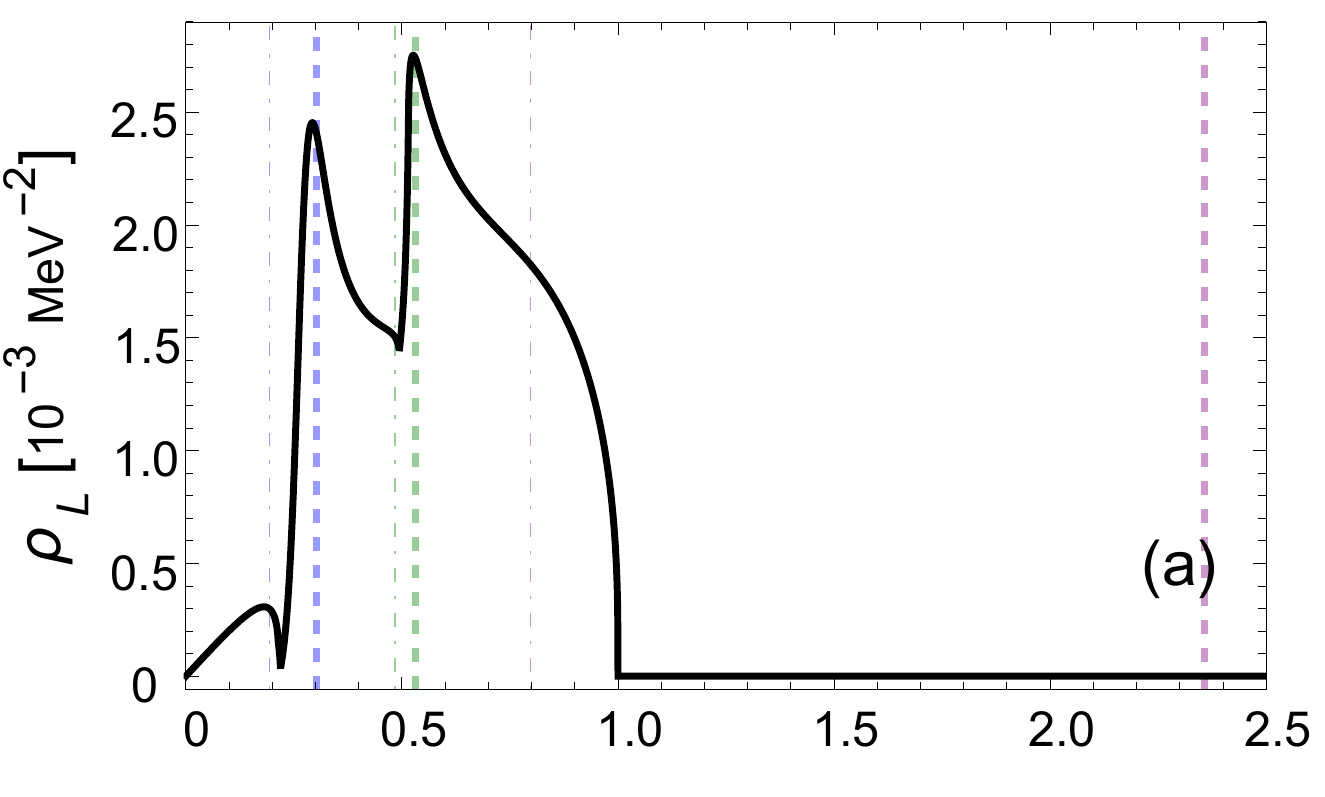}~
\includegraphics[scale=0.486]{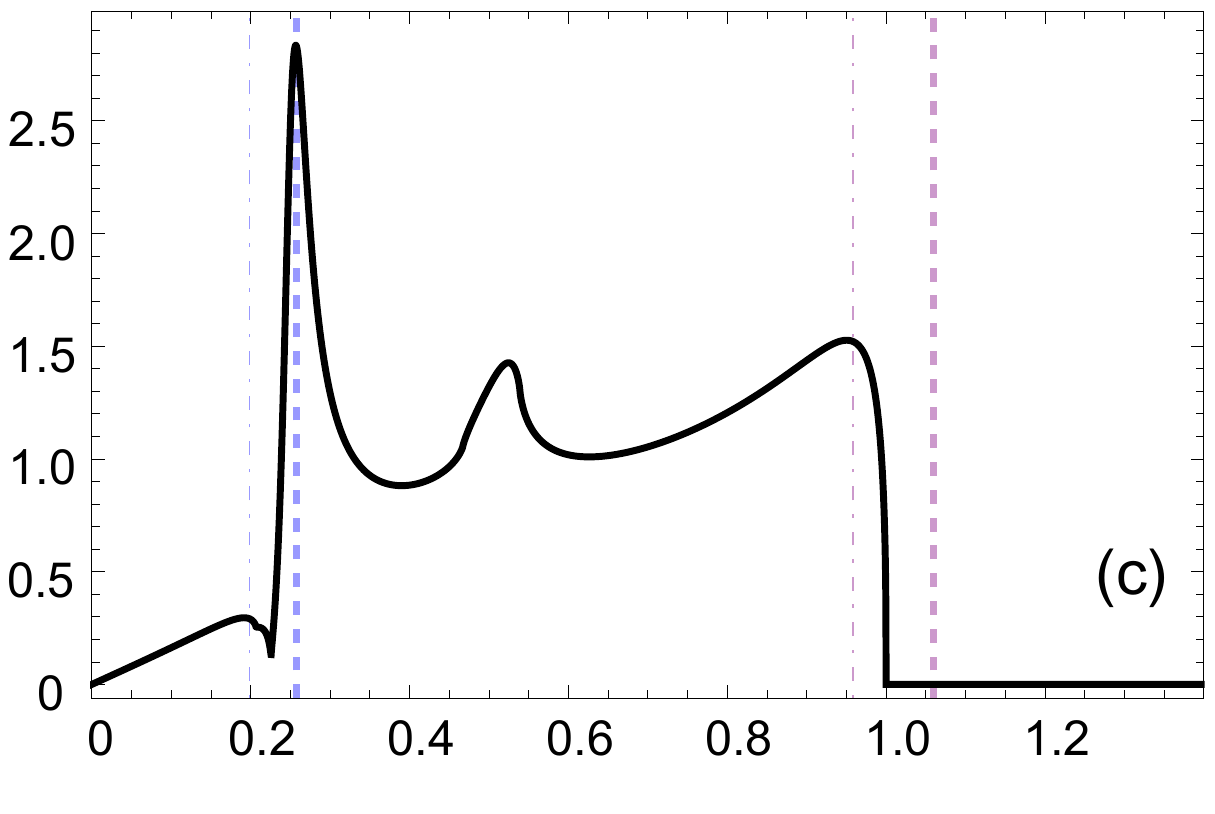}~~~~~
\includegraphics[scale=0.488]{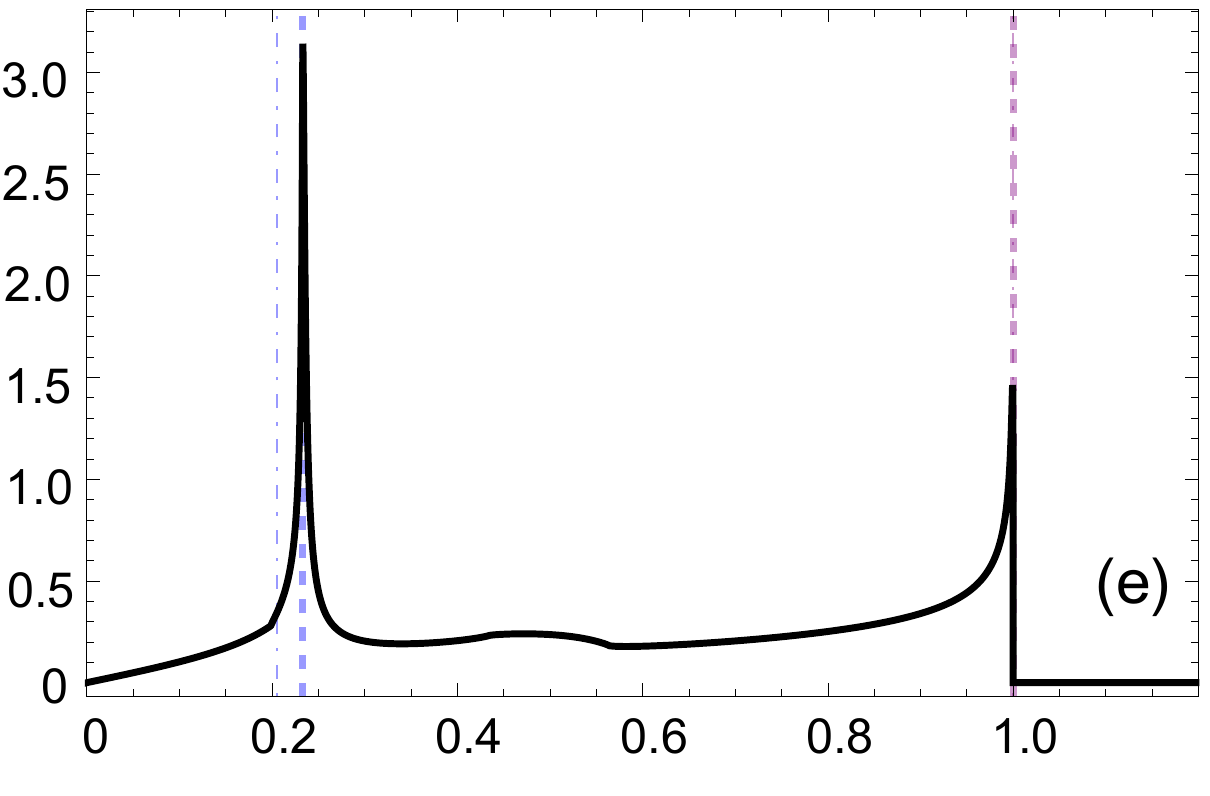}\\
~\includegraphics[scale=0.5]{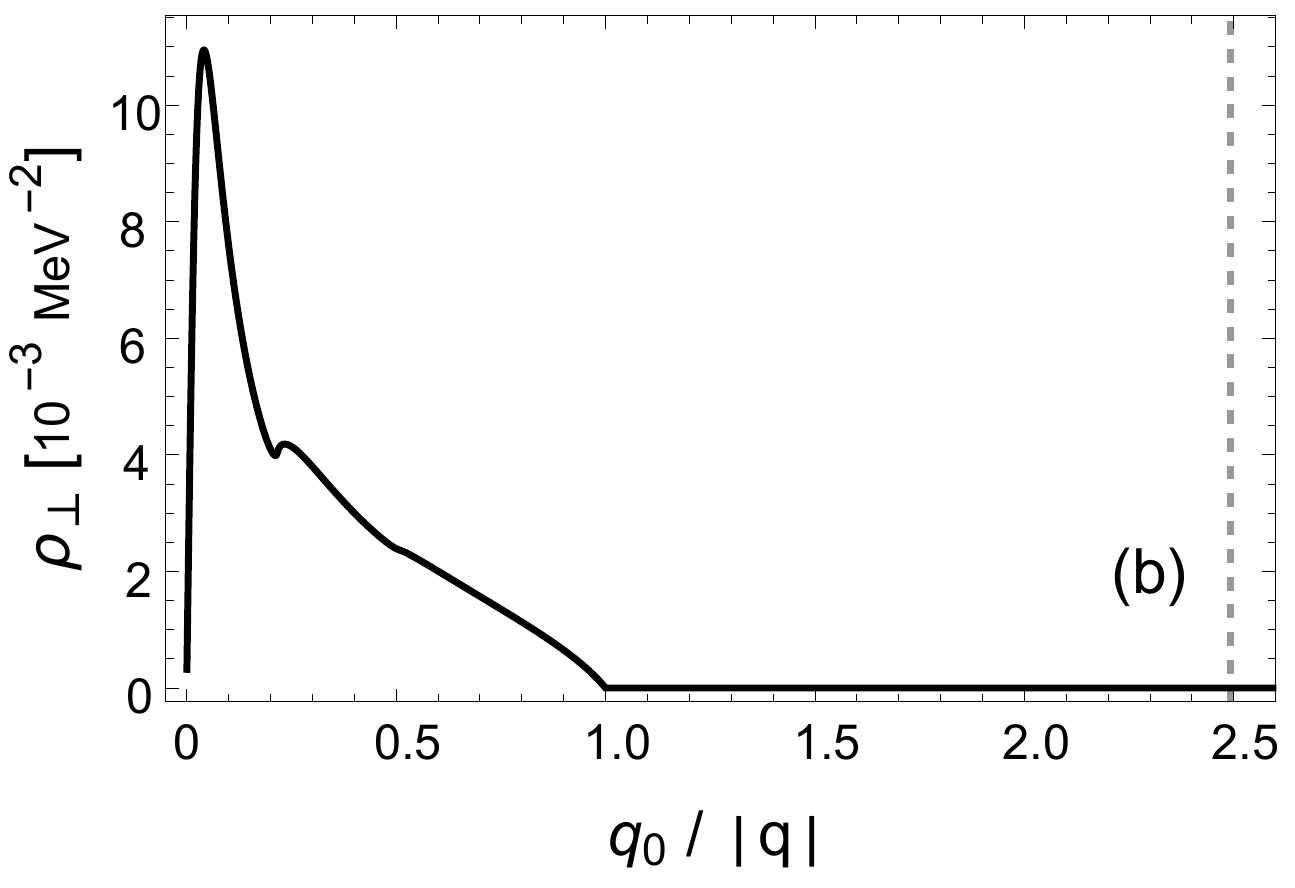}~
\includegraphics[scale=0.5]{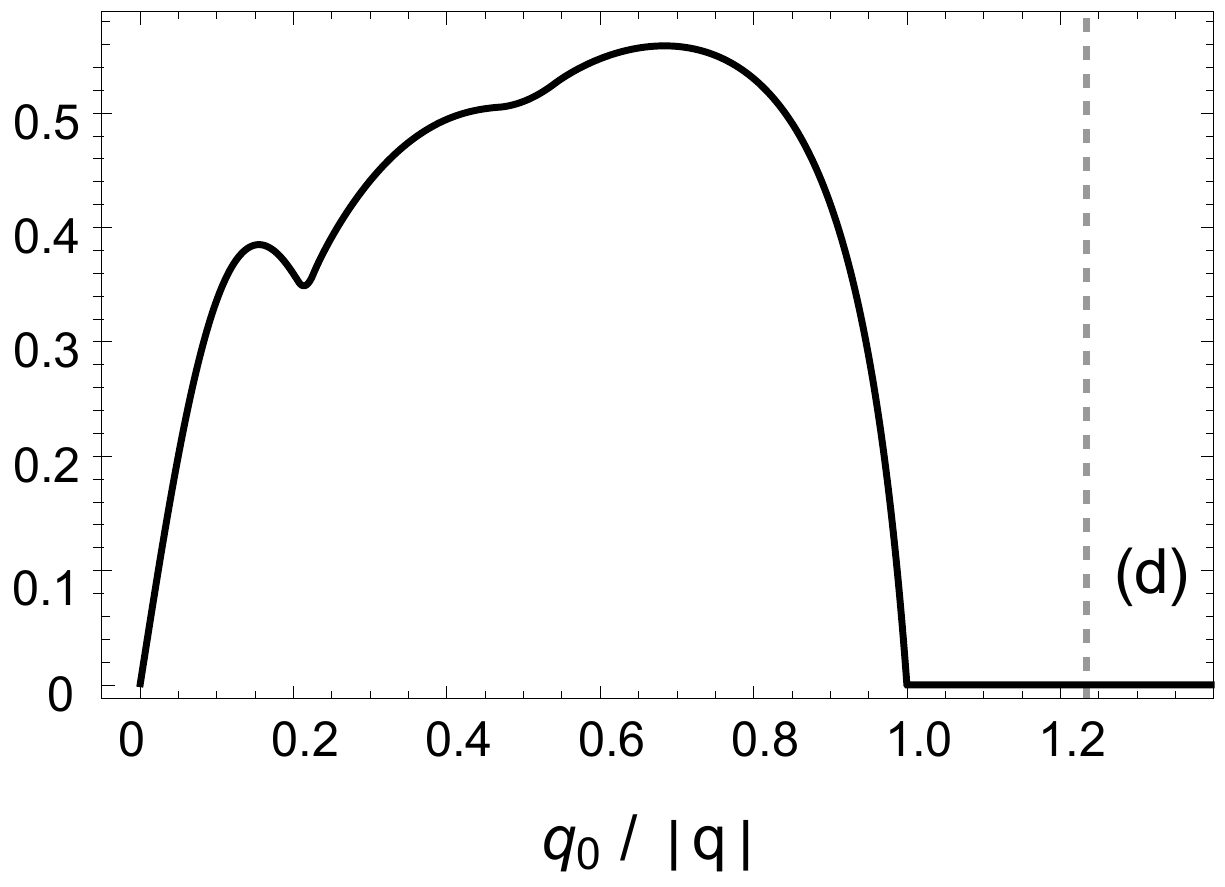}~
\includegraphics[scale=0.5]{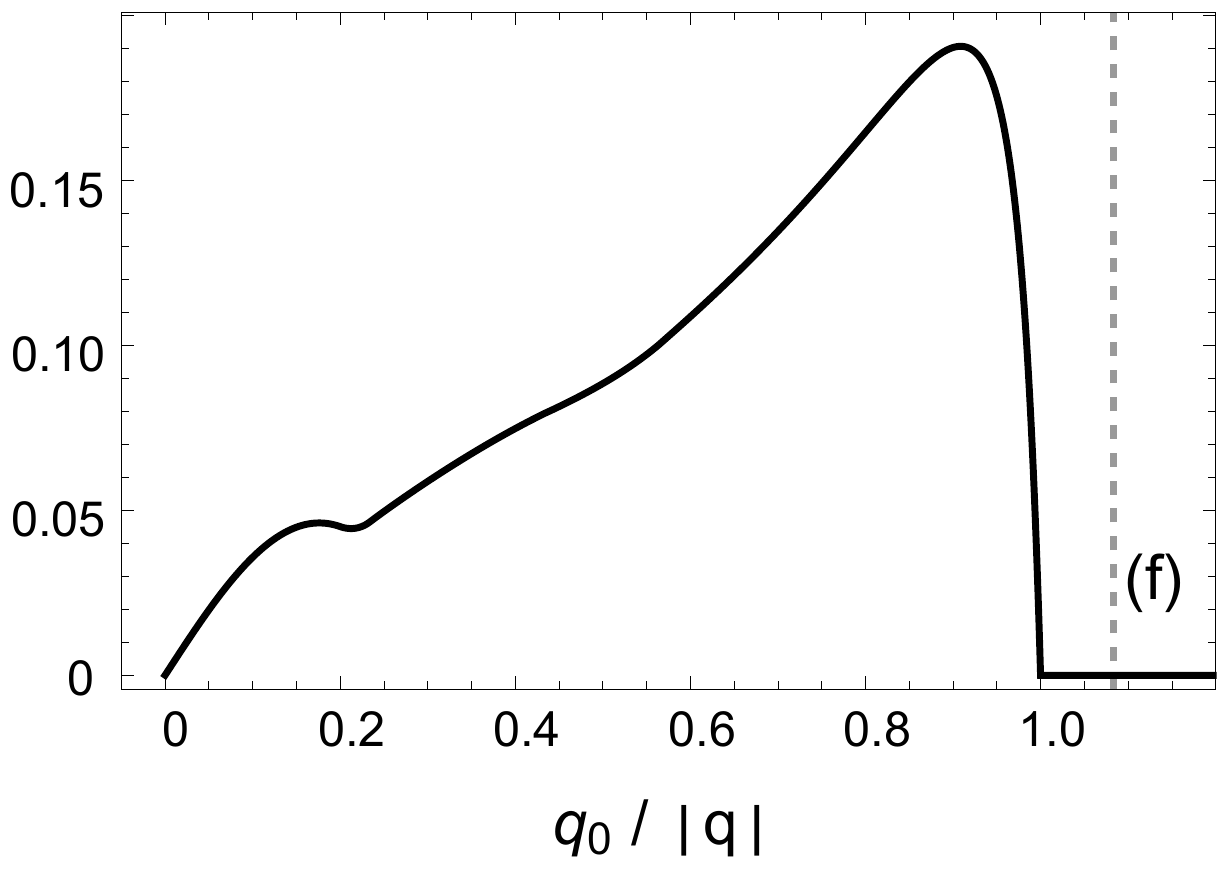}
\par\end{centering}
\setlength{\belowcaptionskip}{-10pt}
\caption{\label{fig:EPMSpectra} Evolution of the longitudinal (a, c, e) and transverse (b, d, f)  low-momentum spectrum of the photon in a degenerate  electron, muon, and proton plasma in $\beta$ equilibrium at saturation density. The spectra are plotted as functions of $q_0$ for fixed momenta; compare as well to Fig. \ref{fig:EPMmodes}. The solutions to Eq. \ref{eq:wLEP} are indicated as follows: thick  dashed lines show the positions of the sound-like solutions $u_{a}$ and the real plasmon mode $\omega_{L}$, and thin dot-dashed lines indicate the positions of the solutions $\omega_{<,\,a}$. While the solutions $u_{a}$ are indeed located at the maxima of the spectrum their corresponding peaks do not stand out too much from the bulk of the spectrum at lower momenta. Only the proton mode develops a well-defined peak at larger momenta.}
\end{figure}
\vspace*{3mm}
\begin{eqnarray}
\rho_{L}(q) & = & -\frac{1}{\pi}\,\text{Im}\,R(q)=\frac{1}{\pi}\frac{\text{Tr}\,(\text{Im}\,\Pi_{00})}{\left[\text{Tr}\,(\text{Re}\,\Pi_{00})- \boldsymbol{q}^{2}\right]^{2}+\left[\text{Tr}\,(\text{Im}\,\Pi_{00})\right]^{2}}\label{eq:RhoMulti}+\textrm{sgn}(q_0/\left|\boldsymbol{q}\right|)\delta\left. \left(\text{Re}\,\Pi^{00} -\boldsymbol{q}^2 \right) \right|_{\omega_{L}}, \\\,\nonumber \\[3ex]
\rho_{\perp}(q) & = & -\frac{1}{\pi}\,\text{Im}\,S(q)=\frac{1}{\pi}\frac{\text{Tr}\,(\text{Im}\,G)}{\left[\text{Tr}\,(\text{Re}\,G)-q^{2}\right]^{2}+\left[\text{Tr}\,(\text{Im}\,G)\right]^{2}}+\textrm{sgn}(q_0 / \left|\boldsymbol{q}\right| )\delta \left(\text{Re}\,G - q^2 \right)\,,\label{eq:RhoMulti2}
\eea

\vspace{4mm}
\noindent where only the real plasmon mode and the transverse photon mode are located in the dissipation-free region. The evolution of the longitudinal and transverse photon spectrum with increasing momentum $\left|\boldsymbol{q}\right|$ is displayed in Fig. \ref{fig:EPMSpectra}. While it is true that the two sound-like solutions $u_{p}$ and $u_{\mu}$ represent maxima, only the proton mode develops a well defined peak at higher momenta; see Fig. \ref{fig:EPMSpectra} (e). The poles $\omega_{<,\,a}$ are located somewhere in the bulk of the spectrum, not necessarily aligned with specific characteristics of the spectrum.     
\subsection{QED and strong interactions: Electrons, Muons, Protons and Neutrons}
\label{subsec:EMPN}
\noindent In the following we include strong interactions in the RPA resummation and account for the induced lepton-neutron scattering. Neglecting the magnetic moment of the neutron its contribution is not directly visible at the level of the resummed photon propagator: Eq. \ref{eq:PropMulti} only knows about electromagnetism and therefore still retains its form. The neutron, however, \textit{does} play a role in the polarization of the medium as described by Eq. \ref{eq:EPMPolResum}. The interaction matrix including electromagnetic and strong interactions reads
\be \label{eq:Vfull}
V^{\mu\nu}(q)=-\frac{1}{q^{2}+i\epsilon}\left(\begin{array}{cccc}
1 & 1 & -1 & 0\\
1 & 1 & -1 & 0\\
-1 & -1 & 1 & 0\\
0 & 0 & 0 & 0
\end{array}\right)G^{\mu\nu}+\left(\begin{array}{cccc}
0 & 0 & 0 & 0\\
0 & 0 & 0 & 0\\
0 & 0 & f_{pp} & f_{pn}\\
0 & 0 & f_{np} & f_{nn}
\end{array}\right)\frac{q^2}{\boldsymbol{q}^2}\, g^{\mu0}g^{\nu0}+\left(\begin{array}{cccc}
0 & 0 & 0 & 0\\
0 & 0 & 0 & 0\\
0 & 0 & \bar{f}_{pp} & \bar{f}_{pn}\\
0 & 0 & \bar{f}_{np} & \bar{f}_{nn}
\end{array}\right)P_\perp^{\mu\nu}\,,
\ee 
where density-density and current-current potentials $f_{ab}$ and $\bar{f}_{ab}$ are determined from \ref{eq:derivatives}. We have introduced the factor $q^2/\boldsymbol{q}^2$ to obtain consistency with $G^{\mu\nu}$; Eq. \ref{eq:PhotonGauge} therefore projects the nuclear interactions onto the vector channel. In Appendix \ref{sub:RMF}, we discuss an instructive example where nucleons interact via the exchange of massive vector mesons as in the RMF model, the static contribution there is $\propto 1/m_{\textit{meson}}^2$, while in Eq. \ref{eq:Vfull} the potentials are extracted from the energy functional \ref{eq:EpsChamel}. This approach to nuclear interactions renders the RPA resummations particularly simple: The evaluation of current-current correlation functions \ref{eq:PIDefinition} is sufficient; axial or mixed correlation functions (which are to be included in a more rigorous treatment of nuclear forces) are not required. We note that our interest here is restricted to  low photon energies and momenta, where induced interactions and collective effects are most pronounced. In this domain, it is reasonable to neglect the momentum dependence of the nuclear interactions as to obtain \ref{eq:Vfull} from Fermi liquid theory. \newline The closed-form solutions of the polarization functions according to Eq. \ref{eq:EPMPolResum} using Eq. \ref{eq:Vfull} are fairly complicated. They can, however, be cast in a simple and intuitive form. To see that, we revisit the expansion of, e.g., the $\tilde{F}_{ee}$ component 
\be \label{eq:ExpandFee}
\tilde{F}_{ee}=F_e + F_e \frac{1}{q^2} F_e + F_e\,\frac{1}{q^2} \left( F_e + F_\mu +\tilde{F}_{p,\,\textrm{nucl}} \right) \frac{1}{q^2} F_e  + \mathcal{O}(\frac{1}{q^6})\,,
\ee
which is identical to expansion \ref{eq:FeeExpand} upon replacing insertions of $F_p$ and $G_p$ with insertions of $\tilde{F}_{p,\,\textrm{nucl}}$ and  $\tilde{G}_{p,\,\textrm{nucl}}$, which read 
\vspace*{1mm}
\bea  \label{eq:ReProtonFull}
\tilde{F}_{p,\,\textrm{nucl}} & = & e^2 \frac{F^{\prime}_{p}\,(1+f_{nn}\,F^{\prime}_{n})}{1+f_{nn}\,F^{\prime}_{n}+f_{pp}\,F^{\prime}_{p}+F^{\prime}_{p}\,\,F^{\prime}_{n}(f_{pp}f_{nn}-f_{np}^{2})}\,,\\[3ex]
\tilde{G}_{p,\,\textrm{nucl}} & = & e^2 \frac{G^{\prime}_{p}\,(1+\bar{f}_{nn}\,G^{\prime}_{n})}{1+\bar{f}_{nn}\,G^{\prime}_{n}+\bar{f}_{pp}\,G^{\prime}_{p}+G^{\prime}_{p}\,\,G^{\prime}_{n}(\bar{f}_{pp}\bar{f}_{nn}-\bar{f}_{np}^{2})}\,. \label{eq:ReProtonFullPerp}
\eea 

\noindent The above expressions contain one global factor of $e^2$,  indicating that \ref{eq:ReProtonFull}  and \ref{eq:ReProtonFullPerp} as a whole replace ``bare" proton loops in the resummation; the polarization functions $F^{\prime}$ and $G^{\prime}$ contain no further factors of $e^2$.\newline
It is easy to understand where Eq. \ref{eq:ReProtonFull} comes from: It is the $(a,b)$ = $(p,p)$ component of the resummed polarization tensor Eq. \ref{eq:EPMPolResum} evaluated in the $2\times2$ subspace spanned by nucleons $\Pi^{\mu\nu}=\textrm{diag}\,(\Pi^{\mu\nu}_p,\, \Pi^{\mu\nu}_n)$ and using only the nuclear interactions contained in matrix $V^{\mu\nu}$, Eq. \ref{eq:Vfull}. Expansion \ref{eq:ExpandFee} consequently shows that nuclear interactions appear ``nested" inside of electromagnetic ones whenever a proton loop appears, which motivates the term \textit{induced} interactions; $f_{pn}$ appears at least quadratic since neutrons need to couple to protons twice to contribute to the electromagnetic response. The photon propagator consequently knows about all direct and induced interactions and neutrons always play a role, even when one considers a scattering process between two leptons. Putting the dressed photon propagator together according to Eq. \ref{eq:PropMulti} leads to a compact result, 
\vspace*{2mm}
\be \label{eq:PropEMPNfull}
D^{\mu\nu}(q)=\frac{q^{2}}{\boldsymbol{q}^{2}}\frac{1}{F_{e}(q)+F_{\mu}(q)+\tilde{F}_{p,\,\textrm{nucl}}(q)-q^{2}}\,g^{\mu0}g^{\nu0}+\frac{1}{G_{e}(q)+G_{\mu}(q)+\tilde{G}_{p,\,\textrm{nucl}}(q)-q^{2}}\,P_{\perp}^{\mu\nu}\,.
\ee

\noindent In summary, expression \ref{eq:PropEMPNfull} resumms contributions from electrons, muons, and protons where protons are themselves polarized by strong interactions; see Fig. \ref{fig:PhotonMulti}. It contains any possible (one-loop) combination of electromagnetic and strong interactions once and only once. We may as well ask for the dressed polarization tensor which is again identical to Eq. \ref{eq:FempFull} upon replacing $F$ with Eq. \ref{eq:ReProtonFull} [and similarly $G$ with \ref{eq:ReProtonFullPerp}]. The effective coupling to neutrons can be expressed in terms of the interaction Lagrangian 
\be 
\mathcal{L}= - e f_{pn}\, \left( \bar{n}\gamma_\mu n \right) \, \Pi^{\mu\nu}_p A_\nu  = - e f_{pn}\, \left( \bar{n}\gamma_\mu n\,\right) \left( F_p P^{\mu\nu}_L + G_p P^{\mu\nu}_\perp \right) A_\nu\,. 
\ee
Modifications of the electromagnetic response due to strong interactions are sizable for longitudinal components but entirely negligible for transverse ones. This is mainly due to the fact that the transverse polarization functions are small compared to their longitudinal counterparts; see, e.g., Figs. \ref{fig:RealPi} and \ref{fig:ImPi}.
\begin{figure}
\begin{centering}
\includegraphics[scale=1]{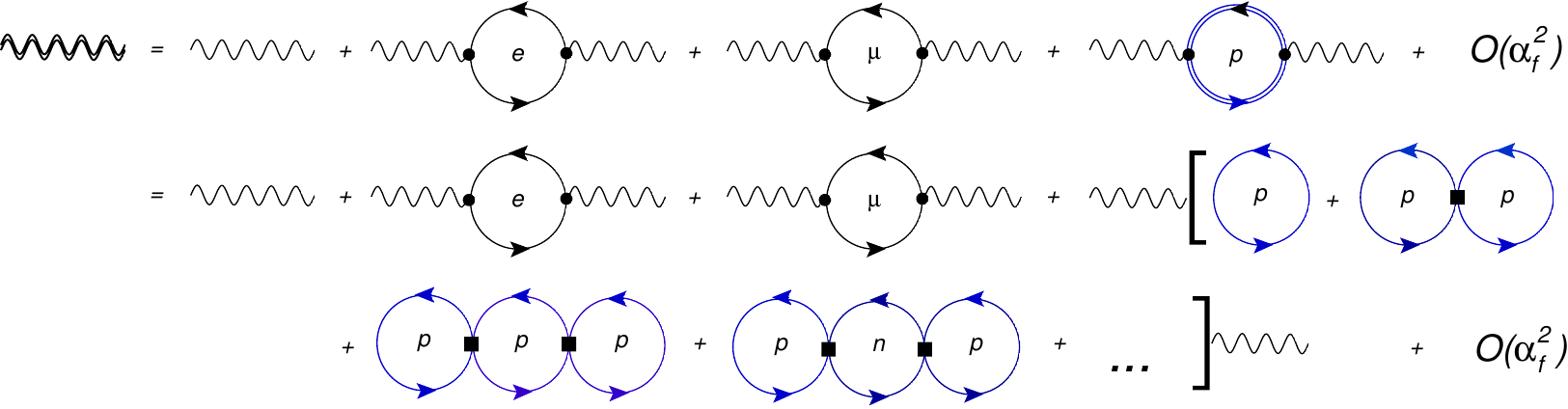}
\par\end{centering}
\caption{\label{fig:PhotonMulti} Photon propagator \ref{eq:PropEMPNfull} expanded in powers of the electromagnetic coupling. QED vertices are indicated by small dots, and short-range strong interactions are shown by small squares. At leading order one obtains the insertions of bare loops corresponding to each particle species. Protons are themselves polarized in the nuclear medium and unfold an infinite series of (proton and neutron) loops whenever they appear. Neutrons appear in the resummation at order $\alpha_f\,f_{pn}^2$.}      
\end{figure}      
\subsubsection{ Screening, collective modes and spectral densities}
\label{subsec:RPempn}
\begin{figure}
\begin{centering}
\includegraphics[scale=0.8]{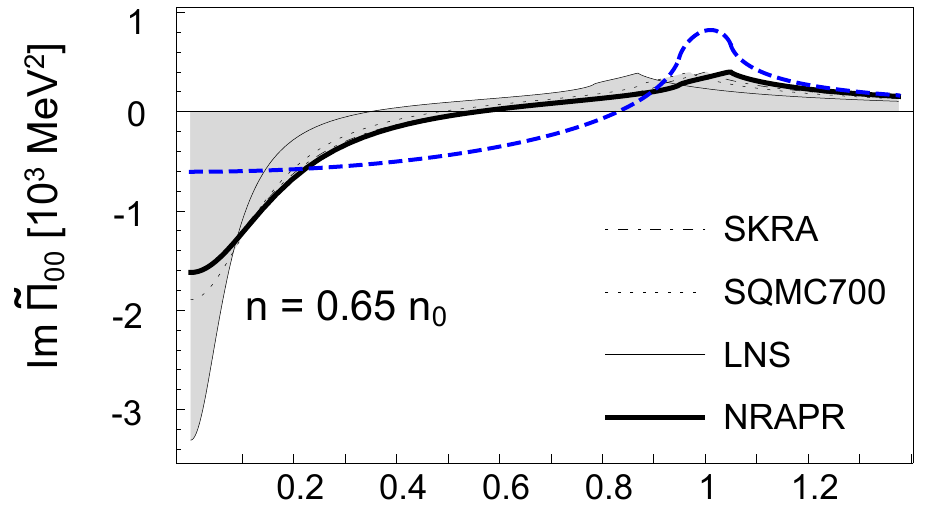}~~~\includegraphics[scale=0.58]{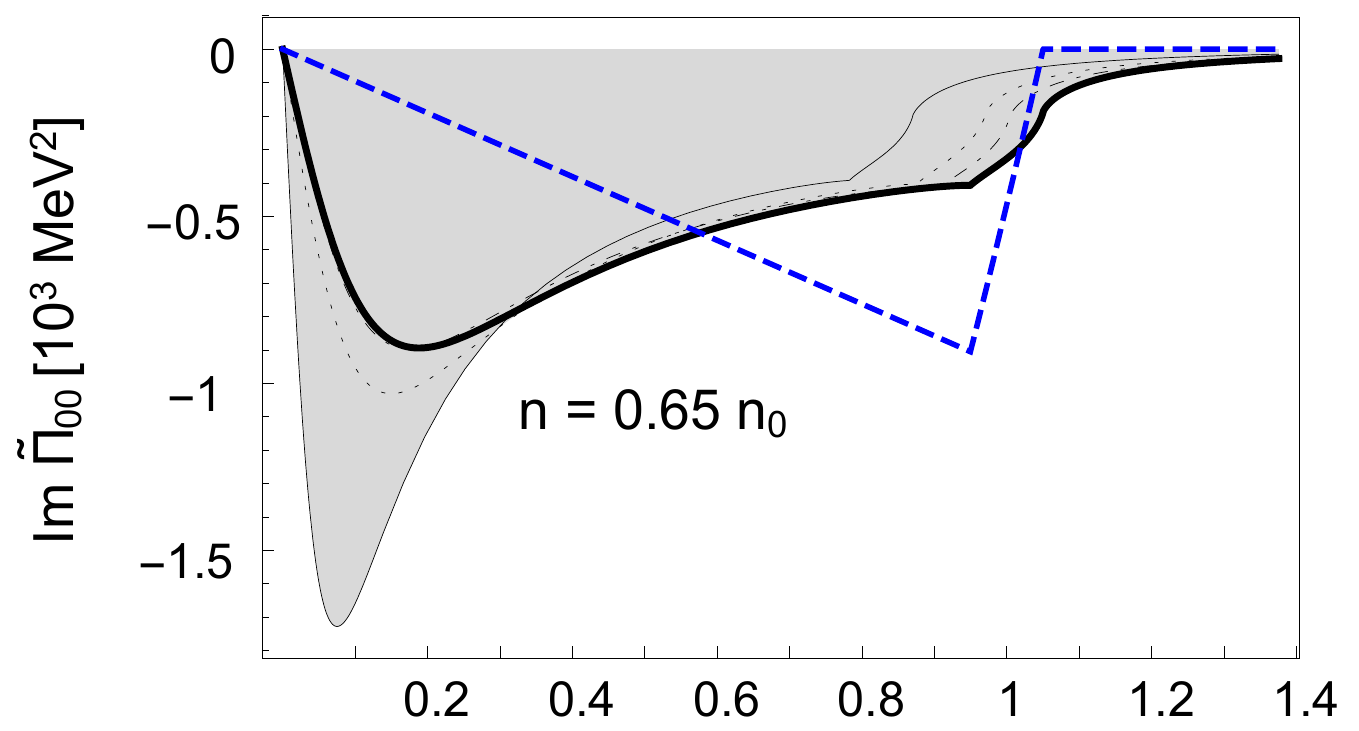}\\
\includegraphics[scale=0.55]{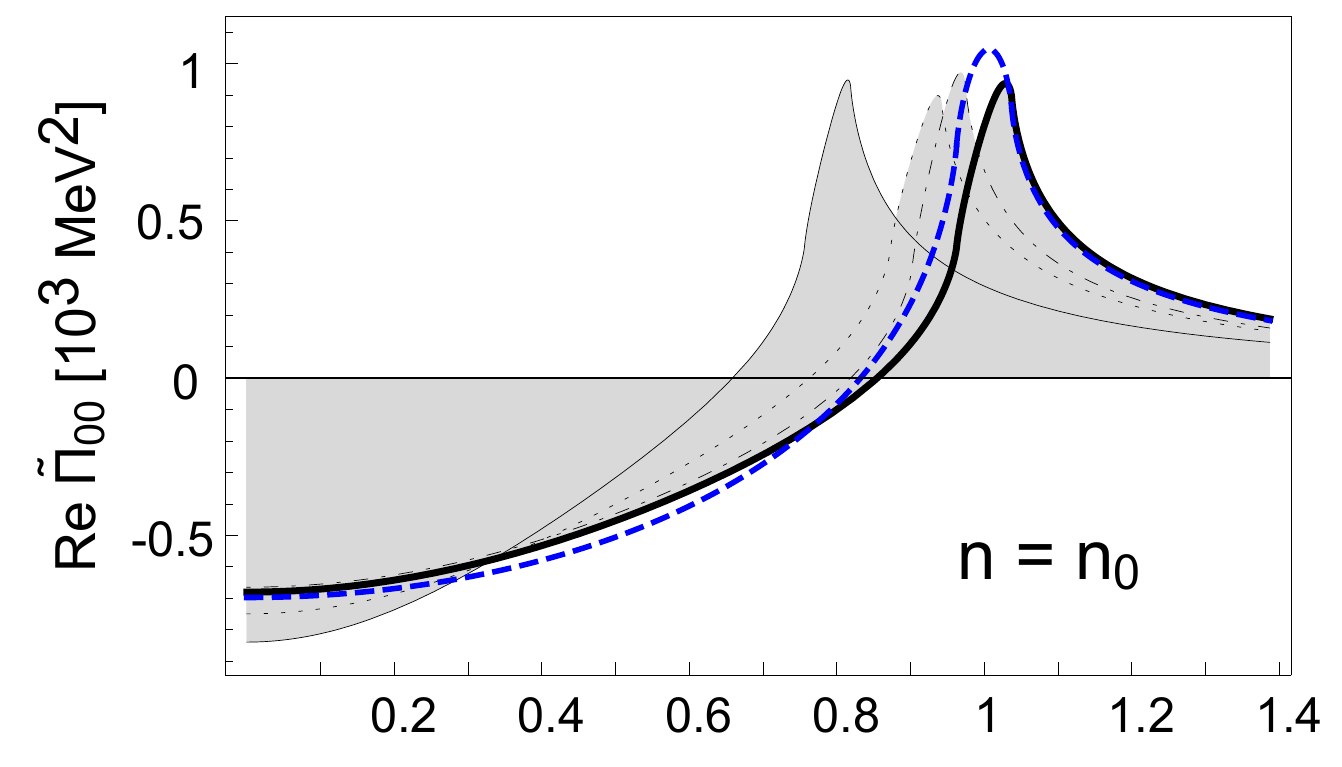}~~~\includegraphics[scale=0.57]{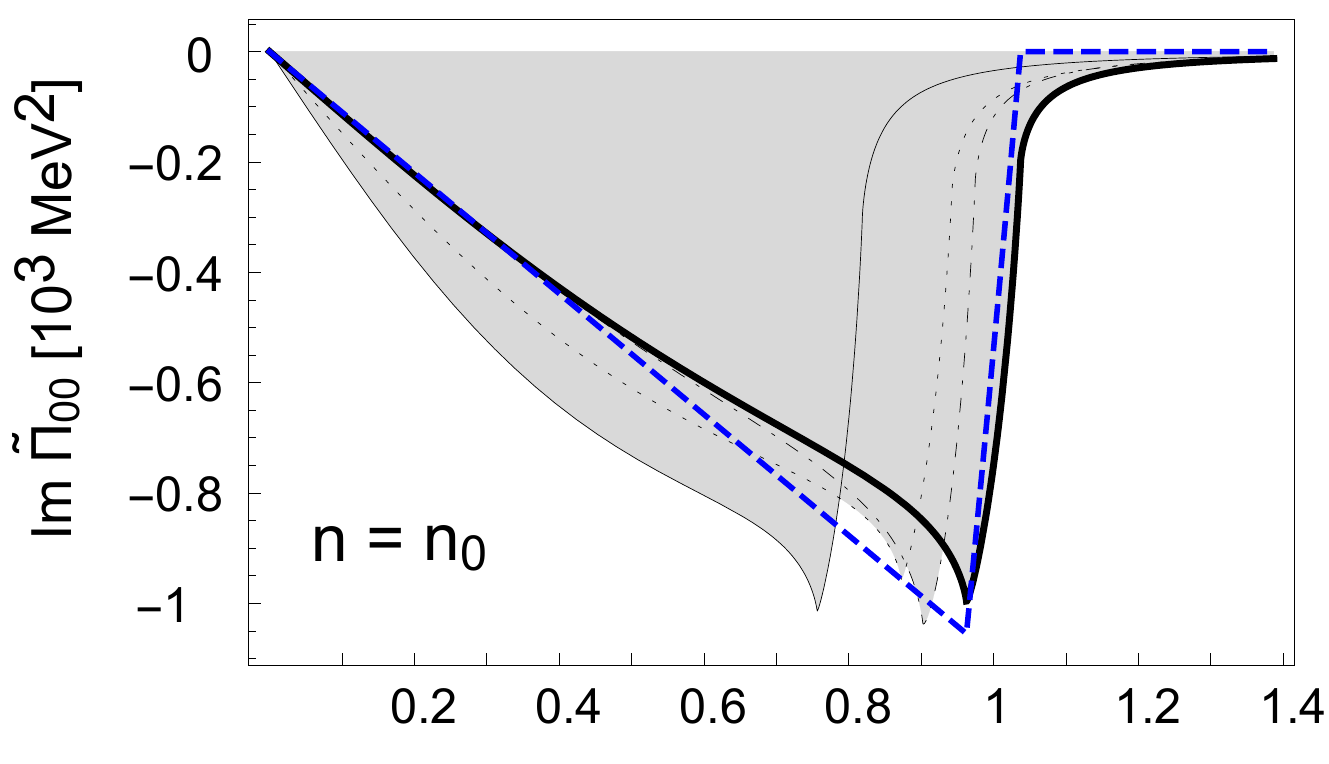}\\
\includegraphics[scale=0.56]{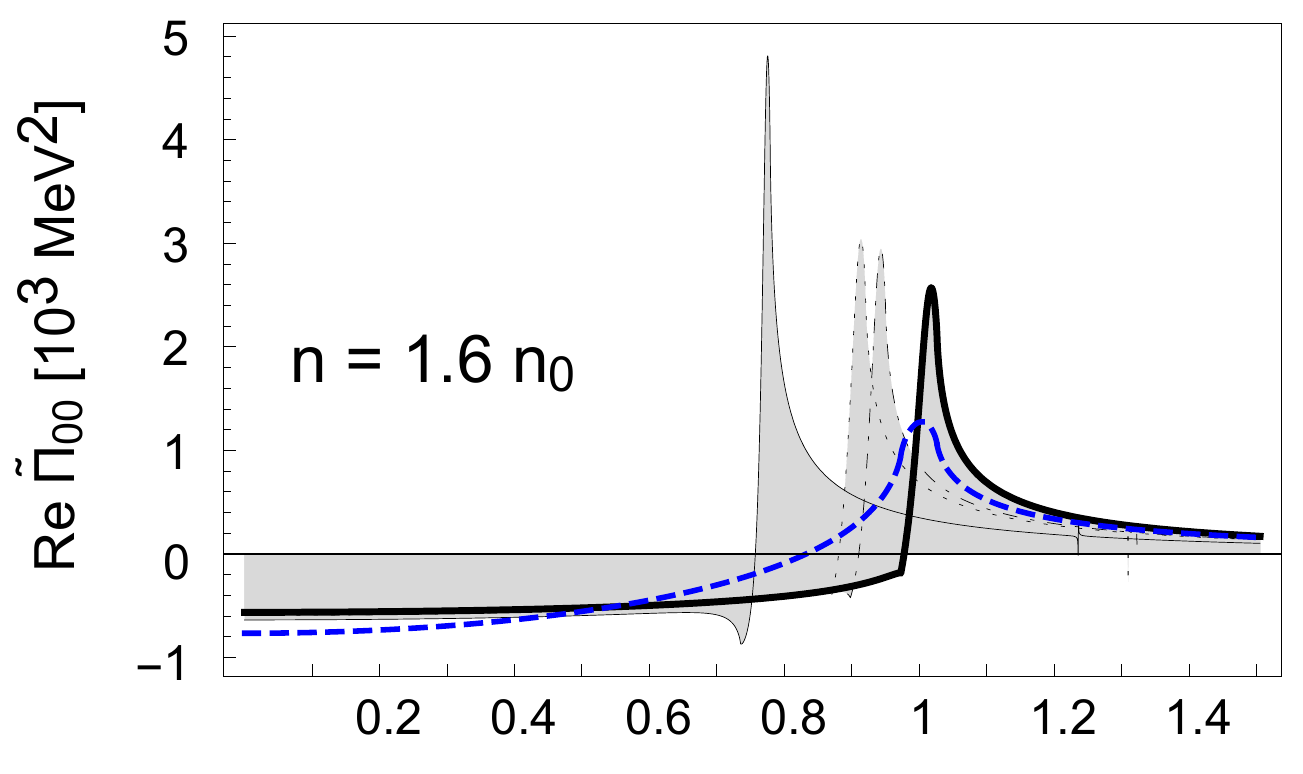}~~~\includegraphics[scale=0.57]{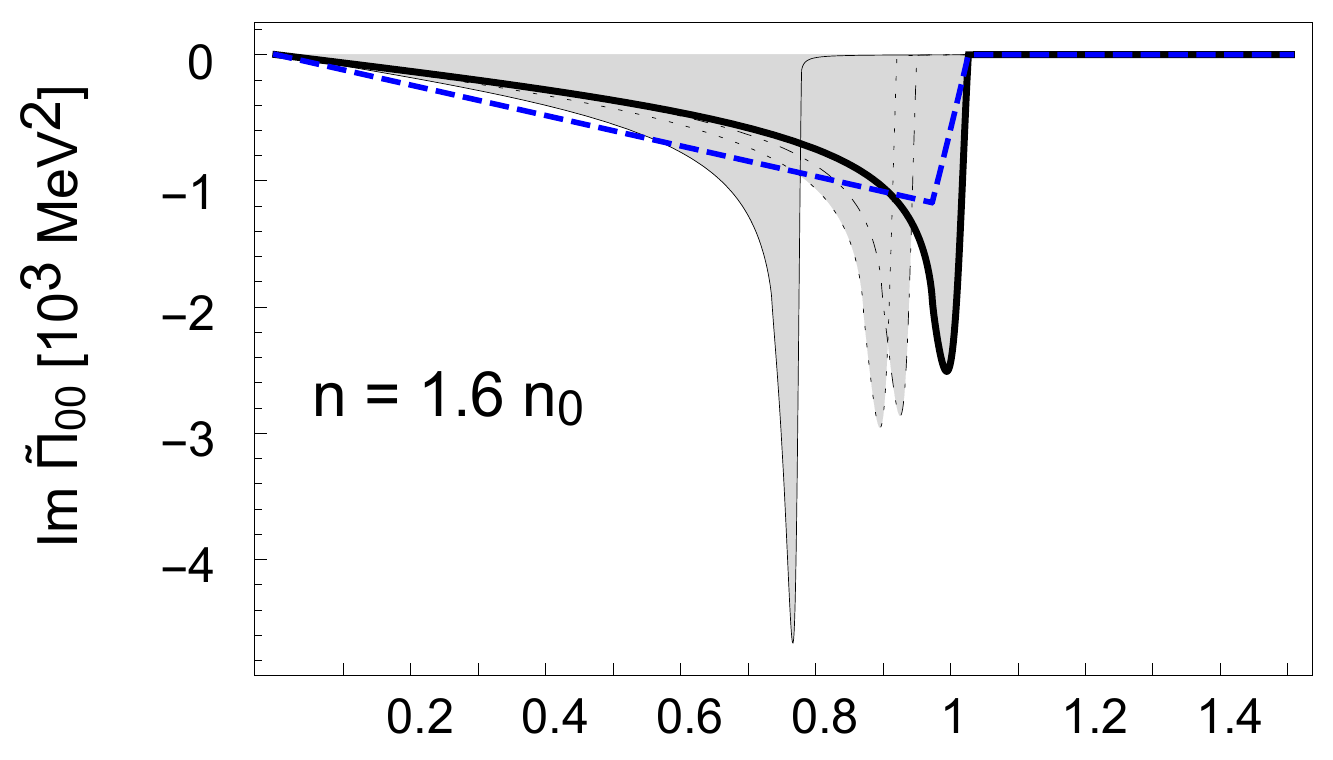}\\
~~~\includegraphics[scale=0.54]{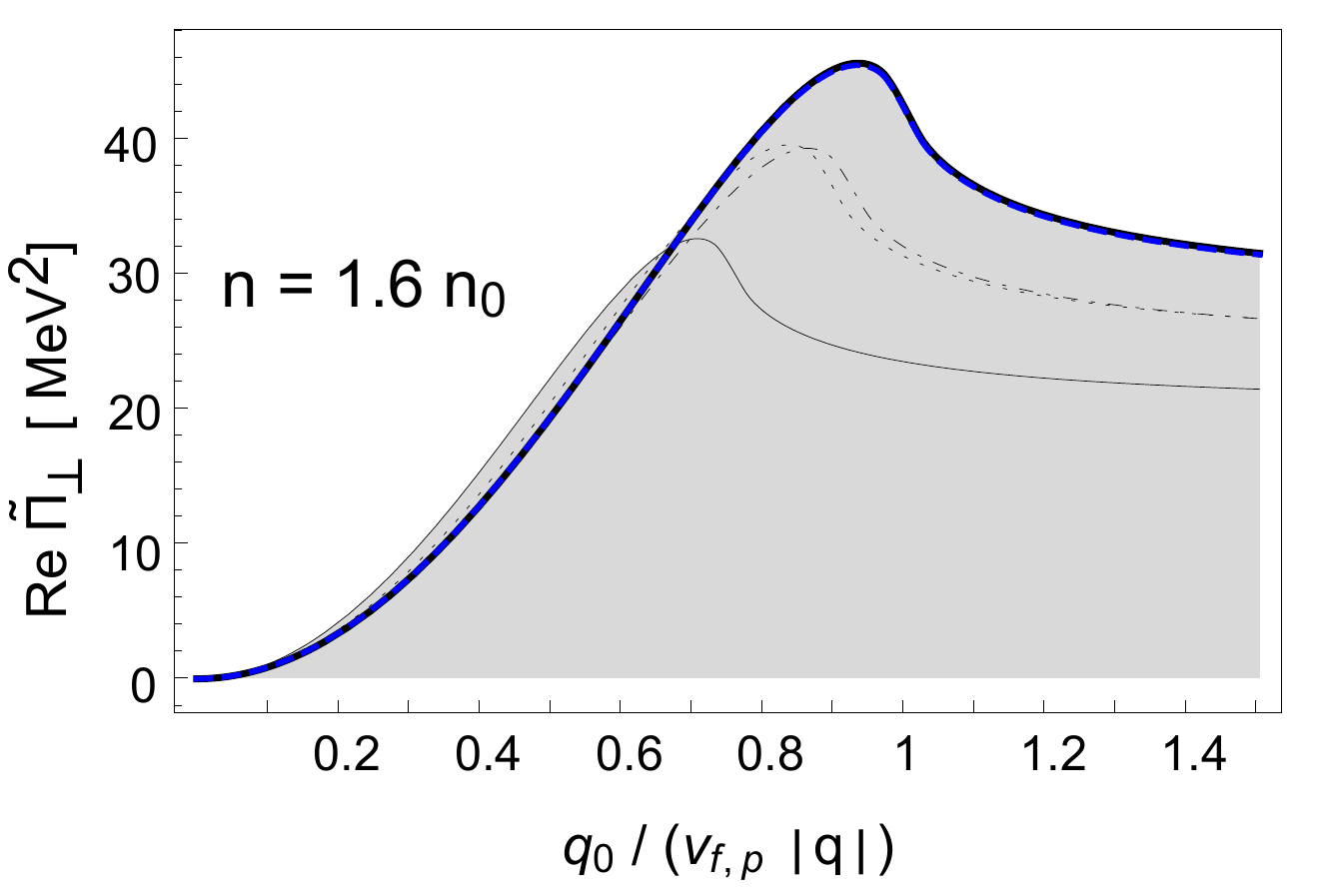}~~\includegraphics[scale=0.57]{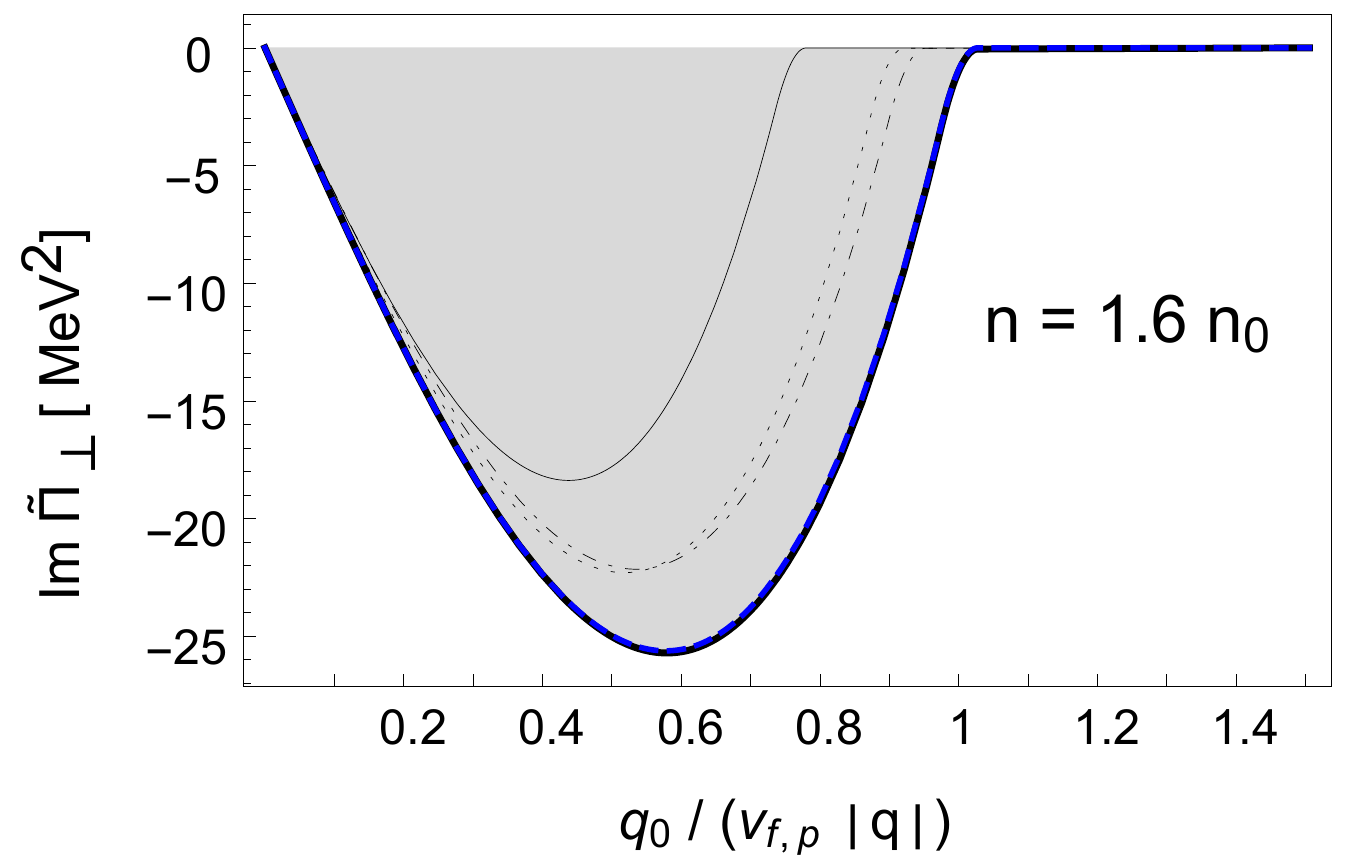}
\par\end{centering}
\caption{\label{fig:ReInduced} Real and imaginary parts of the proton contribution to the photon polarization functions $\tilde{\Pi}_{00}$, Eq. \ref{eq:ReProtonFull}, resummed in the subspace of protons and neutrons interacting via strong forces (see Eq. \ref{eq:ReProtonFull}). In each plot, the momenta $\left| \boldsymbol{q} \right|$ are fixed at 10 MeV. Since $\beta$ equilibrium has to be determined separately for each Skyrme parameter set, the Fermi momenta of protons and neutrons vary in each case. NRAPR parameters (thick black line) which predict the largest proton fraction are chosen as reference and  $q_0$ is normalized over $v_{f,\,p}\,\left| \boldsymbol{q} \right|$ (where the Fermi velocity of the protons consequently also corresponds to the NRAPR set). The blue dashed line displays the polarization effects from protons in the absence of nuclear interactions, again using NRAPR parameters. The peaks of real and imaginary parts of all other parameter sets are hence shifted to the left of the NRAPR peak. Various densities are shown; the last row shows the corresponding transverse polarization at $1.6\,n_0$ for comparison. The impact of induced interactions is completely negligible for transverse components at any density but sizable for longitudinal ones, in particular for densities below $n_0$. At saturation density, the results with and without induced interaction are roughly comparable. At higher densities, results are fairly model dependent. The qualitative tendencies of the impact of induced interactions are, however, consistent in all tested models.}      
\end{figure}
\begin{figure}
\begin{centering}
\textbf{$\hspace{1cm}$ without induced interactions $\hspace{2.5cm}$ with induced interactions}\par\medskip
~~\includegraphics[scale=0.615]{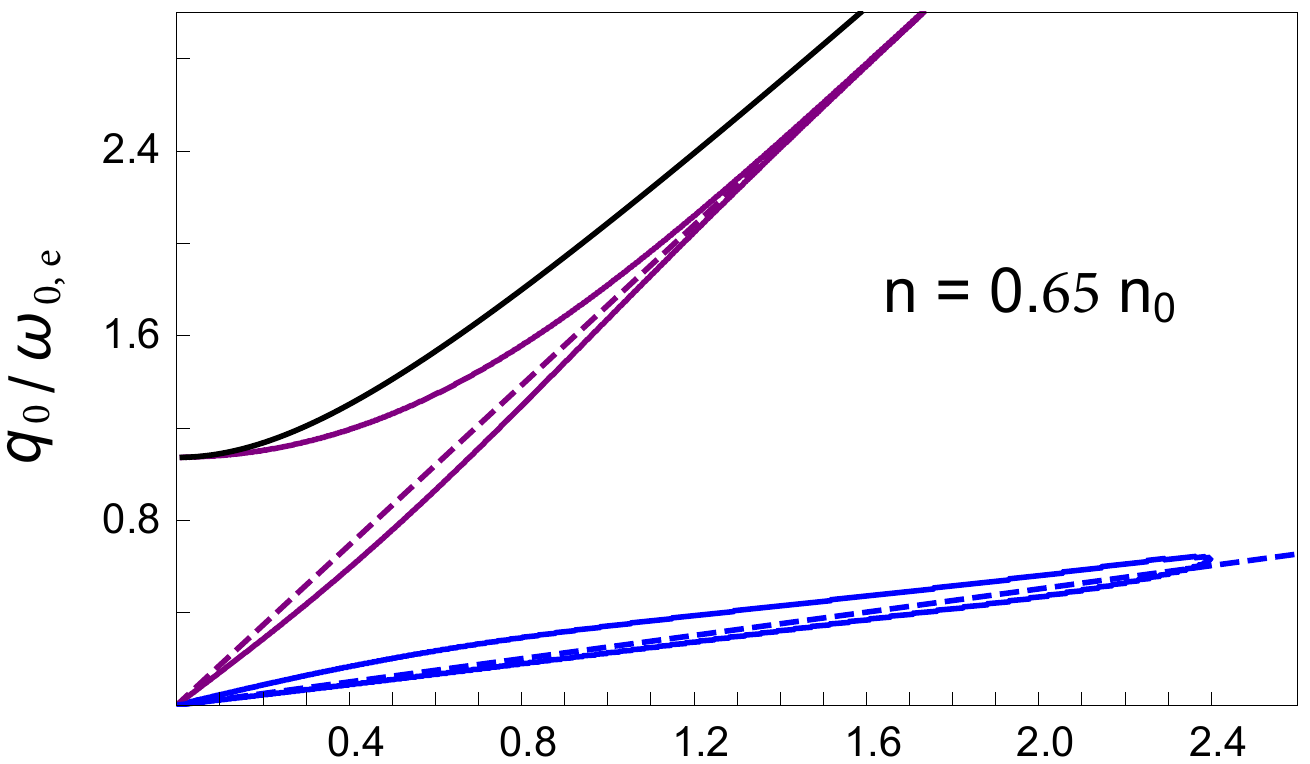}~~~~\includegraphics[scale=0.61]{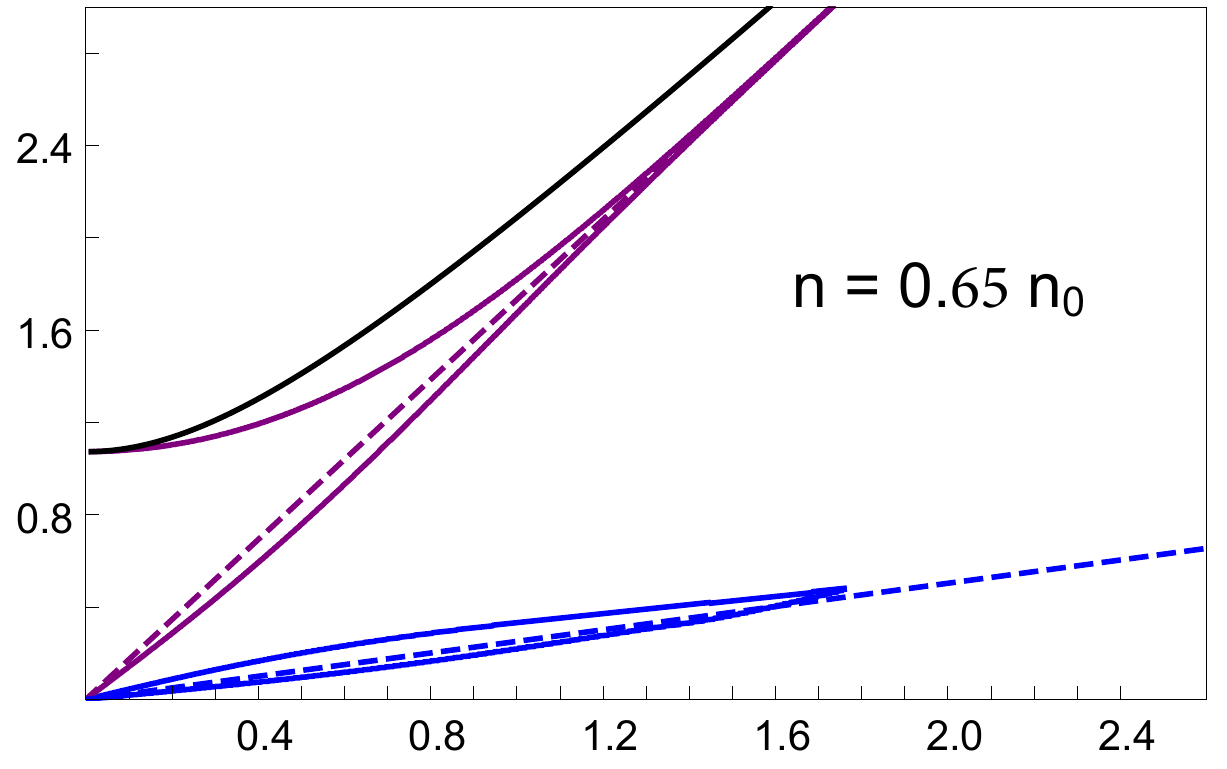}\\
\includegraphics[scale=0.62]{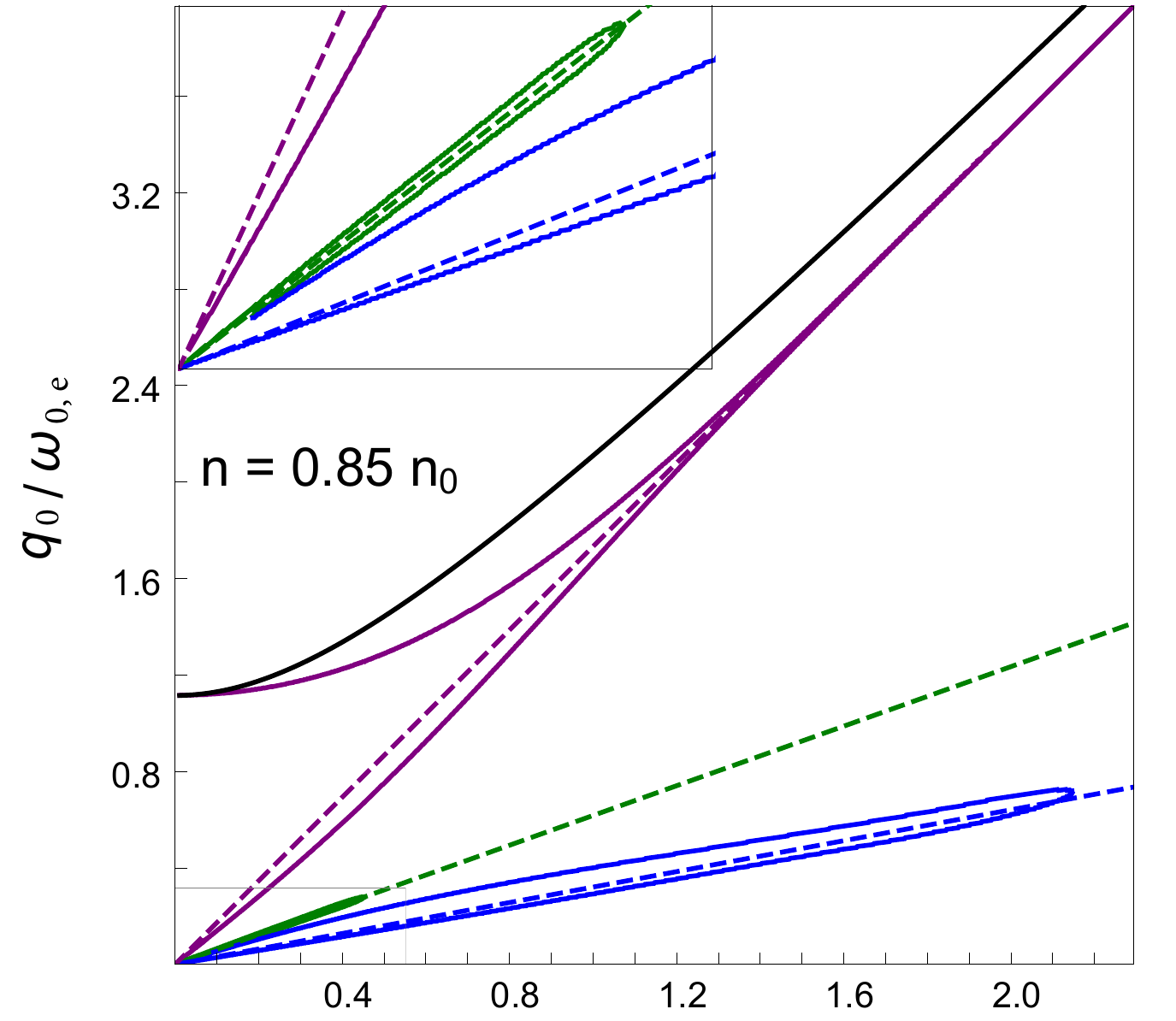}~\includegraphics[scale=0.62]{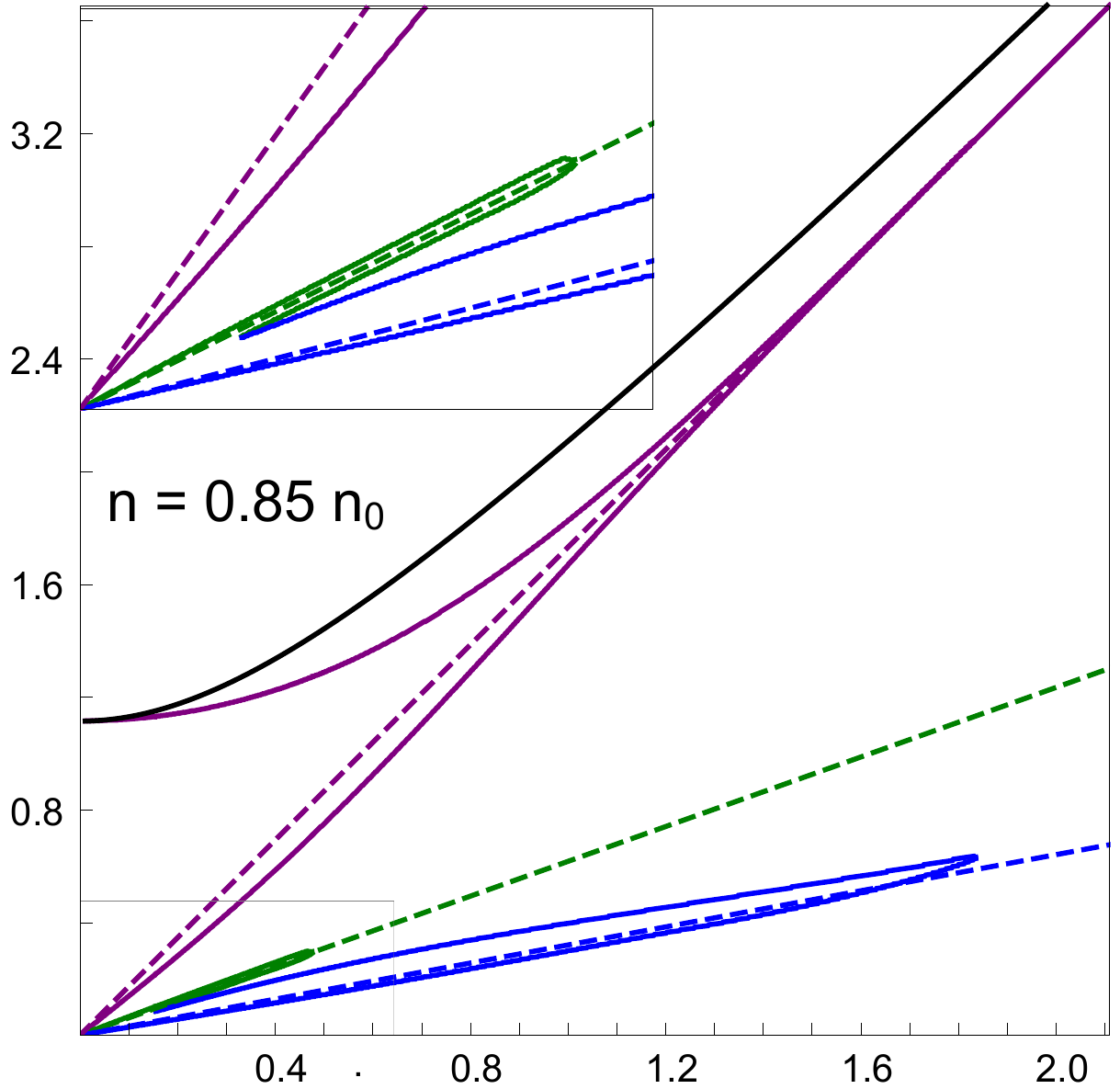}\\
~\includegraphics[scale=0.62]{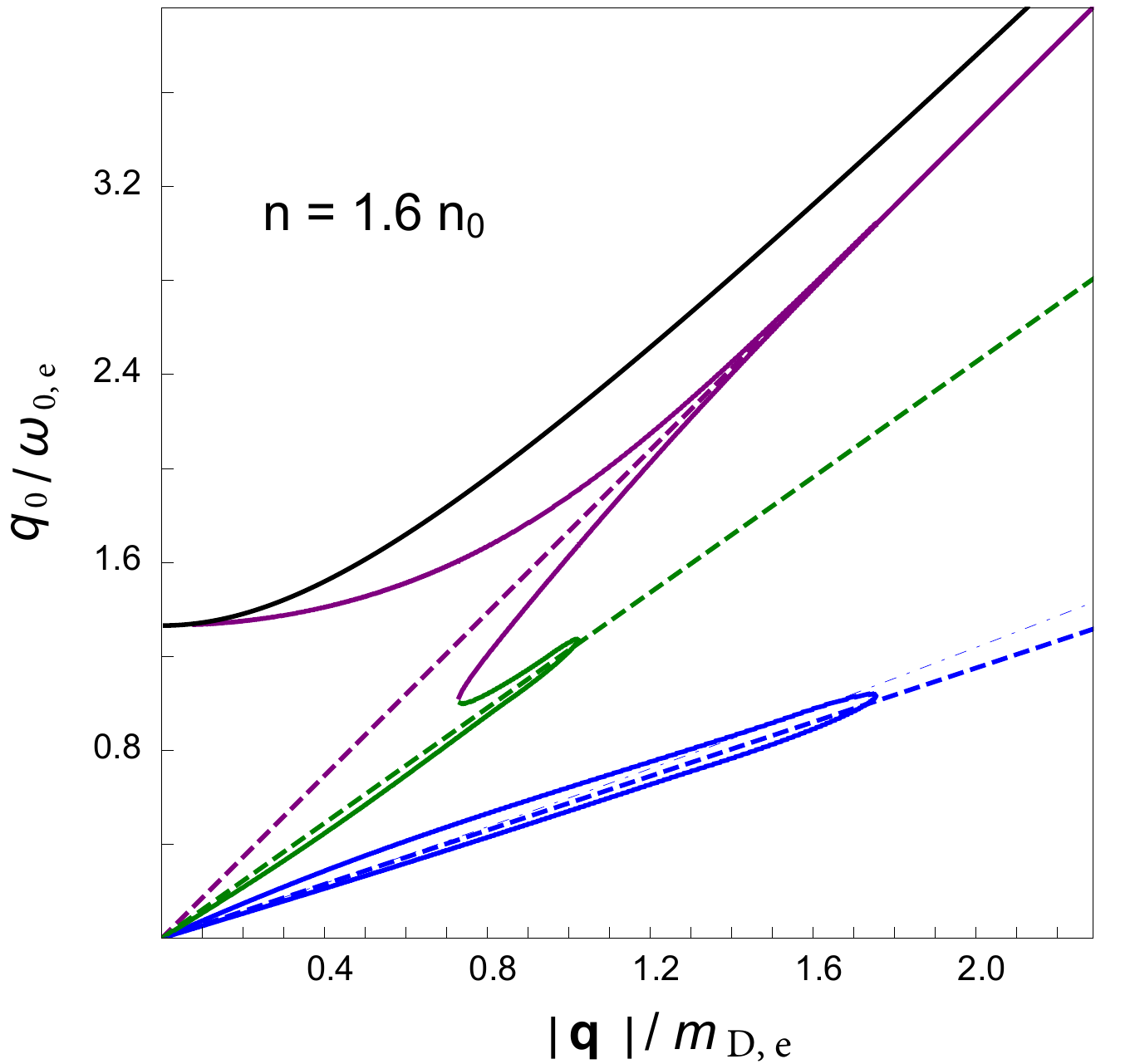}\includegraphics[scale=0.62]{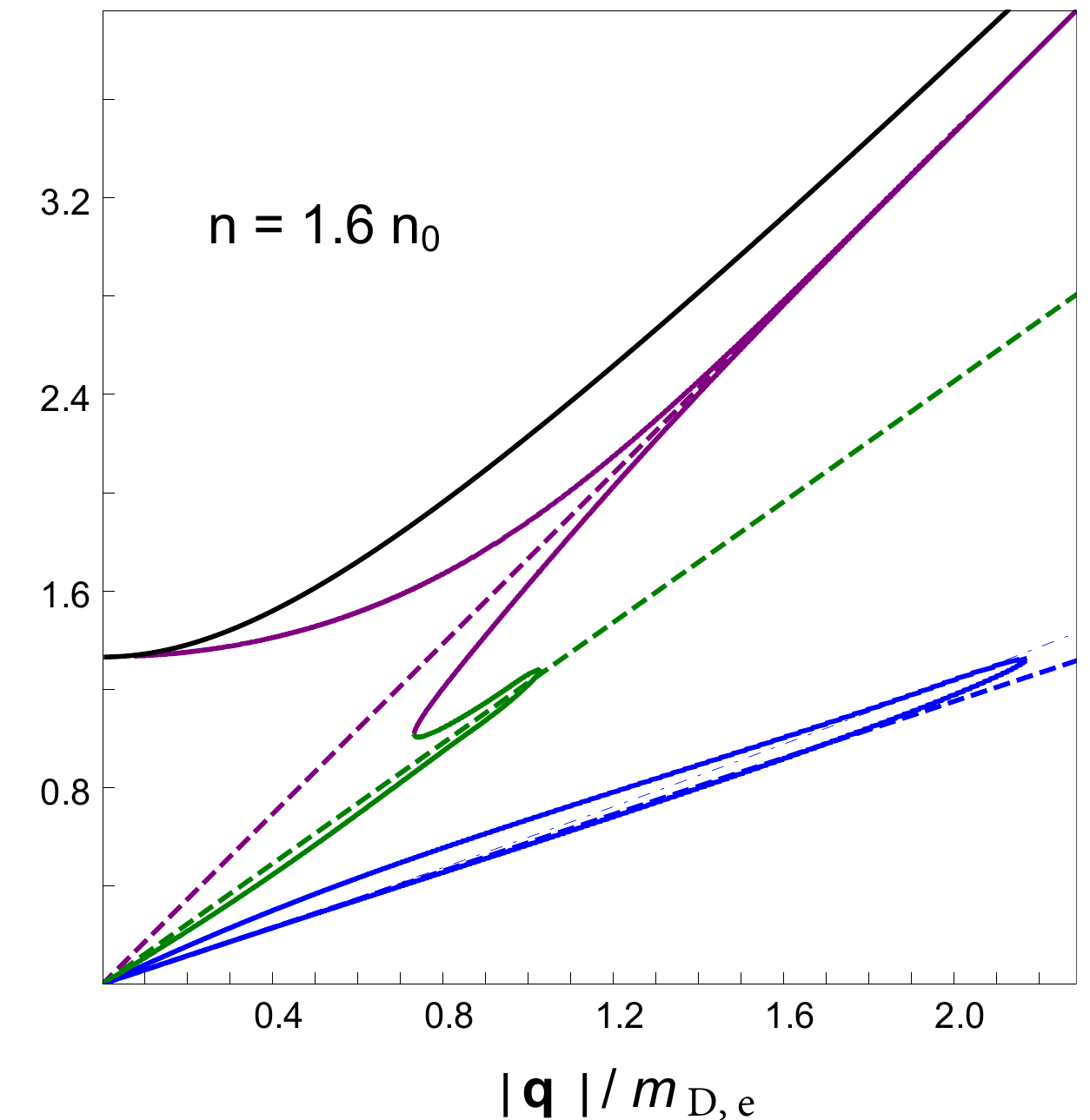}
\par\end{centering}
\caption{\label{fig:AllModes} Poles of the real part of the longitudinal and transverse photon propagator in the usual color coding, using the NRAPR parameter set, with and without induced interactions. At saturation density, both scenarios lead to almost equal results (i.e., to the spectrum displayed in Fig. \ref{fig:EPMmodes}), mainly because there bare and induced screening are roughly of equal magnitude; see Fig. \ref{fig:NucScreening}. Dashed lines correspond to $v_f \left|\boldsymbol{q}\right|$ for each particle species. For comparison, we additionally show the dot-dashed line in the bottom two figures below which the proton contribution to Landau damping sets in. The ``tip of the thumb" is always located below this line.}      
\end{figure}
\noindent The real and imaginary parts of the proton contributions to the polarization functions in the presence of induced interactions are displayed in Fig. \ref{fig:ReInduced}. The densities are fixed at $n=0.65\, n_0$ (where muons are absent), $n=n_0$,  and  $n=1.6 \,n_0$, and the cases with and without induced interactions are compared using various modern Skyrme forces (see Appendix \ref{sub:Skyrme}). Correlations with nuclear interactions induce a sizable change of the longitudinal polarization functions. At sufficiently low densities close to the onset of the spinodal instability, the impact of nuclear interactions is most pronounced: At very low energies $q_0$, the longitudinal polarization is sensitive to the rapid increase of the static screening  displayed in Fig. \ref{fig:NucScreening}. The critical density $n_c$  calculates to a slightly different value in each Skyrme model and the impact of the induced interactions on longitudinal polarization appears to vary strongly  in Fig. \ref{fig:ReInduced}. This is however merely a consequence of the fact that at fixed density some models are closer to the instability while others are farther away. Putting together the spectral function, we shall find that the result at low densities is indeed fairly independent of the chosen Skyrme model. Induced interactions additionally lead to a reduction of the magnitude of the peak around $q_0=v_{f,\,p}\left|\boldsymbol{q}\right|$.\newline  Close to saturation density the cases with and without induced interactions are roughly comparable in magnitude. At $n=1.6\,n_0$ induced interactions reduce the magnitude of longitudinal polarization functions, again following the evolution of the static screening with increasing density. In the transverse channel, the impact of induced interactions is entirely negligible as expected.\newline
Next we turn to the zeros of the real part of the photon propagator \ref{eq:PropEMPNfull}. It is interesting to take another look at the analytical results of $\omega_L$ and $\omega_\perp$ at low momenta. As a result of the (strong) resummation, the plasma frequencies are modified even in the limit $\boldsymbol{q}\rightarrow \boldsymbol{0}$, and the resulting expressions are quite complicated. For longitudinal and transverse modes, one finds
\vspace{1mm}
\bea  \label{eq:wLEMPN}
\omega_{L,\,e}^{2}&=&\tilde{\omega}_{0,\,L}^{2}+\left\{ \frac{5}{3}\left(\omega_{0,\,e}^{2}v_{f,\,e}^{2}+\omega_{0,\,\mu}^{2}v_{f,\,\mu}^{2}\right)+e^{2}\tilde{\omega}_{p,\,L}^{2}\left[1+\left(-1+\frac{5}{3}v_{f,\,p}^{2}\right)\,\frac{\tilde{\omega}_{p,\,L}^{2}}{\omega_{0,\,p}^{\prime2}}+g_{n}^{2}\left(-1+\frac{5}{3}v_{f,\,n}^{2}\right)\frac{\tilde{\omega}_{p\,L}^{2}}{\omega_{0,\,n}^{\prime2}}\right]\right\} \frac{\boldsymbol{q^{2}}}{\bar{\omega}_{0,\,L}^{2}}
\,,\\[4ex] \label{eq:wPerpEMPN}
\omega_{\perp}^{2}&=&\tilde{\omega}_{0,\,\perp}^{2}+\left\{ \left(1+\frac{1}{5}v_{f,\,e}^{2}\right)\omega_{0,\,e}^{2}+\left(1+\frac{1}{5}v_{f,\,\mu}^{2}\right)\omega_{0,\,\mu}^{2}+e^{2}\tilde{\omega}_{p,\perp}^{2}\left[1+\frac{1}{5}\left(v_{f,\,p}^{2}\,\frac{\tilde{\omega}_{p,\,\perp}^{2}}{\omega_{0,\,p}^{\prime2}}+\bar{g}_{n}^2\,v_{f,\,n}^{2}\,\frac{\tilde{\omega}_{p,\,\perp}^{2}}{\omega_{0,\,n}^{\prime2}}\right)\right]\right\} \frac{\boldsymbol{q^{2}}}{\bar{\omega}_{0,\,\perp}^{2}}
\,.
\eea 

\noindent In the above formulae, neutron-proton interactions manifest themselves implicitly in the resummed plasma frequencies $\tilde{\omega}_p$  and explicitely in the dimensonless couplings $g_n=\tilde{\omega}_n^2\,f_{pn}$ and $\bar{g}_n=\tilde{\omega}_n^2\,\bar{f}_{pn}$.  The RPA expressions for $\tilde{\omega}_p$  and $\tilde{\omega}_n$ are  
\vspace{1mm}\label{eq:wresum}
\be
\tilde{\omega}_{p,\,L}^2=\frac{\omega_{0,\,p}^{\prime2}\,(1+f_{nn}\,\omega_{0,\,n}^{\prime2})}{1+f_{pp}\,\omega_{0,\,p}^{\prime2}+f_{nn}\,\omega_{0,\,n}^{\prime2}+\omega_{0,\,p}^{\prime2}\,\omega_{0,\,n}^{\prime2}(f_{pp} f_{nn}-f_{pn}^{2})}\,,\hspace{1cm}\tilde{\omega}_{n,\,L}^2=\frac{\omega_{0,\,n}^{\prime2}}{1+f_{nn}\omega_{0,\,n}^{\prime2}}\,,
\ee 

\vspace{1mm}
\noindent the quantities $\omega_0^{\prime}$ are again defined without a factor of $e^2$, and the perpendicular quantities $\tilde{\omega}_{p,\,\perp}$ and $\tilde{\omega}_{p,\,\perp}$ are obtainable by replacing $f$ with $\bar{f}$. By setting nuclear interactions to zero, one recovers \ref{eq:wLEPM} and \ref{eq:wPEPM}. Since density-density and current-current interactions also affect the plasma frequencies $\omega_0$ one finds a \textit{mode splitting}
\bea 
\tilde{\omega}_{0,\,L}^2 & =& \omega_{0,\,e}^2+\omega_{0,\,\mu}^2+e^2\,\tilde{\omega}_{p,\,L}^2\,,\\[2ex]
\tilde{\omega}_{0,\,\perp}^2 & = & \omega_{0,\,e}^2+\omega_{0,\,\mu}^2+e^2\,\tilde{\omega}_{p,\,\perp}^2\,,
\eea
i.e. the plasmon and photon modes do not coincide in the limit $\boldsymbol{q}\rightarrow\boldsymbol{0}$.  We note that strong interactions appear to distinguish between longitudinal and transverse channels. In our approach this distinction arises from the difference between $l=0$ and $l=1$ Landau parameters. In the simplified RMF model discussed in Appendix \ref{sub:RMF} where only the vector interaction is included, the longitudinal and transverse modes are degengerate, but in a more complete treatment of RMF forces one needs to include scalar and iso-scalar mesons, and this again breaks the degeneracy between longitudinal and transverse modes \cite{Reddy:1998hb}\cite{Lim:1989} \cite{Horowitz1991}. The observed splitting is small because nucleons are essentially non-relativistic and leptons dominate the plasma frequency. Fig. \ref{fig:AllModes} displays to solutions of Eq. \ref{eq:wLEMPN} and \ref{eq:wPerpEMPN} for the three densities $0.65\,n_0$, $0.85\,n_0$ and $1.6\,n_0$ and compares the cases with and without induced interactions. The role of the neutrons is to modify the response of the protons: At lower densities the proton sound mode $u_p$ (together with its overdamped companion $\omega_{<,\,p}$) is restrained to lower momenta $\left|\boldsymbol{q}\right|$ in the presence of induced interactions. This situation is reversed at higher densities. At densities below saturation the Fermi velocities of muons are close to those of protons, leading to a merging of neighboring muon-like and proton-like solutions. As a result, there is no distinct proton sound mode in the spectrum at very low energies; see the second row of Fig. \ref{fig:AllModes}. At densities above saturation, the Fermi velocities of muons are closer to those of electrons such that this time there is no distinct muon sound mode at very low energies (third row of Fig. \ref{fig:AllModes}). This effect persists in the absence of induced interactions and depends mainly on the evolution of the relative densities of the various components in the plasma.  \newline 
The screening mass of the photon subject to strong and electromagnetic interactions calculates to 
\be \label{eq:Screentot} 
m^2_{D,\,\textrm{tot}}=F_e(q_0=0)+F_\mu (q_0=0) + \tilde{F}_{p\,\textrm{nucl}}(q_0=0) = m^2_{D,\,e}+m^2_{D,\,\mu}+e^2\, \tilde{m}^2_{D,\,p}
\ee 
where using the static limit of the resummed polarization tensor [Eq. \ref{eq:ReProtonFull}] one finds that $\tilde{m}^2_{D,\,p}$ indeed agrees with Eq. \ref{eq:DebyePN}. The additional factor of $e^2$ indicates that the screening is of order $\alpha_f$. The static limit of the total (strong and electromagnetic) screening is obtained from Eq. \ref{eq:ScreenEMP} upon replacing $m_{D,\,p}^2$ by $e^2\,\tilde{m}_{D,\,p}^{\prime\,2}$.\newline
Finally, the spectral densities can yet be obtained from \ref{eq:RhoMulti} and \ref{eq:RhoMulti2}, with the usual replacements of $F$ and $G$ with $\tilde{F}$ and $\tilde{G}$. The results are displayed in Fig. \ref{fig:RhoInduced} for the densities $0.65\,n_0$, (a) - (c), $0.85 n_0$, (d) - (f), and $1.6\,n_0$, (g) - (i), for different momenta, using NRAPR Skyrme forces. Induced interactions significantly impact the proton peak (corresponding to the sound-like solution $u_{p}$) in the longitudinal photon spectrum: At lower densities and (moderately) large momenta, it becomes strongly suppressed; see (c) and (f). As a result, there no longer exists a well-defined excitation associated with the collective response of protons in the system. At higher densities, the situation is reversed and the proton peak in the presence of induced interactions outgrows the peak calculated in their absence, albeit by a much smaller margin (i). Since the static (strong) screening in Fig. \ref{fig:NucScreening} assumes a minimum roughly around $n=1.6 \,n_0$, the proton mode will not be promoted to a sharp peak in the spectrum at any higher density, leaving the electron-like plasmon mode $\omega_L$ as the only longitudinal collective mode in the spectrum. Fig. \ref{fig:RhoCompare} compares the results of the proton contribution to the longitudinal spectral function using different Skyrme models. At low densities results are fairly robust, but the model dependence obviously increases with density. \newline 
The considerable modifications of the longitudinal spectrum at lower densities further strengthens the hypothesis of Heiselberg and Pethick  \cite{Heiselberg:1992ha} \cite{Heiselberg:1993cr} that scattering rates of fermions in the plasma (and therefore their corresponding contributions to transport) are dominated by the exchange of transverse photons. The situation changes at higher densities where the induced interactions are in general less relevant. The fact that the low density region is the domain where the impact of nuclear interactions is well under control is fortunate, as it resembles the phenomenologically very relevant crust-core boundary region of neutron stars.    
\begin{figure}
\begin{centering}
\textbf{$\left|\boldsymbol{q}\right|=0.3\,m_{D,\,e}\hspace{4cm}\left|\boldsymbol{q}\right|=\,m_{D,\,e}\hspace{4cm}\left|\boldsymbol{q}\right|=1.8\,m_{D,\,e}$}\par\medskip
\includegraphics[scale=0.49]{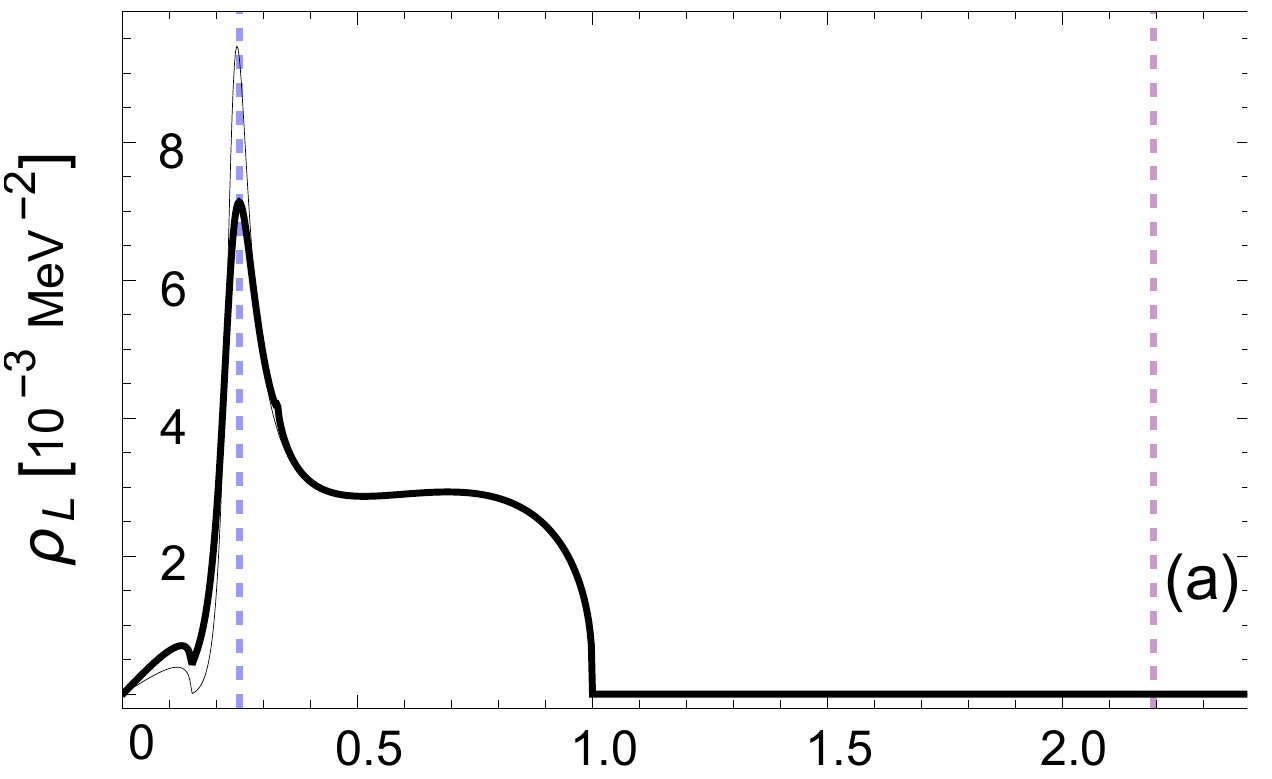}~~\includegraphics[scale=0.49]{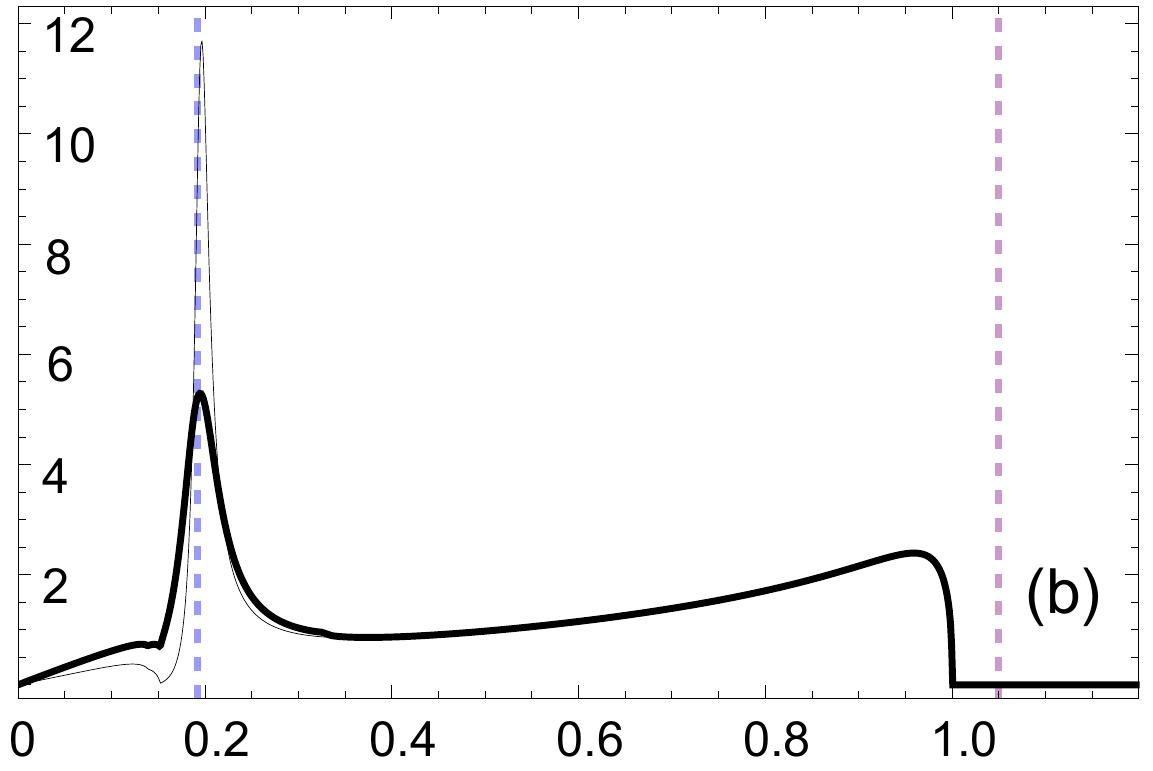}~
\includegraphics[scale=0.48]{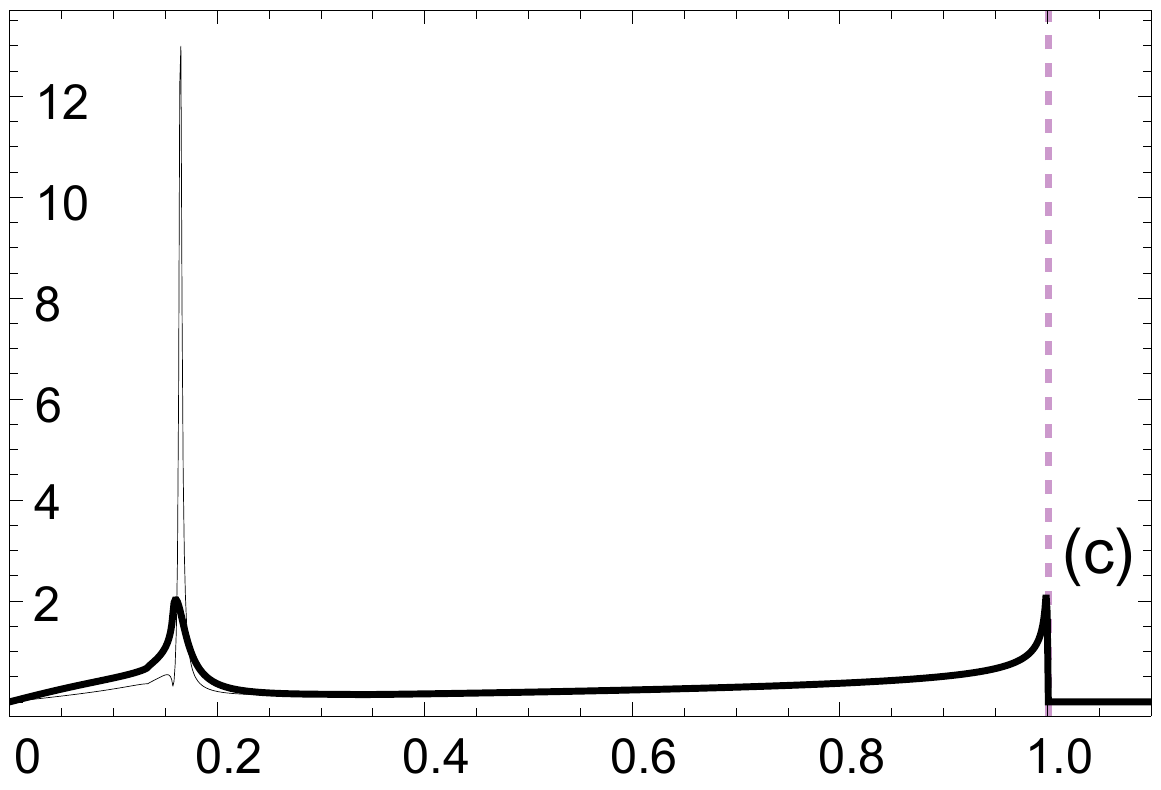}\\
\includegraphics[scale=0.49]{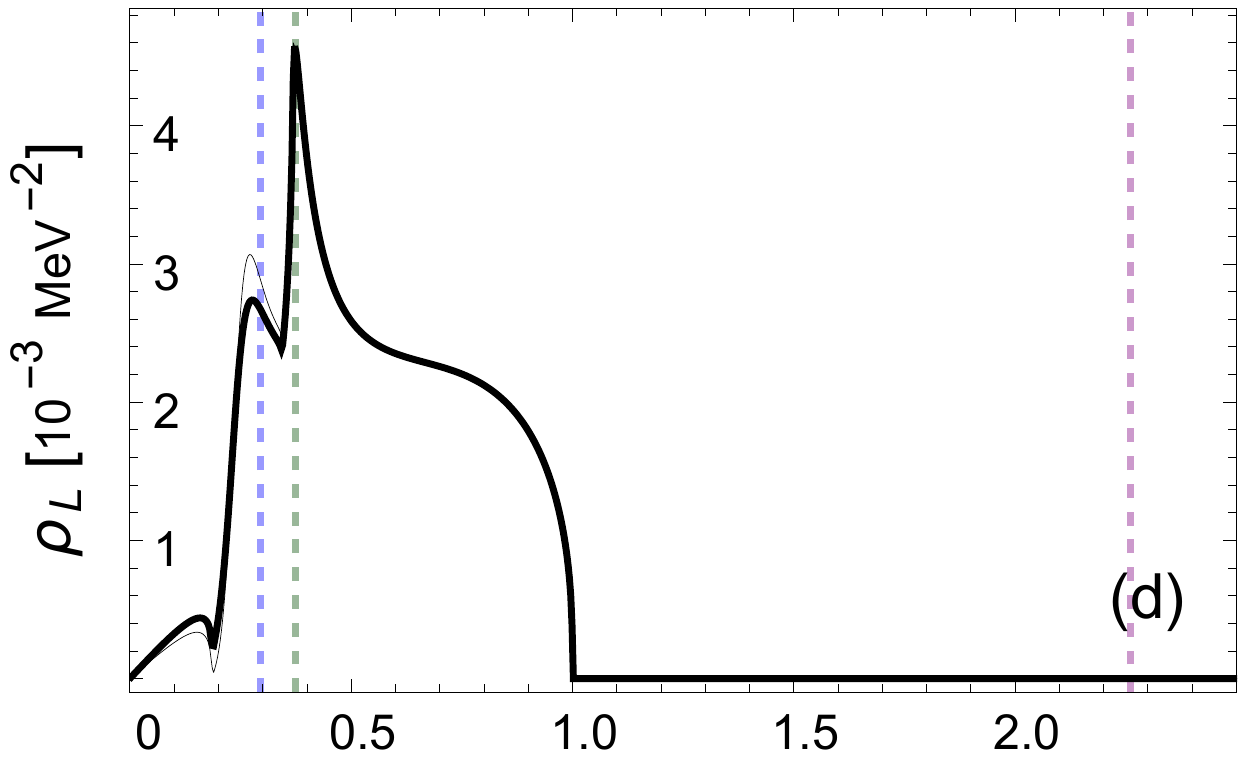}~~~\includegraphics[scale=0.49]{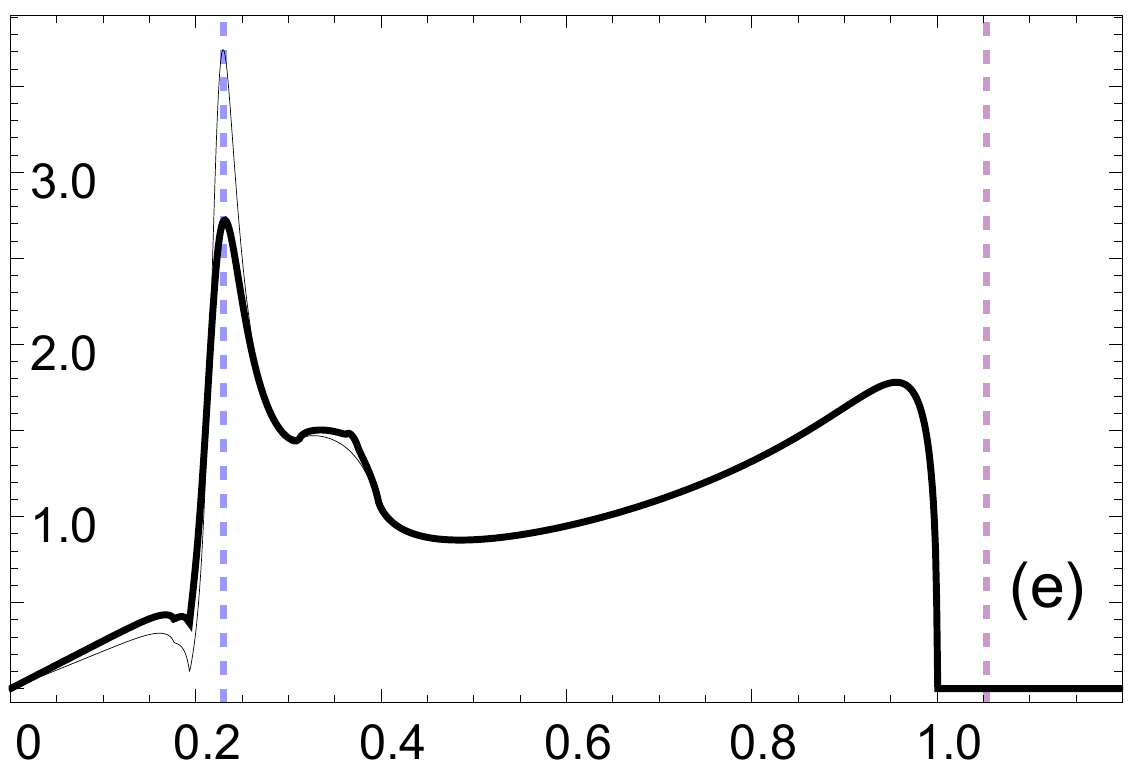}~~~
\includegraphics[scale=0.47]{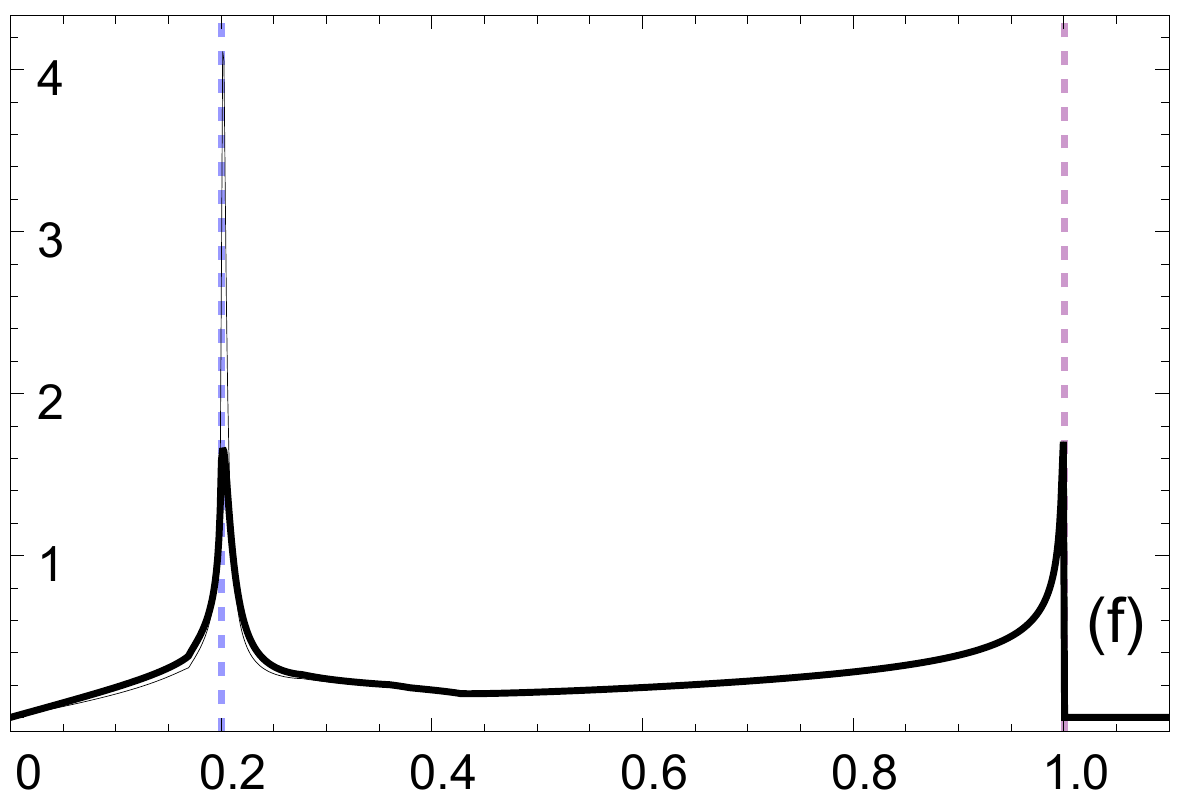}\\
\includegraphics[scale=0.50]
{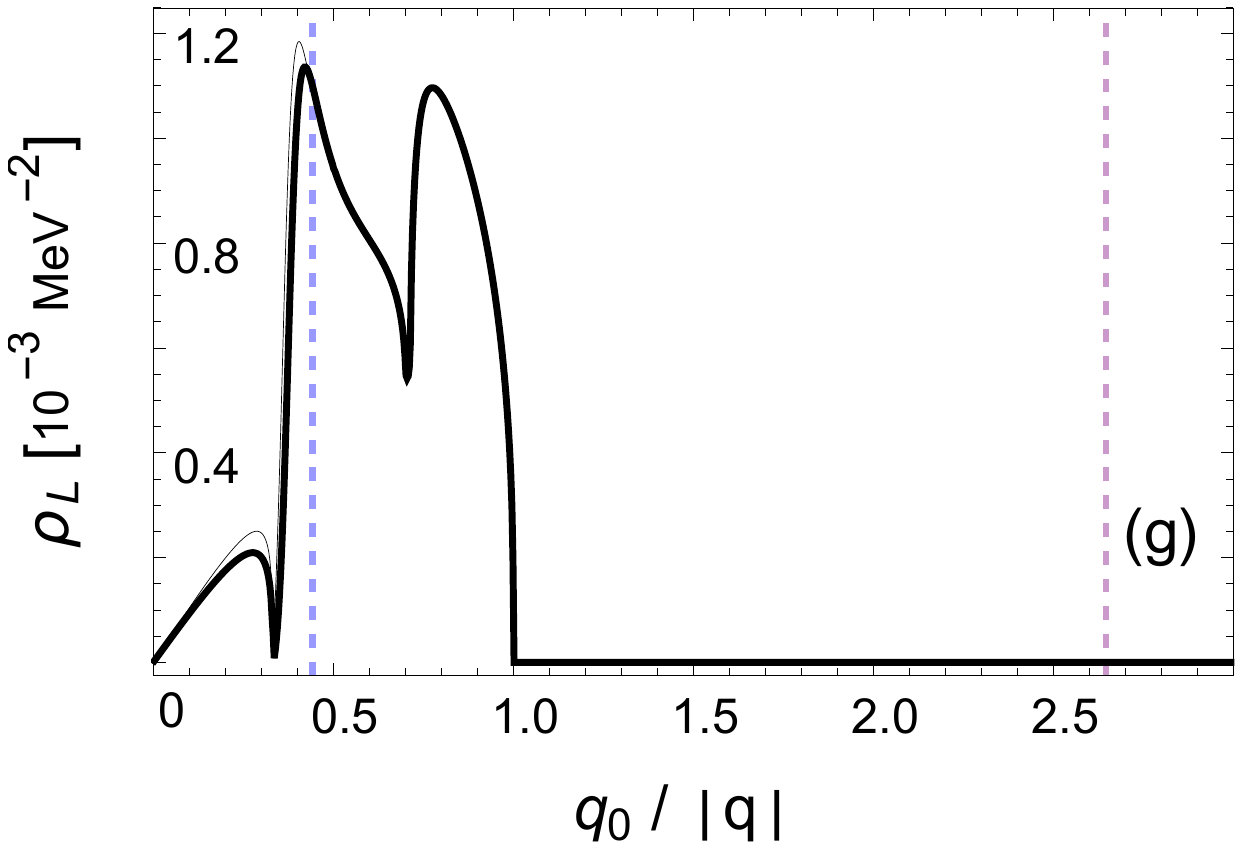}~~
\includegraphics[scale=0.49]
{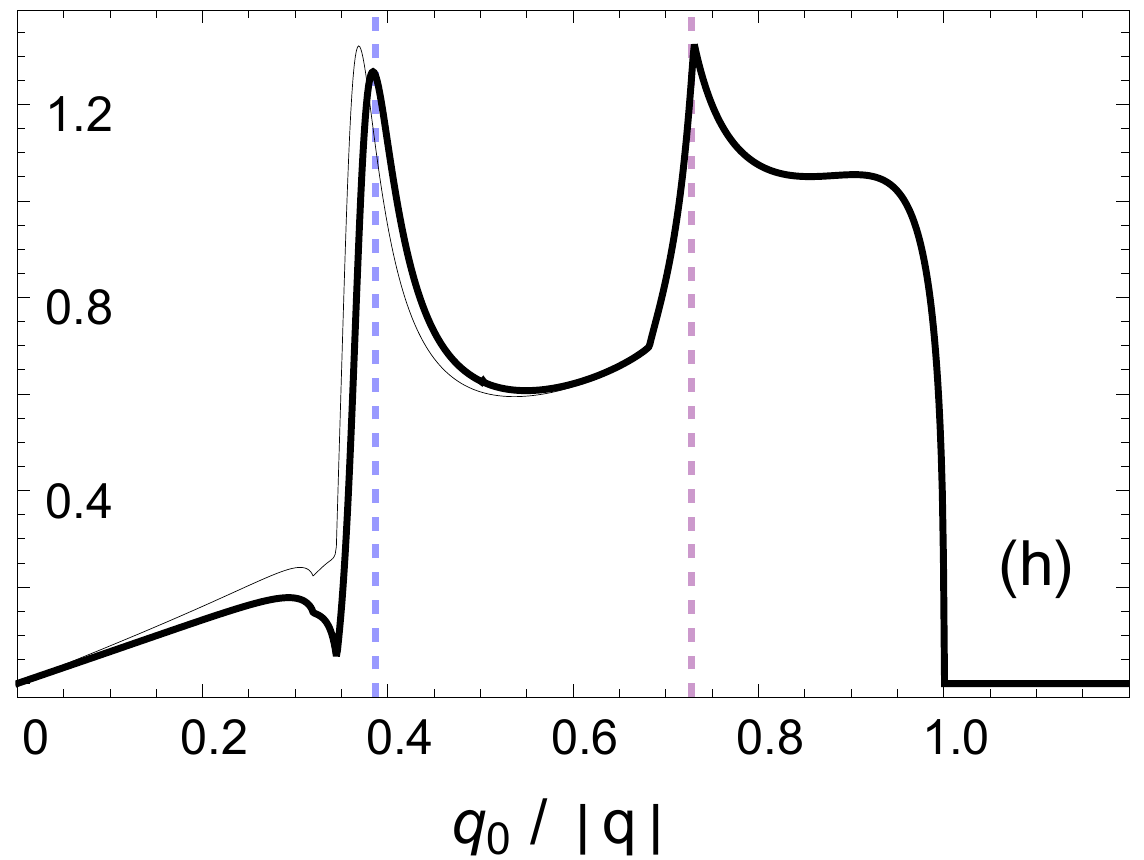}~~
\includegraphics[scale=0.48]
{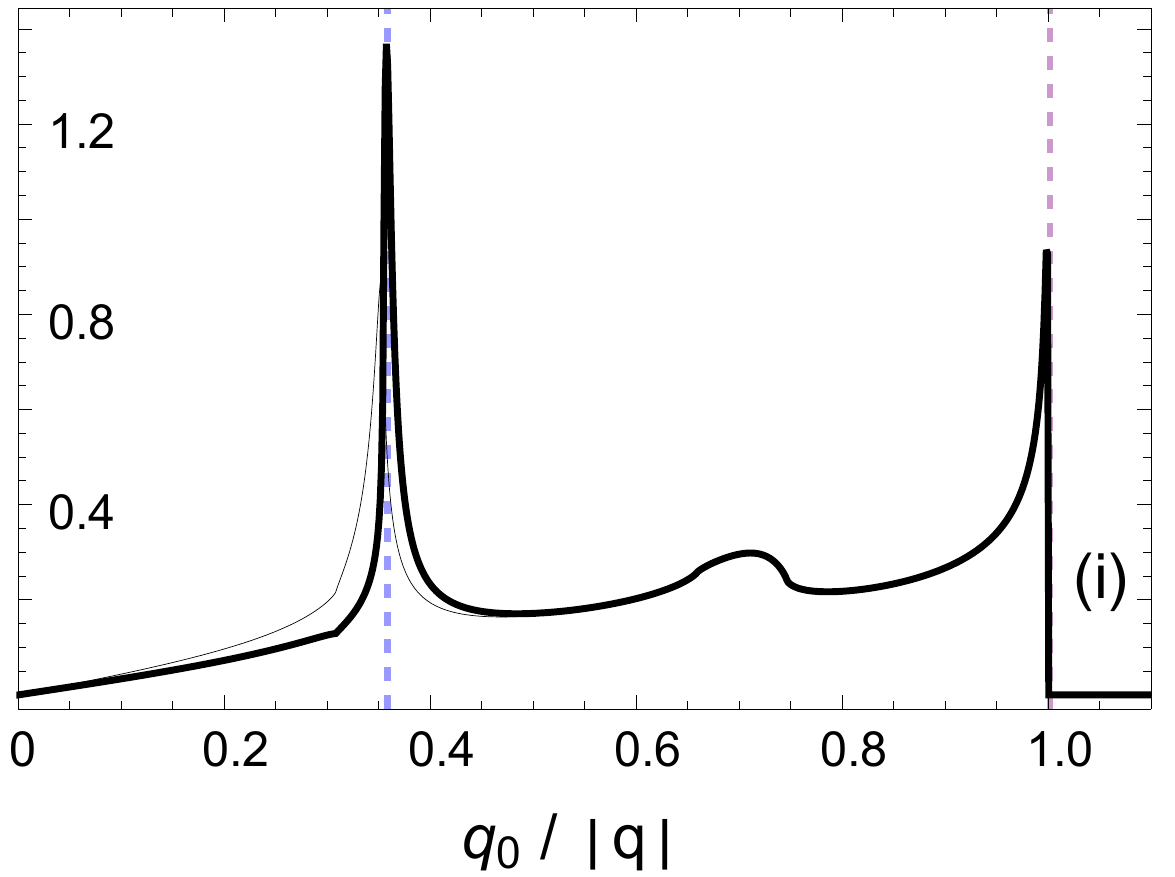}
\par\end{centering}
\setlength{\belowcaptionskip}{-15pt}
\caption{\label{fig:RhoInduced} Evolution of the longitudinal spectral function with increasing momenta using the NRAPR parameter set for $n=0.65\, n_0$ (first row, muons are absent), $n=0.85 n_0$ (second row) and  $n=1.6 \,n_0$ (third row). Thick black lines correspond to the case with induced interactions; thin black lines correspond to the case without. Vertical dashed lines indicate the positions of the electron plasmon mode $\omega_{L}$ and the muon and proton sound modes $u_{p}$  (the overdamped lower branches $\omega_{<}$ are not displayed). }      
\end{figure}

\begin{figure}
\begin{centering}
\includegraphics[scale=0.94]{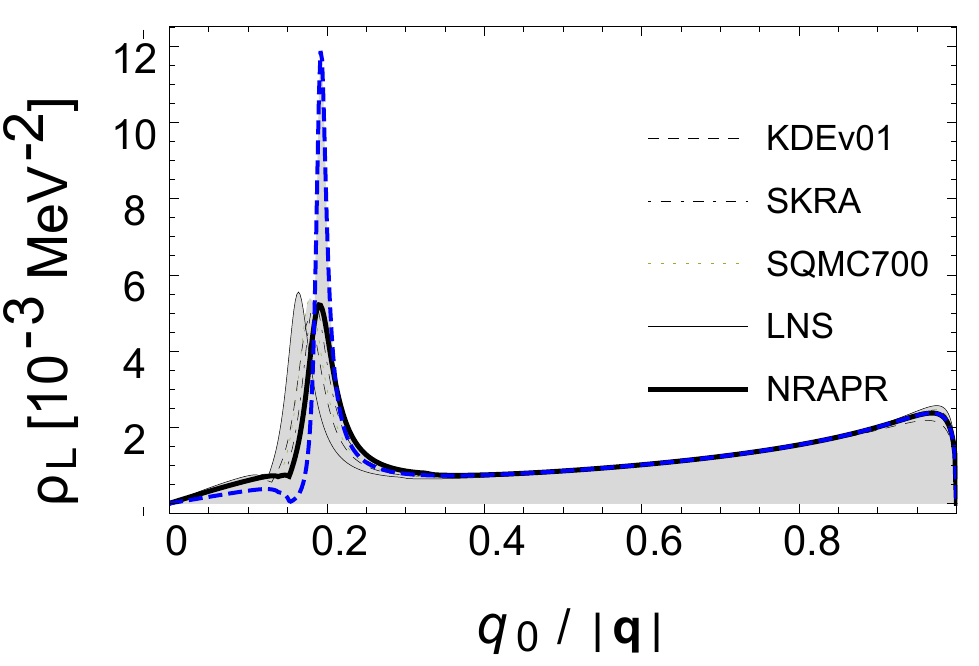}~~\includegraphics[scale=0.67]{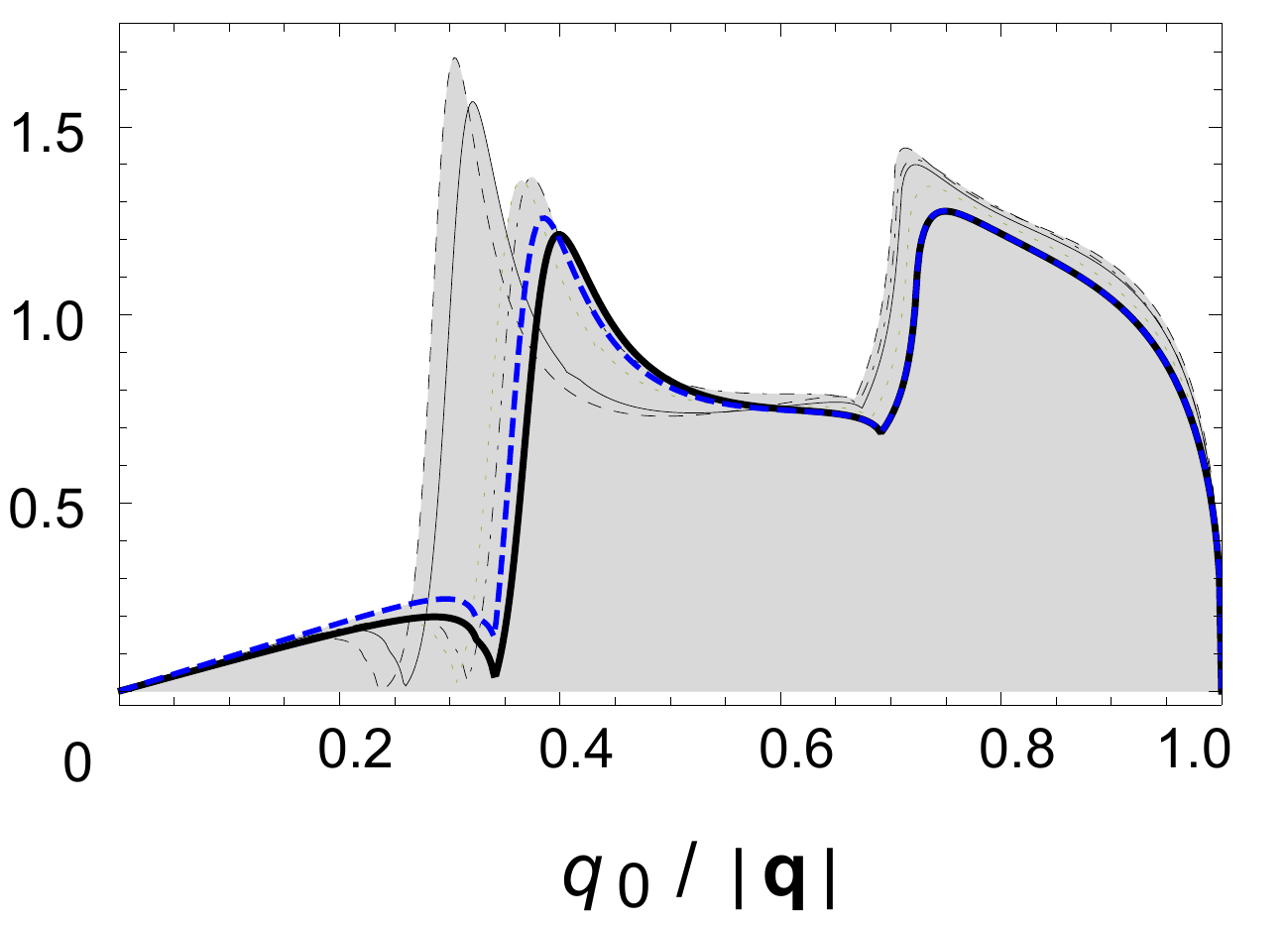}~
\par\end{centering}
\setlength{\belowcaptionskip}{-10pt}
\caption{\label{fig:RhoCompare} Comparison of the various Skyrme parameter sets at $n=0.65\, n_0$ in the absence of muons (left) and  $n=1.6 \,n_0$ (right) at fixed photon momentum $\left| \boldsymbol{q} \right|$ = 10 MeV. Thick black lines correspond to the NRAPR parameter set which we use as a reference. The blue dashed line serves as a reference to the NRAPR result without induced interactions. The suppression of the proton peak at lower densities is to a reasonable degree independent of the chosen parameters; the  (small) enhancement at higher densities is much more model dependent. }      
\end{figure}

\newpage

\section{Conclusions and outlook}
\label{sec:conclusions}
\noindent We have investigated the photon spectrum in a dense relativistic plasma composed of electrons, muons, protons, and neutrons under neutron star conditions using the relativistic random phase approximation (RPA) to account for electromagnetic and strong correlations between them. By incorporating these correlations, we have shown they influence Landau damping and calculated the corresponding damping rates under degenerate conditions for arbitrary energies and momenta. The properties of nuclear matter including the residual quasi-particle interactions have been extracted from a Landau energy functional based on Skyrme forces and matched to the relativistic description. In the following, we summarize the main results and discuss potential applications and improvements. 

For small photon energies, our results are directly relevant for the calculation of transport coefficients. In this regime, collective effects are most pronounced and we show how interactions with neutrons induced by the (strong) polarizability of protons in the nuclear medium are consistently embedded in the resummed photon propagator. We find that dynamical screening effects due to Landau damping give rise to a complicated photon spectrum. Hard dense loop (HDL) approximations often prove insufficient to reproduce the RPA polarization functions, in particular in the presence of nucleons whose large masses violate the condition $\left|\boldsymbol{q}\right|\ll k_f$. Since there is no static screening for transverse polarizations, it is well known that the transverse spectrum is considerably larger than its longitudinal counterpart at low momenta. At densities below nuclear saturation density, we find that the incorporation of short-range strong interactions further enforces this feature: While leaving the transverse spectrum unmodified, they significantly reduce the magnitude of the proton peak in the longitudinal spectrum, located at relatively small momenta $\sim v_{f,\,p} \left|\boldsymbol{q}\right|$. To achieve a more realistic calculation of transport phenomena in degenerate relativistic plasmas, the physics of dynamical screening and induced lepton-neutron scattering should consequently be taken into account. The fact that induced interactions particularly impact the spectrum at densities corresponding to the crust-core boundary region makes them very relevant to the phenomenology of neutron stars. In this regard, the shear viscosity is of high interest as it acts as an important source of damping for hydrodynamic modes and r-modes, which in turn has implications for spin evolution of neutron stars and gravitational wave instabilities.

A qualitative understanding of the spectrum of collective modes is also relevant to study the decay of photons and plasmons into neutrino pairs, which is a potential mechanism for neutron star cooling.  In the most general case, the \textit{real part of the longitudinal propagator} exhibits a total of six distinct poles. Among these, the gapped plasmon mode is a weakly damped (undamped at one-loop order) collective excitation of the system with time-like properties. This mode is termed plasmon and is mainly an excitation of electrons. In addition, one finds two sound-like (gapless) solutions $u_{\mu}$ and $u_{p}$ which manifest themselves as local maxima in the spectrum but are damped by electron and electron-muon particle-hole excitations respectively. The remaining three excitations of electrons, muons, and protons are strongly damped, by particle-hole excitations of all species including their own, and generally do not coincide with any specific peak in the spectrum. The sound-like excitations owe their existence to the presence of lighter charged particles in the plasma which easily follow the motion of the heavier ones, providing efficient screening. Neglecting strong interactions, it may at first look like the proton excitation $u_{p}$ indeed develops a sharp peak in the photon spectrum.  Including strong interactions we find that none of the gapless excitations represents a well defined collective mode, leaving the (mostly) electron excitation $\omega_L$ as the only real longitudinal mode. Upon increasing momenta, all longitudinal solutions eventually cease to exist. The transverse propagator exhibits (two identical) modes which remain undamped in a one-loop calculation and continue to be present for any momentum as expected. 
A shortcoming which warrants further study is our approximate treatment of nuclear interactions which is strictly only valid in the limit where energies and momenta are small compared to typical energy and momentum scales in nuclear matter. While the effective particle-hole interaction in the long wavelength limit can be related to the equation of state and its thermodynamic derivates, its momentum dependence can be also extracted from microscopic approaches using resummation techniques described in Ref. \cite{abrikosov1975methods}. Such effective interactions have been derived recently in Ref. \cite{Benhar:2017oli} and we hope to include these in future work to examine the importance of the momentum dependence. We have further neglected effects due to Cooper pairing between nucleons and our results are not valid at very low temperature $T\le T_c$, where $T_c$ is the critical temperature for superfluidity or superconductivity. The critical temperature is $T_c \ll T_F $, where $T_F$ is the Fermi temperature but its magnitude is poorly known, and our results obtained in the limit when $T \ge T_F$ have a broad range of validity.  It is well known that proton superconductivity will bestow a mass for the transverse photon through the Anderson-Higgs mechanism and fundamentally change its character in the medium. Elementary excitations in a similar setup have recently been studied in Ref. \cite{Baldo:2017qju}. It would desirable to develop a consistent treatment of both particle-particle and particle-hole correlations using realistic nucleon-nucleon interactions, and this work here must be viewed as a first step towards this more ambitious goal.        
\section*{Acknowledgements}
The authors thank Ingo Tews, Jerome Margueron, Aleksey Cherman and Alessandro Roggero for helpful comments and discussions. S. S. was supported by the Schroedinger Fellowship of the FWF, project no.  J3639. S. R. was supported by the U.S. Depart- ment of Energy under Contract No. DE-FG0200ER41132. This work was supported
in part by the Natural Sciences and Engineering
Research Council (NSERC) of Canada, the Canada
Foundation for Innovation (CFI), and the Early Researcher
Award (ERA) program of the Ontario Ministry
of Research, Innovation and Science. Computational resources
were provided by SHARCNET and NERSC.
\newpage
\appendix
\section{Polarization tensor in RTF}
\label{sec:PI}
\noindent In this section, we review the calculation of the retarded photon polarization tensor in RTF where retarded, advanced and symmetric propagators are given by ~\ref{eq:SRetAdv} and ~\ref{eq:SSym} (see also Ref.\cite{Carrington:1997sq}). The full expression of the photon self energy in QED reads
\noindent 
\bea \label{eq:PIformal}
\Pi_{R}^{\mu\nu}(q) & = & -ie^{2}\int\frac{d^{4}k}{(2\pi)^{4}}\left\{ \text{Tr}\left[\gamma^{\mu}S_{S}(k^{\prime})\gamma^{\nu}S_{R}(k)\right]+\text{Tr}\left[\gamma^{\mu}S_{A}(k^{\prime})\gamma^{\nu}S_{S}(k)\right]\right\} \,\\[2ex]
\nonumber 
 &  & -\frac{ie^{2}}{2}\int\frac{d^{4}k}{(2\pi)^{4}}\left\{ \text{Tr}\left[\gamma^{\mu}S_{R}(k^{\prime})\gamma^{\nu}S_{R}(k)\right]+\text{Tr}\left[\gamma^{\mu}S_{A}(k^{\prime})\gamma^{\nu}S_{A}(k)\right]\right\} \nonumber 
\eea
\begin{wrapfigure}{r}{0.32\textwidth}
  \begin{raggedright}
    \includegraphics[width=0.3\textwidth]{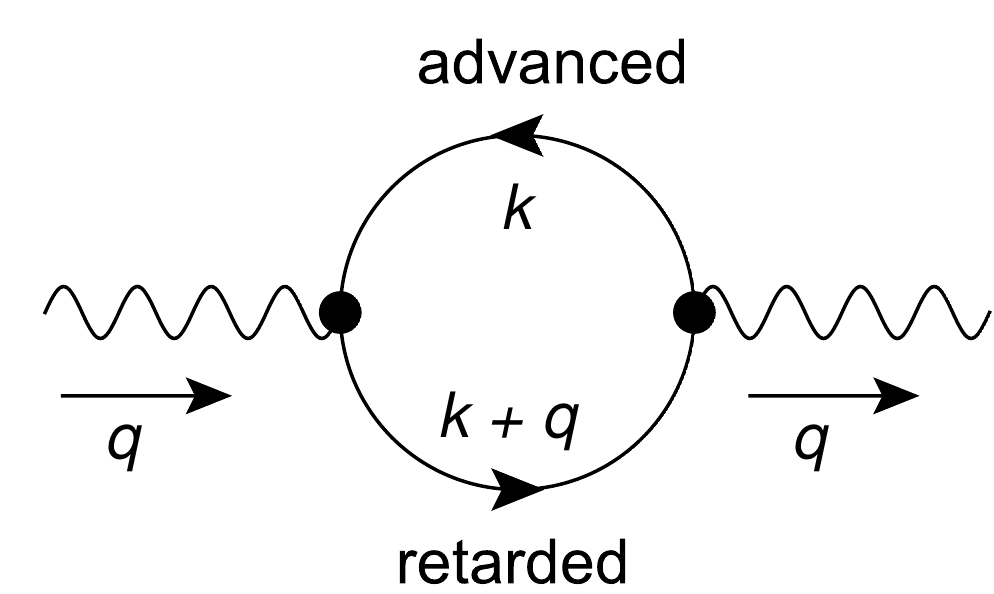}
  \end{raggedright}
  \caption{\label{fig:PIRet} Retarded photon polarization tensor with loop momentum $k$. In the given momentum assignment the propagator carrying momentum $k$ in ~\ref{eq:PIformal} is advanced while the 
propagator carrying momentum $k+q$ is retarded.}
\end{wrapfigure}
where the R/R and A/A propagator combinations equate to zero when the $k_0$ integration is performed. The non vanishing terms each combine an off-shell (retarded or advanced) and an on-shell (symmetric) propagator where only the latter contains a thermal distribution function. To assign momenta to the propagators, one can apply the following ”rule of thumb”: When following a momentum flow of an external leg into a retarded amplitude and tracing through the diagram every momentum momentum flow aligned opposite to the direction we are tracing corresponds to an advanced propagator while every flow in the same direction corresponds to a retarded propagator, see Fig. ~\ref{fig:PIRet}. A simple trace evaluation gives
\bea
T^{\mu\nu}(k,\,q): & = & \text{Tr}\left[\gamma^{\mu}\left(\cancel{k}+m\right)\gamma^{\nu}\left(\cancel{k}^{\prime}+m\right)\right]\nonumber \\[2ex]
 & = & 4\left[\left(k^{\mu}k^{\prime\nu}+k^{\nu}k^{\prime\mu}\right)-g^{\mu\nu}(k_0 k_0^{\prime}-\boldsymbol{k}\cdot\boldsymbol{k}^\prime-m^{2})\right]
\eea 

\noindent By exploiting the delta functions of the symmetric propagators to carry out the $k_{0}$ integral, one ends up with a total of eight contributions:

\bea\label{eq:factorize}
\Pi^{\mu\nu}_{R}(q) & = & -\frac{e^{2}}{2}\int\frac{d^{3}\boldsymbol{k}}{(2\pi)^{3}}(1-2n_{f}^{-}(\epsilon_{\boldsymbol{k}^{\prime}}))\,\frac{1}{4\epsilon_{\boldsymbol{k}}\epsilon_{\boldsymbol{k}^{\prime}}}\left[\frac{1}{q_{0}-\epsilon_{\boldsymbol{k}}-\epsilon_{\boldsymbol{k}^{\prime}}+i\epsilon}-\frac{1}{q_{0}+\epsilon_{\boldsymbol{k}}-\epsilon_{\boldsymbol{k}^{\prime}}+i\epsilon}\right]T^{\mu\nu}(-q_{0}+\epsilon_{\boldsymbol{k}^{\prime}},\,\boldsymbol{k})\nonumber \\[2ex]
 &  & -\frac{e^{2}}{2}\int\frac{d^{3}\boldsymbol{k}}{(2\pi)^{3}}(1-2n_{f}^{+}(\epsilon_{\boldsymbol{k}^{\prime}}))\,\frac{1}{4\epsilon_{\boldsymbol{k}}\epsilon_{\boldsymbol{k}^{\prime}}}\left[\frac{1}{q_{0}-\epsilon_{\boldsymbol{k}}+\epsilon_{\boldsymbol{k}^{\prime}}+i\epsilon}-\frac{1}{q_{0}+\epsilon_{\boldsymbol{k}}+\epsilon_{\boldsymbol{k}^{\prime}}+i\epsilon}\right]T^{\mu\nu}(-q_{0}-\epsilon_{\boldsymbol{k}^{\prime}},\,\boldsymbol{k})\nonumber \\[2ex]
 &  & -\frac{e^{2}}{2}\int\frac{d^{3}\boldsymbol{k}}{(2\pi)^{3}}(1-2n_{f}^{-}(\epsilon_{\boldsymbol{k}}))\,\frac{1}{4\epsilon_{\boldsymbol{k}}\epsilon_{\boldsymbol{k}^{\prime}}}\left[\frac{1}{q_{0}+\epsilon_{\boldsymbol{k}}-\epsilon_{\boldsymbol{k}^{\prime}}+i\epsilon}-\frac{1}{q_{0}+\epsilon_{\boldsymbol{k}}+\epsilon_{\boldsymbol{k}^{\prime}}+i\epsilon}\right]T^{\mu\nu}(\epsilon_{\boldsymbol{k}},\boldsymbol{k})\nonumber\\[2ex]
 &  & -\frac{e^{2}}{2}\int\frac{d^{3}\boldsymbol{k}}{(2\pi)^{3}}(1-2n_{f}^{+}(\epsilon_{\boldsymbol{k}}))\,\frac{1}{4\epsilon_{\boldsymbol{k}}\epsilon_{\boldsymbol{k}^{\prime}}}\left[\frac{1}{q_{0}-\epsilon_{\boldsymbol{k}}-\epsilon_{\boldsymbol{k}^{\prime}}+i\epsilon}-\frac{1}{q_{0}-\epsilon_{\boldsymbol{k}}+\epsilon_{\boldsymbol{k}^{\prime}}+i\epsilon}\right]T^{\mu\nu}(-\epsilon_{\boldsymbol{k}},\boldsymbol{k})\label{eq:PiFullStructure} 
\eea

To calculate $\Pi_{00}$ and $\Pi_{\,\,\,\mu}^{\mu}$, we separate $T^{00}$ and $T^{\mu}_{\,\,\,\mu}$ in equation ~\ref{eq:factorize} into $k_0$-dependent and independent parts:
\be
T^{00}=4\left[k^{0}(k^{0}+q^{0})+(\epsilon_{\boldsymbol{k}}^{2}+\boldsymbol{k}\cdot\boldsymbol{q})\right]\,,\,\,\,\,\,\,\,\,\,\,\,\,T_{\,\,\text{\ensuremath{\mu}}}^{\mu}=-8\left[k^{0}(k^{0}+q^{0})-(\epsilon_{\boldsymbol{k}}^{2}+\boldsymbol{k}\cdot\boldsymbol{q}+m^{2})\right]\,.
\ee
To reorganize the $k_0$-dependent parts, one can derive the following useful identity:
\vspace*{2mm}
\bea
 &  & \sum_{\xi=\pm}\frac{1}{\epsilon_{\boldsymbol{k}}}\frac{1}{\epsilon_{\boldsymbol{k}^{\prime}}}\epsilon_{\boldsymbol{k}^{\prime}}\left(\epsilon_{\boldsymbol{k}^{\prime}}-\xi q_{0}\right)\left[\frac{1}{\left(q_{0}-\epsilon_{\boldsymbol{k}}-\xi\epsilon_{\boldsymbol{k}^{\prime}}\right)}-\frac{1}{\left(q_{0}+\epsilon_{\boldsymbol{k}}-\xi\epsilon_{\boldsymbol{k}^{\prime}}\right)}\right]\,\left[1-2n_{f}^{-\xi}\left(\epsilon_{\boldsymbol{k}^{\prime}}\right)\right]\\[2ex]
 & + & \,\sum_{\xi=\pm}\frac{1}{\epsilon_{\boldsymbol{k}}}\frac{1}{\epsilon_{\boldsymbol{k}^{\prime}}}\epsilon_{\boldsymbol{k}}\,\,\left(\epsilon_{\boldsymbol{k}}+\xi q_{0}\right)\left[\frac{1}{\left(q_{0}+\xi\epsilon_{\boldsymbol{k}}-\epsilon_{\boldsymbol{k}^{\prime}}\right)}-\frac{1}{\left(q_{0}+\xi\epsilon_{\boldsymbol{k}}+\epsilon_{\boldsymbol{k}^{\prime}}\right)}\right]\,\left[1-2n_{f}^{-\xi}\left(\epsilon_{\boldsymbol{k}}\right)\right]\nonumber\\[2ex]
 & = & \sum_{\xi=\pm}\,2\xi\left\{ \,\frac{1}{q_{0}+\xi(\epsilon_{\boldsymbol{k}}+\epsilon_{\boldsymbol{k}^{\prime}})}\left[1-n_{f}^{-\xi}(\epsilon_{\boldsymbol{k}})-n_{f}^{\xi}(\epsilon_{\boldsymbol{k}^{\prime}})\right]-\frac{1}{q_{0}+\xi(\epsilon_{\boldsymbol{k}}-\epsilon_{\boldsymbol{k}^{\prime}})}\left[n_{f}^{-\xi}(\epsilon_{\boldsymbol{k}})-n_{f}^{-\xi}(\epsilon_{\boldsymbol{k}^{\prime}})\right]\right\}.\nonumber 
\eea
\noindent Putting both parts together, one arrives at equation ~\ref{eq:PI0Trace}. For further evaluation, it is often convenient to trade $\left|\boldsymbol{k}\right|$ and the azimuthal angle $\theta$ integrations for integrals over the energies $\epsilon_{\boldsymbol{k}}$ and $\epsilon_{\boldsymbol{k}^{\prime}}$. For $\Pi^{00}$ this results in 
\vspace*{2mm}
\bea
\Pi^{00}_{R}(q) & = & \frac{e^{2}}{8\pi^{2}}\frac{1}{\left|\boldsymbol{q}\right|}\int_{m}^{\infty}d\epsilon_{\boldsymbol{k}}\int_{\epsilon^{-}}^{\epsilon^{+}}d\epsilon_{\boldsymbol{k}^{\prime}}\sum_{\xi=\pm}\,\xi\,\frac{(\epsilon_{\boldsymbol{k}}-\epsilon_{\boldsymbol{k}^{\prime}})^{2}-\boldsymbol{q}^{2}}{(q_{0}+i\epsilon)+\xi(\epsilon_{\boldsymbol{k}}+\epsilon_{\boldsymbol{k}^{\prime}})}\left[1-n_{f}^{-\xi}(\epsilon_{\boldsymbol{k}})-n_{f}^{\xi}(\epsilon_{\boldsymbol{k}^{\prime}})\right]\label{eq:Pi00Final}\\[2ex]
 & + & \frac{e^{2}}{8\pi^{2}}\frac{1}{\left|\boldsymbol{q}\right|}\int_{m}^{\infty}d\epsilon_{\boldsymbol{k}}\int_{\epsilon^{-}}^{\epsilon^{+}}d\epsilon_{\boldsymbol{k}^{\prime}}\sum_{\xi=\pm}\,\xi\,\frac{(\epsilon_{\boldsymbol{k}}+\epsilon_{\boldsymbol{k}^{\prime}})^{2}-\boldsymbol{q}^{2}}{(q_{0}+i\epsilon)+\xi(\epsilon_{\boldsymbol{k}}-\epsilon_{\boldsymbol{k}^{\prime}})}\left[n_{f}^{-\xi}(\epsilon_{\boldsymbol{k}})-n_{f}^{-\xi}(\epsilon_{\boldsymbol{k}^{\prime}})\right]\,,\nonumber 
\eea
\vspace*{1mm}
and for $\Pi_{\,\mu}^{\mu}$ in  
\bea
\Pi_{\,\mu,R}^{\mu}(q) & = & \frac{e^{2}}{4\pi^{2}}\frac{1}{\left|\boldsymbol{q}\right|}\int_{m}^{\infty}d\epsilon_{\boldsymbol{k}}\int_{\epsilon^{-}}^{\epsilon^{+}}d\epsilon_{\boldsymbol{k}^{\prime}}\sum_{\xi=\pm}\,\xi\,\frac{(\epsilon_{\boldsymbol{k}}+\epsilon_{\boldsymbol{k}^{\prime}})^{2}-\boldsymbol{q}^{2}+2m^{2}}{(q_{0}+i\epsilon)+\xi(\epsilon_{\boldsymbol{k}}+\epsilon_{\boldsymbol{k}^{\prime}})}\left[1-n_{f}^{\xi}(\epsilon_{\boldsymbol{k}})-n_{f}^{-\xi}(\epsilon_{\boldsymbol{k}^{\prime}})\right]\label{eq:PiTrFinal}\\[2ex]
 & + & \frac{e^{2}}{4\pi^{2}}\frac{1}{\left|\boldsymbol{q}\right|}\int_{m}^{\infty}d\epsilon_{\boldsymbol{k}}\int_{\epsilon^{-}}^{\epsilon^{+}}d\epsilon_{\boldsymbol{k}^{\prime}}\sum_{\xi=\pm}\,\xi\,\frac{(\epsilon_{\boldsymbol{k}}-\epsilon_{\boldsymbol{k}^{\prime}})^{2}-\boldsymbol{q}^{2}+2m^{2}}{(q_{0}+i\epsilon)+\xi(\epsilon_{\boldsymbol{k}}-\epsilon_{\boldsymbol{k}^{\prime}})}\left[n_{f}^{-\xi}(\epsilon_{\boldsymbol{k}})-n_{f}^{-\xi}(\epsilon_{\boldsymbol{k}^{\prime}})\right]\,,\nonumber 
\eea
\vspace*{1mm}
where the integration boundaries are 
\be
\epsilon^{\pm}=\sqrt{(\sqrt{\epsilon_{\boldsymbol{k}}^{2}-m^{2}}\pm\left|\boldsymbol{q}\right|)^{2}+m^{2}}\,.
\ee
The poles are now simple functions of the integration variables, which renders principal value integrations particularly simple. In the zero-temperature limit both real and imaginary parts can be evaluated analytically. Solutions to the real parts are lengthy,- see e.g. \cite{mcorist_melrose_weise_2007}, a complete evaluation of the imaginary part is provided in Appendix \ref{sec:ImPI}.     
\subsection{Hard dense loop approximation}
\label{subsec:HDL}
\noindent This appendix demonstrates the hard dense loop expansion of $\Pi_{00}$, Eq. \ref{eq:PI0Trace}, and the expansion of $\Pi_{\,\,\mu}^{\mu}$ can be obtained analogously. Assuming $\boldsymbol{q}\ll k_{f}$ and $q_0\ll \mu$, we perform an expansion of the integrand. Since  
\be
\left(1-\frac{K}{\epsilon_{\boldsymbol{k}}\epsilon_{\boldsymbol{k}^{\prime}}}\right)=0+\mathcal{O}(\left|\boldsymbol{q}\right|^{2})\,,\,\,\,\,\,\,\,\,\,\,\,\,\left(1+\frac{K}{\epsilon_{\boldsymbol{k}}\epsilon_{\boldsymbol{k}^{\prime}}}\right)=2+\mathcal{O}(\left|\boldsymbol{q}\right|^{2})\,,
\ee
only the second line of ~\ref{eq:PI0Trace} contributes to the expansion of $\Pi_{00}$. The difference of the dispersion relations approximates to $\epsilon_{\boldsymbol{k}}-\epsilon_{\boldsymbol{k}^{\prime}}\simeq-\left|\boldsymbol{k}\right|\left|\boldsymbol{q}\right|\text{cos}\theta/\epsilon_{\boldsymbol{k}}$, and the expansion of the distribution functions requires some special care: 
\be
n_{f}(\epsilon_{\boldsymbol{k}})-n_{f}(\epsilon_{\boldsymbol{k}^{\prime}})\simeq-\hat{\boldsymbol{k}}\cdot\boldsymbol{q}\,\partial_{\boldsymbol{k}}\,n_{f}(\epsilon_{\boldsymbol{k}})\,,
\ee
where in the zero-temperature limit we formally derive $\partial_{\boldsymbol{k}}\Theta(\mu-\epsilon_{\boldsymbol{k}})=-\delta(\mu-\epsilon_{\boldsymbol{k}})\cdot\left|\boldsymbol{k}\right|/\epsilon_{\boldsymbol{k}}$
such that when all signs and factors considered one obtains
\bea
\Pi_{HDL}^{00} & = & 2e^{2}\int\frac{d^{3}\boldsymbol{k}}{(2\pi)^{3}}\frac{1}{q_{0}-\left|\boldsymbol{k}\right|\left|\boldsymbol{q}\right|\text{cos}\theta/\epsilon_{\boldsymbol{k}}}\cdot\frac{\left|\boldsymbol{k}\right|\left|\boldsymbol{q}\right|\text{cos}\theta}{\epsilon_{\boldsymbol{k}}}\cdot\delta(\mu-\epsilon_{\boldsymbol{k}})\,,\nonumber \\[2ex]
 & = & -2e^{2}\int\frac{d^{3}\boldsymbol{k}}{(2\pi)^{3}}\left(1-\frac{q_{0}}{q_{0}-\boldsymbol{k}\cdot\boldsymbol{q}/\epsilon_{\boldsymbol{k}}}\right)\delta(\mu-\epsilon_{\boldsymbol{k}})\,.
\eea
The angular integrals can now be performed analytically, resulting in 
\vspace*{1mm}
\be
\Pi_{HDL}^{00}=-\frac{e^{2}\mu k_{f}}{\pi^{2}}\left[1-\frac{1}{2}\frac{\mu q_{0}}{k_{f}\left|\boldsymbol{q}\right|}\text{log}\left(\frac{\mu q_{0}+k_{f}\left|\boldsymbol{q}\right|}{\mu q_{0}-k_{f}\left|\boldsymbol{q}\right|}\right)\right]\,.
\ee
Repeating the same steps for $\Pi_{\,\,\mu}^{\mu}$ (where now both lines of ~\ref{eq:PI0Trace} contribute) leads to
\vspace*{1mm}
\be
\Pi_{\,\mu,\,HDL}^{\mu}=-\frac{e^{2}}{\pi^{2}}\left[\mu k_{f}-\frac{1}{2}m^{2}\frac{\mu q_{0}}{k_{f}\left|\boldsymbol{q}\right|}\text{ln}\left(\frac{\mu q_{0}+k_{f}\boldsymbol{q}}{\mu q_{0}-k_{f}\boldsymbol{q}}\right)\right]\,.\label{eq:HDLTrace}
\ee
In the massless limit, the HDL result of the trace returns the Debye mass. 
\section{Imaginary part of polarization tensor}
\label{sec:ImPI}

\noindent In the R/A/S basis the imaginary part of $\Pi^{\mu\nu}$ can easily be obtained from
\begin{equation}
\text{Im}\,\Pi^{\mu\nu}(q)=\frac{1}{2i}\left[\Pi_{R}^{\mu\nu}(q)-\Pi_{A}^{\mu\nu}(q)\right]\,.\label{eq:IMPhoton}
\end{equation}

\noindent where the advanced self-energy is recovered from ~\ref{eq:PIformal} upon replacing $S_{R}$ by $S_{A}$ and vice versa. Again ignoring R/R and A/A combinations, one may use $S_{R}(k^{\prime})-S_{A}(k^{\prime})=-2\pi i\,\text{sign}(k_{0}^{\prime})\delta(k^{\text{\ensuremath{\prime}}2}-m^{2})\left(\cancel{k}^{\prime}+m\right)$, which immediately yields a general expression for the imaginary part:
\vspace{1mm}
\begin{equation}
\text{Im}\,\Pi^{\mu\nu}(q)=-\frac{e^{2}}{4}\int\frac{d^{4}k}{(2\pi)^{2}}\left[\text{sgn}(k_{0})(1-2N_{f}(k_{0}^{\prime}))-\text{sgn}(k_{0}^{\prime})\,(1-2N_{f}(k_{0}))\right]\,\delta(k^{\prime2}-m^{2})\,\delta(k^{2}-m^{2})\,T^{\mu\nu}(k)\,.\label{eq:ImPhotonBare}
\end{equation}

\noindent To evaluate this expression further we again exploit the delta
functions to carry out the $k_0$ integral, resulting in the familiar result:
\begin{eqnarray}\label{eq:ImGeneral}
\text{Im}ß,\Pi^{\mu\nu}(q) & = & -\frac{e^{2}}{2}\int\frac{d^{3}\boldsymbol{k}}{(2\pi)^{2}}\frac{1}{2\epsilon_{\boldsymbol{k}}}\frac{1}{2\epsilon_{\boldsymbol{k}^{\prime}}}\sum_{\xi=\pm}\xi\,\left[1-n_{f}^{-\xi}(\epsilon_{\boldsymbol{k}})-n_{f}^{+\xi}(\epsilon_{\boldsymbol{k}^{\prime}})\right]T^{\mu\nu}(k_{0}=\xi\epsilon_{\boldsymbol{k}})\,\delta(q_{0}+\xi(\epsilon_{\boldsymbol{k}}+\epsilon_{\boldsymbol{k}\prime}))\nonumber \\
\nonumber \\
 &  & -\frac{e^{2}}{2}\int\frac{d^{3}\boldsymbol{k}}{(2\pi)^{2}}\frac{1}{2\epsilon_{\boldsymbol{k}}}\frac{1}{2\epsilon_{\boldsymbol{k}^{\prime}}}\sum_{\xi=\pm}\xi\left[n_{f}^{-\xi}(\epsilon_{\boldsymbol{k}})-n_{f}^{-\xi}(\epsilon_{\boldsymbol{k}^{\prime}})\right]T^{\mu\nu}(k_{0}=\xi\epsilon_{\boldsymbol{k}})\,\delta(q_{0}+\xi(\epsilon_{\boldsymbol{k}}-\epsilon_{\boldsymbol{k}\prime}))\,.
\end{eqnarray}
\noindent from which one easily obtains $\text{Im}\,\Pi_{00}$
and $\text{Im}\,\Pi_{\,\mu}^{\mu}$. Which of the four terms above actually contribute to the imaginary part depends on the kinematic conditions which are enforced by the delta functions. The kinematics for Landau damping are fulfilled by the terms in the second line. After dropping anti-particles, only the first term in the sum remains. Using the same variable transformation as in Eq. \ref{eq:Pi00Final} and \ref{eq:PiTrFinal}, we may exploit the delta functions to carry out the $\epsilon_{\boldsymbol{k^{\prime}}}$ integration, leading to
\vspace{1mm}
\be\label{eq:ImPiFinal}
\text{Im}\,\left.\Pi^{00}(q)\right|_{LD}  =  \frac{1}{\left|\boldsymbol{q}\right|}\,\frac{e^2}{8\pi}\int_{m}^{\infty}\,d\epsilon_{\boldsymbol{k}}\,\left[(2\epsilon_{\boldsymbol{k}}+q_0)^2-\boldsymbol{q}^2)\right]\,\left[n_{f}(q_{0}+\epsilon_{\boldsymbol{k}})-n_{f}(\epsilon_{\boldsymbol{k}})\right]\theta^{\pm}(q_0+\epsilon_{\boldsymbol{k}})
\ee
and
\be\label{eq:ImPiFinal2}
\text{Im}\,\left.\Pi_{\,\,\mu}^{\mu}(q)\right|_{LD}  =  \frac{1}{\left|\boldsymbol{q}\right|}\,\frac{e^2}{8\pi}\int_{m}^{\infty}\,d\epsilon_{\boldsymbol{k}}\,\left[q^2 + 2m^2\right]\,\left[n_{f}(q_{0}+\epsilon_{\boldsymbol{k}})-n_{f}(\epsilon_{\boldsymbol{k}})\right]\theta^{\pm}(q_0+\epsilon_{\boldsymbol{k}})\,.
\ee
\vspace{1mm}
Here we have introduced the symbol $\theta^{\pm}(x)=\theta (x-\epsilon^{-}) \theta (\epsilon^{+}-x)$, which keeps track of kinematic restrictions.   The kinematics for pair creation are fulfilled by the terms in the first line of Eq. \ref{eq:ImGeneral}, which after variable transformation read
\vspace*{1mm}
\bea\label{eq:Pi00PC}\nonumber
\text{Im}\,\left.\Pi^{00}(q)\right|_{PC} & = & -\frac{1}{\left|\boldsymbol{q}\right|}\frac{e^{2}}{8\pi}\int_{m}^{\infty}d\epsilon_{\boldsymbol{k}}\left[(2\epsilon_{\boldsymbol{k}}+q_{0})^{2}-\boldsymbol{q}^{2}\right]\left[1-n_{f}^{-}(\epsilon_{\boldsymbol{k}})\right]\theta^{\pm}(-q_{0}-\epsilon_{\boldsymbol{k}})\\\nonumber
\\
 &  & +\frac{1}{\left|\boldsymbol{q}\right|}\frac{e^{2}}{8\pi}\int_{m}^{\infty}d\epsilon_{\boldsymbol{k}}\left[(2\epsilon_{\boldsymbol{k}}-q_{0})^{2}-\boldsymbol{q}^{2}\right]\left[1-n_{f}^{-}(q_{0}-\epsilon_{\boldsymbol{k}})\right]\theta^{\pm}(q_{0}-\epsilon_{\boldsymbol{k}})
\eea
and
\bea\label{eq:PiTrPC}\nonumber
\text{Im}\,\left.\Pi_{\,\,\mu}^{\mu}(q)\right|_{PC} & = & -\frac{1}{\left|\boldsymbol{q}\right|}\frac{e^{2}}{4\pi}\int_{m}^{\infty}d\epsilon_{\boldsymbol{k}}\left[q_{0}^{2}-\boldsymbol{q}^{2}+2m^{2}\right]\left[1-n_{f}^{-}(\epsilon_{\boldsymbol{k}})\right]\theta^{\pm}(-q_{0}-\epsilon_{\boldsymbol{k}})\\\nonumber
\\
 &  & +\frac{1}{\left|\boldsymbol{q}\right|}\frac{e^{2}}{4\pi}\int_{m}^{\infty}d\epsilon_{\boldsymbol{k}}\left[q_{0}^{2}-\boldsymbol{q}^{2}+2m^{2}\right]\left[1-n_{f}^{-}(q_{0}-\epsilon_{\boldsymbol{k}})\right]\theta^{\pm}(q_{0}-\epsilon_{\boldsymbol{k}})
\eea
\vspace{1mm}
While the ``$1$" in $(1-n_f)$ is a pure vacuum contribution in the calculation of the real parts, it constitutes a (finite) matter contribution in the calculation of the imaginary parts and cannot be neglected.
\subsection{Degenerate limit}
\label{sub:ImPiDegen}
\noindent The degenerate limit allows for an analytical evaluation of the above integrals. To do so, we again replace the distribution functions with step functions. The arguments of these stepfunctions can be combined with the various kinematic restrictions and solved for $q_0$, which results in the domain boundaries of Fig. \ref{fig:Domains}. In the following, we denote these boundaries by 
\be
q_{b}^{a}  =  a\,\mu+\sqrt{\mu^{2}+\left|\boldsymbol{q}\right|^{2}+2 b\,k_{f}\left|\boldsymbol{q}\right|}\,,
\ee
where $a\,,b = \pm$. We may now evaluate the imaginary parts from Landau damping and pair creation in the various sub-domains depicted in Fig. \ref{fig:Domains}. For Landau damping, one finds
\vspace*{1mm}
\bea
\text{Im}\,\left.\Pi^{00}(q_0,\,\left|\boldsymbol{q}\right|)\right|_{LD} & = &- \frac{e^{2}}{24\pi}\frac{q_0}{\left|\boldsymbol{q}\right|}\left(q_0^2-3\boldsymbol{q}^2+12\mu^2\right)\theta(q_{+}^{-}-q_{0})\theta(q_{0}-q_{-}^{-})\,,\\[3ex]
\text{Im}\,\left.\Pi_{\,\,\mu}^{\mu}(q_0,\,\left|\boldsymbol{q}\right|)\right|_{LD} & = & -\frac{e^{2}}{4\pi}\frac{q_0}{\left|\boldsymbol{q}\right|}\left(q^2+2m^2\right)\theta(q_{+}^{-}-q_{0})\theta(q_{0}-q_{-}^{-})\,,\\\nonumber
\eea 
\vspace{1mm}
The above equations readily include the cases $q_0>0$ and $q_0<0$ as well as any value of $\left|\boldsymbol{q}\right|$. Pair creation is slightly more complicated as the various terms in Eqs. \ref{eq:Pi00PC} and \ref{eq:PiTrPC} contain vacuum and matter contributions in (overlapping) intervals of the energy $q_0$. It is sufficient to consider $q_0 > 0$, for $q_0 < 0$ one may take advantage of the fact that the imaginary parts are odd functions in $q_0$. For $q_0 > 0$ we may restrict the calculation to the second lines of \ref{eq:Pi00PC} and \ref{eq:PiTrPC}).  It is further important to distinguish between regions of $\left|\boldsymbol{q}\right|<2k_f$ and $\left|\boldsymbol{q}\right|>2k_f$. We first analyze the case $\left|\boldsymbol{q}\right|<2k_f$. Kinematics allow the terms in the second lines of \ref{eq:Pi00PC} and \ref{eq:PiTrPC} to develop an imaginary part for $q_0^2 > \boldsymbol{q}^2 + 4m^2$ which is the usual threshold for pair creation. In the degenerate limit the Pauli blocking further restricts pair creation to higher values of $q_0$. The actual value of $q_0$ at which a contribution to the imaginary part stemming from pair creation sets in are given by $q_0=q^{+}_{-}$. In total, one obtains for $\left|\boldsymbol{q}\right|<2 k_f$:
\bea 
\text{Im}\,\left.\Pi^{00}(q_0>0,\,\left|\boldsymbol{q}\right|<2k_f)\right|_{PC} & = & -\frac{e^{2}}{16\pi\left|\boldsymbol{q}\right|}\left[A^{-}(q_0)\,\theta(q_{+}^{+}-q_{0})\theta(q_{0}-q_{-}^{+})+B(q_0)\,\theta(q_{0}-q_{-}^{+})\right]\,,\\\nonumber
\\
\text{Im}\,\left.\Pi_{\,\,\mu}^{\mu}(q_0>0,\,\left|\boldsymbol{q}\right|<2k_f)\right|_{PC} & = & -\frac{e^{2}}{8\pi\left|\boldsymbol{q}\right|}\left[C^{-}(q_0)\,\theta(q_{+}^{+}-q_{0})\,\theta(q_{0}-q_{-}^{+})+D(q_0)\,\theta(q_{0}-q_{-}^{+})\right]\,.
\eea
\noindent At $\left|\boldsymbol{q}\right| = 2k_f$ one has $\sqrt{\boldsymbol{q}^2 + 4m^2} = q^{+}_{-} = 2\mu $ and for $\left|\boldsymbol{q}\right|>2k_f$ the imaginary parts become
\bea 
\text{Im}\,\left.\Pi^{00}(q_0>0,\,\left|\boldsymbol{q}\right|>2k_f)\right|_{PC} & = & -\frac{e^{2}}{16\pi\left|\boldsymbol{q}\right|}\left[A^{+}(q_0)\,\theta(q_{+}^{+}-q_0)\,\theta(q_{0}-q_{+}^{-})+B(q_0)\,\theta(q_0-\sqrt{\boldsymbol{q}^2 + 4m^2})\right]\,,\\\nonumber
\\
\text{Im}\,\left.\Pi_{\,\,\mu}^{\mu}(q_0>0,\,\left|\boldsymbol{q}\right|>2k_f)\right|_{PC} & = & -\,\frac{e^{2}}{8\pi\left|\boldsymbol{q}\right|}\,\,\left[C^{+}(q_0)\,\theta(q_{+}^{+}-q_0)\,\theta(q_{0}-q_{+}^{-})+D(q_0)\,\theta(q_0-\sqrt{\boldsymbol{q}^2 + 4m^2})\right]\,.
\eea 
The functions $A^{\pm}$, $B$, $C^{\pm}$, and $D$ are given by (with the abbreviation $\epsilon_q^2 = (q^2-4m^2)/q^2$)

\bea
A^{\pm}(q_0) & = & \epsilon_{\boldsymbol{q}}\left|\boldsymbol{q}\right|^{3}  (1-\frac{\epsilon_{\boldsymbol{q}}^{2}}{3}) \pm (2\mu-q_{0})\left[\left|\boldsymbol{q}\right|^{2}-\frac{1}{3}(2\mu-q_{0})^{2}\right]\,,\hspace{1cm} B(q_0) = \frac{2}{3} \epsilon_{\boldsymbol{q}} \left|\boldsymbol{q}\right| \frac{\boldsymbol{q}^{2}}{q^{2}} \left(q^{2}+2m^{2} \right)\,,
\\[2ex]
C^{\pm}(q_{0}) & = & \left(q^{2}+2m^{2}\right)\left[\left(2\mu-q_{0}\right)\pm\left|\boldsymbol{q}\right|\epsilon_{\boldsymbol{q}}\right]\,,\hspace{3.8cm} D(q_{0})  = -2 \epsilon_{\boldsymbol{q}} \left|\boldsymbol{q}\right|\left(q^{2}+2m^{2}\right).\nonumber
\eea
\subsection{Properties of nuclear matter from Skyrme models}
\label{sub:Skyrme}
\begin{table}[t]
\begin{centering}
\begin{tabular}{|c|c|c|c|c|c|}
\hline 
\textbf{parameter set} & NRAPR & SKRA & SQMC700 & LNS & KDE0v1\tabularnewline
\hline 
\hline 
$n_{c}\,\,[n_{0}]$ (non-rel.) & 0.52 & 0.52 & 0.52 & 0.57 & 0.56\tabularnewline
\hline 
$n_{c,\,\mu}\,[n_{0}]$ (non-rel.) & 0.73 & 0.76 & 0.76 & 0.79 & 0.68\tabularnewline
\hline
\hline
$n_{c}\,\,[n_{0}]$ (rel.) & 0.54 & 0.54 & 0.54 & 0.59 & 0.59\tabularnewline
\hline 
$n_{c,\,\mu}\,[n_{0}]$ (rel.) & 0.75 & 0.77 & 0.77 & 0.80 & 0.70\tabularnewline
\hline 
\end{tabular}
\par\end{centering}
\caption{\label{tab:CritDens}Critical densities for stability of
homogeneous nuclear matter and for the onset of muons in the non-relativistic and relativistic approaches for various parameter sets recommended in Ref. \cite{Dutra:2012mb}. Both values are pushed towards slightly higher values in the relativistic case but the changes are not significant. Stable homogeneous matter is on average achieved for densities larger than $0.56\,n_0$; muons appear on average at densities of about  $0.76\,n_0$ .}
\end{table}

\begin{figure}
\includegraphics[scale=0.6]{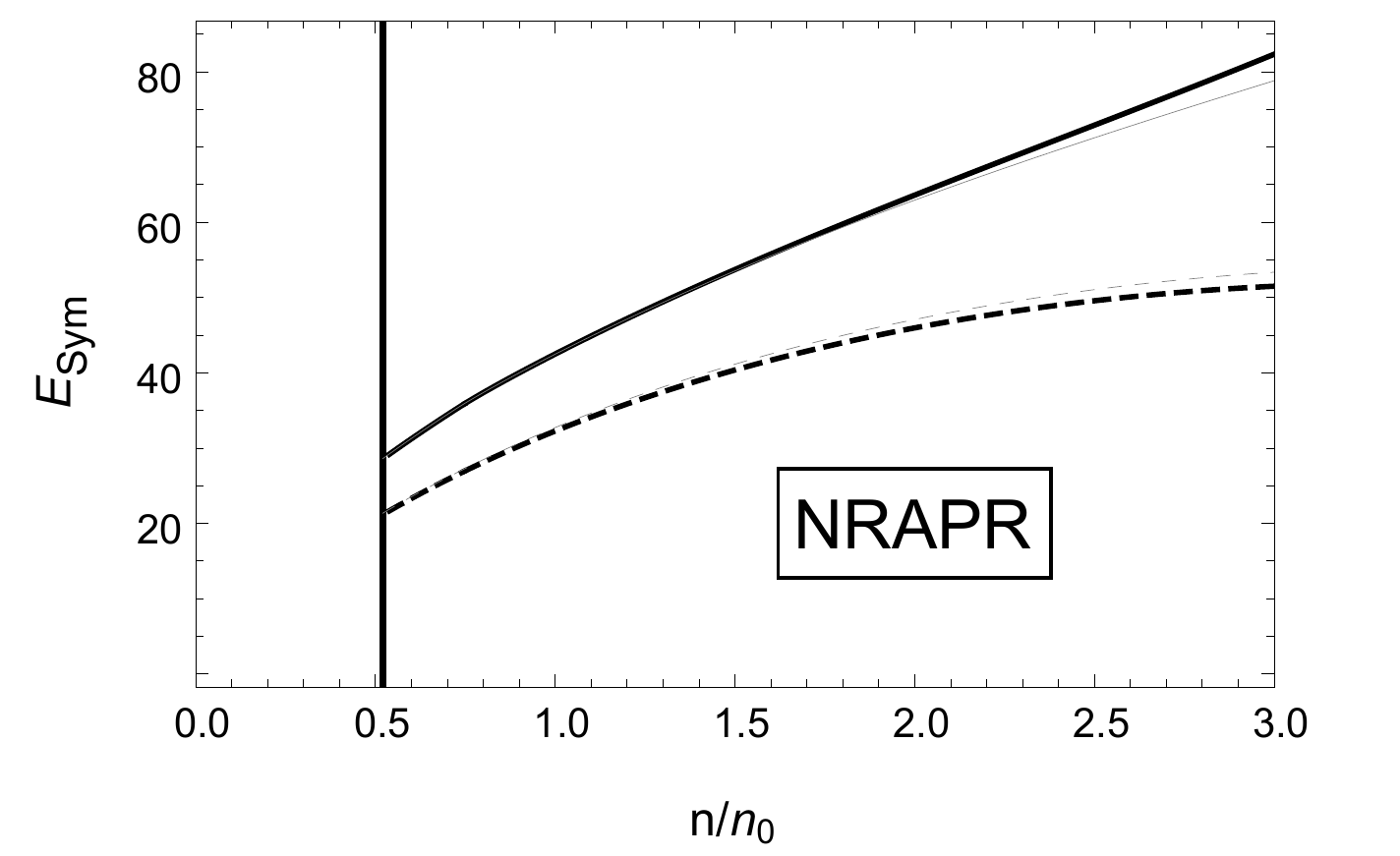}~~\includegraphics[scale=0.61]{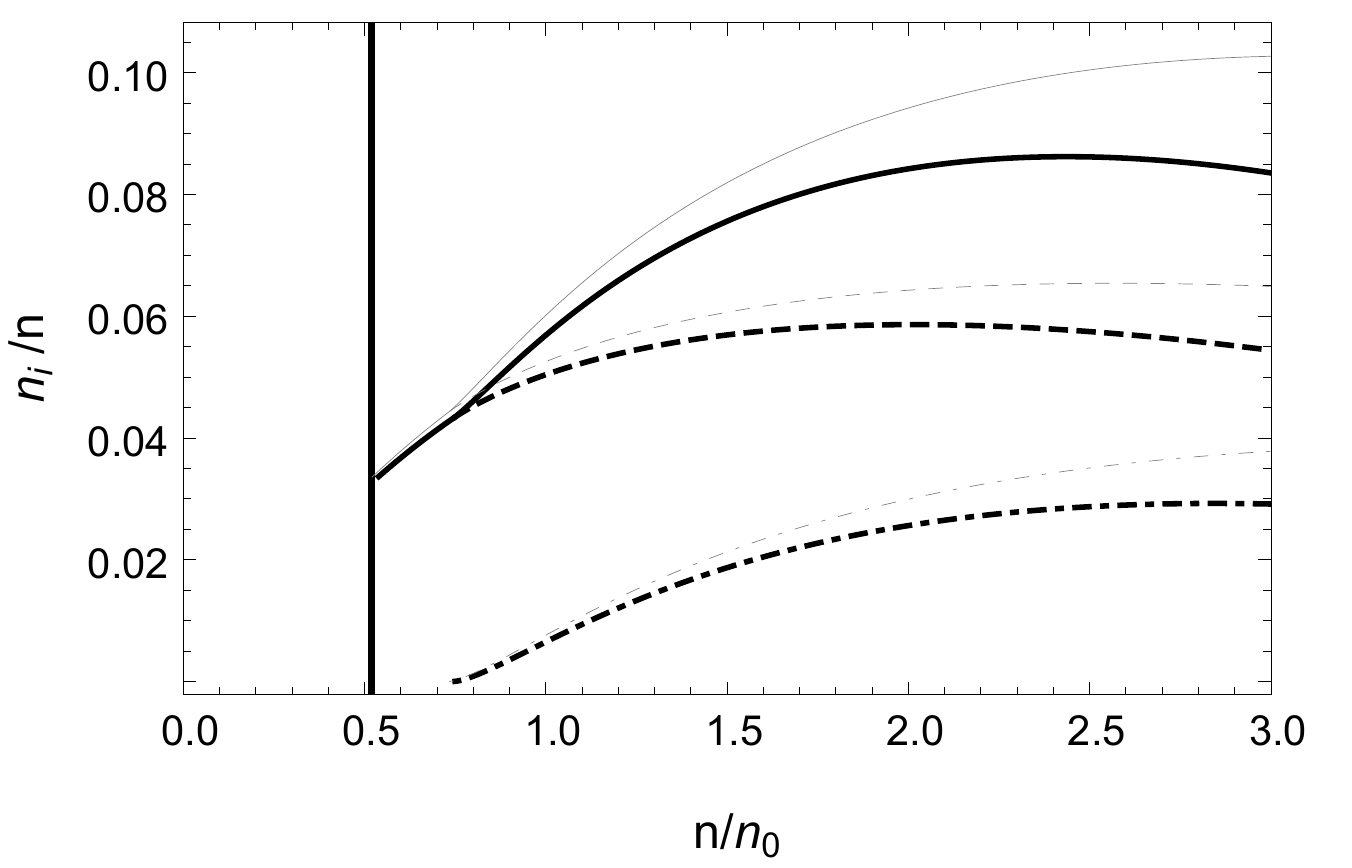}
~
\includegraphics[scale=0.85]{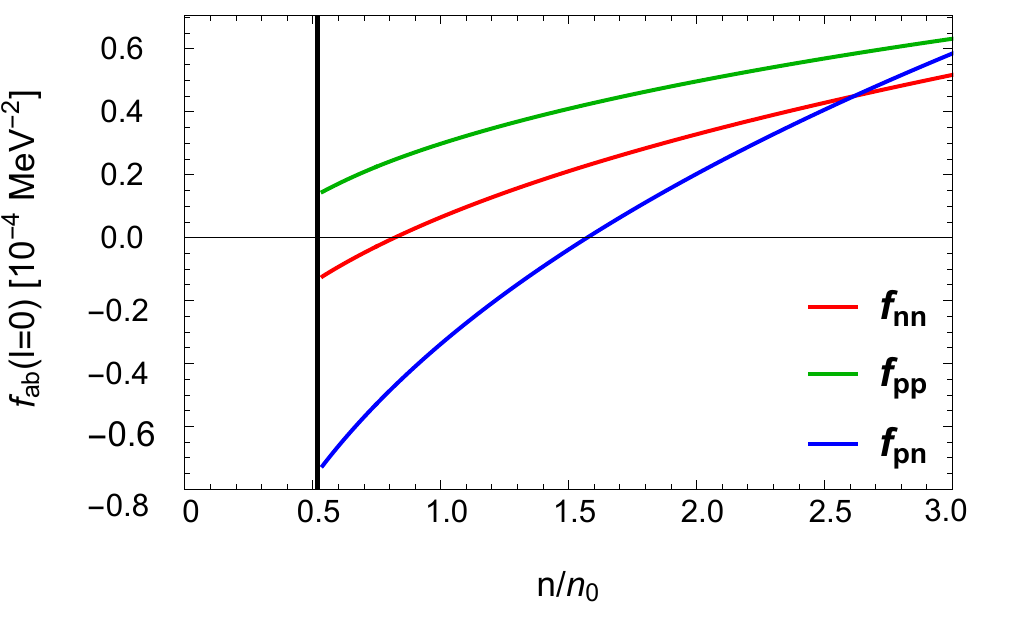}~~\includegraphics[scale=0.62]{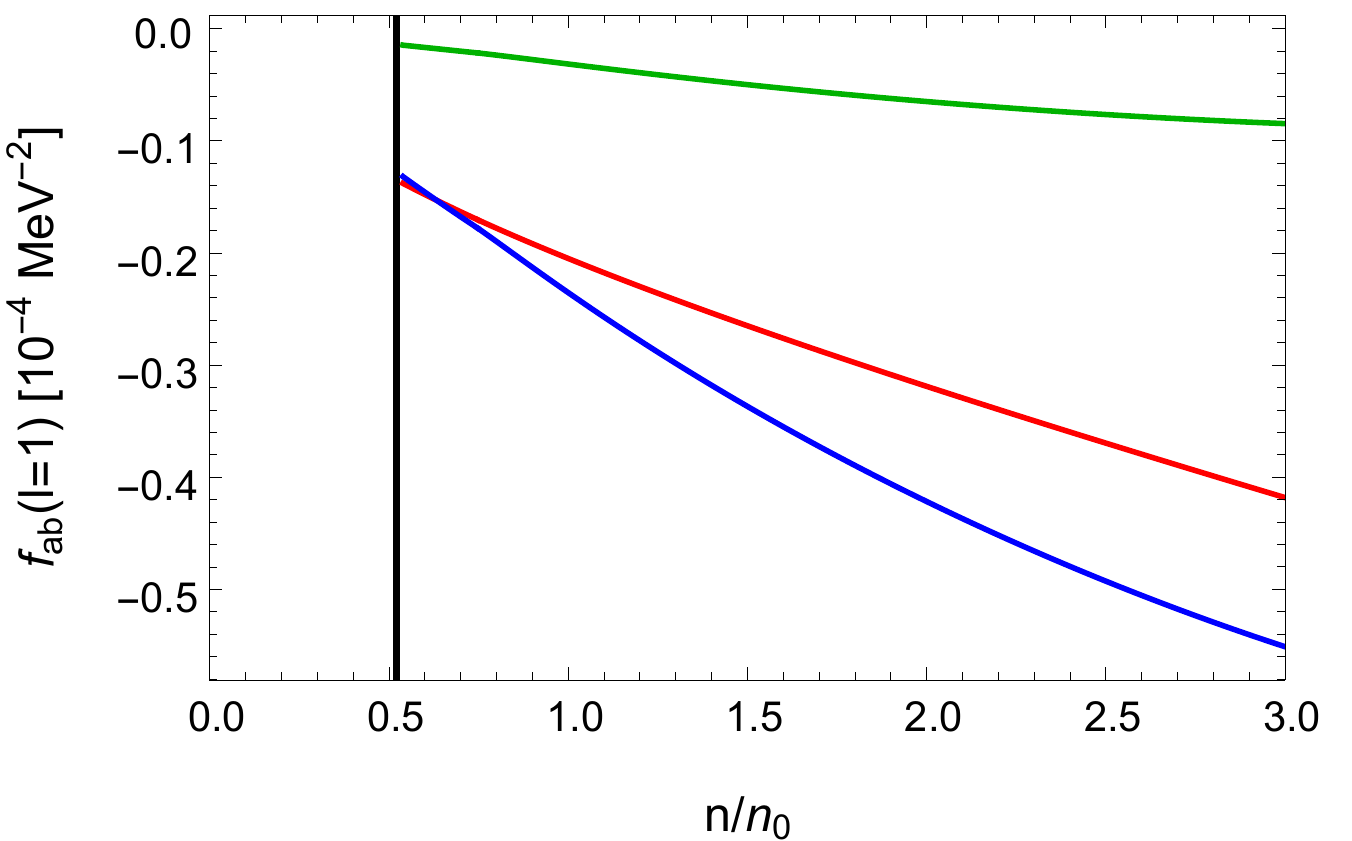}
\setlength{\belowcaptionskip}{-10mm}
\caption{\label{fig:NRAPR} Properties of homogeneous nuclear matter derived from the NRAPR Skyrme parameter set. Homogeneous matter is stable on the right-hand side of the black horizontal line. Top left: comparison of relativistic and original non-relativistic (thin gray lines) symmetry energies in symmetric nuclear matter (dashed) and in $\beta$ equilibrium (solid). In both cases, small deviations due to relativistic corrections appear only at high densities. Top right: proton (solid), electron (dashed), and muon (dot-dashed) fractions. Using NRAPR parameters muons appear around 0.75 $n_0$. Thin gray lines depict the non-relativistic results in $\beta$ equilibrium which indicate a larger proton fraction at higher densities. Bottom left: residual density-density interaction potentials of the quasi-particles. Bottom right: current-current interaction potentials of the quasi-particles.     
 }
\end{figure}

\begin{figure}
\begin{centering}
\includegraphics[scale=0.85]{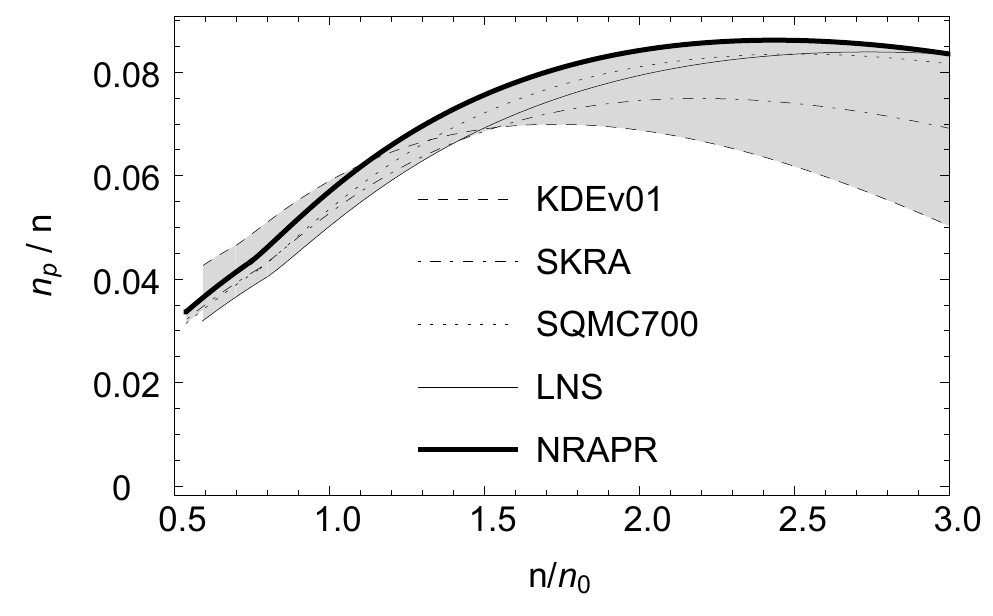}~~~~~~~\includegraphics[scale=0.6]{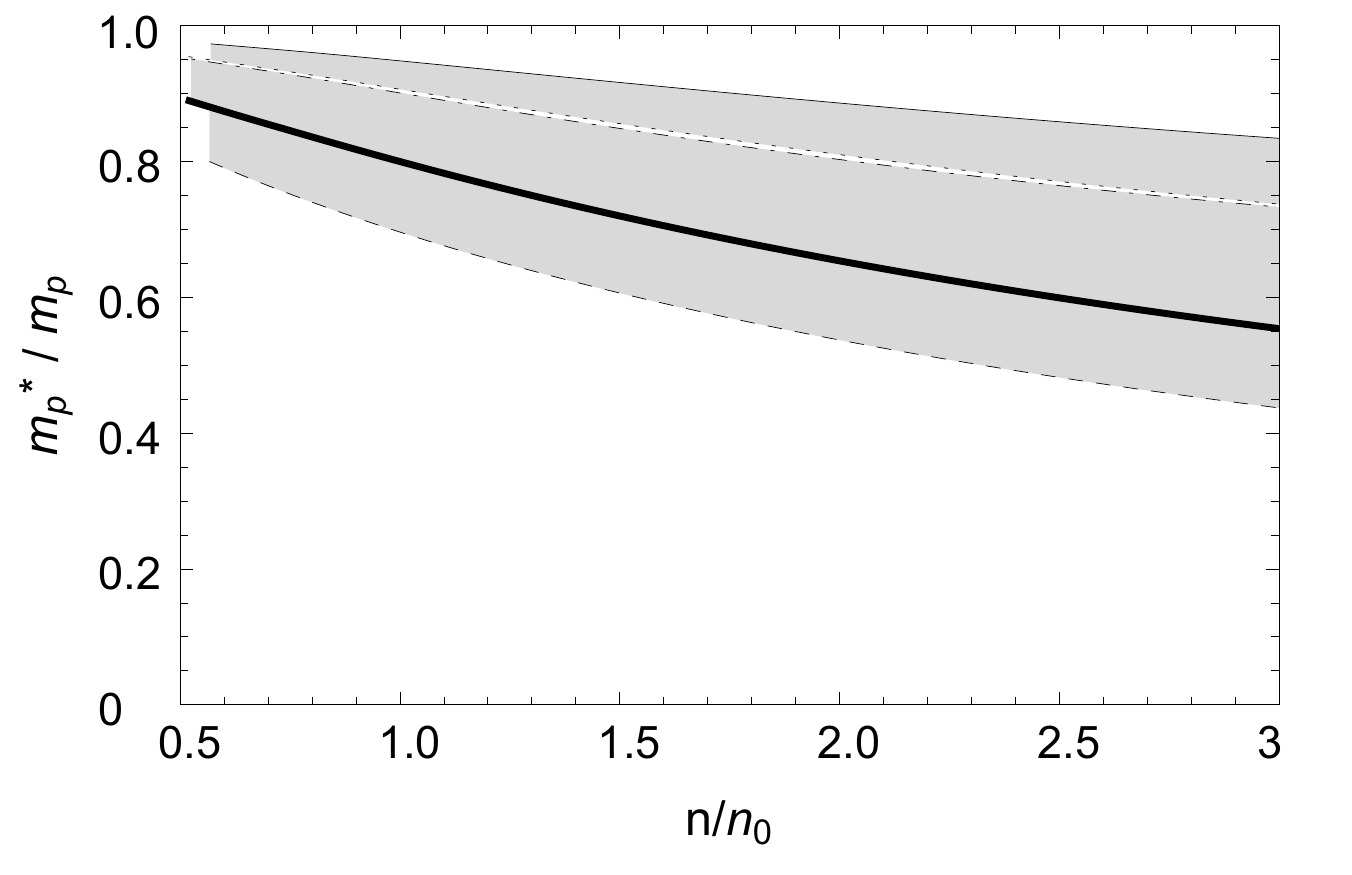}
\par\end{centering}
~
\begin{centering} \includegraphics[scale=0.62]{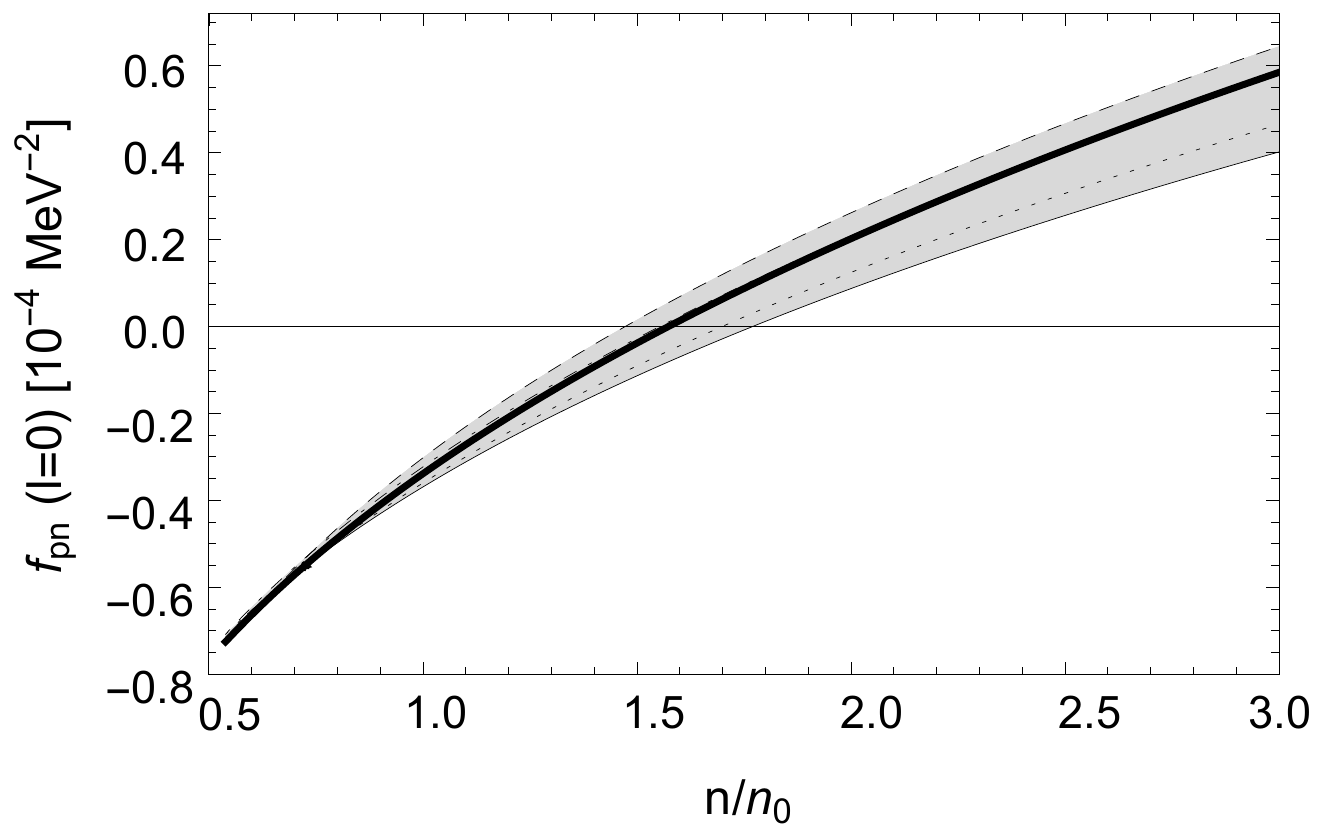}~~~~~\includegraphics[scale=0.62]{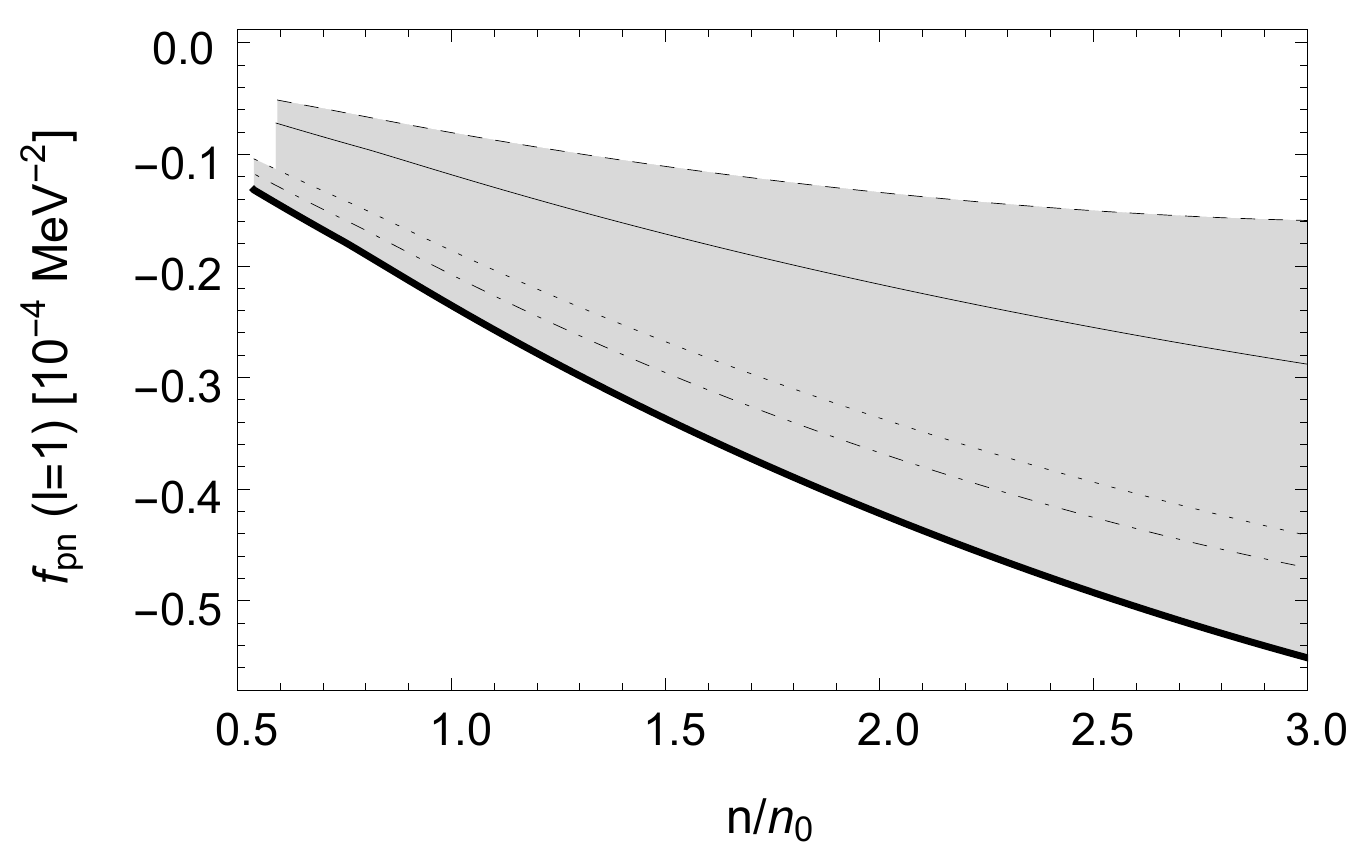}
\end{centering}
\caption{\label{fig:comparison} Comparison of different Skyrme models. The critical densities are slightly different for each model; see table \ref{tab:CritDens}. We are specifically interested in properties relevant to the induced interactions: The proton fractions (top left) agree fairly well at lower densities among the tested models. The reduction of proton the fractions at higher densities is in part due to relativistic corrections; see  Eq.\ref{eq:protonRel}. The same is true for the residual density-density interactions between neutrons and protons (bottom left). Both quantities are derived from $l=0$ contributions of the energy functional \ref{eq:EpsChamel}. The $l=1$ contributions are far less concurring.
 }
\end{figure}

\noindent This appendix demonstrates the derivation of the  basic properties of nuclear matter from a Landau energy functional; see also Ref. \cite{Chamel:2006rc}. The coefficients $C_{T}^{n,\tau,\boldsymbol{j}}$ appearing in the energy functional \ref{eq:EpsChamel} are related to standard Skyrme parameters via 
\bea 
C_{0}^{\tau} & = & -C_{0}^{\boldsymbol{j}}=\frac{3}{16}t_{1}+\frac{1}{4}t_{2}(\frac{5}{4}+x_{2})\,,\,\,\,\,\,\,\,\,\,\,C_{1}^{\tau}=-C_{1}^{\boldsymbol{j}}=-\frac{1}{8}t_{1}(\frac{1}{2}+x_{1})+\frac{1}{8}t_{2}(\frac{1}{2}+x_{2})\,,\label{eq:Constants}\\[2ex]
C_{0}^{n}[n] & = & \frac{3}{8}t_{0}+\frac{3}{48}t_{3}n^{\gamma}\,,\,\,\,\,\,\,\,\,\,\,\,\,\,\,\,\,\,\,\,\,\,\,\,\,\,\,\,\,\,\,\,\,\,\,\,\,\,C_{1}^{n}[n]=-\frac{1}{4}t_{0}(\frac{1}{2}+x_{0})-\frac{1}{24}t_{3}(\frac{1}{2}+x_{3})n^{\gamma}\,.\nonumber 
\eea 
A comparison of modern Skyrme forces can be found in Ref. \cite{Dutra:2012mb}. In this work, we  employ NRAPR, SKRA, SQMC700, LNS, and KDE0v1 parameters. Table \ref{tab:CritDens} lists critical densities $n_c$ below which homogeneous nuclear matter is unstable and $n_{c,\,\mu,}$ above which muons appear with and without the relativistic matching discussed in Sec. \ref{subsec:Fermi}. We further check the symmetry energies for symmetric matter and matter in $\beta$ equilibrium after applying relativistic modifications, obtainable from the energy functional  [i.e., the energy per particle; see Eq. \ref{eq:Eps0}], 
\be 
E_{\textrm{sym}}=\frac{1}{8}\frac{\partial^2}{\partial x_p^2} \frac{E}{A}(x_p)\,.
\ee
Fig. \ref{fig:NRAPR} displays $E_{sym}$ for symmetric and $\beta$-equilibrated matter using NRAPR parameters. The deviations of the symmetry energy from the original non-relativistic result are tiny and appear only at very high densities (this remains true for all other parameter sets is use). We now turn to the calculation of the functional derivatives \ref{eq:derivatives}. One has
\be 
\frac{\delta n_{a}}{\delta n_{\boldsymbol{k},b}}=V\,\delta_{ab}\,,\hspace{1cm}\frac{\delta\tau_{a}}{\delta n_{\boldsymbol{k},b}}=V\,\boldsymbol{k}^{2}\delta_{ab}\,,\hspace{1cm}\frac{\delta^{2}(\boldsymbol{j}_{a}\cdot\boldsymbol{j}_{b})}{\delta n_{\boldsymbol{k},c}\,\delta n_{\boldsymbol{k}^{\prime},d}}=V^{2}\,\boldsymbol{k}\cdot\boldsymbol{k}^{\prime}\left[\delta_{ac}\delta_{bd}+\delta_{bc}\delta_{ad}\right]\,,\label{eq:FuncDerivatives}
\ee 
where $V$ denotes the volume element in momentum space which can be dropped since we are ultimately interested in averaged quasi-particle energies and potentials. Single functional derivatives of $\boldsymbol{j}_{a}\cdot\boldsymbol{j}_{b}$ with respect to $n_{\boldsymbol{k},b}$ equate to zero. Equipped with these relations, it is straight forward to calculate the single-particle energies:
\bea 
e_{\boldsymbol{k},a}=\frac{\delta\mathcal{E}}{\delta n_{a,\boldsymbol{k}}} & = & \sum_{T=0,1}\left[\delta_{T,0}\frac{\hbar^{2}}{2m}+C_{T}^{\tau}n_{T}\right]\frac{\delta\tau_{T}}{\delta n_{a,\boldsymbol{k}}}+\sum_{T=0,1}\left[\frac{\delta}{\delta n_{a,\boldsymbol{k}}}\left(C_{T}^{n}\,n_{T}^{2}\right)+C_{T}^{\tau}\,\tau_{T}\,\frac{\delta}{\delta n_{a,\boldsymbol{k}}}n_{T}\right]\nonumber \\
\nonumber \\
 & = & \frac{\delta\mathcal{E}}{\delta\tau_{a}}\,\boldsymbol{k}^{2}+\frac{\delta\mathcal{E}}{\delta n_{a}}\,.
\eea
\noindent The two resulting terms are usually interpreted as single particle potentials and effective masses
\vspace{1mm}
\bea 
e_{\boldsymbol{k},a} & = & \frac{\hbar^{2}\boldsymbol{k}^{2}}{2m_{a}^{*}}+U_{a}\,,\label{eq:Eparticle}\\[2ex]
\frac{\hbar^{2}}{2m_{a}^{*}} & := & \frac{\delta\mathcal{E}}{\delta\tau_{a}}=\,\frac{\hbar^{2}}{2m_{a}}+\left(C_{0}^{\tau}-C_{1}^{\tau}\right)n+2C_{1}^{\tau}n_{a}\,,\label{eq:mstar}\\[2ex]
U_{a} & = & \frac{\delta\mathcal{E}}{\delta\tau_{a}}=4C_{1}^{n}n_{a}{\color{red}+}2n\left(C_{0}^{n}-C_{1}^{n}\right)+\left(C_{0}^{\tau}-C_{1}^{\tau}\right)\left(\tau_{n}+\tau_{p}\right)+2C_{1}^{\tau}\tau_{a}+\frac{dC_{0}^{n}}{dn}n^{2}+\frac{dC_{1}^{n}}{dn}(2n_{a}-n)^{2}\,.\label{eq:Uparticle}
\eea 
\vspace{1mm}
\noindent The second derivatives yield the residual quasiparticle interactions (see Fig. \ref{fig:NRAPR} for plots using NRAPR parameters and Fig. \ref{fig:comparison} for a comparison of different Skyrme models): 
\bea 
f_{nn,\,pp}^{(0)} & = & 2k_{f,(n,p)}^{2}\left(C_{0}^{\tau}+C_{1}^{\tau}\right)+2(C_{0}^{n}+C_{1}^{n})+4n\frac{d}{dn}C_{0}^{n}\pm4n_{1}\frac{d}{dn}C_{1}^{n}+n^{2}\frac{d^{2}}{dn^{2}}C_{0}^{n}+n_{1}^{2}\frac{d^{2}}{dn^{2}}C_{1}^{n}\,,\label{eq:VL0}\\[2ex]
f_{np}^{(0)} & = & (k_{f,p}^{2}+k_{f,n}^{2})\left(C_{0}^{\tau}-C_{1}^{\tau}\right)+2\left(C_{0}^{n}-C_{1}^{n}\right)+4n\frac{d}{dn}C_{0}^{n}+n^{2}\frac{d^{2}}{dn^{2}}C_{0}^{n}+n_{1}^{2}\frac{d^{2}}{dn^{2}}C_{1}^{n}\,,\label{eq:VNPL0}\\[3ex]
f_{nn,\,pp}^{(1)} & = & 2(C_{0}^{\boldsymbol{j}}+C_{1}^{\boldsymbol{j}})\,k_{f,(n,p)}^{2}\,,\hspace{2cm}f_{pn}^{(1)}=f_{np}^{(1)}=2(C_{0}^{\boldsymbol{j}}-C_{1}^{\boldsymbol{j}})\,k_{f,n}k_{f,p}\,.\label{eq:Vspacial}
\eea 
The superscripts $(0)$ and $(1)$ indicate whether the respective terms are $l=0$ or $l=1$ contributions in a partial wave expansion. The $l=0$ terms \ref{eq:VL0} and \ref{eq:VNPL0} yield the averaged interaction potentials in the static limit $f_{ab}^{(0)}=f_{ab}(\boldsymbol{q}=\boldsymbol{0})$. The $l=1$ terms provide the current-current interactions which alternatively be obtained from
\vspace*{1mm}
\bea
\left(\bar{f}_{pp}\right)^{ij} & = & \left(\bar{f_{nn}}\right)^{ij}=k_{f,(p,n)}^{2}\,\frac{\delta^{2}\mathcal{E}}{\delta j_{(p,n)}^{i}\,\delta j_{(p,n)}^{j}}=2\left(C_{0}^{\boldsymbol{j}}+C_{1}^{\boldsymbol{j}}\right)k_{f,(p,n)}^{2}\,\delta^{ij}\,,\label{eq:VijPPNN}\\[3ex]
\left(f_{pn}^{(1)}\right)^{ij} & = & \left(f_{np}^{(1)}\right)^{ij}=k_{f,p}\,k_{f,n}\,\frac{\delta^{2}\mathcal{E}}{\delta j_{p}^{i}\,\,\delta j_{n}^{j}}=2\left(C_{0}^{\boldsymbol{j}}-C_{1}^{\boldsymbol{j}}\right)k_{f,p}\,k_{f,n}\,\delta^{ij}\,.\label{eq:VijPN}
\eea 

\noindent Unfortunately $\boldsymbol{q}^2$ corrections to the static potentials cannot be obtained from \ref{eq:EpsChamel}. Finally, a comment on units is in order: Because of the Skyrme parametrization, quantities obtained from \ref{eq:EpsChamel} come in various combinations of fm and MeV.  To convert everything to MeV, energy densities are multiplied by $\left[\mathcal{E}_{0}\right] =\text{MeV\,fm}^{-3}=(197)^{3}\,\text{MeV}^{4}$
while single-particle energies (and chemical potentials) are readily obtained in MeV and finally interaction potentials $f$ are to be multiplied by $\left[ f \right] =\text{MeV\,fm}^{3}=(197)^{-3}\text{MeV}^{-2}$. 
\subsection{Vector exchange in a relativistic mean field model }
\label{sub:RMF}
\noindent A simple yet instructive exercise is the derivation of the static screening using strongly simplified relativistic mean field (RMF) forces as a toy model (see e.g., Ref. \cite{Dutra:2014qga} for a review). We include only vector correlations (i.e., the $\rho$ and $\omega$ meson exchange) such that the result serves as a ``blueprint" for the implementation of nuclear forces into the RPA resummation  and as a formal cross-check for the results obtained, in particular the screening mass. To keep things simple, we  neglect self-interactions of $\rho$ and $\omega$ mesons or couplings between the two. This particularly implies that there are no effective masses (which stem from the scalar $\sigma$ meson) and hence $V_{ab}=f_{ab}$. Further more there is no distinction between density and current interactions. One consequently has a very simple isospin $\boldsymbol{\tau}$-dependent Lagrangian 
\be \label{eq:RMFLagrange}
\mathcal{L}=\bar{\psi}(i\gamma^{\mu}\partial_{\mu}-m)\psi-g_{\omega}\bar{\psi}\gamma^{\mu}\omega_{\mu}\psi-\frac{1}{2}g_{\rho}\bar{\psi}\gamma^{\mu}\boldsymbol{\rho}_{\mu}\boldsymbol{\tau}\psi\,.
\ee 
The mean field quantities in infinite nuclear matter are
\begin{equation}
\left\langle \omega_{\mu}\right\rangle =\omega_{0}=\frac{g_{\omega}}{m_{\omega}^{2}}n\,,\,\,\,\,\,\,\,\,\,\,\,\,\,\,\,\left\langle \vec{\rho}_{\mu}\right\rangle =\rho_{0(3)}=\frac{g_{\rho}}{2m_{\rho}^{2}}n_{1}\,.
\end{equation}
with vector and isovector densities as before denoted by
$n=n_{n}+n_{p}$ and $n_{1}=n_{n}-n_{p}$. In this setting the vector interaction matrix describing interactions of protons and neutrons reads in Coulomb gauge  
\bea \label{eq:VecRMF}
V^{\mu \nu}_{V}= \left(\begin{array}{cc}
\chi_{V} & \chi_{I}\\
\chi_{I} & \chi_{V}
\end{array}\right)\left( \frac{q^2}{\boldsymbol{q}^2}\,g^{\mu 0}g^{\nu 0}+P_{\perp}^{\mu\nu} \right)\,, \hspace{1cm} \left(\begin{array}{cc}
\chi_{V} & \chi_{I}\\
\chi_{I} & \chi_{V}
\end{array}\right)=-\left(\begin{array}{cc}
\chi_{\omega}+\frac{1}{4}\chi_{\rho} & \chi_{\omega}-\frac{1}{4}\chi_{\rho}\\
\\
\chi_{\omega}-\frac{1}{4}\chi_{\rho} & \chi_{\omega}+\frac{1}{4}\chi_{\rho}
\end{array}\right)\,,
\eea 
with $\chi_{\omega,\rho}= \, g_{\omega,\rho}^{2}/(q^{2}-m_{\omega,\rho}^{2})$. In the static limit $q_0=0$ and for small momenta $\boldsymbol{q}\rightarrow\boldsymbol{0}$ we thus find
\vspace*{1mm}
\be 
\left(\begin{array}{cc}
\chi_{V} & \chi_{I}\\
\chi_{I} & \chi_{V}
\end{array}\right) \rightarrow
\left(\begin{array}{cc}
\left(g_{\omega}/m_{\omega}\right)^{2}+\frac{1}{4}\left(g_{\rho}/m_{\rho}\right)^{2} & \left(g_{\omega}/m_{\omega}\right)^{2}-\frac{1}{4}\left(g_{\rho}/m_{\rho}\right)^{2}\\[2ex]
\left(g_{\omega}/m_{\omega}\right)^{2}-\frac{1}{4}\left(g_{\rho}/m_{\rho}\right)^{2} & \left(g_{\omega}/m_{\omega}\right)^{2}+\frac{1}{4}\left(g_{\rho}/m_{\rho}\right)^{2}
\end{array}\right) := \left(\begin{array}{cc}
V_{pp} & V_{pn}\\
V_{np} & V_{nn}
\end{array}\right)\,.\label{eq:chiVector}
\ee 
\vspace*{1mm}
To demonstrate that these potentials can alternatively be obtained from  the (interaction part of the) energy density or energy per particle $\mathcal{E}=E/A\cdot n$ we calculate the mean field quantity

\bea
\mathcal{E} & = & \mathcal{E}{}_{\text{kin}}-\frac{1}{2}\left(m_{\omega}^{2}\omega_{0}^{2}+m_{\rho}^{2}\rho_{0}^{2}\right)+g_{\omega}\omega_{0}n_{V}+\frac{1}{2}g_{\rho}\rho_{0}n_{I}\label{eq:EnergyOmRho}\\[2ex]
 & = & \mathcal{E}{}_{\text{kin}}+\frac{1}{2}\left(\frac{g_{\omega}}{m_{\omega}}\right)^{2}n_{V}^{2}+\frac{1}{8}\left(\frac{g_{\rho}}{m_{\rho}}\right)^{2}n_{I}^{2}\,,
\eea 
\noindent and derive it twice with respect to proton and neutron densities, which again yields \ref{eq:chiVector}.  Next, we turn to the resummed polarization functions, Eq. \ref{eq:PolResum}. After contracting Lorentz indices, the closed-form solutions for longitudinal and transverse polarization functions are  
\vspace*{1mm}
\bea 
\tilde{F}_{ab} & = & \left[\delta_{ad}+(V)_{ac}F_{cd}\right]^{-1}F_{db}\,,\label{eq:PiOmegaRhoRPA}\\
\nonumber \\
\tilde{G}_{ab} & = & \left[\delta_{ad}+(V)_{ac}G_{cd}\right]^{-1}G_{db}\,,\label{eq:PiOmegaRhoRPAPerp}
\eea 
where $V$ is the matrix expression in Eq. \ref{eq:VecRMF}. Additional factors of $q^2/\boldsymbol{q}^2$ cancel in the process, as usual. Explicitly expression \ref{eq:PiOmegaRhoRPA} reads
\vspace{1mm}
\begin{equation}
\tilde{F}=\frac{1}{1+\chi_{V}F_{p}+\chi_{V}F_{n}+(\chi_{V}^{2}-\chi_{I}^{2})F_{p}F_{n}}\left(\begin{array}{cc}
F_{p}(1+\chi_{V}F_{n}) & \chi_{I}F_{p}F_{n}\\[2ex]
\chi_{I}F_{p}F_{n} & F_{n}(1+\chi_{V}F_{p})
\end{array}\right)\,.
\end{equation}
\noindent In the static limit and at small momenta, we find that the proton-proton component gives
\begin{equation}
\tilde{F}_{p}(q_{0}=0,\,\boldsymbol{q}\rightarrow\boldsymbol{0})=\tilde{m}_{D,\,p}^{2}=\frac{m_{D,p}^{2}(1+V_{nn}m_{D,n}^{2})}{1+m_{D,p}^{2}V_{pp}+m_{D,n}^{2}V_{nn}+(V_{pp}V_{nn}-V_{np}^{2})m_{D,n}^{2}m_{D,p}^{2}}\,,
\end{equation}
which is exactly Eq. \ref{eq:DebyePN} obtained from thermodynamics, as expected.

\bibliographystyle{apsrev}
\bibliography{RTF.bib}

\begin{thebibliography}{43}
\expandafter\ifx\csname natexlab\endcsname\relax\def\natexlab#1{#1}\fi
\expandafter\ifx\csname bibnamefont\endcsname\relax
  \def\bibnamefont#1{#1}\fi
\expandafter\ifx\csname bibfnamefont\endcsname\relax
  \def\bibfnamefont#1{#1}\fi
\expandafter\ifx\csname citenamefont\endcsname\relax
  \def\citenamefont#1{#1}\fi
\expandafter\ifx\csname url\endcsname\relax
  \def\url#1{\texttt{#1}}\fi
\expandafter\ifx\csname urlprefix\endcsname\relax\def\urlprefix{URL }\fi
\providecommand{\bibinfo}[2]{#2}
\providecommand{\eprint}[2][]{\url{#2}}

\bibitem[{\citenamefont{Altherr et~al.}(1993)\citenamefont{Altherr,
  Petitgirard, and de~Rio~Gaztelurrutia}}]{Altherr:1992jg}
\bibinfo{author}{\bibfnamefont{T.}~\bibnamefont{Altherr}},
  \bibinfo{author}{\bibfnamefont{E.}~\bibnamefont{Petitgirard}},
  \bibnamefont{and}
  \bibinfo{author}{\bibfnamefont{T.}~\bibnamefont{de~Rio~Gaztelurrutia}},
  \bibinfo{journal}{Astropart. Phys.} \textbf{\bibinfo{volume}{1}},
  \bibinfo{pages}{289} (\bibinfo{year}{1993}), \eprint{hep-ph/9212264}.

\bibitem[{\citenamefont{Altherr and Kraemmer}(1992)}]{Altherr:1992mf}
\bibinfo{author}{\bibfnamefont{T.}~\bibnamefont{Altherr}} \bibnamefont{and}
  \bibinfo{author}{\bibfnamefont{U.}~\bibnamefont{Kraemmer}},
  \bibinfo{journal}{Astropart. Phys.} \textbf{\bibinfo{volume}{1}},
  \bibinfo{pages}{133} (\bibinfo{year}{1992}).

\bibitem[{\citenamefont{Litim and Manuel}(2001)}]{Litim:2001mv}
\bibinfo{author}{\bibfnamefont{D.~F.} \bibnamefont{Litim}} \bibnamefont{and}
  \bibinfo{author}{\bibfnamefont{C.}~\bibnamefont{Manuel}},
  \bibinfo{journal}{Phys. Rev.} \textbf{\bibinfo{volume}{D64}},
  \bibinfo{pages}{094013} (\bibinfo{year}{2001}), \eprint{hep-ph/0105165}.

\bibitem[{\citenamefont{Schmitt et~al.}(2004)\citenamefont{Schmitt, Wang, and
  Rischke}}]{Schmitt:2003aa}
\bibinfo{author}{\bibfnamefont{A.}~\bibnamefont{Schmitt}},
  \bibinfo{author}{\bibfnamefont{Q.}~\bibnamefont{Wang}}, \bibnamefont{and}
  \bibinfo{author}{\bibfnamefont{D.~H.} \bibnamefont{Rischke}},
  \bibinfo{journal}{Phys. Rev.} \textbf{\bibinfo{volume}{D69}},
  \bibinfo{pages}{094017} (\bibinfo{year}{2004}), \eprint{nucl-th/0311006}.

\bibitem[{\citenamefont{Heiselberg et~al.}(1992)\citenamefont{Heiselberg, Baym,
  Pethick, and Popp}}]{Heiselberg:1992ha}
\bibinfo{author}{\bibfnamefont{H.}~\bibnamefont{Heiselberg}},
  \bibinfo{author}{\bibfnamefont{G.}~\bibnamefont{Baym}},
  \bibinfo{author}{\bibfnamefont{C.~J.} \bibnamefont{Pethick}},
  \bibnamefont{and} \bibinfo{author}{\bibfnamefont{J.}~\bibnamefont{Popp}},
  \bibinfo{journal}{Nucl. Phys.} \textbf{\bibinfo{volume}{A544}},
  \bibinfo{pages}{569C} (\bibinfo{year}{1992}).

\bibitem[{\citenamefont{Heiselberg and Pethick}(1993)}]{Heiselberg:1993cr}
\bibinfo{author}{\bibfnamefont{H.}~\bibnamefont{Heiselberg}} \bibnamefont{and}
  \bibinfo{author}{\bibfnamefont{C.~J.} \bibnamefont{Pethick}},
  \bibinfo{journal}{Phys. Rev.} \textbf{\bibinfo{volume}{D48}},
  \bibinfo{pages}{2916} (\bibinfo{year}{1993}).

\bibitem[{\citenamefont{Shternin and Yakovlev}(2007)}]{Shternin:2007ee}
\bibinfo{author}{\bibfnamefont{P.~S.} \bibnamefont{Shternin}} \bibnamefont{and}
  \bibinfo{author}{\bibfnamefont{D.~G.} \bibnamefont{Yakovlev}},
  \bibinfo{journal}{Phys. Rev.} \textbf{\bibinfo{volume}{D75}},
  \bibinfo{pages}{103004} (\bibinfo{year}{2007}), \eprint{0705.1963}.

\bibitem[{\citenamefont{Shternin and Yakovlev}(2008)}]{Shternin:2008es}
\bibinfo{author}{\bibfnamefont{P.~S.} \bibnamefont{Shternin}} \bibnamefont{and}
  \bibinfo{author}{\bibfnamefont{D.~G.} \bibnamefont{Yakovlev}},
  \bibinfo{journal}{Phys. Rev.} \textbf{\bibinfo{volume}{D78}},
  \bibinfo{pages}{063006} (\bibinfo{year}{2008}), \eprint{0808.2018}.

\bibitem[{\citenamefont{Schmitt and Shternin}(2017)}]{Schmitt:2017efp}
\bibinfo{author}{\bibfnamefont{A.}~\bibnamefont{Schmitt}} \bibnamefont{and}
  \bibinfo{author}{\bibfnamefont{P.}~\bibnamefont{Shternin}}
  (\bibinfo{year}{2017}), \eprint{1711.06520}.

\bibitem[{\citenamefont{Bertoni et~al.}(2015)\citenamefont{Bertoni, Reddy, and
  Rrapaj}}]{Bertoni:2014soa}
\bibinfo{author}{\bibfnamefont{B.}~\bibnamefont{Bertoni}},
  \bibinfo{author}{\bibfnamefont{S.}~\bibnamefont{Reddy}}, \bibnamefont{and}
  \bibinfo{author}{\bibfnamefont{E.}~\bibnamefont{Rrapaj}},
  \bibinfo{journal}{Phys. Rev.} \textbf{\bibinfo{volume}{C91}},
  \bibinfo{pages}{025806} (\bibinfo{year}{2015}), \eprint{1409.7750}.

\bibitem[{\citenamefont{Rrapaj et~al.}(2018)\citenamefont{Rrapaj, Stetina, and
  Reddy}}]{Reddy2017}
\bibinfo{author}{\bibfnamefont{E.}~\bibnamefont{Rrapaj}},
  \bibinfo{author}{\bibfnamefont{S.}~\bibnamefont{Stetina}}, \bibnamefont{and}
  \bibinfo{author}{\bibfnamefont{S.}~\bibnamefont{Reddy}}, \bibinfo{journal}{In
  preparation}  (\bibinfo{year}{2018}).

\bibitem[{\citenamefont{Horowitz and
  Wehrberger}(1991{\natexlab{a}})}]{Horowitz:1991pg}
\bibinfo{author}{\bibfnamefont{C.~J.} \bibnamefont{Horowitz}} \bibnamefont{and}
  \bibinfo{author}{\bibfnamefont{K.}~\bibnamefont{Wehrberger}},
  \bibinfo{journal}{Phys. Lett.} \textbf{\bibinfo{volume}{B266}},
  \bibinfo{pages}{236} (\bibinfo{year}{1991}{\natexlab{a}}).

\bibitem[{\citenamefont{Reddy et~al.}(1999)\citenamefont{Reddy, Prakash,
  Lattimer, and Pons}}]{Reddy:1998hb}
\bibinfo{author}{\bibfnamefont{S.}~\bibnamefont{Reddy}},
  \bibinfo{author}{\bibfnamefont{M.}~\bibnamefont{Prakash}},
  \bibinfo{author}{\bibfnamefont{J.~M.} \bibnamefont{Lattimer}},
  \bibnamefont{and} \bibinfo{author}{\bibfnamefont{J.~A.} \bibnamefont{Pons}},
  \bibinfo{journal}{Phys. Rev.} \textbf{\bibinfo{volume}{C59}},
  \bibinfo{pages}{2888} (\bibinfo{year}{1999}), \eprint{astro-ph/9811294}.

\bibitem[{\citenamefont{Braaten and Segel}(1993)}]{Braaten:1993jw}
\bibinfo{author}{\bibfnamefont{E.}~\bibnamefont{Braaten}} \bibnamefont{and}
  \bibinfo{author}{\bibfnamefont{D.}~\bibnamefont{Segel}},
  \bibinfo{journal}{Phys. Rev.} \textbf{\bibinfo{volume}{D48}},
  \bibinfo{pages}{1478} (\bibinfo{year}{1993}), \eprint{hep-ph/9302213}.

\bibitem[{\citenamefont{Baldo and Ducoin}(2009)}]{Baldo:2008pb}
\bibinfo{author}{\bibfnamefont{M.}~\bibnamefont{Baldo}} \bibnamefont{and}
  \bibinfo{author}{\bibfnamefont{C.}~\bibnamefont{Ducoin}},
  \bibinfo{journal}{Phys. Rev.} \textbf{\bibinfo{volume}{C79}},
  \bibinfo{pages}{035801} (\bibinfo{year}{2009}), \eprint{0811.0604}.

\bibitem[{\citenamefont{Batell et~al.}(2014)\citenamefont{Batell, deNiverville,
  McKeen, Pospelov, and Ritz}}]{Batell:2014yra}
\bibinfo{author}{\bibfnamefont{B.}~\bibnamefont{Batell}},
  \bibinfo{author}{\bibfnamefont{P.}~\bibnamefont{deNiverville}},
  \bibinfo{author}{\bibfnamefont{D.}~\bibnamefont{McKeen}},
  \bibinfo{author}{\bibfnamefont{M.}~\bibnamefont{Pospelov}}, \bibnamefont{and}
  \bibinfo{author}{\bibfnamefont{A.}~\bibnamefont{Ritz}},
  \bibinfo{journal}{Phys. Rev.} \textbf{\bibinfo{volume}{D90}},
  \bibinfo{pages}{115014} (\bibinfo{year}{2014}).

\bibitem[{\citenamefont{Chang et~al.}(2017)\citenamefont{Chang, Essig, and
  McDermott}}]{Chang:2016ntp}
\bibinfo{author}{\bibfnamefont{J.~H.} \bibnamefont{Chang}},
  \bibinfo{author}{\bibfnamefont{R.}~\bibnamefont{Essig}}, \bibnamefont{and}
  \bibinfo{author}{\bibfnamefont{S.~D.} \bibnamefont{McDermott}},
  \bibinfo{journal}{JHEP} \textbf{\bibinfo{volume}{01}}, \bibinfo{pages}{107}
  (\bibinfo{year}{2017}), \eprint{1611.03864}.

\bibitem[{\citenamefont{Weldon}(1982)}]{Weldon:1982aq}
\bibinfo{author}{\bibfnamefont{H.~A.} \bibnamefont{Weldon}},
  \bibinfo{journal}{Phys. Rev.} \textbf{\bibinfo{volume}{D26}},
  \bibinfo{pages}{1394} (\bibinfo{year}{1982}).

\bibitem[{\citenamefont{Kraemmer and Rebhan}(2004)}]{Kraemmer:2003gd}
\bibinfo{author}{\bibfnamefont{U.}~\bibnamefont{Kraemmer}} \bibnamefont{and}
  \bibinfo{author}{\bibfnamefont{A.}~\bibnamefont{Rebhan}},
  \bibinfo{journal}{Rept. Prog. Phys.} \textbf{\bibinfo{volume}{67}},
  \bibinfo{pages}{351} (\bibinfo{year}{2004}), \eprint{hep-ph/0310337}.

\bibitem[{\citenamefont{Pisarski}(1989)}]{Pisarski:1989cs}
\bibinfo{author}{\bibfnamefont{R.~D.} \bibnamefont{Pisarski}},
  \bibinfo{journal}{Physica} \textbf{\bibinfo{volume}{A158}},
  \bibinfo{pages}{146} (\bibinfo{year}{1989}).

\bibitem[{\citenamefont{Kapusta and Gale}(2011)}]{Kapusta:2006pm}
\bibinfo{author}{\bibfnamefont{J.~I.} \bibnamefont{Kapusta}} \bibnamefont{and}
  \bibinfo{author}{\bibfnamefont{C.}~\bibnamefont{Gale}},
  \emph{\bibinfo{title}{{Finite-temperature field theory: Principles and
  applications}}} (\bibinfo{publisher}{Cambridge University Press},
  \bibinfo{year}{2011}), ISBN \bibinfo{isbn}{9780521173223, 9780521820820,
  9780511222801}.

\bibitem[{\citenamefont{Bellac}(2011)}]{Bellac:2011kqa}
\bibinfo{author}{\bibfnamefont{M.~L.} \bibnamefont{Bellac}},
  \emph{\bibinfo{title}{{Thermal Field Theory}}} (\bibinfo{publisher}{Cambridge
  University Press}, \bibinfo{year}{2011}), ISBN \bibinfo{isbn}{9780511885068,
  9780521654777},
  \urlprefix\url{http://www.cambridge.org/mw/academic/subjects/physics/theoretical-physics-and-mathematical-physics/thermal-field-theory?format=AR}.

\bibitem[{\citenamefont{Chou et~al.}(1985)\citenamefont{Chou, Su, Hao, and
  Yu}}]{Chou:1984es}
\bibinfo{author}{\bibfnamefont{K.-c.} \bibnamefont{Chou}},
  \bibinfo{author}{\bibfnamefont{Z.-b.} \bibnamefont{Su}},
  \bibinfo{author}{\bibfnamefont{B.-l.} \bibnamefont{Hao}}, \bibnamefont{and}
  \bibinfo{author}{\bibfnamefont{L.}~\bibnamefont{Yu}}, \bibinfo{journal}{Phys.
  Rept.} \textbf{\bibinfo{volume}{118}}, \bibinfo{pages}{1}
  (\bibinfo{year}{1985}).

\bibitem[{\citenamefont{Landsman and van Weert}(1987)}]{Landsman:1986uw}
\bibinfo{author}{\bibfnamefont{N.~P.} \bibnamefont{Landsman}} \bibnamefont{and}
  \bibinfo{author}{\bibfnamefont{C.~G.} \bibnamefont{van Weert}},
  \bibinfo{journal}{Phys. Rept.} \textbf{\bibinfo{volume}{145}},
  \bibinfo{pages}{141} (\bibinfo{year}{1987}).

\bibitem[{\citenamefont{Carrington et~al.}(1999)\citenamefont{Carrington, Hou,
  and Thoma}}]{Carrington:1997sq}
\bibinfo{author}{\bibfnamefont{M.~E.} \bibnamefont{Carrington}},
  \bibinfo{author}{\bibfnamefont{D.-f.} \bibnamefont{Hou}}, \bibnamefont{and}
  \bibinfo{author}{\bibfnamefont{M.~H.} \bibnamefont{Thoma}},
  \bibinfo{journal}{Eur. Phys. J.} \textbf{\bibinfo{volume}{C7}},
  \bibinfo{pages}{347} (\bibinfo{year}{1999}), \eprint{hep-ph/9708363}.

\bibitem[{\citenamefont{Peshier et~al.}(1998)\citenamefont{Peshier, Schertler,
  and Thoma}}]{Peshier:1998dy}
\bibinfo{author}{\bibfnamefont{A.}~\bibnamefont{Peshier}},
  \bibinfo{author}{\bibfnamefont{K.}~\bibnamefont{Schertler}},
  \bibnamefont{and} \bibinfo{author}{\bibfnamefont{M.~H.} \bibnamefont{Thoma}},
  \bibinfo{journal}{Annals Phys.} \textbf{\bibinfo{volume}{266}},
  \bibinfo{pages}{162} (\bibinfo{year}{1998}), \eprint{hep-ph/9708434}.

\bibitem[{\citenamefont{Fetter and Walecka}(2003)}]{FetterWalecka}
\bibinfo{author}{\bibfnamefont{A.~L.} \bibnamefont{Fetter}} \bibnamefont{and}
  \bibinfo{author}{\bibfnamefont{J.~D.} \bibnamefont{Walecka}},
  \emph{\bibinfo{title}{Quantum Theory of Many-Particle Systems}}
  (\bibinfo{publisher}{Dover Publications, Inc.}, \bibinfo{year}{2003}).

\bibitem[{\citenamefont{Kobyakov et~al.}(2017)\citenamefont{Kobyakov, Pethick,
  Reddy, and Schwenk}}]{Kobyakov:2017dbl}
\bibinfo{author}{\bibfnamefont{D.~N.} \bibnamefont{Kobyakov}},
  \bibinfo{author}{\bibfnamefont{C.~J.} \bibnamefont{Pethick}},
  \bibinfo{author}{\bibfnamefont{S.}~\bibnamefont{Reddy}}, \bibnamefont{and}
  \bibinfo{author}{\bibfnamefont{A.}~\bibnamefont{Schwenk}},
  \bibinfo{journal}{Phys. Rev.} \textbf{\bibinfo{volume}{C96}},
  \bibinfo{pages}{025805} (\bibinfo{year}{2017}), \eprint{1705.07357}.

\bibitem[{\citenamefont{Cirigliano et~al.}(2011)\citenamefont{Cirigliano,
  Reddy, and Sharma}}]{Cirigliano:2011tj}
\bibinfo{author}{\bibfnamefont{V.}~\bibnamefont{Cirigliano}},
  \bibinfo{author}{\bibfnamefont{S.}~\bibnamefont{Reddy}}, \bibnamefont{and}
  \bibinfo{author}{\bibfnamefont{R.}~\bibnamefont{Sharma}},
  \bibinfo{journal}{Phys. Rev.} \textbf{\bibinfo{volume}{C84}},
  \bibinfo{pages}{045809} (\bibinfo{year}{2011}), \eprint{1102.5379}.

\bibitem[{\citenamefont{Chamel et~al.}(2013)\citenamefont{Chamel, Page, and
  Reddy}}]{Chamel:2012ix}
\bibinfo{author}{\bibfnamefont{N.}~\bibnamefont{Chamel}},
  \bibinfo{author}{\bibfnamefont{D.}~\bibnamefont{Page}}, \bibnamefont{and}
  \bibinfo{author}{\bibfnamefont{S.}~\bibnamefont{Reddy}},
  \bibinfo{journal}{Phys. Rev.} \textbf{\bibinfo{volume}{C87}},
  \bibinfo{pages}{035803} (\bibinfo{year}{2013}), \eprint{1210.5169}.

\bibitem[{\citenamefont{McORIST et~al.}(2007)\citenamefont{McORIST, MELROSE,
  and WEISE}}]{mcorist_melrose_weise_2007}
\bibinfo{author}{\bibfnamefont{J.}~\bibnamefont{McORIST}},
  \bibinfo{author}{\bibfnamefont{D.~B.} \bibnamefont{MELROSE}},
  \bibnamefont{and} \bibinfo{author}{\bibfnamefont{J.~I.} \bibnamefont{WEISE}},
  \bibinfo{journal}{Journal of Plasma Physics} \textbf{\bibinfo{volume}{73}},
  \bibinfo{pages}{495–513} (\bibinfo{year}{2007}).

\bibitem[{\citenamefont{Kobes and Semenoff}(1985)}]{Kobes:1985kc}
\bibinfo{author}{\bibfnamefont{R.~L.} \bibnamefont{Kobes}} \bibnamefont{and}
  \bibinfo{author}{\bibfnamefont{G.~W.} \bibnamefont{Semenoff}},
  \bibinfo{journal}{Nucl. Phys.} \textbf{\bibinfo{volume}{B260}},
  \bibinfo{pages}{714} (\bibinfo{year}{1985}).

\bibitem[{\citenamefont{Das}(1997)}]{Das:1997gg}
\bibinfo{author}{\bibfnamefont{A.~K.} \bibnamefont{Das}},
  \emph{\bibinfo{title}{{Finite Temperature Field Theory}}}
  (\bibinfo{publisher}{World Scientific}, \bibinfo{address}{New York},
  \bibinfo{year}{1997}), ISBN \bibinfo{isbn}{9789810228569, 9789814498234}.

\bibitem[{\citenamefont{Chamel and Haensel}(2006)}]{Chamel:2006rc}
\bibinfo{author}{\bibfnamefont{N.}~\bibnamefont{Chamel}} \bibnamefont{and}
  \bibinfo{author}{\bibfnamefont{P.}~\bibnamefont{Haensel}},
  \bibinfo{journal}{Phys. Rev.} \textbf{\bibinfo{volume}{C73}},
  \bibinfo{pages}{045802} (\bibinfo{year}{2006}), \eprint{nucl-th/0603018}.

\bibitem[{\citenamefont{Baym et~al.}(1971)\citenamefont{Baym, Bethe, and
  Pethick}}]{Baym:1971ax}
\bibinfo{author}{\bibfnamefont{G.}~\bibnamefont{Baym}},
  \bibinfo{author}{\bibfnamefont{H.~A.} \bibnamefont{Bethe}}, \bibnamefont{and}
  \bibinfo{author}{\bibfnamefont{C.}~\bibnamefont{Pethick}},
  \bibinfo{journal}{Nucl. Phys.} \textbf{\bibinfo{volume}{A175}},
  \bibinfo{pages}{225} (\bibinfo{year}{1971}).

\bibitem[{\citenamefont{Muller and Serot}(1995)}]{Muller:1995ji}
\bibinfo{author}{\bibfnamefont{H.}~\bibnamefont{Muller}} \bibnamefont{and}
  \bibinfo{author}{\bibfnamefont{B.~D.} \bibnamefont{Serot}},
  \bibinfo{journal}{Phys. Rev.} \textbf{\bibinfo{volume}{C52}},
  \bibinfo{pages}{2072} (\bibinfo{year}{1995}), \eprint{nucl-th/9505013}.

\bibitem[{\citenamefont{Li and Ko}(1997)}]{Li:1997ra}
\bibinfo{author}{\bibfnamefont{B.-A.} \bibnamefont{Li}} \bibnamefont{and}
  \bibinfo{author}{\bibfnamefont{C.~M.} \bibnamefont{Ko}},
  \bibinfo{journal}{Nucl. Phys.} \textbf{\bibinfo{volume}{A618}},
  \bibinfo{pages}{498} (\bibinfo{year}{1997}), \eprint{nucl-th/9701049}.

\bibitem[{\citenamefont{Dutra et~al.}(2012)\citenamefont{Dutra, Lourenco,
  Sa~Martins, Delfino, Stone, and Stevenson}}]{Dutra:2012mb}
\bibinfo{author}{\bibfnamefont{M.}~\bibnamefont{Dutra}},
  \bibinfo{author}{\bibfnamefont{O.}~\bibnamefont{Lourenco}},
  \bibinfo{author}{\bibfnamefont{J.~S.} \bibnamefont{Sa~Martins}},
  \bibinfo{author}{\bibfnamefont{A.}~\bibnamefont{Delfino}},
  \bibinfo{author}{\bibfnamefont{J.~R.} \bibnamefont{Stone}}, \bibnamefont{and}
  \bibinfo{author}{\bibfnamefont{P.~D.} \bibnamefont{Stevenson}},
  \bibinfo{journal}{Phys. Rev.} \textbf{\bibinfo{volume}{C85}},
  \bibinfo{pages}{035201} (\bibinfo{year}{2012}), \eprint{1202.3902}.

\bibitem[{\citenamefont{Shen and Reddy}(2014)}]{Shen:2013kxa}
\bibinfo{author}{\bibfnamefont{G.}~\bibnamefont{Shen}} \bibnamefont{and}
  \bibinfo{author}{\bibfnamefont{S.}~\bibnamefont{Reddy}},
  \bibinfo{journal}{Phys. Rev.} \textbf{\bibinfo{volume}{C89}},
  \bibinfo{pages}{032802} (\bibinfo{year}{2014}), \eprint{1311.6096}.

\bibitem[{\citenamefont{Bohm and Staver}(1950)}]{Bohm1950}
\bibinfo{author}{\bibfnamefont{D.}~\bibnamefont{Bohm}} \bibnamefont{and}
  \bibinfo{author}{\bibfnamefont{T.}~\bibnamefont{Staver}},
  \bibinfo{journal}{Phys. Rev.} \textbf{\bibinfo{volume}{836}}
  (\bibinfo{year}{1950}).

\bibitem[{\citenamefont{Lim and Horowitz}(1989)}]{Lim:1989}
\bibinfo{author}{\bibfnamefont{K.}~\bibnamefont{Lim}} \bibnamefont{and}
  \bibinfo{author}{\bibfnamefont{C.}~\bibnamefont{Horowitz}},
  \bibinfo{journal}{Nuclear Physics A} \textbf{\bibinfo{volume}{501}},
  \bibinfo{pages}{729 } (\bibinfo{year}{1989}), ISSN \bibinfo{issn}{0375-9474},
  \urlprefix\url{http://www.sciencedirect.com/science/article/pii/0375947489901589}.

\bibitem[{\citenamefont{Horowitz and
  Wehrberger}(1991{\natexlab{b}})}]{Horowitz1991}
\bibinfo{author}{\bibfnamefont{C.}~\bibnamefont{Horowitz}} \bibnamefont{and}
  \bibinfo{author}{\bibfnamefont{K.}~\bibnamefont{Wehrberger}},
  \bibinfo{journal}{Nuclear Physics A} \textbf{\bibinfo{volume}{531}},
  \bibinfo{pages}{665 } (\bibinfo{year}{1991}{\natexlab{b}}), ISSN
  \bibinfo{issn}{0375-9474},
  \urlprefix\url{http://www.sciencedirect.com/science/article/pii/037594749190745R}.

\bibitem[{\citenamefont{Dutra et~al.}(2014)\citenamefont{Dutra, Lourenço,
  Avancini, Carlson, Delfino, Menezes, Providência, Typel, and
  Stone}}]{Dutra:2014qga}
\bibinfo{author}{\bibfnamefont{M.}~\bibnamefont{Dutra}},
  \bibinfo{author}{\bibfnamefont{O.}~\bibnamefont{Lourenço}},
  \bibinfo{author}{\bibfnamefont{S.~S.} \bibnamefont{Avancini}},
  \bibinfo{author}{\bibfnamefont{B.~V.} \bibnamefont{Carlson}},
  \bibinfo{author}{\bibfnamefont{A.}~\bibnamefont{Delfino}},
  \bibinfo{author}{\bibfnamefont{D.~P.} \bibnamefont{Menezes}},
  \bibinfo{author}{\bibfnamefont{C.}~\bibnamefont{Providência}},
  \bibinfo{author}{\bibfnamefont{S.}~\bibnamefont{Typel}}, \bibnamefont{and}
  \bibinfo{author}{\bibfnamefont{J.~R.} \bibnamefont{Stone}},
  \bibinfo{journal}{Phys. Rev.} \textbf{\bibinfo{volume}{C90}},
  \bibinfo{pages}{055203} (\bibinfo{year}{2014}), \eprint{1405.3633}.

\end{thebibliography}
\end{document}